\documentclass[aps,prd,twocolumn,10pt,superscriptaddress,amssymb,amsmath,nofootinbib]{revtex4-2}
\usepackage{url}
\usepackage[utf8]{inputenc}
\usepackage{lmodern}
\usepackage[T1]{fontenc}
\usepackage{graphicx} 
\usepackage[pdftex,breaklinks,colorlinks,
linkcolor=Blue,
citecolor=teal,
anchorcolor=red,
urlcolor=cyan,
pdfencoding=auto]{hyperref}
\usepackage[dvipsnames]{xcolor}
\usepackage[normalem]{ulem}
\usepackage{mathtools}
\usepackage{mathrsfs}
\usepackage{booktabs}
\usepackage[mathscr]{euscript}
\usepackage{multirow}
\usepackage{float}

\usepackage{orcidlink}

\newcommand{\order}{\mathcal{O}}
\usepackage{array}
    \newcolumntype{M}{>{$}c<{$}} 

\newcommand{\be}{\nopagebreak[3]\begin{equation}}
\newcommand{\ee}{\end{equation}}

\newcommand{\ba}{\nopagebreak[3]\begin{eqnarray}}
\newcommand{\ea}{\end{eqnarray}}

\newcommand{\hp}{\hat{p}}
\newcommand{\hv}{\hat{v}}
\newcommand{\htau}{\hat{\tau}}
\newcommand{\hnabla}{\hat{\nabla}}

\newcommand{\hD}{\hat{D}}
\newcommand{\CO}{\mathcal{O}}
\newcommand{\tS}{\hat{S}}
\newcommand{\tJ}{\hat{J}}
\newcommand{\tQ}{\tilde{Q}}
\newcommand{\hR}{\hat{R}}
\newcommand{\e}{\varepsilon}
\newcommand{\beq}{\begin{equation}\begin{aligned}}
\newcommand{\eeq}{\end{aligned}\end{equation}}

\newcommand{\bea}{\begin{eqnarray}}
\newcommand{\eea}{\end{eqnarray}}
\newcommand{\non}{\nonumber}

\newcommand{\bJ}{\mathbf{\mathcal{J}}}
\newcommand{\dJ}{\delta J}
\newcommand{\rO}{r_{0}}
\newcommand{\dM}{\delta M}
\newcommand{\sP}{\Omega, \chi, \dM, \dJ}
\newcommand{\dO}{\dot{\Omega}}
\newcommand{\dchi}{\dot{\chi}}

\newcommand{\ddM}{\dot{\dM}}

\newcommand{\ddJ}{\dot{\dJ}}
\newcommand{\dsP}{\dO, \dchi, \ddM, \ddJ}
\newcommand{\partialsP}{\partial_\Omega, \partial_\chi, \partial_{\dM}, \partial_{\dJ}}
\newcommand{\dbJ}{\dot{\bJ}}
\newcommand{\cF}{\mathcal{F}}

\def\hr{r}

\def\hO{\Omega}

\def\Ang{S_{\ell m}^{a\omega}}
\def\In{\textrm{in}}
\def\up{\textrm{up}}

\def\hr{r}

\def\hO{\Omega}

\def\mode{\ell m \omega}

\def\In{\textrm{in}}
\def\up{\textrm{up}}

\def\RS{\Sigma}
\providecommand{\dif}{\mathrm{d}}
\def\d{\dif}
\def\F{\mathsf{F}}
\begin{document}

\title{Quadrupole and quadratic-in-spin effects in quasicircular, spinning, asymmetric binaries}

\author{Mostafizur Rahman \orcidlink{0000-0003-0904-8548}}
\email{rahman@tap.scphys.kyoto-u.ac.jp}
\affiliation{Department of Physics, Kyoto University, Kyoto 606-8502, Japan}
\affiliation{Astronomical Institute of the Czech Academy of Sciences, Bo\v{c}n\'{i} II 1401/1a, CZ-141 00 Prague, Czech Republic}

\author{Misbah Shahzadi
\orcidlink{0000-0002-3130-1602}}
\email{misbahshahzadi51@gmail.com}
\affiliation{Astronomical Institute of the Czech Academy of Sciences, Bo\v{c}n\'{i} II 1401/1a, CZ-141 00 Prague, Czech Republic}
\affiliation{School of Mathematical Sciences and STAG Research Centre, University of Southampton, Southampton, United Kingdom, SO17 1BJ}

\author{Adam Pound\,\orcidlink{0000-0001-9446-0638}}
\email{a.pound@soton.ac.uk}
\affiliation{School of Mathematical Sciences and STAG Research Centre, University of Southampton, Southampton, United Kingdom, SO17 1BJ}

\author{Josh Mathews
\orcidlink{0000-0002-5477-8470}}
\email{Josh.mat@nus.edu.sg}
\affiliation{
Department of Physics, National University of Singapore, 21 Lower Kent Ridge Rd, 119077, Singapore}

\begin{abstract}
   Next-generation gravitational-wave detectors will require significant improvements in current theoretical waveform models, particularly in the case of asymmetric-mass binaries. Here we provide one such improvement by calculating fully relativistic finite-size effects for small mass ratios---primarily, fluxes of energy---including quadratic-in-secondary-spin terms, spin-induced quadrupole terms, and tidally induced quadrupole terms, for quasicircular inspirals of a small companion into a Kerr black hole. We formulate these calculations within a multiscale waveform-generation framework in self-force theory, which could be used, with an energy-balance law we derive, to develop self-contained waveform models for asymmetric binaries involving stars orbiting black holes. Our results could additionally be used to improve other families of waveform models across all mass ratios. We present results both as complete numerical data sets on a Chebyshev grid and as analytical post-Newtonian expansions (to sixth PN order relative to the leading term in each contribution to the flux).
\end{abstract}

\maketitle

\tableofcontents

\section{Introduction}

The detection of gravitational waves in 2015~\cite{LIGOScientific:2016aoc} has had broad and lasting impact on observational astrophysics. In the decade since, the LIGO-Virgo-KAGRA~(LVK) network has detected over 300 gravitational-wave signals from compact-binary coalescences, illuminating the population of compact objects in the universe~\cite{LIGOScientific:2025slb, LIGOScientific:2026sit}. These observations have helped inform our understanding of how such binaries form and evolve. Moreover, some ``golden'' signals have also allowed us to probe the nature of the compact objects---black holes and neutron stars---that form these binaries. 
In particular, analysis of GW170817, 
GW241011, and GW250114 allowed us to constrain the equation of state of neutron stars and the Kerr nature of black holes~\cite{LIGOScientific:2017vwq, Annala:2017llu, 
LIGOScientific:2025brd,Das:2026tel,LIGOScientific:2025rid}.

Inferring such properties of compact objects largely relies on precise measurements of the objects' masses, spins, and higher moments, relying on the fact that the relationship between these multipole moments is dictated by the composition of the object. A primary goal of next-generation detectors is to sharpen these measurements and more precisely probe the composition of compact stars and black holes~\cite{LISA:2022kgy,LISA:2022yao,LISA:2024hlh,ET:2025xjr}. 
However, this will require improved accuracy of our theoretical waveform models: current models have critical shortcomings that will create systematic biases in such measurements, particularly in the case of asymmetric masses and large spins~\cite{Purrer:2019jcp,
Hu:2022rjq,LISAConsortiumWaveformWorkingGroup:2023arg,Dhani:2024jja, Chandramouli:2024vhw, ET:2025xjr,Mahapatra:2026wsp}. 

To date, waveform models including finite-size effects (primarily, spin and quadrupole moments) have mostly been built from weak-field, post-Newtonian (PN) and post-Minkowskian results (e.g.~\cite{Marsat:2014xea, Bohe:2015ana, Levi:2016ofk, Siemonsen:2017yux, Kastha:2018bcr, Kastha:2019brk, Henry:2019xhg, Henry:2020ski, Cho:2022syn, Blanchet:2023sbv, Blanchet:2023bwj, Mandal:2024iug, Bini:2020flp, Cheung:2020sdj, Kalin:2020lmz, Jakobsen:2023pvx}), which are resummed using effective-one-body (EOB) theory and calibrated to fully nonlinear numerical relativity simulations~\cite{Bernuzzi:2014owa,Steinhoff:2016rfi,Lackey:2018zvw, Nagar:2018plt, Thompson:2020nei, Abac:2023ujg, Williams:2024twp, Haberland:2025luz, Albanesi:2025txj, Schulze:2026ewu, Gamboa:2026jht,RamisVidal:2026ycb}. Such models have achieved high accuracies for the inspiral stage of comparable-mass binaries involving two neutron stars, two black holes, or a star orbiting a black hole. But waveform systematics have proved to be a limiting factor even in the comparable-mass case~\cite{
Owen:2023mid, 
Estelles:2021jnz, Maggio:2022hre, LIGOScientific:2025rsn}, and errors grow significantly for smaller mass ratios~\cite{vandeMeent:2020xgc, Rink:2024swg, Kumar:2026ckp} due to the limited applicability of PN results and the prohibitive cost of numerical simulations in that regime. 

At the same time, self-force theory, which expands the binary problem in powers of the small mass ratio~\cite{Mino:1996nk,Poisson:2011nh,Barack:2018yvs,Pound:2021qin}, has led to self-contained, fast and accurate waveform models tailored to the small-mass-ratio regime~\cite{Wardell:2021fyy,Mathews:2025txc, 
Katz:2021yft,Speri:2023jte,
Nasipak:2023kuf,Chapman-Bird:2025xtd}. In these models the larger, primary object is virtually always taken to be a Kerr black hole, but the smaller, secondary one can be any type of (sufficiently compact) body. However, while there has been a concerted effort to include linear-in-secondary-spin effects in these models~\cite{Ruangsri:2015cvg, Harms:2015ixa, 
Warburton:2017sxk,vandeMeent:2019cam, Witzany:2019nml, Piovano:2020zin, 
Mathews:2021rod, 
Skoupy:2022adh, Drummond:2022efc, 
Witzany:2023bmq, Drummond:2023wqc, Skoupy:2023lih, Kerachian:2023oiw, Ramond:2024sfp, Grant:2024ivt, Piovano:2024yks, Witzany:2024ttz, Skoupy:2024jsi, Mathews:2025nyb, Skoupy:2025nie, Mathews:2025txc, Honet:2025lmk, Drummond:2026haw, Skoupy:2026ewu,Cui:2026qsk}
, higher-order finite-size effects have typically (with a few exceptions~\cite{Steinhoff:2012rw,Rahman:2021eay,Timogiannis:2023pop, Ramond:2024ozy,Shahzadi:2025ebj,Ramond:2026fpi}) been disregarded; this is because these models were initially motivated by extreme-mass-ratio inspirals (EMRIs) with mass ratios $10^{-7}$--$10^{-4}$, and the effects of the secondary's higher moments are suppressed by higher powers of the mass ratio. Yet higher-order effects \emph{can} be significant even in EMRIs if the secondary is a more deformable star such as a white or brown dwarf or an exotic object such as a boson star~\cite{Rahman:2021eay}, and dwarf-star EMRIs could be relevant sources for LISA if they are nearby~\cite{Gourgoulhon:2019iyu,Amaro-Seoane:2020zbo}. Moreover, self-force-based models have been found to be highly accurate for intermediate-mass-ratio inspirals (IMRIs) with mass ratios $10^{-4}$--$10^{-2}$, where finite-size effects will be more significant, and even for mass ratios $\sim 1/10$~\cite{Wardell:2021fyy,Albertini:2022rfe,Ramos-Buades:2022lgf,Mathews:2025txc}. When hybridized with PN information or used to inform EOB models, results from self-force theory can additionally improve models that, in principle, can cover \emph{all} mass ratios~\cite{Damour:2009sm,Nagar:2022fep,vandeMeent:2023ols, Albertini:2024rrs, Leather:2025nhu, Honet:2025gge}. 

While Ref.~\cite{Rahman:2021eay} computed some of the finite-size effects required as inputs for a waveform model and highlighted the potential importance of these effects in EMRIs with highly deformable secondaries through an order-of-magnitude estimation, it did not provide a complete framework that includes these effects alongside self-forces. In particular, the analysis included contributions that formally enter at second post-adiabatic order (2PA) in the mass ratio, such as spin-induced quadrupolar deformations, while neglecting the first post-adiabatic (1PA) self-force corrections. Moreover, the study ignored tidally induced quadrupolar deformations, which can become significant in systems outside the EMRI regime. Finally, it did not compute enough data, or data in an appropriate form, to incorporate into a waveform model.

All of this motivates more complete calculations of finite-size effects within self-force theory. In this paper we take a first step toward that goal by laying out a multiscale self-force framework~\cite{PhysRevD.78.064028,Miller:2020bft,Pound:2021qin,Wei:2025lva,Mathews:2025nyb,Lewis:2025ydo} for a small secondary body on quasicircular orbits around a Kerr primary black hole, with aligned or anti-aligned spins on the two bodies, consistent through 2PA order in the binary's mass ratio and including all finite-size effects through that order. We additionally include the effects of a tidally induced quadrupole moment on the secondary; though those effects enter at 4PA order for compact objects, we expect them to be numerically significant, and even dominate over lower-PA terms, for comparable masses and for non-compact objects such as white dwarfs~\cite{Rahman:2021eay}.

The multiscale expansion leads naturally to a complete waveform generation scheme~\cite{
Miller:2020bft,Mathews:2021rod,Mathews:2025nyb,Honet:2025lmk}. In this scheme, each spherical-harmonic mode of the waveform takes the following simple form in terms of the binary parameters: 
\begin{multline}\label{eq:waveform}
    h_{\ell m} = \Bigl[q \hat h^{1}_{\ell m}(\hat\Omega,\hat a) + q^2 \hat h^{2}_{\ell m}(\hat\Omega,\hat a,\chi) \\
    + q^3 \hat h^{3}_{\ell m}(\hat\Omega,\hat a,\chi,C_Q)+\ldots\Bigr] e^{-im\phi_p},
\end{multline}
where $q:=m/M_{\rm BH}$ is the mass ratio, $\phi_p$ is the orbital phase of the secondary (the ``particle''), and we use a hat to denote normalized, dimensionless quantities. $\hat\Omega:=M_{\rm BH}\Omega$ is a dimensionless orbital frequency, $\hat a:=J_{\rm BH}/M_{\rm BH}^2$ and $\chi$ are the primary and secondary dimensionless spin parameters, and $C_Q$ is the secondary's spin-induced quadrupole deformation. The waveform's time dependence is then governed by the binary's evolution equations, which take the equally simple form
\begin{align}
    \frac{d\phi_p}{dt} &= \Omega,\label{dphi/dt}\\
    \frac{d\Omega}{dt} &= \frac{q}{M^2}\Bigl[ {\hat{\sf F}}^{0}_\Omega(\hat\Omega,\hat a) + q {\hat{\sf F}}^{1}_\Omega(\hat\Omega,\hat a,\chi) \nonumber\\
    &\qquad\qquad + q^2 {\hat{\sf F}}^{2}_\Omega(\hat\Omega,\hat a,\chi,C_Q) + \ldots \Bigr], \label{dOmega/dt}
\end{align}
together with evolution equations for the primary's mass and spin ($\chi$ and $C_Q$ being constant at this order). In this way, the waveform is decomposed into amplitudes $\hat h^{n}_{\ell m}$ that evolve slowly, on the radiation-reaction time scale $\sim M/q$, and the phase factor $e^{-im\phi_p}$ that evolves rapidly, on the orbital time scale $\sim 2\pi/\Omega$.

In this scheme, the accuracy of the orbital phasing is directly determined by the number of orders included in Eq.~\eqref{dOmega/dt}, and the $n$PA counting directly corresponds to the numeric labels on ${\hat{\sf F}}^{n}_\Omega$. Different effects enter at specific orders:
\begin{itemize}
    \item The adiabatic (0PA) term ${\hat{\sf F}}^{0}_\Omega$ depends only on geodesic dynamics and the dissipative piece of the first-order self-force.
    \item The 1PA term ${\hat{\sf F}}^{1}_\Omega$ additionally depends on the first-order conservative self-force, second-order dissipative self-force, and linear-in-$\chi$ effects.
    \item The 2PA term ${\hat{\sf F}}^{2}_\Omega$ additionally depends on the second-order conservative self-force, third-order dissipative self-force, quadratic-in-$\chi$ effects, and the spin-induced quadrupole, along with couplings between lower-order effects.
    \item As mentioned above, the tidally induced quadrupole terms are suppressed in this scheme, appearing in the 4PA term ${\hat{\sf F}}^{4}_\Omega$ (as well as in the amplitudes $\hat h^{5}_{\ell m}$).
\end{itemize}

By deriving an appropriate energy-balance law, we also show that most finite-size effects in Eq.~\eqref{dOmega/dt} can be extracted from the particle's orbital energy together with the fluxes of emitted energy to infinity and down the black hole's horizon. This substantially streamlines the scheme by reducing the number of necessary local calculations at the secondary's position.

Importantly, finite-size effects appear in this construction in a modular way: they can be pre-computed and stored without specifying values of $\chi$ and $C_Q$ and without replacing any of the data for nonspinning, monopolar secondaries. This is because  $h^{2}_{\ell m}$ and ${\sf F}^{1}_\Omega$ are linear in $\chi$, and $h^{3}_{\ell m}$ and ${\sf F}^{2}_\Omega$ are quadratic in $\chi$ and linear in $C_Q$, meaning that one can compute \emph{coefficients} of powers of the spin and quadrupole parameters, without ever needing to specify parameter values. Waveforms can then be generated for any values of the parameters as soon as the coefficients are computed.

Our formulation extends the analogous, 1PA expansions in Refs.~\cite{Mathews:2021rod,Mathews:2025nyb}. However, here we do not directly implement the resulting waveform model. Instead, we focus on using the framework to compute \emph{inputs} for the waveform model (and for other waveform models). Specifically, we compute the quadratic-in-$\chi$ and linear-in-$C_Q$ contributions to the forcing functions ${\sf F}^{n}_\Omega$. These inputs are of two kinds: (i) simple analytical expressions for finite-size contributions to the orbital energy (many but not all of which already exist in equivalent forms in the literature~\cite{Bini:2015zya}), and (ii) the contributions to the emitted energy fluxes. Our main results are numerical values of the energy flux contributions on a grid of $(\hat\Omega,\hat a)$ values, as well as complementary analytical expansions (carried to sixth PN order relative to the leading term in each flux contribution). These results can be immediately incorporated into existing families of self-force~\cite{Mathews:2025txc} and hybrid self-force-PN~\cite{Honet:2025gge,Honet:2025lmk} waveform models that take the form sketched above. Perhaps surprisingly, we find that even high-order PN expansions have very poor accuracy for large primary spins and prograde orbits, suggesting that PN-informed models such as EOB might significantly benefit from utilizing our numerical data. 

The paper is organized as follows. Section~\ref{Sec_2_MPD} reviews the equations of motion of an extended object in a generic stationary, axisymmetric background spacetime, including terms up to quadrupole order (inclusive). In Sec.~\ref{Sec_3_Kerr} we specialize to quasicircular orbits in a Kerr background and formulate the multiscale expansion of the equations of motion and field equations. Section~\ref{Sec_4_BalanceLaws} derives the balance laws that allow one to bypass many calculations of local forces in the multiscale expansion; although our focus is on quasicircular orbits, the results in this section apply for generic, inclined and eccentric orbits. In Sec.~\ref{Sec_5_FieldEquations} we describe our calculation of energy fluxes, along with a comparison between numerical and PN-expanded results. We conclude in Sec.~\ref{Sec_6_Conclusion} with a discussion of future applications of our results. 

	
	\textit{Notation and Convention}: Unless otherwise stated, we work in geometrical units with $c=G=1$ and use the spacetime signature $(-,+,+,+)$. Greek letters $\alpha,~\beta,~\gamma,...$ denotes four-dimensional spacetime indices. Symmetrization and antisymmetrization of indices are denoted by round and square brackets, respectively: $T^{(\mu\nu)}=(T^{\mu\nu}+T^{\nu\mu})/2$, $T^{[\mu\nu]}=(T^{\mu\nu}-T^{\nu\mu})/2$.  

\section{Dynamics of an extended object}\label{Sec_2_MPD}

In this section, we provide a brief overview of the dynamics of an extended spinning object moving in an external universe. Although we ultimately specialize to the case of an object orbiting a Kerr black hole, our summary in this section is valid for any background metric.

    \subsection{Einstein field equations and multipolar approximation method}
    
Within the framework of self-force theory, the small extended object is treated as the source of a small perturbation to the spacetime around it (which we will take to be the Kerr metric in later sections). Accordingly, the physical metric is expressed as a perturbative expansion around the background metric $g_{\mu\nu}$:
\beq
\mathbf{g}_{\mu\nu}= g_{\mu\nu}+\varepsilon h_{\mu\nu}^1+\varepsilon^2 h_{\mu\nu}^2 +\varepsilon^3 h_{\mu\nu}^3 +\mathcal{O}(\e^4),
\eeq
where $h_{\mu\nu}=\sum_{n}\e^n h^{n}_{\mu\nu}$ describes the perturbation to the background spacetime induced by the object. The object's mass $m$ and radius $R$ are both assumed to be much smaller than the length scale $L$ of the external universe (which is the primary mass in the case of a binary). Here $\e$ is a bookkeeping parameter that would typically count powers of $m/L$ . However, to allow flexibility in treating different types of material objects, here we take it to be a more generic counter of powers of $m/L$ \emph{or} $R/L$, in a manner we clarify below. 

There are two general approaches to handling a small object in the Einstein equations and deriving its equations of motion. In the first approach, exemplified by Mathisson, Dixon, and Harte~\cite{Mathisson:2010opl,Dixon1970DynamicsOE,Dixon1973TheDO, Dixon1973TheDO,Dixon1974DynamicsOE,Harte:2011ku,Harte:2025tmd}, one assumes the object is a material body rather than a black hole, with a specified stress-energy tensor $T_{\mu\nu}$. 
If the metric varies slowly in the body's interior, we can employ a multipole expansion, wherein $T_{\mu\nu}$ is expanded in terms of an infinite series of multipole moments that capture the body's internal structure~\cite{Dixon1973TheDO,Harte:2011ku}.  
However, since the body gravitates, the metric inside it actually varies rapidly, on the body's own small scale $R$. Harte showed~\cite{Harte:2011ku,Harte:2025tmd} that Dixon's multipole expansion can be extended to the self-gravitating case by defining an \textit{effective metric} $\hat{g}_{\mu\nu}$ and effective stress-energy tensor $\hat T_{\mu\nu}$ that preserve the body's exact equations of motion (as described in the next section). If $\hat{g}_{\mu\nu}$ is chosen appropriately, such that it varies on the large scale $L$, then the multipole expansion is well behaved. 

Since $R$ is much smaller than the external length scale, the body can be treated as a point particle. Its effective energy momentum tensor can then be expanded explicitly in terms of multipole moments as follows~\cite{Tulczyjew:1959,Steinhoff:2009tk, Steinhoff:2012rw}:
\beq\label{SET}
\hat T^{\alpha \beta }=\e \hat T_{(m)}^{\alpha \beta }+\e^2 \hat T_{(d)}^{\alpha \beta }+\e^3 \hat T_{(q)}^{\alpha \beta }
+\mathcal{O}(\e^4)
,
\eeq
where the first three terms correspond to the monopole, dipole, and quadrupole contributions, respectively. 
Explicitly, these terms can be written as
	\begin{equation}\label{SET1}
		\begin{aligned}
			\hat T_{(m)}^{\alpha \beta }&=\int \text{d$\htau $}\bigg[\frac{\delta ^4(x-x_p(\htau))}{\sqrt{-\hat g}}\hp^{(\alpha }\hv^{\beta)}\bigg],\\
			\hat T_{(d)}^{\alpha \beta }&=-\int \text{d$\htau $}\hnabla _{\gamma }\left[\tS^{\gamma (\alpha }\hv^{\beta)}\frac{\delta ^4(x-x_p(\htau))}{\sqrt{-\hat g}}\right],\\
			\hat T_{(q)}^{\alpha \beta }&=-\frac{1}{3}\int \text{d$\htau $}\bigg[\left(\tJ^{\gamma \delta \epsilon (\alpha }\hR^{\beta )}{}_{\epsilon \gamma \delta }\frac{\delta ^4(x-x_p(\htau))}{\sqrt{-\hat g}}\right)\\&\quad +2\hnabla _{\gamma }\hnabla _{\delta }\left(\tJ^{\delta (\alpha \beta )\gamma }\frac{\delta ^4(x-x_p(\htau))}{\sqrt{-\hat g}}\right)\bigg],
		\end{aligned}
	\end{equation}
where $x^{\mu}_p$ represents the reference worldline of the particle, relative to which the momenta and multipole moments are defined, $\hv^{\mu}:= dx_p^{\mu}/d\htau$ denotes the tangent vector along the worldline, $\hp^{\mu}$ is the object's momentum, $\tS^{\mu\nu}$ is its spin tensor, and $\tJ^{\alpha \beta \gamma \delta}$ is its quadrupole tensor. In this way, the body is represented by its ``skeleton'', a pointlike singularity equipped with the body's multipole moments. Note that here all quantities are defined with respect to the effective metric, not the original metric ${\bf g}_{\mu\nu}$; in the point-particle limit, the original metric becomes singular at the particle's position, such that the above expressions would become ill defined if written in terms of ${\bf g}_{\mu\nu}$, while the effective metric (if chosen correctly) is smooth at the particle's worldline. Concretely, the proper time $\htau$ satisfies the normalization condition $\hat g_{\mu\nu}\hv^{\mu}\hv^{\nu}=-1$, the Riemann tensor $\hR_{\alpha \beta \gamma \delta}$ is that of $\hat g_{\mu\nu}$, and the multipole moments have been implicitly defined in terms of the effective metric~\cite{Harte:2011ku}. Note that the spin tensor is antisymmetric, $\tS^{\mu\nu}=-\tS^{\nu\mu}$, and the quadrupole tensor exhibits all the algebraic symmetries of the Riemann tensor. 



Another way of arriving at the skeletonized stress-energy tensor is to use the second, complementary approach to handling a small object: the method of matched asymptotic expansions~\cite{DEath:1975jps,Thorne:1984mz,Zhang:1985qz,Mino:1996nk,Detweiler:2000gt,Gralla:2008fg,Pound:2012dk,Pound:2015tma,Barack:2018yvs}, which has been the basis for all concrete self-force results beyond linear order in the mass. Unlike the Dixon-Harte approach, this method applies to black holes as well as material bodies. Rather than seeking exact, nonperturbative results, it is formulated in terms of expansions from the beginning. The essential idea is to employ two asymptotic, small-$\e$ expansions: an ``outer expansion'' valid on the scales of the external universe, compared to which the object's scales are all small; and an ``inner expansion'' that zooms in on the object. This method is typically formulated in terms of a one-parameter family of spacetimes, where the parameter $\e$ specifically counts powers of the mass $m/L$, with the assumption that $R\sim m$. In an intermediate region around the object, at distances large compared to $m$ but small compared to external scales, one can solve the vacuum Einstein equation order by order in $m$ and express the solution naturally in terms of a set of multipole moments, which are interpreted as the moments of the object. If this local vacuum solution is analytically extended down to a representative worldline, then one can show that it satisfies a field equation of the form ~\cite{Gralla:2008fg,Pound:2012dk,Upton:2021oxf,Musaeus:InPrep} 
\beq\label{skeleton EFE}
 G_{\mu\nu}[g+h]=8\pi \hat T_{\mu\nu},
\eeq
where $\hat T_{\mu\nu}$ is the same ``skeletonized'' stress-energy tensor~\eqref{SET}. Pointedly, the multipole expansion of the \emph{effective} stress-energy tensor is what appears here, rather than the original $T_{\mu\nu}$. We refer to Refs.~\cite{Upton:2021oxf,Musaeus:InPrep} for a rigorous distributional interpretation of Eq.~\eqref{skeleton EFE}.

A power of this second approach is that it provides the field equation in a solvable perturbative form, and it provides a direct construction of an effective metric that is a smooth vacuum solution at the particle's worldline. The effective metric takes the form~\cite{Pound:2012dk}
\begin{equation}\label{eq:geff}
    \hat g_{\mu\nu} = g_{\mu\nu} + \e h^{1\rm R}_{\mu\nu} + \e^2 h^{2\rm R}_{\mu\nu} + \e^3 h^{3\rm R}_{\mu\nu} + {\cal O}(\e^4),
\end{equation}
where now $\e$ specifically counts powers of $m/L$, again assuming $m\sim R$. Here $h^{\rm R}_{\mu\nu}= h_{\mu\nu} - h^{\rm S}_{\mu\nu}=\sum_n \e^n h^{n{\rm R}}_{\mu\nu}$ is the \emph{regular field}; it is what remains from $h_{\mu\nu}$ after removing the object's \emph{self-field} (or \emph{singular field}) $h^{\rm S}_{\mu\nu}$, which becomes singular on the worldline in the point-particle limit.


As mentioned at the beginning of this section, there is some subtlety in making the order parameter $\e$ count powers of $m/L$. Roughly speaking, the body's $n$th moment scales as $m R^n=m^{n+1}{\cal C}^{-n}$, where ${\cal C}:=m/R$ is the object's compactness. In our expansion in Eq.~\eqref{SET}, the monopole, dipole, and quadrupole terms then scale roughly as $m R^0/L^3$, $m R/L^4$, and $m R^2/L^5$, respectively. These would be treated as $m/L^3$, $m^2/L^4$, and $m^3/L^5$ in the matched-expansions approach in Refs.~\cite{Gralla:2008fg,Pound:2009sm,Pound:2012dk}, for example, even if only implicitly. But in reality the relationship between $R$ and $m$ varies dramatically between different types of objects. For a black hole, $R\propto m$, and the moments scale as $m^{n+1}$, making the expansion in powers of $m/L$ completely natural: one new moment enters the stress-energy tensor at each successive order in $m/L$. For a rigid Newtonian body, $R\propto m^{1/3}$ for fixed density $\rho$, implying the moments scale as $m^{1+n/3}$. In that case, three new moments enter at each successive order in $m/L$. In either case, if we reduce the mass of the object, its size also reduces, and vice versa. However, compact stars (white dwarfs and neutron stars) do not comply with this intuition: their radius \emph{grows} as their mass decreases~\cite{1972ApJ...175..417N,Ozel:2016oaf}, meaning a multipole moment that scales as $mR^n$ gets \emph{larger}. Hence, we cannot freely shrink down such a star in a self-similar way. 

Fundamentally, perturbation theory in general relativity is based on considering a family of spacetimes with one or more parameters that can be taken to zero. We often think of being able to take both the mass and size of the small object to zero simultaneously, while keeping the object's composition fixed, but this is simply not an appropriate description of a family of spacetimes containing a compact star. Instead, the tunable parameter is the external length scale $L$, which can be freely made larger: to take the small-mass-ratio limit for systems containing a compact star, we must consider a family of binaries in which the secondary is always precisely the same star with all its same internal parameters, while the primary's mass grows. In this case, the star's compactness ${\cal C}$ is fixed. The $n$th multipole moment enters the stress-energy tensor with the scaling $\frac{m}{L^3}\left(\frac{m}{L}\right)^n{\cal C}^{-n}$, and it enters the metric perturbation with the scaling $\left(\frac{m}{L}\right)^{n+1}{\cal C}^{-n}$. 

Hence, for compact stars, like black holes, we can continue thinking of $\e$ as counting powers of $m/L$, or of the mass ratio in a binary. However, as mentioned in the Introduction, some quadrupole terms enter at a higher order in this counting, and it will remain convenient to keep a loose interpretation of $\e$. For a Newtonian body, given the scaling $R\propto m^{1/3}$ at fixed density, working with two small parameters, $m/L$ and $R/L$, with two distinct counting parameters, would be most sensible.

As a final comment in this section, we note that for the sake of simplifying expressions, we use $\hat g_{\mu\nu}$ and its inverse $\hat g^{\mu\nu}$ to lower and raise indices on hatted tensors. However, we will later transition to unhatted tensors defined in terms of $g_{\mu\nu}$, and we will use the background metric $g_{\mu\nu}$ and its inverse $g^{\mu\nu}$ to lower and raise indices on those tensors.

\subsection{Mathisson-Papapetrou-Dixon equations}
    Just as for the field equations, there are two approaches to obtaining the equations of motion for the object's momentum $\hat p^\alpha$ and spin $\hat S^{\alpha\beta}$. 
    
    In the Dixon-Harte approach, one first obtains exact equations of motion from conservation of the original (rather than effective) stress-energy tensor in the original metric ${\bf g}_{\alpha\beta}$. In the test-body approximation, where one neglects the gravitational field sourced by the body, the exact equations of motion can be expanded in a multipole series to obtain the well-known Mathisson-Papapetrou-Dixon (MPD) equations~\cite{Mathisson:2010opl,Papapetrou:1951pa,Dixon1974DynamicsOE}. 
    More recently, Harte and collaborators~\cite{Harte:2011ku,Harte:2025tmd} have shown that the test-body result can be extended to account for the body's self-field through some appropriate choice of effective metric and effective stress-energy tensor, with corresponding effective multipole moments. The equations of motion are then the MPD equations in the effective metric:
    \begin{subequations}\label{MPD_eq_hat}
      \begin{align}
			\frac{\hD \hp^{\mu }}{d\htau }&=-\frac{1}{2} \tS^{\rho \sigma } \hv^{\nu } \hR^{\mu }{}_{\nu \rho \sigma }-\frac{1}{6}\tJ^{\alpha \beta \gamma \delta }\hnabla ^{\mu }\hR_{\alpha \beta \gamma \delta }+\mathcal{O}(\e^4)\,,\label{MPD_force} \\
			\frac{\hD \tS^{\mu \nu }}{d \htau}&=2\hp^{[\mu }\hv^{\nu ]}-\frac{4}{3}\hR_{\alpha \beta \gamma }{}^{[\mu }\tJ^{\nu ]\gamma \alpha \beta }+\mathcal{O}(\e^4)\,,\label{MPD_torque}
		\end{align}  
    \end{subequations}
	where $\hD/d\htau:= \hv^{\mu}\hnabla_{\mu}$. These equations are equivalent to conservation of the effective stress-energy tensor in the effective metric, $\hat\nabla_\beta \hat T^{\alpha\beta}=0$.\footnote{Such a conservation equation might appear to be at odds with the Einstein equation~\eqref{skeleton EFE}, since the latter would seem to suggest $\hat T^{\alpha\beta}$ should be conserved in the full metric ${\bf g}_{\alpha\beta}$, not in $\hat g_{\alpha\beta}$. However, the (perhaps surprising) consistency of Eq.~\eqref{skeleton EFE} with $\hat\nabla_\beta \hat T^{\alpha\beta}=0$ can be established as in Refs.~\cite{Upton:2021oxf,Musaeus:InPrep}.}

In the matched-expansions approach, the equations of motion are instead derived from the field equations \emph{outside} the extended object. This approach has been used to derive the MPD equations~\cite{Thorne:1984mz,Zhang:1985qz}, but without a precise specification of the effective metric, essentially defining it loosely as the metric in which the object would be in free-fall if it had no moments beyond its monopole; see Ref.~\cite{Pound:2015tma} for a historical perspective. The equations of motion \emph{with} a complete specification of $\hat g_{\alpha\beta}$ have only been derived through second perturbative order in the mass for a nonspinning, spherical object~\cite{Pound:2012nt,Pound:2017psq}\footnote{The equation of motion in Ref.~\cite{Gralla:2012db} is expected to be equivalent to this on short timescales, though this equivalence has never been verified.}---i.e., $\hD\hp^\alpha/d\htau={\cal O}(\e^4)$ with a known $\hat g_{\alpha\beta}=g_{\alpha\beta}+\e h^{1\rm R}_{\alpha\beta}+\e^2 h^{2\rm R}_{\alpha\beta}$---and to one order lower for a spinning object~\cite{Mino:1997wh,Gralla:2008fg,Pound:2009sm}.

This situation is not entirely satisfying because no scheme has consistently derived Eq.~\eqref{MPD_eq_hat} with a specification of the effective metric in which the equation is formulated: the MPD equations have not been shown to hold in the effective metric defined in Refs.~\cite{Pound:2012nt,Pound:2012dk,Pound:2017psq} once coupling between the spin and regular field arise or once quadrupole moments enter; and while Harte's work shows there should exist effective metrics for which the MPD equations hold, it does not provide a prescription for constructing them (moreover, as pointed out in Ref.~\cite{Mathews:2021rod}, the example effective metric in Ref.~\cite{Harte:2011ku} actually becomes singular in the point-particle limit at second order in $\e$, unlike the effective metric in Ref.~\cite{Pound:2012dk}, which is smooth to all orders).  

However, in this paper, we will only require that there exists at least one appropriate choice of effective metric. Harte's work suffices for that purpose, at least for material bodies. Additionally, we note that there is no ambiguity in the test-body limit, where both the matched-expansions and Dixon-Harte approach yield the same MPD equations. Put another way, any uncertainties arise from couplings to the regular field. Such couplings will be less relevant in cases where finite-size effects are large (i.e., when $R\gg m$). So, for the rest of this paper, we take the MPD equations~\eqref{MPD_eq_hat} as a given. 

References~\cite{Steinhoff:2009tk, Steinhoff:2012rw, Hinderer:2013uwa} provide a modern introduction to the MPD equations, which we draw from in the rest of this section. We first note that the system of MPD equations is underdetermined as the number of unknowns exceeds the number of equations. To have a closed system of equations, it is necessary to impose some additional condition, known as the \textit{spin supplementary condition}, which specifies the reference worldline $x^\alpha_p$ along which the MPD equations are defined. 
In this work, we employ the Tulczyjew-Dixon spin supplementary condition~\cite{Tulczyjew:1959,Dixon1970DynamicsOE},
	\begin{equation}\label{SSC}
		\hp_{\mu}\tS^{\mu\nu}=0~,
	\end{equation}
	which fixes the reference worldline to be the centre of mass of the object. 
    
    The Tulczyjew-Dixon condition 
    also provides a relation between the 4-velocity $\hv^{\mu}$ and the 4-momentum $\hp^{\mu}$. Differentiating Eq.~\eqref{SSC} and substituting Eq.~\eqref{MPD_eq_hat}, we obtain~\cite{Bohe:2015ana}
	\begin{multline}\label{rel_v_p}
		\hp^{\mu}=m_d\hv^{\mu}-\frac{4}{3}\hR_{\alpha\beta\gamma}{}^{[\mu}\tJ^{\nu]\gamma\alpha\beta}\hv_{\nu}\\
        -\frac{1}{2m_d}\hR_{\nu\gamma\alpha\beta}\tS^{\alpha\beta}\tS^{\mu\nu}\hat{v}^{\gamma}+\mathcal{O}(\e^4)~,
	\end{multline}	
    where $m_d:=\sqrt{-\hp_{\mu}\hp^{\mu}}$ is the dynamical mass of the object. Note the mass $m$ we used previously could stand for this dynamical mass or for a conserved mass we introduce below. 
    
    As discussed before, the quadrupole tensor $\tJ^{\mu\nu\alpha\beta}$ contains information about the deformation due to spin and tidal forces.  Thus, we choose the following form of the quadrupole tensor~\cite{Steinhoff:2012rw, Hinderer:2013uwa, Bini:2015zya, Bohe:2015ana, Dones:2024odv}:
    \begin{equation}\label{quadrupole}
		\tJ^{\alpha \beta \gamma \delta }=\tJ^{\alpha \beta \gamma \delta }_{Q}+\tJ^{\alpha \beta \gamma \delta }_{\mathcal{G}}+\tJ^{\alpha \beta \gamma \delta }_{\mathcal{H}},
	\end{equation}
    where $\tJ^{\alpha \beta \gamma \delta }_{Q}$, $\tJ^{\alpha \beta \gamma \delta }_{\mathcal{G}}$, and $\tJ^{\alpha \beta \gamma \delta }_{\mathcal{H}}$ represent the deformation of the secondary induced by spin, gravito-electric, and gravito-magnetic tidal forces. These quantities are defined as 
	\beq\label{quadrupole1}
		\tJ^{\alpha \beta \gamma \delta }_{Q}&=-3c_{Q}\,\hv^{[\alpha }\tQ^{\beta ][\gamma }\hv^{\delta ]},\\
        \tJ^{\alpha \beta \gamma \delta }_{\mathcal{G}}&=-3\mu_2\,\hv^{[\alpha }\mathcal{G}^{\beta ][\gamma }\hv^{\delta ]},\\
        \tJ^{\alpha \beta \gamma \delta }_{\mathcal{H}}&=\sigma_2\left[\epsilon^{\alpha\beta}{}_{\mu\nu}\hv^{\mu}\mathcal{H}^{\nu[\gamma}\hv^{\delta]}+\epsilon^{\gamma\delta}{}_{\mu\nu}\hv^{\mu}\mathcal{H}^{\nu[\alpha}\hv^{\beta]}\right],
	\eeq
where $\tQ^{\alpha\beta}$ and $\mathcal{G}^{\alpha\beta}$ represent the mass quadrupole tensor due to spin and gravito-electric tidal effect, respectively, and $\mathcal{H}^{\alpha\beta}$ is the current quadrupole (gravito-magnetic) tidal tensor. These terms can be written as follows \cite{Steinhoff:2012rw, Bohe:2015ana, Dones:2024odv}:
\beq
\tQ^{\alpha\beta} &= \tS^{\alpha}{}_{\mu}\tS^{\beta\mu}\,,\quad
\mathcal{G}_{\alpha\beta} =-\hR_{\alpha\gamma\beta\delta}\hv^{\gamma}\hv^{\delta}\,,\\
\mathcal{H}_{\alpha\beta} &=2 \hR^{*}_{(\alpha\underline{\rho}\beta)\sigma}\hv^{\rho}\hv^{\sigma},
\eeq
where $\hR^{*}_{\alpha\rho\beta\sigma}$ is the Hodge dual of the Riemann tensor $\hR_{\alpha\rho\beta\sigma}$. The parameter $c_Q$ characterizes the spin-induced quadrupolar polarization, whereas 
$\mu_2$ and $\sigma_2$ 
 describe the mass-type and current-type tidal polarizabilities. 
\begin{table*}[tb] 
\centering
\renewcommand{\arraystretch}{1.25}
\begin{ruledtabular}
\begin{tabular}{c c  c c c c}
&$\mathcal{C}=m_d/R$ & $\chi_{\rm max}$  & $C_Q$ & $k_2$ & $-j_2$\\
\midrule
Black hole & 1/2& 1 & 1~\cite{Hansen:1974zz, PhysRevD.57.5287} & 
0~\cite{Binnington:2009bb, Damour:2009vw, LeTiec:2020spy,Chia:2020yla} &0~\cite{Binnington:2009bb, Damour:2009vw, LeTiec:2020spy,Chia:2020yla} \\
Neutron star & 0.1--0.25 \cite{Hinderer:2007mb} &0.7~\cite{1994ApJ...424..823C, Lo:2010bj} & 2--14~\cite{Laarakkers:1997hb, Harry:2018hke, Pappas:2012qg} &  \begin{tabular}{@{}c@{}}0.05--0.15  \\[-3pt] ~\cite{Hinderer:2007mb, Hinderer:2009ca, Piekarewicz:2018sgy} \end{tabular}& \begin{tabular}{@{}c@{}}0.001--0.01  \\[-3pt] \cite{Binnington:2009bb,Damour:2009vw, Gagnon-Bischoff:2017tnz, Landry:2015cva} \end{tabular}\\ \begin{tabular}{@{}c@{}}Massive \\[-3pt]  boson star \end{tabular}& $\lesssim ~0.158$ \cite{Sennett:2017etc}&$\gtrsim1$~\cite{PhysRevD.55.6081, Siemonsen:2020hcg, Vaglio:2022flq} & 10--150~\cite{PhysRevD.55.6081,Vaglio:2022flq} & $\tilde{\Lambda}_E\gtrsim 280$ \cite{Sennett:2017etc}  &-- \\
White dwarf & $10^{-4}$--$10^{-3}$ \cite{1972ApJ...175..417N, Hartl:2002ig}& 20--30 \cite{Hartl:2002ig,2000ApJ...534..359G, Rahman:2021eay} & $10^3$--$10^5$~\cite{Taylor:2019hle} & 0.05--0.30~\cite{Taylor:2019hle}&-- \\
\end{tabular}
\end{ruledtabular}
\caption{Representative values of the compactness parameter $\mathcal{C}=m_d/R$, maximum value of the dimensionless spin parameter $\chi_{\rm max}$, spin-induced quadrupole coefficient $C_Q$, electric quadrupolar tidal Love number $k_2$, and magnetic quadrupolar tidal Love number $j_2$ for various astrophysical compact objects. }
%
\label{tab:param_astro_object}
\end{table*}
 One can also define a set of dimensionless parameters $C_Q$, $\tilde{\Lambda}_E$, and $\tilde{\Lambda}_M$, which quantify the spin-induced quadrupolar, mass-type (``electric'') tidal, and current-type (``magnetic'') tidal deformations, respectively. These are defined as \cite{Steinhoff:2012rw}
 \beq\label{deformation parameters}
C_Q:=c_Q m_d\,,\quad \tilde{\Lambda}_E:=\frac{\mu_2}{m_d^5}\,,\quad \tilde{\Lambda}_M:=\frac{\sigma_2 }{ m_d^5}.
\eeq
Furthermore, the parameters $\tilde{\Lambda}_E$ and $\tilde{\Lambda}_M$ are related to the dimensionless tidal Love numbers $k_2$ and $j_2$ through the relation
$k_2=3\tilde{\Lambda}_E \mathcal{C}^5/2$ and $j_2=48\tilde{\Lambda}_M \mathcal{C}^5$,
where $\mathcal{C}= m_d/R$ is the object's compactness. 

The dimensionless quantities $C_Q$, $\tilde\Lambda_E$, and $\tilde\Lambda_M$ are specifically adimensionalised using powers of the mass $m_d$. \emph{If} these quantities and the dimensionless spin $\hat S^{\alpha\beta}/m_d^2$ are all of order unity, then we can see that the spin-induced quadrupole moment scales as $(m_d)^3$, while the tidally induced quadrupole moments scale as $(m_d)^5$. We could then imagine a one-parameter family of binaries in which the mass ratio is varied while holding $C_Q$, $\tilde\Lambda_E$, and $\tilde\Lambda_M$ (or $C_Q$, $k_2$, and $j_2$) all fixed. In the limit of small mass ratios, the spin-induced quadrupole moment then enters the equations of motion~\eqref{MPD_eq_hat} as a second-order-in-mass-ratio acceleration---the same (2PA) order as the conservative second-order self-force (generated by $h^{\rm R(2)}_{\alpha\beta}$)---while the tidally induced quadrupole moments enter at two orders higher (4PA). This is the PA counting used in the Introduction. As foreshadowed in the previous section, this picture can be misleading because if we consider a specific type of material body, such as a neutron star or white dwarf, its dimensionless parameters $C_Q$, $\tilde\Lambda_E$, and $\tilde\Lambda_M$ are not independent of its mass~\cite{Rodriguez:2026iot,Chakraborty:2026qru}. Instead, as explained in the previous section, the appropriate family of binaries to consider is one in which the large mass varies and the secondary's intrinsic parameters, including its mass, are all held fixed. In the limit as the large mass becomes much larger than $m_d$, we then obtain the PA counting above. 

If the secondary is a Kerr black hole, then we do not need to consider that subtlety because the parameters $C_Q$, $\tilde\Lambda_E$, and $\tilde\Lambda_M$ are all universal for Kerr black holes: the no-hair theorem implies $C_Q=1$~\cite{PhysRevD.57.5287} and $k_2=0=j_2$~\cite{Rodriguez:2026iot,Chakraborty:2026qru}. For stars, the values of these parameters vary substantially depending on the assumed equation of state, and they also vary substantially with $m_d$ for a given equation of state~\cite{Rodriguez:2026iot,Chakraborty:2026qru}. We summarize predicted ranges of values for various types of stars in Table~\ref{tab:param_astro_object}. For a neutron star, most values appear moderate, and $k_2$ appears numerically small. However, $\tilde\Lambda_E$, which is what actually appears in the stress-energy tensor and MPD equations, is several orders of magnitude larger than $k_2$ due to the factor ${\cal C}^5$ between them. For less compact stars such as white dwarfs, the spin-induced quadrupole can reach enormous values, but the tidal deformability is even more extreme, with $\tilde\Lambda_E\sim 10^{14}$; since this term in the equation of motion is suppressed by four orders in the mass ratio (relative to 0PA), the impact would still only be moderate for an EMRI, but potentially comparable to a 1PA term (consistent with Ref.~\cite{Rahman:2021eay}'s estimates for the impact of a spin-induced quadrupole). Notably, $C_Q$ can also assume relatively large values for certain exotic compact objects such as boson stars, as shown in the table.

    It is worth noting that the dynamical mass and the monopole mass of the object are not conserved. Using the MPD equations~\eqref{MPD_force} and~\eqref{MPD_torque} and Bianchi identities, we can show that 
	\begin{equation}\label{dyn_mass}
	\frac{dm_d}{d\htau}=-\frac{\hp_\mu}{m_d}\frac{\hD \hp^{\mu}}{d\htau}=\frac{1}{6}\frac{\hD \hR_{\rho\alpha\beta\gamma}}{d\htau}\tJ^{{\rho\alpha\beta\gamma}}+\mathcal{O}(\e^{4}),
	\end{equation}
    which indicates that the dynamical mass of the object generically evolves with time. However,
    we can define a mass-like quantity $m_s$, so-called the \textit{renormalized mass}, given by the following expression~\cite{Steinhoff:2012rw, Bini:2015zya}:
	\begin{equation}\label{con_mass}
		m_s:= m_d-\frac{1}{6}\hR_{\rho\alpha\beta\gamma}\tJ_Q^{{\rho\alpha\beta\gamma}}+\frac{\mu_2}{4}\mathcal{G}_{\alpha\beta}\mathcal{G}^{\alpha\beta}+\frac{\sigma_2}{6}\mathcal{H}_{\alpha\beta}\mathcal{H}^{\alpha\beta},
    \end{equation}
    which \emph{is} conserved at quadrupolar order. Furthermore,  the spin supplementary condition also ensures that the spin magnitude $\tS:=\sqrt{\tS_{\mu\nu}\tS^{\mu\nu}/2}$ is conserved at quadrupole order, as can be seen from
    \begin{equation}\label{spin_length}
		\begin{aligned}
			\tS\frac{d\tS}{d\htau}=\tS_{\mu\nu}\Bigl(\hp^{\mu }\hv^{\nu}-\frac{2}{3}\hR_{\alpha \beta \gamma }{}^{\mu }\tJ^{\nu \gamma \alpha \beta }\Bigr)=\mathcal{O}(\e^5)~.
		\end{aligned}
	\end{equation}
This implies $d\tS/d\htau={\cal O}(\e^3)$.

If $\hat g_{\alpha\beta}$ were a stationary, axisymmetric spacetime, the energy and angular momentum of the object,
	\beq\label{Conserved}
		E &=-\Bigl(\hp^{\mu}\hat\xi_{\mu}^{t}+\frac{1}{2}\tS^{\mu\nu}\hnabla_{\mu }\hat\xi_{\nu}^{t}\Bigr),\\
        L_z &=\hp^{\mu}\hat\xi_{\mu}^{\phi}+\frac{1}{2}\tS^{\mu\nu}\hnabla_{\mu }\hat\xi_{\nu}^{\phi},
	\eeq  
would also be conserved. Here, $\xi^{\mu}_t=(\partial_t)^{\mu}$ and $\xi^{\mu}_\phi=(\partial_\phi)^{\mu}$ are the timelike and azimuthal Killing vector, respectively, and $\hat\xi^t_\mu:= \hat g_{\mu\nu}\xi^\nu_t$ and $\hat\xi^\phi_\mu:= \hat g_{\mu\nu}\xi^\nu_\phi$. Note that, unlike the renormalized mass $m_s$ and the spin magnitude $S$, the energy and angular momentum of the object would be conserved even at higher multipole orders. However, in our context, $\xi^{\mu}_t=(\partial_t)^{\mu}$ and $\xi^{\mu}_\phi=(\partial_\phi)^{\mu}$ are only Killing vectors of the background spacetime, $g_{\alpha\beta}$, and the $E$ and $L_z$ defined above are only approximately conserved. We will derive flux-balance law equations for their evolution in Sec.~\ref{Sec_4_BalanceLaws}.


Substituting Eq.~\eqref{con_mass} in Eq.~\eqref{rel_v_p}, we obtain another relation between the 4-velocity and 4-momentum, which can be written as follows: 
	\begin{equation}\label{rel_v_p_1}	\hp^{\mu}=m_s\hv^{\mu}+\hp_{Q}^{\mu}+\hp_{\mathcal{G}}^{\mu}+\hp_{\mathcal{H}}^{\mu}+\mathcal{O}(\e^4),
	\end{equation}	
where
\begin{subequations}
    \begin{align}
     \hp_{Q}^{\mu} &= -\frac{4}{3}\hR_{\alpha\beta\gamma}{}^{[\mu}\tJ_Q^{\nu]\gamma\alpha\beta}\hv_{\nu}-\frac{1}{2m_s}\hR_{\nu\gamma\alpha\beta}\tS^{\alpha\beta}\tS^{\mu\nu}\hat{v}^{\gamma} \nonumber\\&\quad 
     +\frac{1}{6}\hR_{\rho\alpha\beta\gamma}\tJ_Q^{{\rho\alpha\beta\gamma}}\hv^{\mu}\,,\\
     \hp_{\mathcal{G}}^{\mu} &= \mu_2\left[-\hR^{\mu}{}_{\alpha\gamma\beta}\hv^{\gamma}\mathcal{G}^{\alpha\beta}+\frac{3}{4}\hv^{\mu}\mathcal{G}^{\alpha\beta}\mathcal{G}_{\alpha\beta}\right]\,,\\
     \hp_{\mathcal{H}}^{\mu} &= \sigma_2\left[\hR^{(\mu\underline{\alpha}\gamma)\beta}_{*}\hv_{\gamma}\mathcal{H}_{\alpha\beta}+\frac{1}{2}\hv^{\mu}\mathcal{H}^{\alpha\beta}\mathcal{H}_{\alpha\beta}\right]\,.
    \end{align}
\end{subequations}
We use Eq.~\eqref{rel_v_p_1} to reformulate the equations of motion in the next section.	

\subsection{Equations of motion with respect to the background spacetime}	

As discussed above, the multipolar terms presented in Eq.~\eqref{SET1} are defined   in terms of the \textit{effective metric} $\hat{g}_{\mu\nu}=g_{\mu\nu}+ h^{R}_{\mu\nu}$. However, it is customary to rewrite the equations of motion~\eqref{MPD_force} and~\eqref{MPD_torque} in terms of the background metric $g_{\mu\nu}$. 

The relationship between the proper time in the two metrics is given by  $d\htau/d\tau=\sqrt{1- h^{R}_{\rho\sigma}v^{\rho}v^{\sigma}}$, where $\tau$ is the proper time with respect to the background spacetime and $v^{\mu}=dx_p^{\mu}/d\tau$ is the four-velocity. The Riemann tensor is also expanded as 
\begin{eqnarray}
    \hat R_{\alpha\beta\gamma\delta} = R_{\alpha\beta\gamma\delta} + \delta R_{\alpha\beta\gamma\delta}[h^R] + {\cal O}[(h^R)^2],
\end{eqnarray}
and implicit appearances of $\hat g_{\mu\nu}$ or $\hat g^{\mu\nu}$ in lowered or raised indices are additionally expanded around $ g_{\mu\nu}$ or $ g^{\mu\nu}$. 

In terms of the background metric, Eq.~\eqref{MPD_force} for the object's trajectory can be written as 
\begin{equation}\label{self_force1}
     \begin{aligned}
       \frac{D^2 x_p^{\mu}}{d\tau^2}= F^{\mu}= F^{\mu}_{m}+F^{\mu}_{S}+F^{\mu}_{Q}+F^{\mu}_{\mathcal{G}}+F^{\mu}_{\mathcal{H}}+\CO(\e^3),
     \end{aligned}
 \end{equation}
 where we have used the relation~\eqref{rel_v_p_1} between the 4-velocity and 4-momentum and 
\begingroup\allowdisplaybreaks%
 \begin{subequations}\label{SF_terms}
     \begin{align}
       F^{\mu}_{m}&= -\frac{1}{2} P^{\mu\nu}(g_\nu{}^\lambda - h^{\mathrm{R}\, \lambda}_\nu)\left(2 h_{\lambda \rho ; \sigma}^{\mathrm{R}}-h_{\rho \sigma ; \lambda}^{\mathrm{R}}\right) v^{\rho} v^{\sigma}\,,\\
F^{\mu}_{S}&=  -\frac{m_s}{2}R^{\mu}{}_{\alpha \beta \gamma}\left(1- \frac{1}{2}h^{\mathrm{R}}_{\rho\sigma}v^\rho v^\sigma\right)v^{\alpha} S^{\beta \gamma}\nonumber\\*
&\quad +\frac{m_s}{2} P^{\mu\nu}\bigl(2h^{\mathrm{R}}_{\nu(\alpha;\beta)\gamma}-h^{\mathrm{R}}_{\alpha\beta;\nu\gamma}\bigr)v^{\alpha} S^{\beta \gamma}\,, \\
F^{\mu}_{Q}&= -\frac{m_s^2}{6}J_Q^{\alpha\beta\gamma\delta}\nabla^{\mu}R_{\alpha\beta\gamma\delta}-
         \frac{m_s^2}{6}v^{\mu}\frac{DR_{\rho\alpha\beta\gamma}}{d\tau}J_Q^{\rho\alpha\beta\gamma}\nonumber\\*
         &\quad +\frac{4m_s^2}{3}\frac{DR_{\alpha \beta \gamma }{}^{[\mu }}{d\tau}J_Q^{\nu]\gamma\alpha\beta}v_{\nu}\nonumber\\*
         &\quad +\frac{m_s^2}{2}\frac{DR_{\nu\gamma\alpha\beta}}{d\tau}S^{\mu\nu}S^{\alpha\beta}v^{\gamma}\,,\\
     F^{\mu}_{\mathcal{G,H}}&=  -\frac{1}{6m_s}\tJ_{\mathcal{G,H}}^{\alpha\beta\gamma\delta}\nabla^{\mu}R_{\alpha\beta\gamma\delta}   -\frac{1}{m_s}\frac{D\hp^\mu_{\mathcal{G,H}}}{d\tau}\,,
     \end{align}
 \end{subequations}%
 \endgroup
 with the projection tensor $P^{\mu\nu}:= g^{\mu\nu}+v^\mu v^\nu$, and the normalized quantities
 \begin{equation}
 S^{\mu\nu}:= \frac{\tS^{\mu\nu}}{m_s^2}, \qquad J_Q^{\alpha \beta \gamma \delta}:=\frac{\tJ_Q^{\alpha \beta \gamma \delta}}{m_s^3}.
 \end{equation}
The expressions for $F^\mu_m$ and $F^\mu_S$ are reproduced from Ref.~\cite{Mathews:2021rod}, though in that reference there was no distinction between $m_d$ and $m_s$; here we have explicitly written the results in terms of $m_s$.
 
 Similarly, Eq.~\eqref{MPD_torque} for the object's spin evolution can be written as 
 \begin{equation}\label{ST_terms}
     \begin{aligned}
         \frac{D S^{\mu\nu}}{d \tau} &:= N^{\mu\nu}=N_{S}^{\mu\nu}+N_{Q}^{\mu\nu}
         +\mathcal{O}(\e^2)\,,
     \end{aligned}
 \end{equation}
where the torques are 	
 \begin{subequations}\label{ST_terms1}
     \begin{align}
N_{S}^{\mu\nu} &=  v^{(\rho} S^{\sigma)[\mu}g^{\nu]\lambda}\left(2 h_{\lambda \rho ; \sigma}^{\mathrm{R}}-h_{\rho \sigma ; \lambda}^{\mathrm{R}}\right)  + \mathcal{O}(\e^2),\\
N_{Q}^{\mu\nu}&=2m_s v^{[\mu }{p}_Q^{\nu ]}-\frac{4m_s}{3}R_{\alpha \beta \gamma }{}^{[\mu }J_Q^{\nu ]\gamma \alpha \beta }+\mathcal{O}(\e^2),
     \end{align}
 \end{subequations}
    with $p^{\mu}_Q:=  \hp^{\mu}_Q/m_s^3$. The ``self-torque'' $N^{\mu\nu}_S$ is the same as appeared in Ref.~\cite{Mathews:2021rod} but also in earlier work such as Ref.~\cite{Akcay:2019bvk}.

We can also write the spin evolution in terms of the spin vector  
\begin{equation}\label{eq:spin vector def}
    {S}^{\mu} := -\frac{1}{2} \epsilon^{\mu}{}_{\alpha \beta \gamma} v^{\alpha}{S}^{\beta \gamma}.
\end{equation}
Substituting ${S}^{\mu \nu}=-\epsilon^{\mu \nu}{}_{\alpha \beta}{S}^{\alpha}v^{\beta}$ into Eq.~\eqref{ST_terms} and contracting the equation with $P^\alpha_{\ \beta}$, we find
\beq\label{self_torque1}
P^\alpha_{\ \beta} \frac{D S^\beta}{d\tau} = N^\alpha  =N_F^\alpha+ N_N^\alpha, 
\eeq
where 
\begin{subequations}
\begin{align}
   N_F^\alpha &:= -\frac{1}{2}P^\alpha_{\ \mu}\epsilon^\mu{}_{\nu\rho\sigma}F^\nu S^{\rho\sigma},\\
   N_N^\mu&:= -\frac{1}{2}\epsilon^\mu{}_{\nu\rho\sigma}v^\nu N^{\rho\sigma}.
\end{align}
\end{subequations}
Note that the spin vector $S^\mu$ defined here agrees with the one in Ref.~\cite{Mathews:2021rod} but differs from the one in Ref.~\cite{Mathews:2025nyb}, which was defined in terms of $\hat\epsilon^\mu{}_{\alpha\beta\gamma}$ and $\hat v^{\mu}$.

 \section{Multiscale expansion for extended bodies in a Kerr background}\label{Sec_3_Kerr}
 
We now specialize to the motion of an extended object orbiting a Kerr black hole with background mass $M$ and spin $J=Ma$. The Kerr background is given in terms of the following line element
\beq\label{Kerr_BH}
\d s^2 = g_{tt} \d{t}^2 + g_{rr} \d{r}^2 + g_{\theta\theta}\d\theta^2 + g_{\phi\phi}\d\phi^2 +2g_{t\phi}\d{t}\d \phi, 
\eeq
where the nonzero components of the metric tensor $g_{\mu\nu}$, expressed in standard Boyer-Lindquist coordinates, are given by
\begin{align}
g_{tt} &= -\left(\frac{\Delta-a^{2}\sin^{2}\theta}{\RS}\right), \quad
g_{rr} = \frac{\RS}{\Delta}, \quad g_{\theta\theta} = \RS,\nonumber\\
g_{\phi\phi} &= \frac{\sin^{2}\theta}{\RS}\left[(r^{2}+a^{2})^{2}-\Delta
a^{2}\sin^{2}\theta\right], \nonumber\\
g_{t\phi}&=\frac{a\sin^{2}\theta}{\RS}\left[\Delta-(r^{2}+a^{2})\right] ,
\label{eq:MetricCoef}
\end{align}
where 
\beq
\Delta = r^2 - 2 Mr + a^2,\quad  \RS = r^2 + a^2 \cos^2\theta.
\eeq
Note that the outer horizon of the black hole is situated at 
\beq
r_{+} = M + \sqrt{ M^2 -a^2 }.
\eeq

In what follows, we 
assume that the secondary inspirals along an equatorial, quasicircular orbit. To keep the motion equatorial, the spins of both the primary and the secondary are taken to be aligned or anti-aligned with each other and oriented perpendicular to the equatorial plane. Our treatment in this section follows Ref.~\cite{Mathews:2021rod}, which carried out the same analysis at 1PA order (and therefore limited to linear in spin) in a Schwarzschild background.
 \subsection{Fixed-frequency expansion in the test-body approximation}\label{Sec:FF_energy}
 
 As discussed in the previous section, the motion of the secondary is modified by the presence of self-force, spin, and quadrupolar effects, as described in Eq.~(\ref{self_force1}) and Eq.~(\ref{self_torque1}). However, before proceeding to solve these equations, we first neglect the self-force contributions by setting $h_{\mu\nu}^R=0$  and focus solely on the effects of spin and quadrupolar deformation on the motion of the secondary. 
 
 To describe the resulting shifted trajectory, we adopt the \textit{fixed-frequency} formalism, in which the perturbed (circular) orbit with $\e>0$ has the same frequency $\Omega$ as the zeroth-order, Kerr-geodesic orbit at $\e=0$, while the particle's perturbed orbital radius is shifted to 
 \begin{multline}
 r_p=\rO+\e r_{\chi} \\+\e^2 \bigg(r_{\chi\chi} + C_Q r_{C_Q} + \mu_2 r_{\mu_{2}} + \sigma_2 r_{\sigma_2} \bigg) + {\cal O}(\e^3)\,.
 \end{multline}
  Here, $\rO$ is the orbital radius of a test mass:
 \beq\label{radius_shift}
 \rO = \left(\frac{\sqrt{M} (1 - a \Omega )}{\Omega }\right)^{2/3}.
 \eeq
  Throughout this work, we assume $\Omega>0$, meaning the orbital angular momentum is always pointing in the $+z$ direction. With this consideration, $a>0$ corresponds to prograde orbits, whereas $a<0$ corresponds to retrograde orbits. The corrections $r_{\chi}$ and $r_{\chi\chi}$ denote the contributions from linear-in-spin and quadratic-in-spin terms, respectively, while $r_Q$, $r_{\mu_{2}}$, and $r_{\sigma_2}$ correspond to the spin-induced quadrupole deformation, electric tidal quadrupole, and magnetic tidal quadrupole effects. Consistent with the discussion in the previous section, we formally include the tidal terms at order $\varepsilon^2$ even though they are actually of order $\varepsilon^4$.
  
  With these conventions, we write the particle's trajectory as 
 \beq\label{ff_trajectory}
x_p^\mu(t,\e)&=x_0^\mu(t)+\e x_1^\mu+\e^2 x_2^\mu+\CO(\e^3)\,,
 \eeq
 where $x_0^\mu(t)$ denotes the geodesic trajectory and $x_{1,2}^\mu$ represent the shift due to finite-size effects, which for a circular orbit is given by
\begin{subequations}
\begin{align}
x_0^i(t)&=\left(t, \rO,\pi/2, \Omega t\right)\,,\\
x_1^\mu&=r_{\chi}\delta^{\mu}_r\,,\\
x_2^\mu&= 
\left(r_{\chi\chi}+C_Q r_{C_Q}+\mu_2 r_{\mu_2} +\sigma_2 r_{\sigma_2} \right)\delta^{\mu}_r\,.
\end{align}
\end{subequations}
Here, the orbital frequency is defined through 
\beq\label{frequency}
\frac{d\phi_p}{dt} :=\Omega\,. 
\eeq

We emphasize that expanding in powers of $\e$ at fixed $\Omega$ rather than fixed orbital radius (for example) simply corresponds to a choice of coordinates on the orbital phase space. A spatial trajectory in the equatorial plane naturally has four (not necessarily canonically conjugate) phase-space coordinates, $(r_p,\phi_p,\hat p_r,\hat p_\phi)$. Specializing to circular orbits reduces the phase-space trajectory to two dimensions, with natural coordinates $(\phi_p,\tilde p_\phi)$ and a constraint $r_p=r_p(\tilde p_\phi)$. We adopt instead $(\phi_p,\Omega)$ as our coordinates and expand in powers of $\varepsilon$ at fixed values of these coordinates. A fixed-$r_p$ expansion would instead adopt coordinates $(\phi_p,r_p)$ and expand at fixed values of \emph{those} coordinates. Since the two expansions represent different choices of phase-space coordinates, we refer to them as choices of the phase-space gauge. This choice of phase-space gauge is distinct from but related to the choice of coordinates on spacetime:  $(r_p,\phi_p,\hat p_r,\hat p_\phi)$ all transform under a change in background coordinates and under a change in the perturbative gauge on spacetime, while $\Omega$ is an invariant feature of the dynamics (and very closely related to observable waveform frequencies~\cite{Warburton:2024xnr}). See discussions in Ref.~\cite{Lewis:2025ydo}, for example.

As alluded to above, restricting the orbit to the equatorial plane requires specializing to spin-aligned (or anti-aligned) configurations. Thus, the rescaled spin vector of the secondary is orthogonal to the equatorial plane,
 \beq\label{ff_spin_param}
S^{\alpha}=-\frac{\chi}{r_p}\delta_{\theta}^{\alpha}\,,
 \eeq
where the secondary's dimensionless spin is defined as
\begin{equation}\label{eq:chi def}
\chi:=\pm\sqrt{S^{\mu}S_{\mu}} = \pm\sqrt{S_{\mu\nu}S^{\mu\nu}/2}\,,   \end{equation}
with the + (-) sign corresponding to a spin vector pointing up (down), parallel (anti-parallel) to the orbital angular momentum.
In this section, the quantity $\chi$  agrees with the quantity  $\pm\tS/m_s^2 = \pm\sqrt{\hat g_{\mu\alpha}\hat g_{\nu\beta} S^{\alpha\beta}S^{\mu\nu}/2}$ since here we neglect $h^R_{\mu\nu}$ in $\hat g_{\mu\nu}$. However, the two quantities will differ in the next section. 

 Substituting Eqs.~\eqref{ff_trajectory}--\eqref{ff_spin_param} into Eq.~\eqref{self_force1}, we can obtain the expressions for $x^{\mu}_{1,2}$. 
 To present these expressions (along with those for orbital energy and angular momentum) in a compact form, we introduce the following set of helper functions: 
 \begingroup%
\allowdisplaybreaks%
\begin{subequations}\label{helper_functions}
     \begin{align}
         \mathbb{A}&=\sqrt{M r_0}- a\,,\\ \mathbb{B}&=r_0^{3/2}+a\sqrt{M}\,,\\
\mathbb{C}&= r_0 (r_0 -3M)+ 2a\sqrt{M r_0}\,,\\ \mathbb{F}&=\left(1-\frac{2M\sqrt{r_0}}{\mathbb{B}}\right)\,,\\
\mathbb{D}&=\left(r_0^2+a^2-2 a \sqrt{M r_0}\right)\,,\\
\mathbb{H}&=r_0^4+3 a^4-6 a^3 \sqrt{M r_0} +a^2 M r_0 +3 a^2 r_0^2\notag\\&\quad-2 a \sqrt{M} r_0^{5/2}+3 M^2 r_0^2-3 M r_0^3\,,\\
\mathbb{K} &=-6 a^5+9 a^4 \sqrt{M r_0} +a^3 r_0 (5 M-9 r_0)+\sqrt{M} r_0^{9/2}\notag\\&\quad+2 a^2 \sqrt{M} r_0^{3/2} (5 r_0-2 M)-3 a r_0^2 \left(2 M^2-M r_0 + r_0^2\right)\,.
     \end{align}
 \end{subequations}
\endgroup
Note that the orbital frequency and the helper function $\mathbb{B} $ satisfy the relation $\mathbb{B}=\sqrt{M}/\Omega$. 

With the aid of the helper functions, we can write $x^{\mu}_{1,2}$ in the following manner:
\begingroup%
\allowdisplaybreaks%
\begin{subequations}\label{ff_rad}
    \begin{align}
  r_{\chi} &= - m_{s}  \frac{\mathbb{A}}{\rO}  \chi\,,\label{rchi}\\
r_{\chi\chi} &=  3\, a\, m_{s}^2  \frac{\mathbb{A}}{\rO^3}  \chi ^2,\label{rchi2}\\
r_{C_Q} &=  m_{s}^2 \left( \frac{\Delta_0-4a \mathbb{A} }{2 \rO^3}\right) \chi^2,\label{rCq}\\
r_{\mu_{2}} &= \frac{3 M (v^t_0)^2}{\mathbb{B}^2 m_s r_0^5}\left(2 a\mathbb{A} \left(2a\mathbb{A}-r_0(r_0-M)\right)+\Delta_0^2\right)\,,\label{rmu}\\
r_{\sigma_2} &=\frac{8 M \mathbb{A}(v^t_0)^2}{\mathbb{B}^2 m_s r_0^5}\big(\mathbb{A}\left(-5 M r_0+3 a^2+2 r_0^2\right)-2a \Delta_0\big)\label{rsigma}\,,      
    \end{align}
\end{subequations}
\endgroup%
where $\Delta_0=\Delta(r_0)$.
Note that the expression for $r_{\chi}$, $r_{\chi\chi}$ and $r_{C_Q}$ matches with Eq.~(4.27) in \cite{Bini:2015zya}.
The expressions for $r_{\mu_2}$ and $r_{\sigma_2}$ involve the four-velocity $v^{\mu} = (v^{t}, 0, 0, v^{\phi})$, which satisfies the circular orbit condition $v^{\phi} = \Omega v^{t}$ together with the normalization condition $v^{\mu} v_{\mu} = -1$. 
Using this, we find that the four-velocity admits the following expansion: 
\beq\label{v0_fff}
v^{t}=v^{t}_{0}+\e^2 v^{t}_{2}+\mathcal{O}(\e^3)\,,
\eeq
 where 
\beq \label{v0_fff1}
v^{t}_{0} =\frac{\mathbb{B}}{\sqrt{\rO \mathbb{C}}}\,,\qquad
v^{t}_{2} =\frac{3 M r_{\chi}^2}{2 \mathbb{B}^2}(v^{t}_{0})^3\,.
\eeq

Finally, we compute the conserved energy and angular momentum within the fixed-frequency parametrization using Eq.~\eqref{Conserved}. These quantities can be expanded in a manner analogous to Eq.~(\ref{radius_shift}) as
\begin{multline}\label{eq:E test body}
E = E_0 + \e E_\chi \\+ \e^2\left(E_{\chi\chi} + C_Q E_{C_Q} + \mu_2 E_{\mu_2} + \sigma_2 E_{\sigma_2}\right) +{\cal O}(\e^3),
\end{multline}  
and 
\begin{multline}
L_z = L_0 +\e  L_\chi \\+ \e^2 \left(L_{\chi\chi} + C_Q L_{C_Q} +\mu_2 L_{\mu_2} + \sigma_2 L_{\sigma_2}\right) +{\cal O}(\e^3) .
\end{multline}

The explicit forms of these quantities can be expressed in terms of the helper functions introduced above. The geodesic contributions to the orbital parameters are given by
\beq\label{energy_geo}
E_0=\mathbb{F}v^{t}_{0}\,,\quad
      L_0=\frac{\mathbb{D}}{\mathbb{B}} \sqrt{M}v^{t}_{0}.
\eeq
The leading-order corrections to the orbital energy and angular momentum due to the spin of the secondary can be written as
\begin{subequations}\label{energy_chi}
    \begin{align}
      E_\chi &=\frac{M r_{\chi}  v^{t}_{0}}{\mathbb{B}\sqrt{r_0}}\,,\\
L_\chi&=2r_{\chi}v^{t}_{0}\sqrt{\frac{M}{\rO}}+m_s\chi v^{t}_{0} \left[1-\frac{M}{\rO}-\frac{\mathbb{D} M}{\mathbb{B}r_0^{3/2}}\right],
    \end{align}
\end{subequations}
whereas the quadratic-in-spin contributions are
\begingroup%
\allowdisplaybreaks%
\begin{subequations}\label{energy_chi2}
    \begin{multline}
E_{\chi\chi}=\mathbb{F}v^{t}_{2}+\frac{v^t_0 r_{\chi}^2}{\mathbb{A}\mathbb{B}} \bigg(\frac{5 a M}{\rO^{3/2}}-\frac{M^{3/2}}{\rO}\\*-\frac{3 \Delta_0 M^{3/2} (v^t_0)^2}{\mathbb{B}^2}\bigg), 
    \end{multline}
    \vspace{-5mm}
    \begin{multline}
L_{\chi\chi}=\frac{\mathbb{D}}{\mathbb{B}} \sqrt{M}v^{t}_{2}+\frac{v^t_0 r_{\chi}^2}{\mathbb{A}} \bigg(\frac{5 a \sqrt{M}}{\rO^{3/2}}-\frac{M}{\rO}\\*-\frac{3 \Delta_0 M (v^t_0)^2}{\mathbb{B}^2}\bigg). 
    \end{multline}
\end{subequations}
\endgroup
The contribution to orbital energy and angular momentum from the spin-induced quadrupolar deformation can be written as follows:
\begingroup%
\allowdisplaybreaks%
\begin{subequations}\label{energy_Cq}
    \begin{align}  
    E_{C_Q}&=\frac{2r_Q M v^{t}_{0}}{\mathbb{B}\sqrt{\rO}}- r_{\chi}^2\frac{\mathbb{F} (v^t_0)^3
M (\mathbb{D}- 2a  \mathbb{A})}{2 \mathbb{A}^2\mathbb{B}^2}\notag\\*&\quad+ r_{\chi}^2\frac{3  \Delta_0 M^{3/2} (v^t_0)^3}{\mathbb{A}\mathbb{B}^3}\,,\\
     L_{C_Q}&=2r_Q v^{t}_{0}\sqrt{\frac{M}{\rO}}-\frac{\mathbb{D} r_{\chi}^2 (v^t_0)^3 M^{3/2} (\mathbb{D}-2 \mathbb{A} a)}{2 \mathbb{A}^2 \mathbb{B}^3}\notag\\*&\quad +\frac{3 \Delta_0 r_{\chi}^2 (v^t_0)^3 M}{\mathbb{A}\mathbb{B}^2}\,.
    \end{align}
\end{subequations}
\endgroup
We have compared our expressions for the orbital energy and angular momentum, including the geodesic, linear-in-spin, quadratic-in-spin, and spin-induced quadrupolar contributions, with those presented in Ref.~\cite{Bini:2015zya}, specifically Eq.~(4.28) therein. We find complete agreement between the two results.

The terms $(E_{\mu_2} ,L_{\mu_2})$ and $(E_{\sigma_2}, L_{\sigma_2})$ denote the corrections to the orbital energy and angular momentum arising from gravito-electric and gravito-magnetic tidal deformations of the secondary, respectively. The gravito-electric tidal contributions are given by 
\begingroup%
\allowdisplaybreaks%
\begin{subequations}\label{energy_mu}
    \begin{align}  
    E_{\mu_2}&=\frac{2r_{\mu_{2}} M v^{t}_{0}}{\mathbb{B}\sqrt{\rO}}-\frac{3 M^2 (v^t_0)^3}{\mathbb{B}^3 m_s \rO^{11/2}}\bigg[2\mathbb{H}-\mathbb{A} \sqrt{M\rO} \notag\\*&\quad\times\left(3 a^2-4 a \sqrt{M\rO} +\rO^2\right)\bigg]
+\frac{9 \mathbb{F} M^2 (v^t_0)^5}{2 \mathbb{B}^4 m_s \rO^4}\notag\\*&\quad\times\bigg[\left(\mathbb{A}^2 (\rO (\rO-M)-2 a \mathbb{A})+\Delta_0^2\right)\bigg]\,\,,\\
     L_{\mu_2}&= 2r_{\mu_{2}} v^{t}_{0}\sqrt{\frac{M}{\rO}}+\frac{3 M^{2} \mathbb{K} (v^t_0)^3}{\mathbb{B}^3 m_s \rO^{11/2}}\notag\\*&\quad+\frac{9 \mathbb{D} \mathbb{H} M^{5/2}(v^t_0)^5}{2 \mathbb{B}^5 m_s \rO^4}\,,
    \end{align}
\end{subequations}
\endgroup%
whereas gravito-magnetic tidal contributions are given by 
\begingroup%
\allowdisplaybreaks%
\begin{subequations}\label{energy_sigma}
    \begin{align}  
    E_{\sigma_2} &=\frac{2r_{\sigma_2} M v^{t}_{0}}{\mathbb{B}\sqrt{\rO}}+\frac{24 \mathbb{A}M^2 (v^t_0)^3 \Delta_0\left(2a-\sqrt{M r_0}\right)}{\mathbb{B}^3 m_s \rO^{11/2}}\notag\\*&\quad +\frac{36 \mathbb{A}^2 \Delta_0 \mathbb{F} M^2 (v^t_0)^5}{\mathbb{B}^4 m_s \rO^4}\,,\\
     L_{\sigma_2}&= 2r_{\mu_{2}} v^{t}_{0}\sqrt{\frac{M}{\rO}}+\frac{36 \mathbb{A}^2 \Delta_0 \mathbb{D} M^{5/2} (v^t_0)^5}{\mathbb{B}^5 m_s \rO^4}\notag\\*
&\quad+\frac{24 \mathbb{A}M^{2} \Delta_0 (v^t_0)^3}{\mathbb{B}^3 m_s \rO^{11/2}}\bigg[2a^2-a \sqrt{M \rO}+\rO^2\bigg]\,.
    \end{align}
\end{subequations}
\endgroup%
These equations for the tidal terms appear here for the first time. 
 \subsection{Multiscale expansion}
In this section, we incorporate the contributions arising from self-force effects, following the approach in Ref.~\cite{Mathews:2021rod}. The dissipative component of the self-force drives the secondary into a slow inspiral motion. As a result, the system is characterized by two distinct timescales: the fast orbital timescale of the secondary, $T_o= \order(M)$, over which the orbital phase $\phi_p$ evolves, and the slow inspiral timescale, $T_i= \order(M/\e)$, over which the set of state parameters $\bJ^a=(\sP)$ evolves. Here, $m_s\delta M$ and $m_s\delta J$ represents the correction to the mass and angular momentum of the primary due to absorption of energy and angular momentum through its horizon. Note that the parameter $\chi$ remains constant up to $\order({\e^3})$ residuals for circular orbits, as we show later in this section. Moreover, we have preemptively omitted $m_s$, $C_Q$, $\mu_2$, and $\sigma_2$ from the state vector because they are constant to even higher order.


\subsubsection{Expansion of the trajectory}

To develop the split between fast and slow variables, we adopt a $3+1$ decomposition into space and time, and we assume that the particle's spatial trajectory and the spacetime metric only depend on time $t$ through the time dependence of the set of mechanical parameters $(\phi_p, \bJ^a)$. We then write the particle's worldline as  $x_p^{\mu}(t,\e)=(t,x_p^i(\phi_p(t,\e), \bJ^a(t,\e),\e))$, where its spatial trajectory $x_p^i(\phi_p, \bJ^a,\e)$ can be expanded as
 \beq\label{trajectory}
x_p^i(\phi_p, \bJ^a)&=x_0^i(\phi_p, \Omega)+\e x_1^i(\bJ^a)+\e^2 x_2^i(\bJ^a)+\CO(\e^3)
 \eeq
in analogy with Eq.~\eqref{ff_trajectory} and suprressing the functional dependence on the constants $\{M,a,m_s,C_Q,\mu_2,\sigma_2\}$. Unlike in Eq.~\eqref{ff_trajectory}, here even the leading term, $x_0^i$, ultimately depends on $\e$ because $\phi_p=\phi_p(t,\e)$ and $\bJ^a=\bJ^a(t,\e)$. Moreover, the 
subleading terms, $x_{1,2}^{\mu}$, include corrections due to self-force as well as spin and quadrupolar effects. Here again, we assume that the secondary is inspiralling around the primary in an equatorial, quasi-circular orbit. Thus, the explicit expression of $x_{0,1,2}^i$ can be written as 
\begin{subequations}\label{tts_trajectory}
\begin{align}
x_0^i(\phi_p, \Omega)&=\left(r_0(\Omega),\pi/2,\phi_p\right),\\
x_{1,2}^i(\phi_p, \bJ^a)&=r_{1,2}(\bJ^a)\,\delta^i_r,
\end{align}
\end{subequations}
 where $\rO(\Omega)$ is as given in  Eq.~(\ref{radius_shift}). 
 The frequency remains defined as $\frac{d\phi_p}{dt}:= \Omega$, as for a test body.
 Furthermore, we are interested in the scenario where the spin vector of the secondary remains orthogonal to the equatorial plane, meaning
 \beq\label{spin_param}
S^{\alpha}(\bJ^a,\varepsilon)=-\frac{\chi}{r_p}\delta^\alpha_\theta,
 \eeq
with $\chi$ defined by Eq.~\eqref{eq:chi def}. On the face of it, this equation is identical to Eq.~\eqref{ff_spin_param}, but the interpretation is different. First, $r_p$ itself differs (as a function of ${\cal J}^a$ and $\varepsilon$) due to the self-force. Second, $\chi$ is no longer equal to $\hat S/m_s^2 = \sqrt{\hat g_{\mu\alpha}\hat g_{\nu\beta} S^{\alpha\beta}S^{\mu\nu}/2}$, as the two differ by $h^R_{\mu\nu}$ terms:
\begin{align}
    \tS = m_s^2\chi\sqrt{1 + \frac{h^R_{\beta\gamma}S_{\alpha}{}^\beta S^{\alpha\gamma}}{\chi^2} + \frac{h^R_{\mu\alpha}h^R_{\beta\nu}S^{\alpha\beta}S^{\mu\nu}}{2\chi^2}}.
\end{align}

The slow evolution of the state parameters can likewise be expanded in powers of $\e$ at fixed values of $(\phi_p,\bJ^a)$: 
\begin{subequations}\label{state_evo}
\begin{align}
    \dO &= \e \F_{\Omega}^{0}(\Omega) + \e^2 \F_{\Omega}^{1}(\bJ^a) +\e^3 \F_\Omega^{2}(\bJ^a)+ {\cal O}(\e^4),\\        
    \dchi &= \e \F_\chi^{1}(\bJ^a) +\e^2 \F_\chi^{2}(\bJ^a) + {\cal O}(\e^4),\\
    \delta\dot{\mathsf{M}} &= \e \F_{\delta \mathsf{M}}^{1}(\Omega) + \e^2 \F_{\delta \mathsf{M}}^{2}(\bJ^a) + {\cal O}(\e^3),
\end{align}
\end{subequations}
where $\delta\mathsf{M}\in (\delta M\,,\delta J)$. Here, the overdot represents a total derivative with respect to $t$. The forcing functions ${\sf F}^{n}_\Omega$ are the same, up to normalization, as those appearing in Eq.~\eqref{dOmega/dt}. There, to make the dependence on all parameters (including constant ones) manifest, we defined dimensionless functions $\hat {\sf F}^{n}_\Omega$ of dimensionless variables and did away with the bookkeeping parameter $\e$. We also implicitly re-expanded functions of the background mass and spin in terms of the primary's physical mass and spin $M_{\rm BH}=M+m_s \delta M$ and $J_{\rm BH} = a M+m_s \delta J$, which cancels the dependence on $\delta {\sf M}$~\cite{Warburton:2024xnr}.

To obtain the expressions for the forcing functions~$\mathsf{F}^{n}$ along with the expressions for $r_{1,2}$ in Eq.~(\ref{tts_trajectory}), we employ the orbital and spin equations given in Eq.~(\ref{self_force1}) and Eq.~(\ref{self_torque1}), respectively. 
It is more convenient to rewrite these equations in terms of $t$, as 
\begin{subequations}
\begin{align}
 \ddot{x}_p^\mu &+\frac{\dot{v}^t}{v^t}\dot{x}_p^{\mu}+\Gamma^{\mu}_{\nu\alpha}\dot{x}_p^{\nu}\dot{x}_p^{\alpha}=\frac{1}{(v^t)^2}F^{\mu}~\,,\label{force_eqn_na} 
 \\\dot{S}^{\mu} &+\Gamma^{\mu}_{\nu\alpha}\dot{x}_p^{\nu}S^{\alpha}=\frac{1}{v^t}N^{\mu}\,,\label{Torque_eqn_na}
\end{align} 
\end{subequations}
where $v^t=dt/d\tau$ and $\dot{x}_p^\mu=d{x}_p^\mu/dt$ is the coordinate velocity. Using Eq.~(\ref{trajectory}) and Eq.~(\ref{partial_t}), we can write the coordinate velocity as 
\beq\label{coord_vel}
 \dot{x}_p^{\mu}&=\dot x_0^{\mu}+\e \dot x_1^{\mu}+\e^2 \dot x_2^{\mu}+\CO(\e^3),
 \eeq	
where $\dot x_0^{\mu}= (1,0,0,\Omega)$ and    
\begin{subequations}
\begin{align}\label{coord_vel1}
\dot x_1^{\mu}&= \left(r_{0}'(\Omega)\F_{\Omega}^{0}(\Omega)\right)\delta^{\mu}_r\,,\\
\dot x_2^{\mu}&= \left(r_{0}'(\Omega)\F_{\Omega}^{1}(\Omega)+\dbJ_1^{a} \partial_{\bJ^{a}}r_1\right)\delta^{\mu}_r\,.
\end{align}
\end{subequations}
The four-velocity is then given by $v^{\mu }=v^t \dot{x}_p^{\mu }$. Using the expression for the coordinate velocity, along with Eq.~(\ref{trajectory}) and the normalization condition $v^{\mu } v_{\mu }=(v^t)^2 \dot{x}^{\mu } \dot{x}_{\mu }=-1$, we find the $t$-component of the four-velocity can be expanded as $v^t=v^t_0+\e^2 v^t_2+\CO(\e^3)$, where 
\beq\label{v00_v02_tts}
v^{t}_{0} =\frac{\mathbb{B}}{\sqrt{\rO \mathbb{C}}}\,,\qquad
v^{t}_{2} =\frac{3 r_{1}^2M}{2 \mathbb{B}^2}(v^{t}_{0})^3\,,
\eeq
in analogy with Eq.~\eqref{v0_fff}; no dissipative terms appear at these orders. 

We can now obtain the expressions for the radial shifts $r_{1,2}$ appearing in Eq.~\eqref{tts_trajectory}. To do so, we substitute Eqs.~\eqref{trajectory} and \eqref{v00_v02_tts} into the conservative sector of the self-force equation (i.e., the radial component of Eq.~\eqref{force_eqn_na}) and solve the resulting equation order by order in powers of $\e$, while keeping the state parameters $\bJ^{a}$ fixed. This leads to
\begin{subequations}
\begin{align}\label{r0_r1}
  r_1 =&-F^r_1\,\mathbb{T}^2\,,
  \\\non
  r_2=&-F^r_2\,\mathbb{T}^2+\left(\frac{3}{r_0}-\frac{\Delta'_0}{\Delta_0}\right)r_1^2+\dot{r}_0\left(\frac{\dot{\F}^{0}_{\Omega}}{\F^{0}_{\Omega}}+\frac{\dot{v}^t_0}{{v}^t_0}\right)(v^t_0 \mathbb{T})^2\\ &+\left[\frac{\dot{r}_0^2}{\Delta_0} \left(M -\frac{a^2}{r_0}\right)-\left(\F^{0}_{\Omega}\right)^2 \frac{d^2 r_0}{d\Omega^2}\right](v^t_0 \mathbb{T})^2,
 \end{align}
 \end{subequations}
 where $\Delta'_{0}=2(r_0-M)$, and $\dot{f}=\F^{0}_{\Omega}df/d\Omega$ for any function $f(\Omega)$. Here we also introduced another helper function, $\mathbb{T}=1/\sqrt{3 (v^t_0 \Omega)^2 \Delta_0/r_0^2}$. If we neglect dissipation and self-forces, then these expressions for $r_{1,2}$ reduce to Eq.~\eqref{ff_rad} from the previous section.

Next, we determine the evolution of the state parameters on the inspiral timescale, as described by Eq.~\eqref{state_evo}. We start by utilizing the evolution equation~\eqref{Torque_eqn_na} for the spin. Specifically, we substitute Eqs.~\eqref{trajectory} and \eqref{v00_v02_tts} into the $\theta$ component of Eq.~\eqref{Torque_eqn_na} and solve the resulting equation order by order in powers of $\e$ at fixed $\bJ^{a}$. This leads to
 \beq\label{chidot}
 \F_\chi^{1} =0\,,\quad  \F_\chi^{2} =-\frac{N^{\theta} r_0}{v^t_0}\,,\quad N^{\mu}\propto v^{r}\delta_\theta^\mu=0.
\eeq
Here, we use the fact that for circular orbits $v^r=0$.
 Thus, for circular orbits,  Eq.~\eqref{Torque_eqn_na} dictates that the spin parameter of the secondary remains constant up to $\mathcal{O}(\e^3)$ residuals. This simplifies the calculation significantly.
 
Similarly, we can obtain the expression for each $\F^{n}_\Omega$ by substituting Eqs.~\eqref{trajectory} and~\eqref{v00_v02_tts} in the $t$ component of the orbital evolution equation~\eqref{force_eqn_na}, and solving the equation order by order in $\e$ at fixed $\bJ^a$. Particularly, at orders $\e$ and $\e^2$, we obtain
\begin{subequations}
\begin{align}\label{Omegadot0}
\F_\Omega^{0} =&\frac{F^t_1}{\mathbb{W}},\\\non
\F_{\Omega}^{1} = &\frac{F^t_2}{\mathbb{W}}-\frac{2 M \mathbb{D}(v^t_0)^2 }{\mathbb{W}\mathbb{B} \Delta_0 \sqrt{r_0}}\dot{r}_1+\frac{4M(v^t_0)^2 }{\mathbb{W}\mathbb{B} \Delta_0^2 r_0^{3/2}}r_1\dot{r}_0\times\\\non&\bigg[a^2 r_0 (2 r_0-3 M)+2 a \left((M r_0)^{3/2}-\sqrt{M} r_0^{5/2}\right)\\&+r_0^3 (r_0-M)+a^4\bigg],
\end{align}
\end{subequations}
    where we introduce another helper function $\mathbb{W}$, which can be expressed as
\beq
\mathbb{W}=v^t_0\left(\dfrac{dv^t_0}{d\Omega}+\frac{2 M \mathbb{D}v^t_0 }{\mathbb{B} \Delta_0 \sqrt{r_0}}\dfrac{dr_0}{d\Omega}\right).
\eeq

As reviewed in the Introduction, up to 1PA order (inclusive), only the contributions from $\F^{0}_{\Omega}$ and $\F^{1}_{\Omega}$ are required, while the 2PA corrections enter through $\F^{2}_{\Omega}$. These 2PA terms include the spin-squared and quadrupole terms. An expression for $\F^{2}_{\Omega}$ can be obtained in a similar way as for $\F^{0}_{\Omega}$ and $\F^{1}_{\Omega}$ by substituting Eqs.~\eqref{trajectory} and \eqref{v00_v02_tts} into the $t$ component of the force equation~\eqref{force_eqn_na} and solving the resulting equation at order $\e^3$, while keeping the $\bJ^a$s fixed. However, the resulting expression is rather lengthy. More importantly, in the end we will not require such expressions for $\F^{n}_{\Omega}$ in terms of local forces. Instead, in Sec.~\ref{Sec_4_BalanceLaws} below we will derive more useful expressions for finite-size contributions to the orbital evolution (i.e., to $\F^{2}_{\Omega}$) from a flux-balance law. 

\subsubsection{Expansion of the metric, computation of forcing functions, and waveform generation}

Calculating $\dot\Omega$, whether from balance laws or self-forces, requires solving the Einstein equations (or some subset of them such as the Teukolsky equation). Generating the waveform~\eqref{eq:waveform} also requires the mode amplitudes $h^{(n)}_{\ell m}$, which are likewise obtained from the solution to the Einstein equations.

Similar to Eq.~(\ref{trajectory}), we can expand the spacetime metric and stress-energy tensor as
\begin{align}
\mathbf{g}_{\mu\nu}&= g_{\mu\nu}(x^i)+\varepsilon h_{\mu\nu}^1(x^i,\phi_p,{\cal J}^a)+\varepsilon^2 h_{\mu\nu}^2\left(x^i,\phi_p, \bJ^a\right)\non\\
&\quad+\varepsilon^3 h_{\mu\nu}^3\left(x^i,\phi_p, \bJ^a\right)+\mathcal{O}(\e^4),\label{eq:g multiscale expansion}\\
\hat T^{\mu\nu}&= \varepsilon T^{\mu\nu}_1(x^i,\phi_p,\Omega)+\varepsilon^2 T^{\mu\nu}_2\left(x^i,\phi_p, \bJ^a\right)\non\\
&\quad +\varepsilon^3 T^{\mu\nu}_3\left(x^i,\phi_p, \bJ^a\right)+ \mathcal{O}(\e^4),\label{eq:T multiscale expansion}
\end{align}
where first-order terms are independent of the secondary's spin, second-order terms are at most linear in it, and third-order terms are at most quadratic in it (and at most linear in the secondary's quadrupole moments). 

Given that the linear perturbation $h_{\mu\nu}^1$ is linear in $\delta M$ and $\delta J$, it is more conveniently written as 
\begin{multline}
h_{\mu\nu}^1(x^i,\phi_p,\Omega,\delta M,\delta J)=h_{\mu\nu}^{1,\rm pp}(x^i,\phi_p,\Omega)+\delta M h_{\mu\nu}^{\delta M}(x^i)\\+ \delta J h_{\mu\nu}^{\delta J}(x^i),
\end{multline}
where the first term,  $h_{\mu\nu}^{1,\rm pp}(x^i,\phi_p,\Omega)$, represents the linear perturbation to the background Kerr spacetime due to a point mass on a circular orbit; this would be equal to the standard first-order ``test-particle'' perturbation if we replaced $\phi_p$ with~$\Omega t$. 
The terms $\delta M h_{\mu\nu}^{\delta M}$ and $\delta J h_{\mu\nu}^{\delta J}$ then represent the linear perturbation toward a Kerr black hole with mass $M+ m_s \delta M$ and angular momentum $a M+ m_s \delta J$. 

Since we assume that the particle's trajectory and the spacetime metric do not have explicit time dependence, we write partial $t$ derivatives in the Einstein equation as
 \beq\label{partial_t}
\partial_t=\Omega \partial_{\phi_p}+\dbJ^{a} \partial_{\bJ^{a}},
 \eeq
 where $\dbJ^{a} =(\dsP)$ represents the evolution of the state parameters, and $\partial_{\bJ^a}=(\partialsP)$. Substituting the chain rule~\eqref{partial_t} and the expansions~\eqref{state_evo} into the Einstein field equations then leads to a sequence of equations for the metric perturbations $h^n_{\mu\nu}(x^i,\phi_p,\bJ^a)$, as described in Refs.~\cite{Miller:2020bft,Pound:2021qin,Mathews:2021rod,Miller:2023ers} (for example).

The form of the resulting field equations will not be our focus here. Instead, we highlight the main idea of how they are used to build the inputs for a waveform model: (i) the field $h^1_{\mu\nu}$ is solved for without knowledge of the forcing functions in Eqs.~\eqref{state_evo}, since the forcing functions only enter the field equations at higher orders; from $h^1_{\mu\nu}$, one computes the forcing functions $\F^{0}_\Omega$ and $\F^{0}_{\delta{\sf M}}$; these are used to construct the source terms in the field equation for $h^2_{\mu\nu}$; from the solution for $h^2_{\mu\nu}$, one computes the next-order forcing functions and waveform amplitudes; and so on.

The outputs of this procedure are hence precisely the inputs for the waveform model described by Eqs.~\eqref{eq:waveform}--\eqref{dOmega/dt}. In Sec.~\ref{Sec_5_FieldEquations} we describe the subset of the field equations that we solve to obtain (most of) the finite-size contributions to that model.

Before discussing the flux-balance law from which we extract ${\sf F}^{2}_\Omega$, we note that the forcing functions for the primary's mass and spin evolution,  $\F^{1}_{\delta \sf M}$, cannot be determined directly from the orbital and spin equations~\eqref{force_eqn_na} and~\eqref{Torque_eqn_na}. Rather, they also come from balance laws. Specifically, the second-order Einstein field equations imply that $\F^{1}_{\delta \sf M}$ is given by the standard energy or angular momentum flux carried by $h^{pp}_{\alpha\beta}$ down through the black hole horizon~\cite{Miller:2020bft}.

    \section{Balance laws}\label{Sec_4_BalanceLaws}

Our formulas for the multiscale evolution in the preceding section were written in terms of the local self-force. However, it is generally far more efficient to utilize balance laws that relate the evolution of the system's mechanical energy to the emitted fluxes of gravitational radiation. In this section we derive appropriate balance laws and apply them to determine the finite-size contributions to the forcing functions ${\sf F}^{(n)}_\Omega$ appearing in $d\Omega/dt$.

In the first two subsections, we leave the orbit generic, not restricting to quasicircular orbits until Sec.~\ref{sec:point particle forcing functions}.

\subsection{Conserved charges}

To obtain the balance laws, we will work with conserved currents and charges. 
We first define the currents
\begin{equation}
j^\alpha_\xi := \hat T^\alpha_{\ \beta} \xi^\beta, 
\end{equation}
using with the effective metric $\hat g_{\alpha\beta}$, with $\hat T^\alpha_{\ \beta} := \hat g_{\beta\gamma}\hat T^{\alpha\gamma}$. Here $\xi^\alpha$ can be either of the two background Killing vectors. The MPD equations in the effective spacetime imply $\hat\nabla_\beta T^{\alpha\beta}=0$ and hence
\begin{equation}
    \hat\nabla_\alpha j^\alpha_\xi =\hat T^{\alpha\beta}\hat\nabla_\alpha \hat \xi_\beta = \frac{1}{2}\hat T^{\alpha\beta}{\cal L}_\xi \hat g_{\alpha\beta},
\end{equation}
where $\hat\xi_\beta:=\hat g_{\alpha\beta}\xi^\alpha$. Since $\xi^\alpha$ is a Killing vector of the background, this reduces to
\begin{equation}\label{eq:quasi-conservation eqn}
    \hat\nabla_\alpha j^\alpha_\xi = \frac{1}{2}\hat T^{\alpha\beta}{\cal L}_\xi h^R_{\alpha\beta}.
\end{equation}
In plain language, this equation shows that the current associated with $\xi^\alpha$ is not locally conserved unless the metric perturbation possesses the same symmetry as the background spacetime. The nonzero flux emissions and global balance laws we obtain will hence be directly linked to $\xi^\alpha$ being only an \emph{approximate} Killing vector of the full spacetime.

Next we define a charge $Q_\xi(\Sigma)$ on any given spacelike surface $\Sigma$:
\begin{equation}\label{eq:Q def}
    Q_\xi(\Sigma) := \int_{\Sigma} j_\xi^\alpha d\hat\Sigma_\alpha.
\end{equation}
We specifically define $\Sigma_t$ as surfaces of constant $t$ (which need not be Boyer-Lindquist time), choosing each such surface such that the particle's worldline intersects its interior. Integrating Eq.~\eqref{eq:quasi-conservation eqn} over a spacetime volume $V_t$ bounded by spacelike surfaces $\Sigma_{t+T}$ and $\Sigma_{t-T}$ and by a (possibly disconnected) timelike surface $\Gamma_t$, appealing to Stokes' theorem, and noting that $j^\alpha_\xi$ vanishes on $\Gamma_t$, we then find that the net change in the charge between the two slices is
\begin{equation}
    Q_\xi(\Sigma_{t+T}) - Q_\xi(\Sigma_{t-T}) = \frac{1}{2}\int_{V_t}\hat T^{\alpha\beta}{\cal L}_\xi h^R_{\alpha\beta}d\hat V.
\end{equation}
Note that to apply Stokes' theorem we have used the volume element associated with $\hat g_{\alpha\beta}$, and we take $d\hat\Sigma^\alpha$ to be future-directed. The charge's instantaneous rate of change is hence
\begin{subequations}
\begin{align}
    \frac{dQ_\xi}{dt} &= \lim_{T\to0}\frac{Q_\xi(\Sigma_{t+T}) - Q_\xi(\Sigma_{t-T})}{2T} \\
    &= \frac{1}{2}\int_{\Sigma_t}\hat T^{\alpha\beta}{\cal L}_\xi h^R_{\alpha\beta}d\hat \Sigma,
\end{align}
\end{subequations}
and its average rate of change over a large time interval is\footnote{There is considerable subtlety in utilizing such infinite time averages in a multiscale expansion, as the system ultimately evolves by a large amount away from the state at time $t$. Averages should more generally be defined as averages over the system's phase variables~\cite{Hinderer:2008dm,Lewis:2025ydo} (meaning, e.g., an average over $0\leq\phi_p<2\pi$ for quasicircular orbits, holding all other variables fixed). However, for the test-body effects we assess here, the infinite time average directly provides the relevant torus average. This is discussed in detail in Sec.~VIA2 of Ref.~\cite{Mathews:2025nyb}.}
\begin{subequations}\label{eq:<dQdt>}
\begin{align}
    \left\langle\frac{dQ_\xi}{dt}\right\rangle &= \lim_{T\to\infty}\frac{Q_\xi(\Sigma_{t+T}) - Q_\xi(\Sigma_{t-T})}{2T} \\
    &= \lim_{T\to\infty}\frac{1}{4T}\int_{V_t}\hat T^{\alpha\beta}{\cal L}_\xi h^R_{\alpha\beta}d\hat V.
\end{align}
\end{subequations}
This will be the starting point for our balance law. 

An advantage of the form~\eqref{eq:<dQdt>} is that it is manifestly invariant under coordinate transformations of interest. The Leibniz integral formula tells us that, under a small coordinate shift $\delta x^\mu$, the integral in Eq.~\eqref{eq:<dQdt>} changes by an amount
\begin{equation}
        \delta\left\langle\frac{dQ_\xi}{dt}\right\rangle = \lim_{T\to\infty}\frac{1}{4T}\int_{\partial V_t}\hat T^{\alpha\beta}{\cal L}_\xi h^R_{\alpha\beta}\delta x^\mu d\hat\Sigma_\mu.
\end{equation}
Due to the division by $T$, this vanishes so long as the coordinate shift $\delta x^\alpha$ grows less rapidly than $t$, given that $\hat T^{\alpha\beta}{\cal L}_\xi h^R_{\alpha\beta}$ has bounded variation in time. The restriction to transformations that grow less rapidly than $t$ is natural in the multiscale expansion, analogous to the usual restriction to transformations that preserve frequencies~\cite{Barack:2018yvs,Shah:2015nva}.


In Appendix~\ref{app:charges} we relate the conserved charges $Q_\xi$ to the quantities 
\begin{equation}\label{eq:P def}
P_\xi:= -\left(\hp_\alpha\xi^\alpha + \frac{1}{2}\hat S^{\alpha\beta}\hat\nabla_\alpha\hat\xi_\beta\right),
\end{equation}
which represent the particle's energy and angular momentum~\eqref{Conserved} with $P_{\xi_t} = E$ and $P_{\xi_\phi} = -L_z$. The punchline of that analysis is
\begin{equation}
    Q_\xi = P_\xi + {\cal O}(\varepsilon^4).
\end{equation}
Hence, our balance law for $Q_\xi$ will also be a balance law for~$P_\xi$.

\subsection{Test-body balance laws}

To establish a balance law, we must express the right-hand side of Eq.~\eqref{eq:<dQdt>} in terms of gravitational-wave fluxes. This is challenging because of the complexity of $h^R_{\alpha\beta}$ at nonlinear orders in perturbation theory. It is obtained from the solution to the perturbative Einstein equation expanded to cubic order in $\e$: 
\begin{align}
    \delta G^{\alpha\beta}[h] &= 8\pi \hat T^{\alpha\beta} - \delta^2 G^{\alpha\beta}[h,h] \nonumber\\
    &\quad - \delta^3 G^{\alpha\beta}[h,h,h] -\ldots,
\end{align}
where $\delta G^{\alpha\beta}$ is the linearized Einstein tensor, $\delta^2 G^{\alpha\beta}$ is quadratic in the metric perturbation, $\delta^3 G^{\alpha\beta}$ is cubic, and so on. $h^R_{\alpha\beta}$ is a specific, smooth piece of the solution to this equation. The complicated nonlinearities on the right-hand side (and the strong singularities they introduce) make application of traditional methods~\cite{Galtsov:1982hwm,Sago:2005fn,Isoyama:2018sib,Akcay:2019bvk,Grant:2024ivt} difficult, and this difficulty is further exacerbated by the additional source terms that arise in the multiscale expansion through the application of Eq.~\eqref{partial_t}.  To date, no complete balance laws have appeared even at second order in $\e$. 

The challenge is eliminated if we limit our attention to \emph{linear} effects. Specifically, we neglect all source terms arising from interactions between the secondary and the regular field, defining a linear ``test body'' field satisfying\footnote{Here we follow common practice by referring to this as a ``test body'' field, even though a test body, by definition, does not generate a field.} 
\begin{equation}
    \delta G^{\alpha\beta}[h^{\rm test}] = 8\pi T^{\alpha\beta}_{\rm test},
\end{equation}
with the skeletonized test-body stress-energy tensor
\begin{align}\label{eq:Tpp def}
	T_{\rm test}^{\alpha \beta } &= \int d\tau\biggl\{p^{(\alpha }v^{\beta)}\delta^4(x,x_p)-\nabla_\gamma\Bigl[\tS^{\gamma (\alpha }v^{\beta)}\delta^4(x,x_p)\Bigr]\non\\
	&\quad -\frac{1}{3}\tJ^{\gamma \delta \epsilon (\alpha }R^{\beta )}{}_{\epsilon \gamma \delta }\delta^4(x,x_p)\non\\
    &\quad -\frac{2}{3}\nabla_{\gamma }\nabla_{\delta }\Bigl[\tJ^{\delta (\alpha \beta )\gamma }\delta^4(x,x_p)\Bigr]\biggr\}.
\end{align}
Here $\tau$, $\nabla$, $R_{\alpha\beta\gamma\delta}$, and $\delta^4(x,x_p):=\frac{\delta^4(x^\mu-x^\mu_p(\tau))}{\sqrt{-g}}$ are all defined with respect to the background metric. 

We write the regular field associated with the skeletonized stress-energy tensor as
\begin{equation}\label{eq:pp R field}
    h^{\rm R,test}_{\alpha\beta} := \int G^{\rm R}_{\alpha\beta\alpha'\beta'} T^{\alpha'\beta'}_{\rm test}dV',
\end{equation}
where $G^{\rm R}_{\alpha\beta\alpha'\beta'}$ is the Detweiler-Whiting regular two-point function~\cite{Detweiler:2002mi,Poisson:2011nh}. Substituting Eq.~\eqref{eq:pp R field} into Eq.~\eqref{eq:<dQdt>} yields the rate of change due to $h^{\rm R,test}_{\alpha\beta}$, which we write in the symmetrical form
\begin{equation}\label{eq:Qdot average pp}
    \left\langle\frac{dQ^{\rm test}_\xi}{dt}\right\rangle = \lim_{\substack{T\to\infty\\T'\to\infty}}\frac{1}{4T}\int_{V_t}\int_{V_t'}T^{\alpha\beta}_{\rm test}{\cal L}_\xi G^{\rm R}_{\alpha\beta\alpha'\beta'} T^{\alpha'\beta'}_{\rm test}dV'dV.
\end{equation}
Here we have defined $Q^{\rm test}_\xi$ by discarding $h^{\rm R}_{\alpha\beta}$ terms in the definition of $Q_\xi$.

The right-hand side of Eq.~\eqref{eq:Qdot average pp} can be converted into asymptotic fluxes by first expressing it in terms of the radiative part of the field~\cite{Galtsov:1982hwm,Sago:2005fn,Drasco:2005is,Isoyama:2018sib}. To achieve that, we split the regular Green's function into its symmetric and radiative parts: 
\begin{equation}
    G^{\rm R}_{\alpha\beta\alpha'\beta'} = G^{\rm R, Sym}_{\alpha\beta\alpha'\beta'} + G^{\rm Rad}_{\alpha\beta\alpha'\beta'},
\end{equation}
where the two pieces have the (anti)symmetry properties
\begin{align}
G^{\rm R, Sym}_{\alpha\beta\alpha'\beta'}(x,x') &= G^{\rm R, Sym}_{\alpha'\beta'\alpha\beta}(x',x),\\
G^{\rm Rad}_{\alpha\beta\alpha'\beta'}(x,x') &= -G^{\rm Rad}_{\alpha'\beta'\alpha\beta}(x',x).
\end{align}
All of these two-point functions satisfy\footnote{Note the two-point functions are tensors on two copies of the spacetime manifold, ${\cal M}\times {\cal M}$. We use ${\cal L}_\xi$ to denote the Lie derivative associated with a flow generated by $\xi^\alpha$ on the first copy (reducing to $\partial_t$ or $\partial_\phi$ in Boyer-Lindquist coordinates), and ${\cal L}_{\xi'}$ to denote the Lie derivative associated with a flow on the second copy (reducing to $\partial_{t'}$ or $\partial_{\phi'}$). This contrasts with the definition in Ref.~\cite{Harte:2011ku}, for example, where the Lie derivative would correspond to $({\cal L}_{\xi}+{\cal L}_{\xi'})$ in our notation.} 
\begin{equation}
{\cal L}_\xi G_{\alpha\beta\alpha'\beta'} = -{\cal L}_{\xi'} G_{\alpha\beta\alpha'\beta'},
\end{equation}
implying
\begin{align}
{\cal L}_\xi G^{\rm R, Sym}_{\alpha\beta\alpha'\beta'} = -{\cal L}_{\xi'} G^{\rm R, Sym}_{\alpha'\beta'\alpha\beta},\label{Lie GSym}\\
{\cal L}_\xi G^{\rm Rad}_{\alpha\beta\alpha'\beta'} = +{\cal L}_{\xi'} G^{\rm Rad}_{\alpha'\beta'\alpha\beta}.\label{Lie GRad}
\end{align}
Using these identities, we find the contribution from the symmetric part is
\begin{subequations}
\begin{align}
    I^{\rm Sym}&:=\int_{V_t}\int_{V_t'} T^{\alpha\beta}_{\rm test} {\cal L}_\xi G^{\rm R, Sym}_{\alpha\beta\alpha'\beta'} T^{\alpha'\beta'}_{\rm test} \, dV' \, dV \\
    &\hphantom{:}= \int_{V_t'}\int_{V_t} T^{\alpha'\beta'}_{\rm test} {\cal L}_{\xi'} G^{\rm R, Sym}_{\alpha'\beta'\alpha\beta} T_{\rm test}^{\alpha\beta} \, dV \, dV'\\
    &\hphantom{:}= -\int_{V_t'}\int_{V_t} T_{\rm test}^{\alpha'\beta'} {\cal L}_{\xi} G^{\rm R, Sym}_{\alpha\beta\alpha'\beta'} T_{\rm test}^{\alpha\beta} \, dV \, dV'\\
    &\hphantom{:}= - I^{\rm Sym},
\end{align}
\end{subequations}
where we swapped the integration dummy variable labels in the first equality and used Eq.~\eqref{Lie GSym} in the second equality. Hence, we conclude $I^{\rm Sym}=0$, and so only the radiative piece contributes:
\begin{multline}\label{Qdot pp rad}
\left\langle \frac{dQ^{\rm test}_\xi}{dt}\right\rangle \\
=\lim_{T\to\infty}\frac{1}{2T}\int_{V_t}\int T_{\rm test}^{\alpha\beta} {\cal L}_\xi G^{\rm Rad}_{\alpha\beta\alpha'\beta'} T_{\rm test}^{\alpha'\beta'}\, dV' \, dV.
\end{multline}

The right-hand side of Eq.~\eqref{Qdot pp rad} can be expressed in the standard form of an asymptotic (physical, retarded) flux by using a factorized form of the radiative Green's function~\cite{Chrzanowski:1974nr,Galtsov:1982hwm,Sago:2005fn}. We can then write
\begin{equation}\label{eq:test body balance law}
    \left\langle \frac{dQ^{\rm test}_\xi}{dt}\right\rangle = - {\cal F}^{\rm test}_\xi,
\end{equation}
where ${\cal F}^{\rm test}_{\xi_t}$ is the flux of energy out to infinity and down the primary's horizon, and ${\cal F}^{\rm test}_{\xi_\phi}$ is (minus) the flux of azimuthal angular momentum.

We emphasize that the balance law~\eqref{eq:test body balance law} only neglects $h^R_{\alpha\beta}$ contributions to $Q_\xi$ and nonlinear couplings in the field equations. It does \emph{not} neglect self-forces and torques in the dynamics; those self-forces and torques are precisely what drive the local evolution of $Q_\xi$. The above balance law establishes that these changes in $Q_\xi$ due to local self-forces and torques are equal (on average) to the energy and angular momentum carried off by emitted fluxes.

Going beyond the result here, by accounting for \emph{all} self-interactions and nonlinear effects, will be the subject of future work.

\subsection{Test-body contributions to forcing functions}
\label{sec:point particle forcing functions}

The balance law~\eqref{eq:test body balance law} is valid for generic (eccentric and inclined) orbits in Kerr spacetime. If we specialize to quasicircular orbits, we can discard the average (since there are no oscillations to average over), and we can focus on energy balance alone (since angular momentum balance becomes redundant). We now use the energy-balance law to obtain the test-body terms in the forcing functions $\F_{\Omega}^{n}$ that appear in the evolution equation~\eqref{state_evo}.

Defining ${\cal F}_{\rm test}:={\cal F}_{\xi_t}^{\rm test}$, we write the energy-balance law~\eqref{eq:test body balance law} as 
 \beq\label{balance_law}
 \dot{E}_{\rm test}=-\cF_{\rm test}=-(\e\cF^{\rm test}_0+\e^2\cF^{\rm test}_1+\e^3\cF^{\rm test}_2).
 \eeq
Here $E_{\rm test}$ is the \emph{test-body} expression~\eqref{eq:E test body}, which neglects $h^{\rm R}_{\alpha\beta}$ contributions in Eq.~\eqref{Conserved}. $\cF^{\rm test}_0 $ corresponds to the leading-order gravitational wave flux of a point mass, which is equal to the 0PA flux, $\cF_0$. The correction $\cF^{\rm test}_1=\cF^{\rm test}_{1|\chi}$ 
is the linear-in-spin contribution. The term $\cF_2$ accounts for 
quadratic-in-spin effects ($\cF_{2|\chi\chi}$), the quadrupolar deformation of the secondary due to rotation ($\cF_{2|C_Q}$), electric tidal effects ($\cF_{2|\mu_2}$), and magnetic tidal effects $(\cF_{2|\sigma_2})$.


Written more explicitly, the test-body orbital energy~\eqref{eq:E test body} has the form 
\begin{multline}\label{energy}
 E_{\rm test}(\bJ)=E_0(\Omega)+\e \chi E_{\chi}(\Omega,\chi)
 +\e^2 C_Q E_{C_Q}(\Omega,\chi)\\
 +\e^2 \mu_2 E_{\mu_2}(\Omega)+\e^2 \sigma_2 E_{\sigma_2}(\Omega),
 \end{multline}
where the explicit expressions for $E_0$, $E_{\chi}$, $E_{\chi\chi}$, $E_{C_Q}$, $E_{\mu_2}$, and $E_{\sigma_2}$ are presented in Sec.~\ref{Sec:FF_energy}. 
If we differentiate this expression with respect to time, apply the chain rule $d/dt = \dot\Omega\partial_\Omega$ (noting all other parameters are constant at the orders we work to), and appeal to the balance law~\eqref{balance_law}, then we find
\begin{equation}
    \left(\frac{d\Omega}{dt}\right)_{\rm test} = -\frac{{\cal F}_{\rm test}}{\partial E_{\rm test}/\partial\Omega}.\label{eq:dOmegadt test body}
\end{equation}

Substituting the expansions of the flux and energy in the right-hand side of Eq.~\eqref{eq:dOmegadt test body}, and expanding the fraction in powers of $\varepsilon$, we obtain the test-body contributions to the forcing functions $\F^n_\Omega$. At leading order, we find the adiabatic evolution equation,
 \beq\label{adiabatic}
 \F_\Omega^{0}=-\frac{\cF_0}{\partial_\Omega E_0}.
 \eeq
At order $\varepsilon^2$, we obtain the linear-in-spin contribution,
 \beq\label{lin_spin}
 \F_\Omega^{1|\chi}=-\frac{\cF_{1|\chi}+\F_\Omega^{0} \partial_\Omega E_{\chi}}{\partial_\Omega E_0}.
 \eeq
Similarly, at order $\e^3$, the quadratic in spin contribution can be expressed as 
 \beq\label{quad_spin}
 \F_\Omega^{2|\chi\chi}=-\frac{\cF_{2|\chi\chi}+\F_{\Omega}^{1} \partial_\Omega E_{\chi}+
 \F_\Omega^{0} \partial_\Omega E_{\chi\chi}}{\partial_\Omega E_0}.
 \eeq
Finally, the quadrupolar contributions can be expressed as   
 \beq\label{quadrupolar_F}
 \F_\Omega^{2|I}=-\frac{\cF_{2|I}+\F_\Omega^{0} \partial_\Omega E_{I}}{\partial_\Omega E_0}.
 \eeq
where $I\in(C_Q,\mu_2,\sigma_2)$.

Rather than reading off the contributions to $\F^n_\Omega$ in this way, we can also work with a balance law for the entire system, including all self-interactions and nonlinearities. Following Refs.~\cite{Mathews:2025txc,Trestini:2026tky}, we can choose $t$ to be a coordinate such that slices of constant $t$ foliate the primary's horizon on one end and future null infinity on the other. On each slice, we can define the system's binding energy as $E_{\rm bind}:=M_{\rm Bondi} - M_{\rm BH} - m_s$. We can then appeal to exact balance laws that hold fully nonlinearly: the Bondi mass-loss formula $dM_{\rm Bondi}/dt=-{\cal F}_\infty$~\cite{Compere:2019gft}; and the horizon energy-absorption formula $dM_{\rm BH}/dt={\cal F}_{\cal H}$~\cite{Ashtekar:2004cn,Ashtekar:2023kei,Chandrasekaran:2018aop}.\footnote{Although Ref.~\cite{Chandrasekaran:2018aop} notes that there is no black hole mass associated with a symmetry of its event horizon, the black hole's mass can be \emph{defined} from its area and spin using the Kerr relationship~\cite{Ashtekar:2004cn,Ashtekar:2023kei}, and it then satisfies a balance law. Here the black hole's dynamical horizon or its event horizon can be used.} In the multiscale expansion, the binding energy can always be written as a function of the system's parameters, $E_{\rm bind}=E_{\rm bind}(\Omega,\delta M,\delta J)$ (suppressing dependence on constant parameters). Differentiating $E_{\rm bind}$ with respect to $t$, applying the chain rule, appealing to the exact balance laws, and rearranging, we then obtain
\begin{equation}
    \frac{d\Omega}{dt} = -\frac{{\cal F} + \dot{\delta M}\partial E_{\rm bind}/\partial\delta M + \dot{\delta J}\partial E_{\rm bind}/\partial\delta J}{\partial E_{\rm bind}/\partial\Omega},\label{eq:Omega dot full balance}
\end{equation}
with ${\cal F}:={\cal F}_\infty+{\cal F}_{\cal H}$. Here for simplicity we have treated the secondary's mass and spin as constants at the orders of interest; inclusion of their rates of change, if significant, is straightforward. As pointed out in Ref.~\cite{Mathews:2025txc}, the accuracy of a self-force waveform model can be significantly improved for comparable masses by leaving the fraction intact rather than fully expanding the right-hand side of Eq.~\eqref{eq:Omega dot full balance}. 

An important uncertainty in Eq.~\eqref{eq:Omega dot full balance} is that the binding energy defined from the Bondi mass differs, in general, from the energy defined from the local mechanics of the two-body system; the two will generally differ by Schott terms~\cite{Pound:2019lzj,Trestini:2025nzr,Lewis:2025ydo}. We refer to Refs.~\cite{Trestini:2026tky,Lewis:2025ydo} for further discussion of that point. However, the balance law we have derived in this paper ensures that we \emph{can} safely use Eq.~\eqref{energy} as the test-body contribution to $E_{\rm bind}$. The test-body contributions to the orbital evolution then comprise these terms in $E_{\rm bind}$ in the denominator of Eq.~\eqref{eq:Omega dot full balance} and the corresponding test-body terms in the fluxes in the numerator.

 \section{Flux calculations in the fixed-frequency expansion}\label{Sec_5_FieldEquations}

To compute the fluxes that enter the balance law (and therefore the binary evolution), we solve the Teukolsky equation with the skeletonized stress-energy tensor as a source. In this section, we outline that calculation and present our numerical and analytical (PN expanded) results. As highlighted in the previous sections, our calculations here neglect contributions from $h^R_{\alpha\beta}$ and $\dot{\cal J}^a$ that would appear in the complete 2PA fluxes. Future work should account for those contributions.

\subsection{Teukolsky equation}

In ordinary linear perturbation theory, when the four-dimensional Teukolsky equation ${\cal O}_4\psi_4=S_4$ is separated into frequencies and spheroidal modes, the radial and angular Teukolsky equations are given by~\cite{Pound:2021qin} 
\begin{subequations}
\begin{align}
&\Delta^2\frac{d}{d\hr}\left(\frac{1}{\Delta}\frac{dR_{\mode}(r)}{d\hr}\right)- V(\hr)R_{\mode}(r) = \mathcal{J}_{\mode}(r),\label{rad_Teuk}\\
& \Bigg[\frac{1}{\sin\theta} \frac{d}{d\theta}\left(\sin\theta \frac{d}{d\theta} \right) - a^2 \omega^2 \sin^2\theta - \left(\frac{m - 2\cos\theta}{\sin\theta}\right)^2  \notag\\ & \quad + \, 4 a \omega \cos\theta - 2 + 2 a m \omega + \lambda_{\ell m \omega} \Bigg] \Ang(\theta) = 0,
\label{ang_Teuk}
\end{align}
\end{subequations}
where $\Ang(\theta)$ is the spin-weighted spheroidal harmonic function with spin weight $-2$, satisfying the normalization condition
\begin{equation}\label{ang_norm}
\int \sin\theta d\theta d\phi |\Ang e^{im\phi}|^2=1,
\end{equation} 
and the potential $V(r)$ is given by
\begin{subequations}
\begin{align}\label{Potential}
V(\hr)&=-\frac{K^2+4i(\hr-M)K}{\Delta} + 8i\omega \hr + \lambda_{\mode},\\
K&= (r^2 + a^2)\omega - a m.
\end{align}
\end{subequations}
Here $\omega$ is the mode frequency and $\lambda_{\mode}$ represents the eigenvalues. The source term $\mathcal{J}_{\mode}(r)$ is constructed from the stress-energy tensor that sources the linear perturbation. 

The radial Teukolsky equation ordinarily assumes that the stress-energy tensor and linear metric perturbation are written as inverse Fourier transforms: $T^{\alpha\beta}=\int d\omega e^{-i\omega t}T^{\alpha\beta}_\omega$ and $h_{\alpha\beta}=\int d\omega e^{-i\omega t}h^\omega_{\alpha\beta}$ (or at least that the four-dimensional Weyl scalar $\psi_4$, if not the metric perturbation, is expressed in that form). In our multiscale expansion, $T^{\alpha\beta}$ and $h_{\alpha\beta}$ are functions of slow and fast variables rather than~$t$, and Fourier transforms are no longer used. Instead, because~$\phi_p$ is a periodic variable, the coefficients in Eqs.~\eqref{eq:g multiscale expansion} and \eqref{eq:T multiscale expansion} are expanded in discrete Fourier series:
\begin{align}
    T^{\alpha\beta}_n &= \sum_{m=-\infty}^\infty T^{\alpha\beta}_{n,m}(r,\theta,\bJ^a)e^{im(\phi-\phi_p)},\\
    h^n_{\alpha\beta} &= \sum_{m=-\infty}^\infty h_{\alpha\beta}^{n,m}(r,\theta,\bJ^a)e^{im(\phi-\phi_p)},
\end{align}
where we have used the fact that the functions can only depend on $\phi_p$ in the combination $(\phi-\phi_p)$ due to the form of the stress-energy tensor~\eqref{SET1} and the axial symmetry of the background spacetime. We refer to Refs.~\cite{Miller:2020bft,Pound:2021qin,Mathews:2021rod,Mathews:2025nyb,Lewis:2025ydo} for extensive discussion of these points.

We can recover the Teukolsky equation~\eqref{rad_Teuk} by appealing to (i) the above discrete Fourier series, (ii) the chain rule~\eqref{partial_t} for $t$ derivatives, and (iii) our neglect of $\dbJ^a$ and $h^R_{\alpha\beta}$ terms in the field equations. The source ${\cal J}_{\mode}$ then corresponds to an expansion of the four-dimensional Teukolsky source $S_4$ in the form~\cite{Pound:2021qin} 
\begin{equation}
    S_4 = \frac{\rho^4}{16\pi\Sigma}\sum_{\ell=2}^\infty\sum_{m=-\ell}^{+\ell} {\cal J}_{\mode}(r,\bJ^a) S^{a\omega}_{\ell m}(\theta)e^{im(\phi-\phi_p)},
\end{equation}
with the discrete frequencies $\omega=\omega_m:= m\Omega$ [and having defined $\rho:=(r-ia\cos\theta)^{-1}$]. Here the source mode coefficients ${\cal J}_{\mode}$ are constructed from the background test-body/bare point-particle stress-energy tensor $T^{\alpha\beta}_{\rm pp}$ defined in Eq.~\eqref{eq:Tpp def}, where $\dbJ^a$ and $h^R_{\alpha\beta}$ are set to zero both in $T^{\alpha\beta}$ and in the calculation of derivatives acting on $T^{\alpha\beta}$ (which arise in the Teukolsky source). The detailed calculation of the source term is presented in Appendix~\ref{App:source_term}. 

Setting $h^R_{\alpha\beta}=0=\dbJ^a$ makes our calculations closely analogous to numerous flux calculations for linear ``test spins'' (e.g.,~\cite{Tanaka:1996ht, Piovano:2020zin, Piovano:2021iwv, Skoupy:2021asz, Skoupy:2022adh, Skoupy:2023lih, Piovano:2024yks, Skoupy:2024jsi, Skoupy:2025nie, Skoupy:2026ewu}) and to our own calculation of quadratic and quadrupole corrections~\cite{Rahman:2021eay}. Our computations are more limited than many of those in that we restrict to quasicircular binaries, but we go beyond them by including tidal quadrupole contributions and by computing a complete data set that can be immediately used in waveform models. 

Moreover, our calculations are formulated differently than most earlier ones in our consistent use of the multiscale expansion, following our earlier Refs.~\cite{Mathews:2021rod,Mathews:2025txc}. In this formulation, as emphasized in earlier sections, we consistently expand all quantities, including the test-body source, in powers of $\e$ at fixed $\Omega$. If one did not consistently expand in this way, instead working with some unexpanded, $\e$-dependent relationship $\Omega(r_p,a,m_s,\chi,C_Q,\mu_2,\sigma_2)$, then one would need to solve the field equations for each value of the parameters, computing and storing flux data on a high-dimensional grid of $(r_p,a,m_s,\chi,C_Q,\mu_2,\sigma_2)$ values. This would be at odds with the underlying expansion in powers of $\e$ and with the natural modularity of the multiscale waveform construction, which (by design) allows one to immediately add new physical effects without enlarging the dimensionality of one's grid. But such an approach, solving field equations with unexpanded (or only partially expanded) test-spin sources, is in fact how most linear-in-spin calculations have been done; in that case, very often, the linear-in-spin coefficients have been extracted by numerically fitting data, as in Ref.~\cite{Akcay:2019bvk}, for example. 

The main benefits of the multiscale expansion are not specifically tied to using the frequency as a phase-space coordinate. For example, one could expand for small~$\e$ at fixed~$r_p$ (or fixed energy) rather than at fixed~$\Omega$. But expanding at fixed $\Omega$ has several advantages over other choices: $\Omega$ closely tracks the waveform frequency~\cite{Warburton:2024xnr,Honet:2025gge}, leading to accurate waveforms for comparable masses~\cite{Wardell:2021fyy}; $\Omega$ is gauge invariant, allowing simple hybridization of results that might have used different gauge choices or entirely different formalisms~\cite{Honet:2025gge,Trestini:2026tky}; and $\Omega$ appears directly in the field equations, through the chain rule $\partial_t = \Omega \partial_{\phi_p} + \cdots$, such that if~$\Omega$ is expanded at fixed $r_p$ (for example), then this leads to noncompact source terms for $h^2_{\mu\nu}$ involving products of $h^1_{\mu\nu}$ with $\Omega_1$ (for example), which are more difficult to deal with than point-particle sources.

\begin{figure*}
\begin{center}
\includegraphics[width=\textwidth,trim={15pt 32.5pt 15pt 10pt},clip]{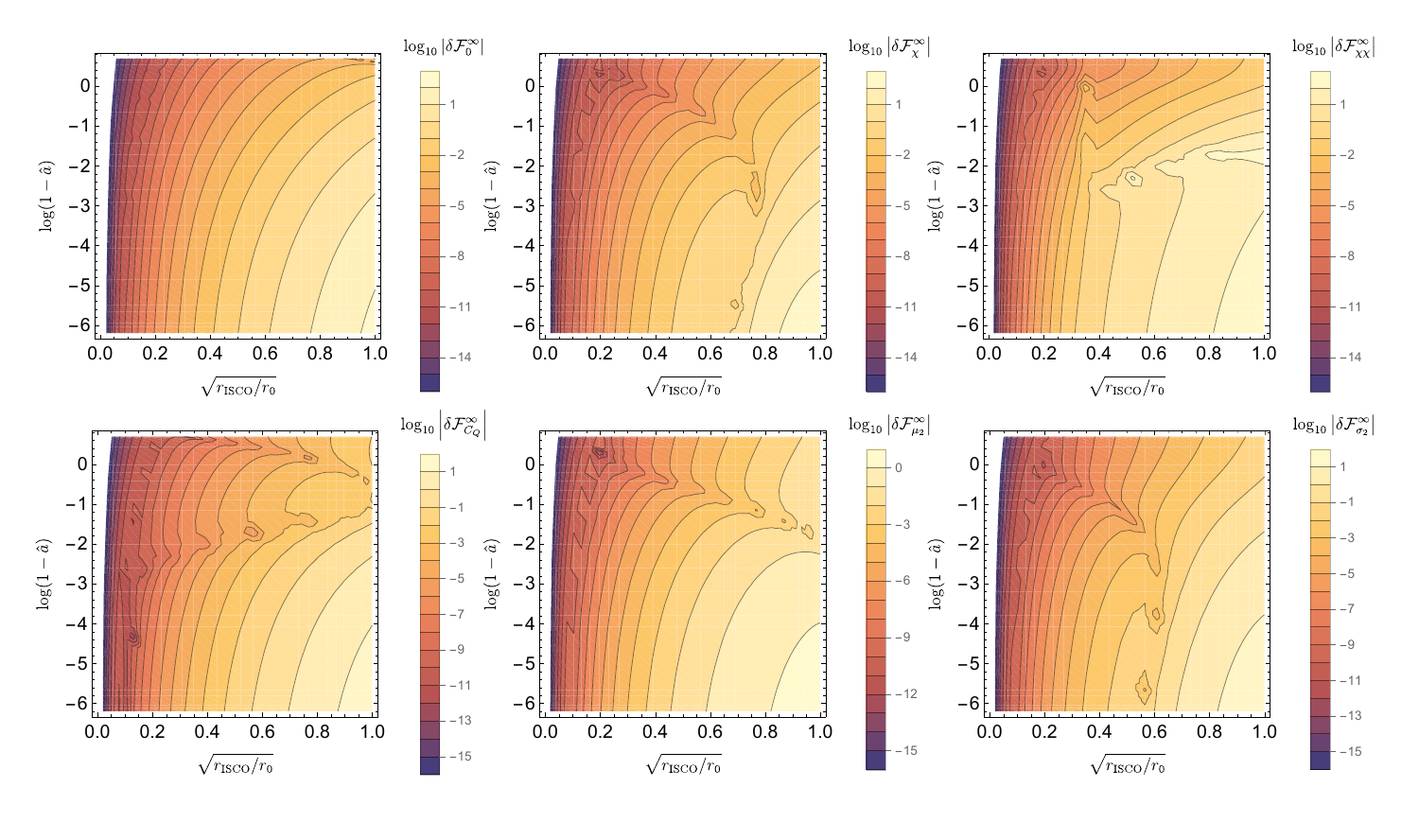}
\end{center}
\caption{Relative difference between numerical fluxes at infinity and our highest-order corresponding analytical PNSF expressions on the two-dimensional $(\sqrt{r_{\rm ISCO}/r_0}, \, \log(1- \hat a) )$ space. }\label{fig_RelErrInfFlux}
 \end{figure*}

\begin{figure*}
\begin{center}
\includegraphics[width=\textwidth,trim={15pt 32.5pt 15pt 10pt},clip]{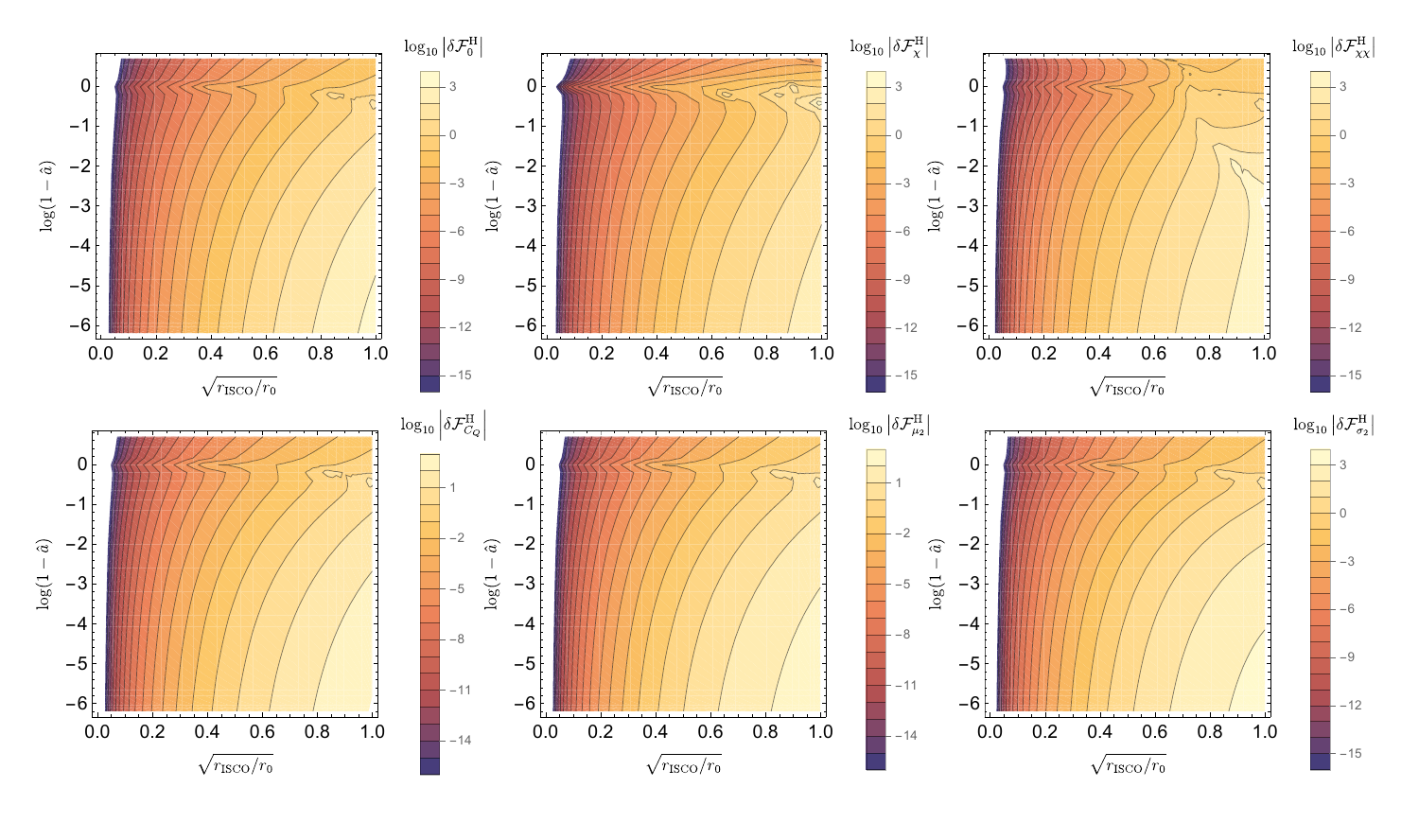}
\end{center}
\caption{Relative difference between numerical fluxes at the horizon and our highest-order corresponding analytical PNSF expressions on the two-dimensional $(\sqrt{r_{\rm ISCO}/r_0}, \, \log(1-\hat a) )$ parameter space.
}\label{fig_RelErrHorFlux}
 \end{figure*}

\begin{figure}
\begin{center}
\includegraphics[width=\hsize]{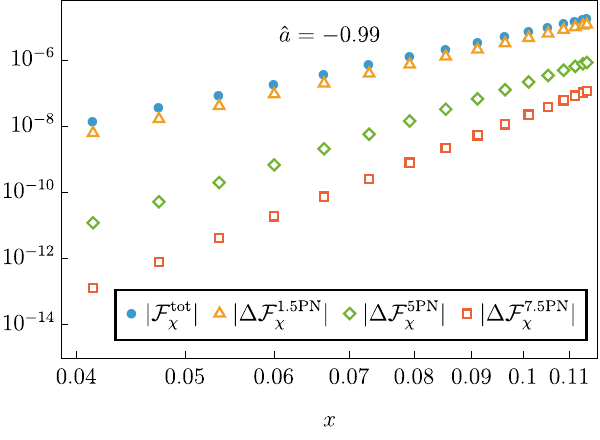}
\includegraphics[width=\hsize,trim={0 5pt 0pt 0pt},clip]{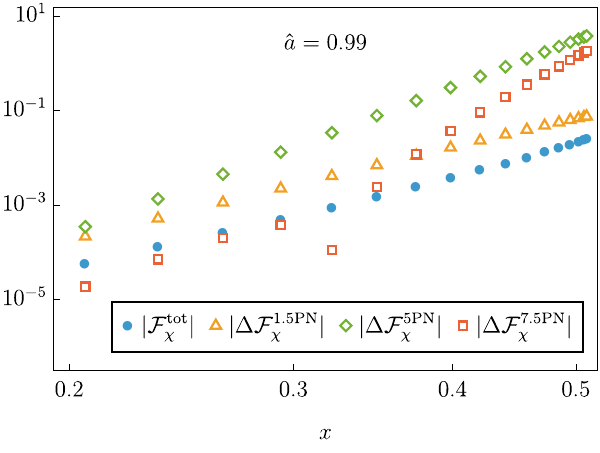}
\end{center}
\caption{Comparison of the total numerical linear-in-spin flux correction $\mathcal{F}_{\chi}^{\rm tot}$ with the corresponding PN approximations at $1.5$PN (leading spin-dependent order), $5$PN, and $7.5$PN (our highest PN order) for retrograde ($\hat a=-0.99$, upper panel) and prograde ($\hat a = 0.99$, lower panel) orbits. The plotted quantities $|\Delta \mathcal{F}_{\chi}^{\rm nPN}| = | \mathcal{F}_{\chi}^{\rm tot} - \mathcal{F}_{\chi}^{\rm nPN} |$, with $n=1.5,\, 5,\, 7.5$, measure the absolute deviation of each PN approximation from the total numerical flux. For the retrograde case, PN errors grow with $x$ but decrease systematically with increasing PN order. In contrast, for the prograde orbits, the errors are not only severe but also do not improve monotonically with PN order (though we stress that for sufficiently small values of $x$, not displayed here, the errors become small and do monotonically decrease with increasing order). 
}\label{fig3}
 \end{figure}

\subsection{Test-body fluxes}

We employ the Green function method to obtain the radial solution $R_{\mode}(r)$ to Eq.~\eqref{rad_Teuk}. The retarded Green function is written in terms of the linearly independent solutions of the homogeneous radial Teukolsky equation, $R^{\In}_{\mode}(\hr)$ and $R^{\up}_{\mode}(\hr)$, that represent purely incoming boundary conditions at the horizon and purely outgoing boundary conditions at infinity, respectively. In terms of these functions, the solution of Eq.~\eqref{rad_Teuk} can be written as 
\begin{align}\non
R_{\mode}(\hr) &= \frac{1}{\mathcal W} \Biggl\{R^{\up}_{\mode}\int_{\hr_+}^{\hr}d\hr\frac{R^{\In}_{\mode}\mathcal{J}_{\mode}}{\Delta^2} \\ &\quad + R^{\In}_{\mode}\int_{\hr}^{\infty}d\hr\frac{R_{\mode}^{\up}\mathcal{J}_{\mode}}{\Delta^2}\Biggr\},\label{sol_Teuk}
\end{align}
where $\mathcal W=\left(R^{\In}_{\mode}\partial_{\hr} R^{\up}_{\mode}-R^{\up}_{\mode}\partial_{\hr} R^{\In}_{\mode}\right)/\Delta$ is the constant Wronskian. The asymptotic behavior of the radial function~\eqref{sol_Teuk} is given as follows~\cite{Pound:2021qin}:
\begin{equation}\label{asymp_behave}
		\begin{aligned}
			R_{\mode}=
			\begin{cases}
				\mathcal{Z}_{\mode}^{\infty}\hr^3~e^{i\omega\hr_{*}}~, & \hr\to \infty\\
				\mathcal{Z}_{\mode}^{\rm H}\Delta^2~e^{-i\omega\hr_{*}}~, & \hr\to \hr_+
			\end{cases}
		\end{aligned}
\end{equation}
where $\hr_{*}$ is the tortoise coordinate of the Kerr metric, defined as
\begin{align}\non
\hr_{*} (r) &= r + \frac{2 M r_{+}}{r_{+} - r_{-}} \ln \left( \frac{r - r_{+}}{2 M} \right) \\ &\quad - \frac{2 M r_{-}}{r_{+} - r_{-}} \ln \left( \frac{r - r_{-}}{2 M} \right).
\end{align}
The solution~\eqref{asymp_behave} satisfies the physical boundary conditions, being purely ingoing at the horizon and purely outgoing at infinity. 

The amplitudes $ \mathcal{Z}_{\mode}^{\rm H,\infty}$ in the solution are given by the relation
\begin{equation}\label{amp_def}
\begin{aligned}
\mathcal{Z}_{\mode}^{\rm H,\infty}=\mathcal{C}_{\mode}\int_{\hr_+}^{\infty}d\hr\frac{R_{\mode}^{\up,\In}\mathcal{J}_{\mode}}{\Delta^2}\,,
\end{aligned}
\end{equation}
where the constant $\mathcal{C}_{\mode}$ is defined as
\begin{equation}\label{constant_term}
\mathcal{C}_{\mode}=\frac{1}{2i \omega B^{\textrm{in}}_{\mode}}.
\end{equation}
These amplitudes fully determine the asymptotic gravitational-wave fluxes at infinity and at the horizon, respectively. The factor $\mathcal{C}_{\mode}$ depends on the normalization of the homogeneous solutions; the value~\eqref{constant_term} assumes $R^{\rm in}_{\mode}\to \Delta^2 e^{-i\omega r_*}$ at the horizon and $R^{\rm up}_{\mode}\to r^3 e^{+i\omega r_*}$ at infinity.

For the energy-momentum tensor of a spinning deformed secondary object presented in Eq.~\eqref{SET}, the amplitudes take the following form:
\begin{equation}\label{eq:Teukolsky amplitudes}
\mathcal{Z}_{\ell m\omega}^{\rm H,\infty}            = \mathcal{C}_{\ell m\omega} \, \mathcal{I}_{\ell m\omega}^{\rm H,\infty}\,,
\end{equation}
where the explicit expression for $\mathcal{I}_{\ell m\omega}^{\rm H,\infty}$ is provided in Appendix~\ref{App:source_term}. The time-averaged energy flux at infinity and the horizon can be written as~\cite{Pound:2021qin}
\begin{equation}\label{flux_inf/H}
	\bigg\langle\frac{dE}{dt}\bigg\rangle^{\!A} =\sum_{\ell=2}^{\infty}\sum_{m=1}^{\ell}\frac{\alpha_{\ell m}^{A}|\mathcal Z^{A}_{\mode}|^2}{2\pi( m\hO)^2},
\end{equation}
where $A=\infty$ or $A={\rm H}$. The coefficients at infinity and the horizon are $\alpha^{\infty}_{\ell m}=1$ and 
\begin{equation}
\alpha^{\rm H}_{\ell m} = \frac{256\, \kappa \left(2 M r_{+} \right)^5 \left(\kappa^2 + 4 \epsilon^2 \right) \left(\kappa^2 + 16 \epsilon^2 \right) \left(m \Omega \right)^3}{ \left|\mathcal{C}_{\ell m} \right|^2},
\end{equation}
with $\epsilon = \sqrt{M^2 - a^2}/(4 M r_{+})$, and 
\begin{align}\non
\left|\mathcal{C}_{\ell m} \right|^2 &= \left[\left(2 + \lambda_{\ell m \omega} \right)^2 - 4 a^2 \left(m \Omega \right)^2 + 4 a m \left(m \Omega \right) \right] \\\non &\quad\times
\left[\lambda^{2}_{\ell m \omega} -36 a^2 \left(m \Omega \right)^2 + 36 m a \left(m \Omega \right) \right] \\\non &\quad +
\left[96 a^2 \left(m \Omega \right)^2 -48 m a \left(m \Omega \right) \right]\\ &\quad \times
\left( 2 \lambda_{\ell m \omega} + 3 \right) + 144 \left( 1 - a^2 \right) \left(m \Omega \right)^2.
\end{align}

Expanding the amplitudes in powers of the small parameter $\varepsilon$ yields
\begin{equation}
  \bigg\langle\frac{dE}{dt}\bigg\rangle^{\!A} =  \e^2 \mathcal{F}^A_{0}+\e^3 \mathcal{F}^A_{1}+\e^4 \mathcal{F}^A_{2}.
\end{equation}
As discussed in Sec.~\ref{sec:point particle forcing functions}, the leading coefficient, $\mathcal{F}^A_{0}$, corresponds to the flux generated by a point mass. It can be expressed as
\begingroup%
\allowdisplaybreaks%
\begin{subequations}\label{0pa_flux}
\begin{align}
    \mathcal{F}^A_{0} &= \ \sum_{\ell=2}^{\infty}\sum_{m=1}^{\ell}\frac{\alpha_{\ell m}^A|\mathcal{Z}^A_{0}|^2}{2\pi( m\hO)^2}.
\end{align}
\end{subequations}%
\endgroup%
The coefficient $\mathcal{F}^A_{1}$ contains contributions from both the subleading point-particle corrections and the linear-in-spin effects of the secondary object. The contribution arising from the spin of the secondary can be written as 
\begingroup%
\allowdisplaybreaks%
\begin{subequations}\label{0pa_1pa_flux}
\begin{align}
\mathcal{F}^{A}_{1|\chi} &= \ \sum_{\ell=2}^{\infty}\sum_{m=1}^{\ell}\frac{2 \alpha_{\ell m}^{A}\, \mathrm{Re}\left(\mathcal{Z}^{A}_{0} \overline{\mathcal{Z}^{A}_{\chi}}\right)}{2\pi( m\hO)^2}.
\end{align}
\end{subequations}%
\endgroup%
For the sake of brevity, hereafter we denote $\mathcal{F}^{A}_{1|\chi}$ by $\mathcal{F}^{A}_{\chi}$.
\par
The coefficient $\mathcal{F}^{A}_{2}$ contains several contributions associated with the internal structure of the secondary object, as well as higher-order self-force corrections associated with the point-mass sector. In the work, we neglect the latter contributions and retain only the finite-size effects. Thus, $\mathcal{F}^{A}_{2}$  can be decomposed as
\begin{equation}
    \mathcal{F}^{A}_{2} = \mathcal{F}^{A}_{2|\chi \chi} + C_Q \mathcal{F}^{A}_{2|C_Q} +  \mu_2 \mathcal{F}^{A}_{2|\mu_2} + \sigma_2 \mathcal{F}^{A}_{2|\sigma_2},
\end{equation}
where $\mathcal{F}^{A}_{2|\chi \chi}$ denotes the pure quadratic-in-spin contribution, $\mathcal{F}^{A}_{2|C_Q}$ corresponds to the quadratic-in-spin associated with the spin-induced quadrupole moment, while $\mathcal{F}^{A}_{2|\mu_2}$ and $\mathcal{F}^{A}_{2|\sigma_2}$ represent the electric and magnetic tidal quadrupole deformation contributions, respectively. For notational simplicity, hereafter we denote these terms as $\mathcal{F}^{A}_{\chi \chi}$, $\mathcal{F}^{A}_{C_Q}$, $\mathcal{F}^{A}_{\mu_2}$ and $\mathcal{F}^{A}_{\sigma_2}$, respectively.  The explicit expressions of these terms can be written as
\begin{subequations}\label{2pa_flux}
\begin{align}
    \mathcal{F}^{A}_{\chi \chi} &= \ \sum_{\ell=2}^{\infty}\sum_{m=1}^{\ell} \alpha_{\ell m}^{A}\frac{2 \, \mathrm{Re}\left(\mathcal{Z}^{A}_{0} \overline{\mathcal{Z}^{A}_{ \chi \chi}}\right) + |\mathcal{Z}^{A}_{\chi}|^2}{2\pi( m\hO)^2},\\
    \mathcal{F}^{A}_{C_Q} &= \ \sum_{\ell=2}^{\infty}\sum_{m=1}^{\ell} \frac{2\alpha_{\ell m}^{A} \, \mathrm{Re}\left(\mathcal{Z}^{A}_{0} \overline{\mathcal{Z}^{A}_{C_Q}}\right) }{2\pi( m\hO)^2},\\
    \mathcal{F}^A_{\mu_2} &= \ \sum_{\ell=2}^{\infty}\sum_{m=1}^{\ell} \frac{2\alpha_{\ell m}^{A} \, \mathrm{Re}\left(\mathcal{Z}^{A}_{0} \overline{\mathcal{Z}^{A}_{\mu_2}}\right) }{2\pi( m\hO)^2},\\
\mathcal{F}^{\infty}_{\sigma_2} &= \ \sum_{\ell=2}^{\infty}\sum_{m=1}^{\ell} \frac{2 \alpha_{\ell m}^{A}\, \mathrm{Re}\left(\mathcal{Z}^{A}_{0} \overline{\mathcal{Z}^{A}_{\sigma_2}}\right) }{2\pi( m\hO)^2}.
\end{align}
\end{subequations}
We have presented the expressions of the mode amplitudes $\mathcal{Z}_{0}^{A}$, $\mathcal{Z}_{\chi}^{A}$, $\mathcal{Z}_{\chi\chi}^{A}$, $\mathcal{Z}_{C_Q}^{A}$, $\mathcal{Z}_{\mu_2}^{A}$ and $\mathcal{Z}_{\sigma_2}^{A}$ in Eq.~\eqref{mode_amplitudes}.

\begin{table}[tb] 
\centering
\renewcommand{\arraystretch}{1.35}
\begin{ruledtabular}
\begin{tabular}{c c c c}
Flux  & \begin{tabular}{@{}c@{}}Leading \\[-3pt] PN order\end{tabular}  & \begin{tabular}{@{}c@{}} Max. \\[-3pt] PN order\end{tabular} & \begin{tabular}{@{}c@{}} Max. \\[-3pt] PNSF order\end{tabular} \\
\hline
\multicolumn{4}{c}{Infinity flux} \\
\hline
$\mathcal F^{\infty}_{0}$ & $0$PN & \ $4$PN~\cite{
Marsat:2014xea,Siemonsen:2017yux,Cho:2022syn,Blanchet:2023bwj} & $20$PN~\cite{Shah:2014tka}\\
$\mathcal F^{\infty}_{\chi}$ & $1.5$PN & $4$PN~\cite{
Bohe:2013cla,Marsat:2014xea,Siemonsen:2017yux,Cho:2022syn} & $\mathbf{7.5}$\textbf{PN}
\\
$\mathcal F^{\infty}_{\chi\chi}$ & $2$PN & $4$PN~\cite{
Marsat:2014xea,Siemonsen:2017yux,Cho:2022syn} & $\mathbf{8}$\textbf{PN} \\
$\mathcal F^{\infty}_{C_Q}$ & $2$PN & $3.5$PN~\cite{
Marsat:2014xea} & $\mathbf{8}$\textbf{PN} \\
$\mathcal F^{\infty}_{\mu_{2}}$ & $5$PN & $ 6.5$PN~\cite{Abdelsalhin:2018reg}  & $\mathbf{11}$\textbf{PN} \\
$\mathcal F^{\infty}_{\sigma_{2}}$ & $6$PN & $6.5$PN~\cite{Abdelsalhin:2018reg} & $\mathbf{12}$\textbf{PN} \\
\hline
\multicolumn{4}{c}{Horizon flux} \\
\hline
$\mathcal F^{\rm H}_{0}$ & $2.5$PN & 4PN~\cite{
Saketh:2022xjb} & $20$PN~\cite{Shah:2014tka}\\
$\mathcal F^{\rm H}_{\chi}$ & $4$PN & 4PN~\cite{
Saketh:2022xjb} & $\mathbf{10}$\textbf{PN} \\
$\mathcal F^{\rm H}_{\chi\chi}$ & $4.5$PN & - & $\mathbf{10.5}$\textbf{PN} \\
$\mathcal F^{\rm H}_{C_Q}$ & $4.5$PN & - & $\mathbf{10.5}$\textbf{PN} \\
$\mathcal F^{\rm H}_{\mu_{2}}$ & $7.5$PN & - & $\mathbf{13.5}$\textbf{PN} \\
$\mathcal F^{\rm H}_{\sigma_{2}}$ & $8.5$PN & - & $\mathbf{14.5}$\textbf{PN} \\
\end{tabular}
\end{ruledtabular}
\caption{Post-Newtonian results for different contributions to the gravitational-wave energy flux to infinity and through the horizon. Orders are counted relative to the 0PN quadrupole formula. 
The second column indicates the PN order at which each contribution first enters the flux. The third column lists the highest PN order available in full PN theory, and ``-'' indicates that no PN result is currently available in the literature. The fourth column indicates the highest PN order obtained from PNSF expansions, with entries in bold indicating our new results. 
Note that since the definition of $\mathcal{F}_0^{\infty}$ includes all orders in primary spin, its PN accuracy is limited by progress on the spinning sector --- for non-spinning binaries the result is available to 4.5PN~\cite{Blanchet:2023bwj}. Furthermore, for non-spinning binaries, tidal deformation results ($\mathcal{F}_{\mu_2}^{\infty}$ and $\mathcal{F}_{\sigma_2}^{\infty}$) are available up to $7.5$PN~\cite{Henry:2020ski,Mandal:2024iug}. Higher-order PNSF results are also available for a nonspinning primary, such as 22PN~\cite{Fujita:2012cm} and even 30PN~\cite{Cipriani:2025ikx} 0PA energy fluxes. $\mathcal F^{\infty}_{C_Q}$ is known to 4PN for black holes binaries~\cite{Siemonsen:2017yux}.
}
\label{tab:PN_orders}
\end{table}

\begin{table*}[htb!]
\centering
\def\arraystretch{1.4}      	
\setlength{\tabcolsep}{.9em}
\begin{tabular}{M M M M M M M}
\hline\hline
\multicolumn{1}{c|}{$\rO$} & \multicolumn{1}{c|}{$\mathcal{F}^{\rm tot}_0$}        
& \multicolumn{1}{c|}{$\mathcal{F}^{\rm tot}_\chi$}   
& \multicolumn{1}{c|}{$\mathcal{F}^{\rm tot}_{\chi\chi}$}          
& \multicolumn{1}{c|}{$\mathcal{F}^{\rm tot}_{C_Q}$}
& \multicolumn{1}{c|}{$\mathcal{F}_{\mu_2}^{\textrm{tot}}$}
& \multicolumn{1}{c}{$\mathcal{F}_{\sigma_2}^{\textrm{tot}}$}   \\ 
\hline\hline

\multicolumn{7}{c}{$\hat a = -0.5$} \\ \hline
8.0 & 2.42491 \times 10^{-4} & -2.86551 \times 10^{-5} & -2.93089 \times 10^{-6} &  1.02932 \times 10^{-5} & 2.14698 \times 10^{-7} & 2.05340 \times 10^{-7} \\
8.5 & 1.72046 \times 10^{-4} & -1.81487 \times 10^{-5} & -1.66075 \times 10^{-6} & 6.30488 \times 10^{-6} & 1.07819 \times 10^{-7} & 9.51348 \times 10^{-8} \\
9.0 & 1.25191 \times 10^{-4} & -1.18658 \times 10^{-5} & -9.79365 \times 10^{-7} & 4.00353 \times 10^{-6} & 5.68985 \times 10^{-8} & 4.65526 \times 10^{-8} \\
9.5 & 9.30720 \times 10^{-5} & -7.97342 \times 10^{-6} & -5.97575 \times 10^{-7} &  2.62135 \times 10^{-6} & 3.13239 \times 10^{-8} & 2.38715 \times 10^{-8} \\
10.0 & 7.04835 \times 10^{-5} & -5.48735 \times 10^{-6} & -3.75540 \times 10^{-7} & 1.76238 \times 10^{-6} & 1.7887 \times 10^{-8} & 1.27485 \times 10^{-8} \\

\hline
\multicolumn{7}{c}{$\hat a = 0$} \\ \hline
6.1 & 8.56192 \times 10^{-4} & -1.26327 \times 10^{-4} & -1.57070 \times 10^{-5} & 6.06110 \times 10^{-5} & 3.03579 \times 10^{-6} & 2.90414 \times 10^{-6} \\
6.5 & 6.00306 \times 10^{-4} & -7.82198 \times 10^{-5} & -8.68042 \times 10^{-6} & 3.65008 \times 10^{-5} & 1.47785 \times 10^{-6} & 1.29701 \times 10^{-6} \\
7.0 & 4.00172 \times 10^{-4} & -4.51590 \times 10^{-5} & -4.40557 \times 10^{-6} & 2.04555 \times 10^{-5} & 6.48669 \times 10^{-7} & 5.15891 \times 10^{-7} \\
7.5 & 2.76157 \times 10^{-4} & -2.72959 \times 10^{-5} & -2.36739 \times 10^{-6} & 1.20476 \times 10^{-5} & 3.05218 \times 10^{-7} & 2.21879 \times 10^{-7} \\
8.0 & 1.96104 \times 10^{-4} & -1.71438 \times 10^{-5} & -1.33361 \times 10^{-6} & 7.39293 \times 10^{-6} & 1.52131 \times 10^{-7} & 1.01826 \times 10^{-7} \\

\hline
\multicolumn{7}{c}{$\hat a = 0.5$} \\ \hline
4.3 &  3.78644 \times 10^{-3} & -6.69031 \times 10^{-4} & -8.26459 \times 10^{-5} &  4.92966 \times 10^{-4} & 7.75638 \times 10^{-5} & 6.6897 \times 10^{-5} \\
5.0 & 1.72640 \times 10^{-3} & -2.22559 \times 10^{-4} & -2.16030 \times 10^{-5} & 1.60280 \times 10^{-4} & 1.50740 \times 10^{-5} & 1.06093 \times 10^{-5} \\
5.5 & 1.06337 \times 10^{-3} & -1.13347 \times 10^{-4} & -9.55174 \times 10^{-6} & 8.02666 \times 10^{-5} & 5.49867 \times 10^{-6} & 3.43705 \times 10^{-6} \\
6.0 & 6.86535 \times 10^{-4} & -6.18544 \times 10^{-5} & -4.59174 \times 10^{-6} & 4.30210 \times 10^{-5} & 2.21593 \times 10^{-6} & 1.24918 \times 10^{-6} \\
6.5 & 4.60277 \times 10^{-4} & -3.56698 \times 10^{-5} & -2.35702 \times 10^{-6} & 2.43449 \times 10^{-5} & 9.67299 \times 10^{-7} & 4.97602 \times 10^{-7} \\

\hline\hline
\end{tabular}
\caption{Sample of our numerical results for the gravitational-wave energy fluxes corresponding to the point-mass (monopole) contribution $\mathcal{F}_{0}^{\textrm{tot}}$, the linear-in-spin contribution $\mathcal{F}_{\chi}^{\textrm{tot}}$, the quadratic-in-spin contribution $\mathcal{F}_{\chi \chi}^{\textrm{tot}}$, the quadratic-in-spin contribution associated with the spin-induced quadrupole $\mathcal{F}_{C_Q}^{\textrm{tot}}$, and the electric and magnetic tidal quadrupole contributions $\mathcal{F}_{\mu_2}^{\textrm{tot}}$ and $\mathcal{F}_{\sigma_2}^{\textrm{tot}}$. Results are shown for three values of the dimensionless primary spin $\hat a=a/M$ and are accurate to all significant figures displayed.
}
\label{tab:flux0}
\end{table*}

\subsection{Results}
We compute the gravitational-wave fluxes~\eqref{flux_inf/H} (i) numerically on a two-dimensional Chebyshev parameter grid to facilitate interpolation and integration into waveform models with complete strong-field self-force information and (ii) as an analytic PN series expansion (following a standard approach~\cite{10.1143/PTP.95.1079,Sasaki:2003xr,Tanaka:1996ht, Akcay:2019bvk, Nagar:2019wrt,Skoupy:2024jsi}). 

We employ the same two-dimensional Chebyshev parameter grid $(36 \times 36)$ as in Ref.~\cite{Honet:2025gge}, containing 1296 data points, and spanning $r_0 \in [r_{\rm ISCO} ,\, 4 \times 10^{7} M]$, $\hat a \in [-0.993, \, 0.998]$, where $r_{\rm ISCO}$ corresponds to the innermost stable circular orbit (ISCO) of a geodesic in Kerr spacetime. The Chebyshev distribution provides enhanced resolution near the boundaries of the parameter space, where the fluxes exhibit the strongest variation, namely for orbits near the ISCO and for near-extremal black hole spins. The complete numerical dataset is provided as Supplemental Material~\cite{SupplementalMaterial}. 

We refer to the analytic expansion as post-Newtonian self-force (PNSF) to stress the difference from full PN theory, which does not involve expansions in the small mass ratio. Using our PNSF expansions, we assess convergence of the PN series to the numerical fluxes in the strong field, and we validate our calculations by analytically establishing agreement with the PN literature (see Table~\ref{tab:PN_orders}).

We first solve the homogeneous radial Teukolsky equation to obtain the radial solutions $R^{\rm in}_{\mode}$ and $R^{\up}_{\mode}$ and solve Eq.~\eqref{ang_Teuk} for the eigenvalue $\lambda_{\mode}$ and spheroidal harmonics. For this purpose, we employ the Black Hole Perturbation Toolkit's \texttt{Teukolsky} and \texttt{SpinWeightedSpheroidalHarmonics} packages~\cite{BHPT,bhpt_teukolsky, bhpt_swsh}; for numerical results we compute $R^{\rm in}_{\mode}$, $R^{\up}_{\mode}$ using the default Mano-Suzuki-Takasugi (MST) method~\cite{10.1143/PTP.95.1079}, while for analytical PNSF calculations we use the new ``PN'' functionality of the \texttt{Teukolsky} package, which computes the radial modes as analytic small frequency/large radii series expansions via a separate implementation of the MST method. The \texttt{Teukolsky} package interfaces directly with the \texttt{SpinWeightedSpheroidalHarmonics} package to compute $\lambda_{\mode}$ either numerically or as a PN expansion. The remaining ingredients for the fluxes in Eq.~\eqref{flux_inf/H} are analytic functions that we evaluate numerically or expand as PNSF expressions as required. In the mode sums, we sum over all $m$ and truncate the sum over $\ell$ at some $\ell_{max}$. In the PNSF expressions, we set $\ell_{max}=8$ as every $\ell+1$ mode is suppressed by a full PN order in comparison to the preceding $\ell$ mode and we keep six PN corrections to the leading PN term (which enters at $\ell=2$) in every flux expression. In Table~\ref{tab:PN_orders}, we list the PN orders of each expression relative to the leading order total energy flux. In the numeric flux calculations, $\ell_{max}$ is chosen to meet a fixed tolerance which we elucidate next.

Concretely, we evaluate twelve distinct expressions via Eq.~\eqref{0pa_1pa_flux} and Eq.~\eqref{2pa_flux} (six in the horizon fluxes and six in the fluxes at infinity): the point-mass term ($\mathcal{F}_{0}^{ A}$), the linear-in-spin correction ($\mathcal{F}_{\chi}^{A}$), the quadratic-in-spin term ($\mathcal{F}_{\chi \chi}^{A}$), the quadratic-in-spin contribution associated with the spin-induced quadrupole ($\mathcal{F}_{C_Q}^{A}$), and the electric ($\mathcal{F}_{\mu_2}^{A}$) and magnetic ($\mathcal{F}_{\sigma_2}^{A}$) quadrupole deformations. The $\ell$-mode summation is terminated when the maximum relative change between successive partial sums across all twelve terms satisfies
\begin{equation}
\max \left| 1-\frac{\mathcal{F}^{A}_{\ell-1,i}} {\mathcal{F}^{A}_{\ell,i}} \right| <10^{-7},
\end{equation}
where the subscript $\ell$ denotes the partial sum over all mode contributions up to and including $\ell$, and $i$ runs over the six flux contributions considered in this work for each $A=\infty$ or $A={\rm H}$. 
A small sample of the corresponding total fluxes ($\mathcal{F}^{\rm tot} = \mathcal{F}^{\rm H} + \mathcal{F}^{\infty}$) for each contribution is presented in Table~\ref{tab:flux0}. We have checked that our results for the point-mass and the linear-in-spin cases agree with the fluxes in Refs.~\cite{Mathews:2021rod,Honet:2025lmk}, the latter of which are computed from adapted codes from Refs.~\cite{vandeMeent:2015lxa,Piovano:2021iwv, Piovano:2024yks}. Meanwhile, the PNSF expressions agree with all available terms in the PN literature as summarized in Table~\ref{tab:PN_orders}. We have made all new PNSF results available in the \texttt{PostNewtonianSelfForce} package~\cite{bhpt_postnewtonianselfforce} of the Black Hole Perturbation Toolkit~\cite{BHPToolkit}.


To quantify the accuracy and convergence of the PNSF expansions ($\mathcal{F}_{\rm PN}$) against the numerical fluxes ($\mathcal{F}_{\rm Num}$) across the full parameter space, we compute the relative difference
\begin{equation}
   | \delta \mathcal{F} | = \Big| 1 - \frac{\mathcal{F}_{\rm PN}}{\mathcal{F}_{\rm Num}} \Big|.
\end{equation}
We display this relative difference in Figs.~\ref{fig_RelErrInfFlux} and \ref{fig_RelErrHorFlux} for the fluxes at infinity and at the horizon, respectively. We perform the comparisons over a wide two-dimensional parameter space with coordinates $(\sqrt{r_{\rm ISCO}/r_0}, \, \log(1-\hat a) )$ (recall $\hat a:= a/M$). 
This allows us to identify the regions where the analytical model (PNSF) performs best and where deviations become significant. The agreement is naturally strongest in the weak-field regime, corresponding to small values of $\sqrt{r_{\rm ISCO}/r_0}$, where the PN expansion converges rapidly and accurately captures both the orbital and field dynamics. Moving toward the ISCO, the relative difference gradually increases as strong-field effects become more important. More broadly, the relative differences remain reasonably small throughout a large portion of the parameter space, with our PNSF approximation achieving 2 digits of agreement with the numerics for all $r_0\gtrsim 6 r_{\rm ISCO}$ for the fluxes to infinity and for all $r_0\gtrsim 11 r_{\rm ISCO}$ for the fluxes through the horizon.  
   
However, the most striking feature of Figs.~\ref{fig_RelErrInfFlux} and \ref{fig_RelErrHorFlux} is the strong dependence on the sign of the primary spin. For retrograde orbits, corresponding to $\log(1- \hat a)>0$, our PNSF approximation agrees with the numerics to 2 digits all the way down to the ISCO for many of the flux contributions.  
This is again to be expected because the ISCO lies in a weaker-field, larger-radii regime for retrograde orbits, while the ISCO radius approaches the event horizon (which itself approaches $r_+=M$) for a near-extremal black hole ($\hat a \rightarrow 1$), suggesting a weak-field PNSF approximation should be liable to much smaller errors near a rapidly spinning black hole's retrograde ISCO than near its prograde ISCO.

\begin{figure}
\begin{center}
\includegraphics[width=\hsize]{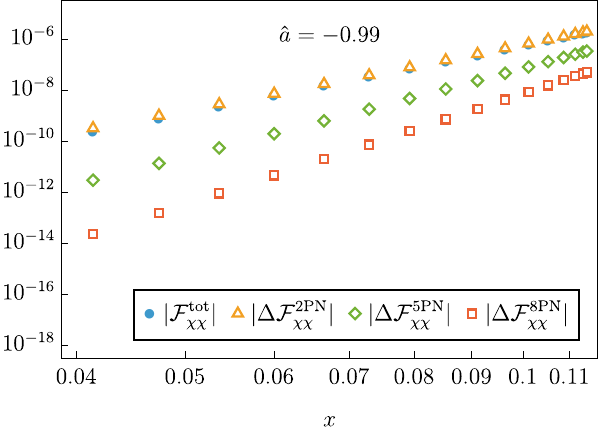}
\includegraphics[width=\hsize,trim={0 5pt 0pt 0pt},clip]{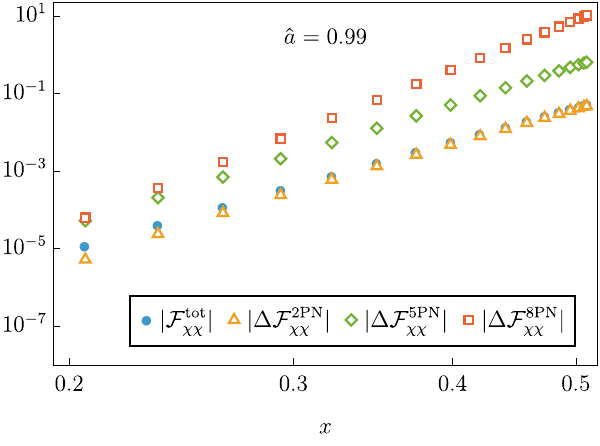}
\end{center}
\caption{
Comparison of the total numerical quadratic-in-spin flux correction, $\mathcal{F}_{\chi \chi}^{\rm tot}$, with the corresponding PN approximations at $2$PN (leading quadratic-in-spin order), $5$PN, and $8$PN (highest known PN order) for retrograde ($\hat a=-0.99$, upper panel) and prograde ($\hat a=0.99$, lower panel) orbits. The plotted quantities $|\Delta \mathcal{F}_{\chi \chi}^{\rm nPN}| = | \mathcal{F}_{\chi \chi}^{\rm tot} - \mathcal{F}_{\chi \chi}^{\rm nPN} |$, with $n=2,\, 5,\, 8$, measure the absolute deviation of each PN approximation from the total numerical flux.   
}\label{fig4}
 \end{figure}

\begin{figure}
\begin{center}
\includegraphics[width=\hsize]{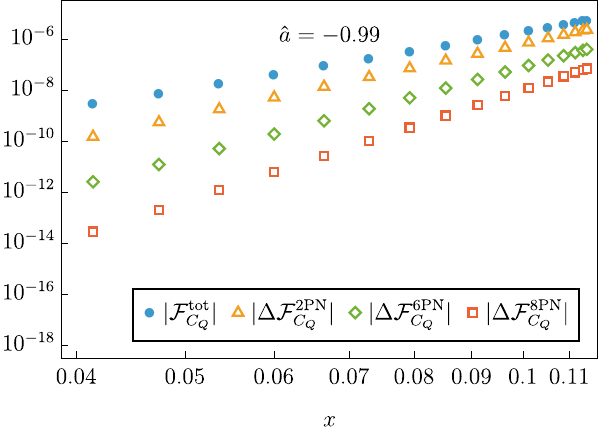}
\includegraphics[width=\hsize,trim={0 5pt 0pt 0pt},clip]{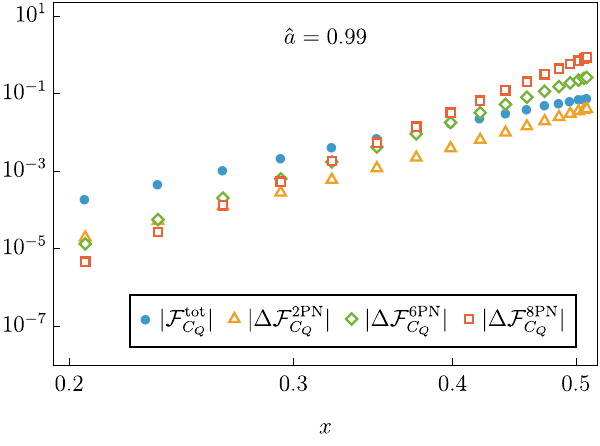}
\end{center}
\caption{Comparison of the total numerical quadratic-in-spin with spin-induced quadrupole flux correction, $\mathcal{F}_{C_Q}^{\rm tot}$, with the corresponding PN approximations at $2$PN (leading quadratic-in-spin with spin-induced quadrupole order), $6$PN, and $8$PN (highest available PN order) for retrograde ($\hat a=-0.99$, upper panel) and prograde ($\hat a=0.99$, lower panel) orbits. The plotted quantities $|\Delta \mathcal{F}_{C_Q}^{\rm nPN}| = | \mathcal{F}_{C_Q}^{\rm tot} - \mathcal{F}_{C_Q}^{\rm nPN} |$, with $n=2,\, 6,\, 8$, measure the absolute deviation of each PN approximation from the total numerical flux.
}\label{fig5}
 \end{figure}

\begin{figure}
\begin{center}
\includegraphics[width=\hsize]{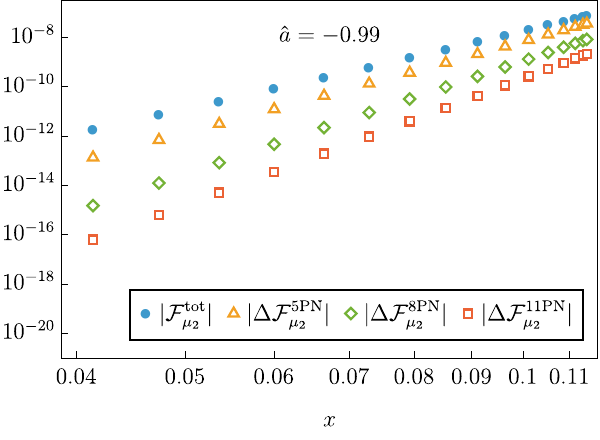}
\includegraphics[width=\hsize,trim={0 5pt 0pt 0pt},clip]{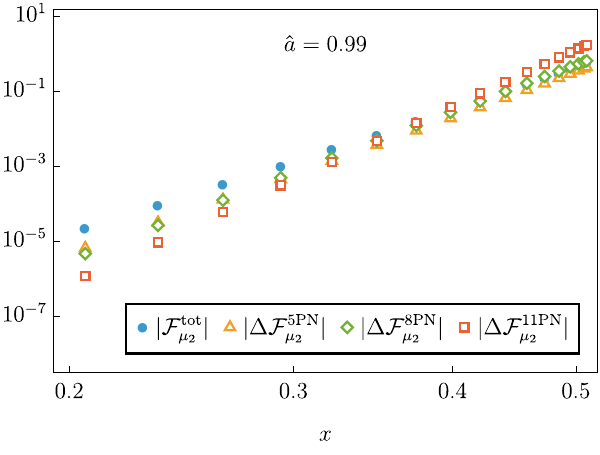}
\end{center}
\caption{Comparison of the total numerical electric tidal flux correction, $\mathcal{F}_{\mu_2}^{\rm tot}$, with the corresponding PN approximations at $5$PN (leading order), $8$PN, and $11$PN (highest known PN order) for retrograde ($\hat a=-0.99$, upper panel) and prograde ($\hat a=0.99$, lower panel) orbits. The plotted quantities 
$|\Delta \mathcal{F}_{\mu_2}^{\rm nPN}| = | \mathcal{F}_{\mu_2}^{\rm tot} - \mathcal{F}_{\mu_2}^{\rm nPN} |$, with $n=5,\, 8,\, 11$, measure the absolute deviation of each PN approximation from the total numerical flux.
}\label{fig6}
 \end{figure}

\begin{figure}
\begin{center}
\includegraphics[width=\hsize]{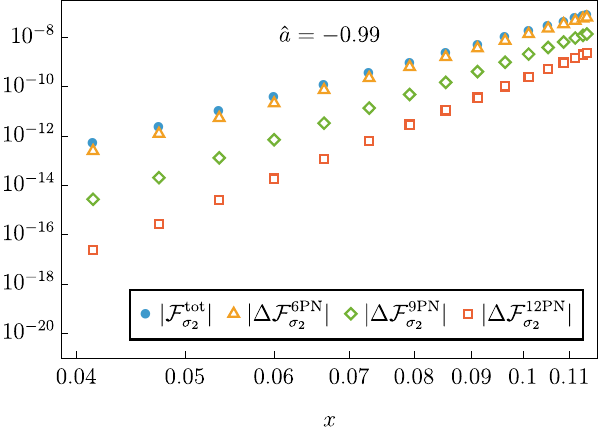}
\includegraphics[width=\hsize,trim={0 5pt 0pt 0pt},clip]{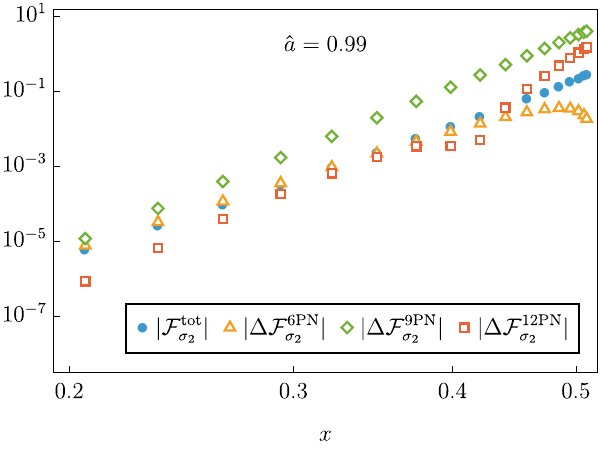}
\end{center}
\caption{Comparison of the total numerical magnetic tidal flux correction, $\mathcal{F}_{\sigma_2}^{\rm tot}$, with the corresponding PN approximations at $6$PN (leading order), $9$PN, and $12$PN (highest known PN-order) for retrograde ($\hat a=-0.99$, upper panel) and prograde ($\hat a=0.99$, lower panel) orbits. The plotted quantities 
$|\Delta \mathcal{F}_{\sigma_2}^{\rm nPN}| = | \mathcal{F}_{\sigma_2}^{\rm tot} - \mathcal{F}_{\sigma_2}^{\rm nPN} |$, with $n=6,\, 9,\, 12$, measure the absolute deviation of each PN approximation from the total numerical flux.
}\label{fig7}
 \end{figure}

To further illustrate the behavior of the PNSF expressions in the prograde and retrograde cases, we compare our total numerical fluxes with the corresponding PNSF expressions at successive PN orders, focusing on the strong-field regime at fixed values of the primary spin. This comparison is shown in Figs.~\ref{fig3}--\ref{fig7} for the linear-in-spin, quadratic-in-spin, spin-induced quadrupole, electric, and magnetic tidal deformation contributions, respectively. We specifically compare the total numerical flux with the leading-order PN term, an intermediate PN approximation, and our highest-order PN approximation as a function of the PN expansion parameter $x = (M \Omega)^{2/3} =1/[(r_{0}/M)^{3/2}+a/M]^{2/3}$.  

In all cases, the PN truncation errors grow with the PN parameter $x$, indicating the expected loss of accuracy of finite-order PN expansions as the binary approaches the strong-field regime. But the distinction between retrograde and prograde orbits is even clearer in Figs.~\ref{fig3}--\ref{fig7} than in the contour plots~\ref{fig_RelErrInfFlux} and \ref{fig_RelErrHorFlux}. For retrograde configurations with $\hat a=-0.99$, the PN approximations exhibit a systematic convergence pattern, with higher-order truncations yielding progressively smaller deviations from numerical results, even at the ISCO. In contrast, for prograde configurations with $\hat a=0.99$, the higher-order PN terms do not lead to a reduction in the truncation error, and in many cases, sufficiently close to the ISCO, higher-order approximations are consistently worse than lower-order ones.       

We emphasize that for sufficiently small $x$, we always find the expected behavior that higher-order PNSF expansions are better approximations than lower-order ones. And as described above, at small $x$ we find perfect order-by-order agreement between the numerics and the PNSF expansions. But our strong-field results do suggest a breakdown of the PN expansions in the strong-field regime that is accessible to prograde orbits. This has bearing on the long-standing question of whether PN expansions are genuinely convergent~\cite{Yunes:2008tw,Sun:2026lfd} (i.e., the PN expansion summed to infinite order at a given $x$ converges to a definite number) or only asymptotic (i.e., in the present context, $\displaystyle\lim_{x\to0}\frac{|\Delta{\cal F}^{n\rm PN}|}{x^{(n+5)}}=0$ for all $n\geq0$). Although it is little known in the literature, there \emph{are} formulations of PN expansions that are convergent~\cite{Oliynyk:2007uno,Oliynyk:2008vx}, and for nonspinning quasicircular binaries, there is no indication of non-convergence of the 0PA energy flux even when carried to 22PN and all the way to the ISCO~\cite{Fujita:2012cm}. But just as we have found here, previous studies of 0PA fluxes have observed an apparent failure of convergence in more complex strong-field scenarios~\cite{Fujita:2017wjq,Sago:2026gxb}. This does not strictly establish failure to converge (nor even a finite radius of convergence), which would require knowing the behavior to all PN orders. But it does clearly show the potential hazards of applying finite-order PN expansions in the strong-field regime.

%

\section{Conclusion}\label{Sec_6_Conclusion}

We have achieved several objectives:
\begin{enumerate}
    \item extended the multiscale expansion of Refs.~\cite{Mathews:2021rod} to 2PA order, including all quadratic-in-$\chi$ and spin-induced quadrupole effects. Like in Refs.~\cite{Mathews:2021rod}, this expansion is formulated in a fixed-frequency gauge that simplifies many calculations.
    \item incorporated the effects of a tidally induced quadrupole on the secondary. Though this is a 4PA term, it involves new physical deformability parameters, making it qualitatively distinct from the 3PA and other 4PA terms we neglect.
    \item derived a balance law relating the emitted gravitational-wave fluxes to the secondary's loss of mechanical energy and angular momentum. In practice this substantially simplifies waveform generation as it circumvents the need for calculations of local forces.
    \item analytically derived the fixed-frequency expansions of the conserved energy and angular momentum with a tidally induced quadrupole, completing the earlier calculations for the spin-induced quadrupole in Ref.~\cite{Bini:2015zya}.
    \item numerically computed the energy fluxes (to infinity and through the primary's horizon) proportional to $\chi^2$ and to each of the secondary's quadrupole parameters $C_Q$, $\mu_2$, and $\sigma_2$.
    \item analytically calculated the same fluxes in a weak-field (PNSF) expansion through 6PN order relative to the leading order of each individual term in the total flux. These expressions identically recover and extend the 4PN results in the literature.
\end{enumerate}
We have made our analytical and numerical results available in the Black Hole Perturbation Toolkit \texttt{PostNewtonianSelfForce} package~\cite{bhpt_postnewtonianselfforce}. Our numerical data is provided on the same Chebyshev grid used in Ref.~\cite{Honet:2025gge}.

Together, these results provide all the needed inputs to extend the waveform models in Refs.~\cite{Mathews:2025txc,Honet:2025gge,Honet:2025lmk} to the case of a star orbiting a black hole, while also improving the strong-field accuracy of the spin-squared terms in Refs.~\cite{Honet:2025gge,Honet:2025lmk} (which relied on a PN approximation for those terms). We will carry out that implementation in future work.

One of our main conclusions is that even high-order PN expansions can suffer from very low accuracy and extremely poor (or perhaps nonexistent) convergence in the strong-field regime near the ISCO for rapidly spinning primary black holes. We hence expect our results can be used to improve the treatment of finite-size effects in models that rely on PN inputs.

\begin{acknowledgments}
We would like to thank Soichiro Isoyama, Takahiro Tanaka, Paul Ramond, Viktor Skoup\'{y}, Vojt\v{e}ch Witzany, Georgios Lukes-Gerakopoulos, Gabriel Andres Piovano, and Loïc Honet for helpful discussions. We thank David Trestini particularly for guidance in navigating PN results. 
JM thanks Jakob Neef for assistance with the post-Newtonian functionality of the \texttt{Teukolsky} package~\cite{bhpt_teukolsky}. This work makes use of the \texttt{XAct} package~\cite{xACT}, Black Hole Perturbation Toolkit~\cite{BHPToolkit} and PNpedia~\cite{PNpedia}. MR has been partially supported by the JSPS KAKENHI Grant No. JP23KF0233. MS acknowledges support from the Czech Science Foundation (GA\v{C}R) under Grant number 25-15272I. AP acknowledges the support of a Royal Society Research Grant for Research Fellows and the UKRI Frontier Research Grant GWModels (as selected by the ERC and funded by UKRI under grant number EP/Y008251/1). JM acknowledges support from the NUS Faculty of Science, under the research grant 22-5478-A0001.

\end{acknowledgments}

\appendix

%


\section{Relationship between conserved charges}\label{app:charges}

In this appendix we relate the conserved charge $Q_\xi$ to the energy or angular momentum $P_\xi$, as defined in Eqs.~\eqref{eq:Q def} and \eqref{eq:P def}. 

Referring to the definition of $Q_\xi$, we consider spatial surfaces $\Sigma_t$ of constant time $t$, where $t$ need not be Boyer-Lindquist time. The future-directed surface element on each surface is then $d\hat\Sigma_\alpha = - t_\alpha \sqrt{-\hat g}\,d^3y$ with $t_\alpha:=\partial_\alpha t$ and with coordinates $y^i$ on $\Sigma_t$. To conveniently evaluate the integral over~$\Sigma_t$, we write the stress-energy tensor~\eqref{SET} as
\begin{multline}
    \hat T^{\alpha\beta}(x) = \int dt \Bigl[\hat t^{\alpha\beta}(x,x_p) + \hat\nabla_\gamma \hat t^{\alpha\beta\gamma}(x,x_p) \\
    + \hat\nabla_\gamma\hat\nabla_\delta \hat t^{\alpha\beta\gamma\delta}(x,x_p)\Bigr],
\end{multline}
where we have used $d\hat\tau = \frac{dt'}{\hat v^t(x_p(t'))}$. We then define a charge density $j^\alpha_\xi t_\alpha = T^{\alpha\beta}\hat\xi_\beta t_\alpha$, which can be written as
\begin{align}
    j^\alpha_\xi t_\alpha 
    &= \int dt'\, \bigl(\hat t^{\alpha\beta} t_\alpha  - \hat t^{\alpha\beta\gamma}\hat\nabla_\gamma t_\alpha + \hat t^{\alpha\beta\gamma\delta}\hat\nabla_\delta\hat\nabla_\gamma t_\alpha \bigr) \hat\xi_\beta  \non\\
    &\quad + \int dt' \,\bigl(-\hat t^{\alpha\beta\gamma}t_\alpha + 2\hat t^{\alpha\beta(\gamma\delta)}\hat\nabla_\delta t_\alpha\bigr)\hat\nabla_\gamma\hat\xi_\beta 
   \nonumber\\
    &\quad  + \int dt' \,\hat t^{\alpha\beta\gamma\delta}t_\alpha\hat\nabla_\delta\hat\nabla_\gamma\hat\xi_\beta \non\\
    &\quad + \int dt'\, \hat \nabla_\gamma\bigl(\hat t^{\alpha\beta\gamma}t_\alpha\hat\xi_\beta  + \hat\nabla_\delta \hat t^{\alpha\beta\gamma\delta}t_\alpha\hat\xi_\beta \non\\
    &\qquad\qquad\qquad - \hat t^{\alpha\beta\delta\gamma}t_\alpha\hat\nabla_{\delta}\hat\xi_\beta \bigr). 
\end{align}
Expressing $\hat t^{\alpha\beta\cdots}$ as $\hat t^{\alpha\beta\cdots} = \hat j^{\alpha\beta\cdots}\frac{\delta^4(x-x_p(t'))}{\sqrt{-\hat g}}$ to evaluate the integral $-\int j^\alpha_\xi t_\alpha \sqrt{-\hat g}d^3y$, we immediately find the charge at time $t$ reduces to 
\begin{align}\label{Q_xi}
    Q_\xi(\Sigma_t) &= -\bigl(\hat j^{\alpha\beta} t_\alpha  - \hat j^{\alpha\beta\gamma}\hat\nabla_\gamma t_\alpha + \hat j^{\alpha\beta\gamma\delta}\hat\nabla_\delta\hat\nabla_\gamma t_\alpha \bigr) \hat\xi_\beta  \non\\
    &\quad - \bigl(-\hat j^{\alpha\beta\gamma}t_\alpha + 2\hat j^{\alpha\beta(\gamma\delta)}\hat\nabla_\delta t_\alpha\bigr)\hat\nabla_\gamma\hat\xi_\beta 
   \nonumber\\
    &\quad  - \hat j^{\alpha\beta\gamma\delta}t_\alpha\hat\nabla_\delta\hat\nabla_\gamma\hat\xi_\beta ,
\end{align}
where the right-hand side is evaluated at $x=x_p(t)$. Note that the divergence term in $j^\alpha_\xi t_\alpha$ does not contribute to $Q_\xi$ because it integrates to boundary terms at $t'=\pm\infty$; these vanish due to the delta function $\delta^4(x-x_p(t'))$, which has support only at $t'=t$.

The first term in $Q_\xi$ evaluates to
\begin{align}
    \hat j^{\alpha\beta} t_\alpha \hat\xi_\beta &= \frac{\hat p^{\alpha}\hat v^\beta}{2\hat v^t} ( t_\alpha \hat\xi_\beta +  t_\beta \hat\xi_\alpha) -\frac{1}{3\hat v^t}\tJ^{\gamma \delta \epsilon (\alpha }\hR^{\beta )}{}_{\epsilon \gamma \delta } t_\alpha \hat\xi_\beta \non\\
    &= \hat p^\alpha\xi_\alpha + \frac{2}{3}\hR_{\alpha\mu\gamma}{}^{[\beta}\tJ^{\nu]\gamma\alpha\mu}\hv_{\nu}\hat\xi_\beta \non\\
    &\quad +\frac{1}{4m_d}\hR_{\nu\gamma\alpha\mu}\tS^{\alpha\mu}\tS^{\beta\nu}\hat{v}^{\gamma}\hat\xi_\beta \non\\
    &\quad -\frac{2}{3\hat v^t}\hR_{\mu\beta\gamma}{}^{[\alpha}\tJ^{\nu]\gamma\mu\beta}\hv_{\nu}t_\alpha \hv^\rho\hat\xi_\rho \non\\
    &\quad -\frac{1}{4m_d\hat v^t}\hR_{\nu\gamma\mu\beta}\tS^{\mu\beta}\tS^{\alpha\nu}\hat{v}^{\gamma} t_\alpha \hv^\rho\hat\xi_\rho\non\\
    &\quad -\frac{1}{3\hat v^t}\tJ^{\gamma \delta \epsilon (\alpha }\hR^{\beta )}{}_{\epsilon \gamma \delta } t_\alpha \hat\xi_\beta,\label{eq:jtxi}
\end{align}
where we made use of Eq.~\eqref{SET} in the first line and the relationship~\eqref{rel_v_p} between the four-velocity and the momentum in obtaining the second equality.

To simplify the other terms in $Q_\xi$, we note the identities
\begin{align}
    \hat\nabla_\gamma t_\alpha &=\hat\nabla_\gamma\hat\nabla_\alpha t = \hat\nabla_\alpha t_\gamma,\label{eq:Dt}\\
    \hat\nabla_\delta\hat\nabla_\gamma t_\alpha &= \hat\nabla_\gamma\hat\nabla_\delta t_\alpha + \hat R^\beta{}_{\alpha\gamma\delta}t_\beta,\label{eq:DDt}\\
    \hat\nabla_\gamma\hat\xi_\beta &= -\hat\nabla_\beta\hat\xi_\gamma + {\cal L}_\xi h^R_{\beta\gamma},\label{eq:Dxi}\\
    \hat\nabla_\delta\hat\nabla_\gamma\hat\xi_\beta &= \hat R_{\alpha\gamma\delta\beta}\xi^\alpha + {\cal O}(\e).\label{eq:DDxi}
\end{align}
The last equality would be an exact, standard identity if $\xi^\alpha$ were an exact Killing field; the omitted order-$\e$ terms all involve $2\hat\nabla_{(\alpha}\hat\xi_{\beta)} = {\cal L}_\xi h^R_{\alpha\beta}$. 
Using these identities, Eq.~\eqref{SET}, and the symmetries of $\hat S^{\alpha\beta}$ and $\hat J^{\alpha\beta\gamma\delta}$, we obtain
\begin{align}
    - &\hat j^{\alpha\beta\gamma}\hat\nabla_\gamma t_\alpha\hat\xi_\beta -\hat j^{\alpha\beta\gamma}t_\alpha \hat\nabla_\gamma\hat\xi_\beta \non\\
    &= \frac{\hat S^{\gamma\alpha}\hat v^{\beta}}{\hat v^t}(\hat\nabla_\gamma t_{(\alpha}\hat\xi_{\beta)} +  \hat\nabla_\gamma\hat\xi_{(\alpha} t_{\beta)})\non\\
    &= \frac{1}{2}\hat S^{\gamma\alpha}\hat\nabla_\gamma\hat\xi_{\alpha} + \frac{\hat S^{\gamma\alpha}\hat v^{\beta}}{2\hat v^t}[2\hat\nabla_\beta (t_{\gamma}\hat\xi_{\alpha}) - t_\gamma\hat\nabla_\beta \hat\xi_{\alpha} \non\\
    &\qquad\qquad\qquad\qquad\qquad + \hat\xi_\gamma\hat\nabla_\beta t_\alpha+ t_\alpha{\cal L}_\xi h^R_{\beta\gamma}].\label{eq:jdtxi}
\end{align}
Similarly, 
\begin{align}\label{eq:jddtxi}
     \hat j^{\alpha\beta\gamma\delta}&(\hat\nabla_\delta\hat\nabla_\gamma t_\alpha \hat\xi_\beta +t_\alpha\hat\nabla_\delta\hat\nabla_\gamma\hat\xi_\beta) +2\hat j^{\alpha\beta(\gamma\delta)}\hat\nabla_\delta t_\alpha\hat\nabla_\gamma\hat\xi_\beta \non\\
    &=-\frac{2}{3v^t} \hat J^{\gamma(\alpha\beta)\delta}(\hat\nabla_\delta\hat\nabla_\gamma t_\alpha \hat\xi_\beta +t_\alpha\hat\nabla_\delta\hat\nabla_\gamma\hat\xi_\beta) \non\\
    &\quad -\frac{4}{3v^t} \hat J^{\gamma(\alpha\beta)\delta}\hat\nabla_\delta t_\alpha\hat\nabla_\gamma\hat\xi_\beta \non\\
     &= \frac{1}{3\hat v^t}\hat R_{\epsilon\mu\gamma}{}^\beta(\hat J^{\epsilon\alpha\gamma\mu}t_\beta\hat\xi_\alpha - 2\hat J^{\gamma(\epsilon\alpha)\mu}t_\alpha\hat\xi_\beta) + {\cal O}(\varepsilon^4). 
\end{align}
In obtaining the third line, we made use of Eqs.~\eqref{eq:Dt}, \eqref{eq:DDt}, and \eqref{eq:DDxi}, along with the symmetries of $\hat J^{\alpha\beta\gamma\delta}$, to eliminate second derivatives. We also noted that the final term on the second line vanishes because $\hat J^{\gamma(\alpha\beta)\delta}\hat\nabla_\delta t_\alpha$ is symmetric in $(\gamma,\beta)$ while $\hat\nabla_\gamma\hat\xi_\beta$ is antisymmetric in $(\gamma,\beta)$ (at leading order in $\varepsilon$).

All of the above is valid for generic slices $\Sigma_t$. $Q_\xi$ is conserved for any choice of slicing, but we can simplify its relationship with $P_\xi$ by choosing these slices such that their normal vector becomes parallel to $\hat p^\alpha$ at the point where they intersect the particle's worldline: 
\begin{equation}\label{eq:p=-mNt}
\hat p^\alpha = -\frac{m_d\hat g^{\alpha\beta}t_\beta}{\sqrt{-\hat g^{\mu\nu} t_\mu t_\nu} } := - m_d\hat N \hat t^\alpha, 
\end{equation}
implying 
\begin{equation}\label{eq:vt}
\hat v^t = -\displaystyle\frac{\hat v^\alpha\hat p_\alpha}{m_d\hat N} = \displaystyle\frac{m_0}{m_d\hat N},
\end{equation}
where $\hat N = 1/\sqrt{-\hat g^{\mu\nu} t_\mu t_\nu}$ is the lapse. Our calculation below is largely identical (and our conclusion is precisely identical) if instead we choose the slicing such that the normal vector is parallel to $\hat v^\alpha$.

With our choice of slicing, $\hat v_{\nu}t_\alpha=\hat v_{(\nu}t_{\alpha)} + {\cal O}(\varepsilon^2)$, $\hat S^{\gamma\alpha}t_\gamma = 0$, and 
\begin{align}
\hat S^{\gamma\alpha}\hat v^\beta\hat\nabla_\beta t_\alpha &= - t_\alpha \hat v^\beta\hat\nabla_\beta\hat S^{\gamma\alpha}.
\end{align}
Given these relationships along with Eqs.~\eqref{eq:p=-mNt} and~\eqref{eq:vt}, Eq.~\eqref{eq:jtxi} reduces to
\begin{align}\label{term_pxi}
    \hat j^{\alpha\beta} t_\alpha \hat\xi_\beta 
    &= \hat p^\alpha\xi_\alpha + \frac{1}{6}\hR_{\epsilon\mu\gamma}{}^{\beta}\tJ^{\alpha\gamma\epsilon\mu}(3\hv_{\alpha}\hat\xi_\beta - \hv_{\beta}\hat\xi_\alpha) \non\\
    &\quad +\frac{1}{4m_d}\hR_{\nu\gamma\alpha\mu}\tS^{\alpha\mu}\tS^{\beta\nu}\hat{v}^{\gamma}\hat\xi_\beta +{\cal O}(\e^5).    
\end{align}
Appealing to the same identities and the MPD equation~\eqref{MPD_torque}, we similarly reduce Eq.~\eqref{eq:jdtxi} to
\begin{align}\label{term_sxi}
        - &\hat j^{\alpha\beta\gamma}\hat\nabla_\gamma t_\alpha\hat\xi_\beta -\hat j^{\alpha\beta\gamma}t_\alpha \hat\nabla_\gamma\hat\xi_\beta \non\\
        &= \frac{1}{2}\hat S^{\gamma\alpha}\hat\nabla_\gamma\hat\xi_{\alpha} -\frac{1}{4m_d}\hat R_{\nu\gamma\alpha\mu}\hat S^{\alpha\mu}\hat S^{\beta\nu}\hat v^\gamma\hat\xi_{\beta}  + {\cal O}(\e^4).
\end{align}

At this stage, it is easily seen that the spin-squared terms cancel when we sum Eqs.~\eqref{eq:jddtxi}, \eqref{term_pxi}, and \eqref{term_sxi} to obtain $Q_\xi$. However, it is not so easily seen that the quadrupole terms cancel. To establish that they do in fact cancel, we use the cyclic symmetry of $\hat J^{\alpha\beta\gamma\delta}$, along with Eq.~\eqref{eq:p=-mNt}, to rewrite Eq.~\eqref{eq:jddtxi} as
\begin{align}\label{term_jvxi}
         \hat j^{\alpha\beta\gamma\delta}&(\hat\nabla_\delta\hat\nabla_\gamma t_\alpha \hat\xi_\beta +t_\alpha\hat\nabla_\delta\hat\nabla_\gamma\hat\xi_\beta) +2\hat j^{\alpha\beta(\gamma\delta)}\hat\nabla_\delta t_\alpha\hat\nabla_\gamma\hat\xi_\beta \non\\
         &= -\frac{1}{3}{\sf B}^{\beta\alpha}\hv_\beta \hat\xi_\alpha -\frac{1}{3}({\sf B}^{\beta\alpha}+2{\sf A}^{\beta\alpha})\hv_\alpha\hat\xi_\beta + {\cal O}(\varepsilon^4).
\end{align}
Here we have defined the helper quantities
\begin{align}
\mathsf{A}^{\alpha\beta}&:=\hR_{\epsilon\mu\gamma}{}^{\alpha}\hat J^{\beta\gamma\epsilon\mu}, \label{Aab}\\
\mathsf{B}^{\alpha\beta}&:=\hR_{\epsilon\mu\gamma}{}^{\alpha}\hat J^{\beta\epsilon\mu\gamma}.
\end{align}
Again using the cyclic symmetry of $\hat J^{\alpha\beta\gamma\delta}$, we can readily show
$\mathsf{B}^{\alpha\beta}
=-\mathsf{B}^{\alpha\beta}-\mathsf{A}^{\alpha\beta}$.
Hence,
\beq\label{Bab}
\mathsf{B}^{\alpha\beta}=-\frac{1}{2}\mathsf{A}^{\alpha\beta}.
\eeq
Using this to replace $\mathsf{B}^{\alpha\beta}$ in Eq.~\eqref{term_jvxi}, we obtain 
\begin{align}\label{term_jvxi1}
     &\hat j^{\alpha\beta\gamma\delta}(2\hat\nabla_\delta\hat\nabla_\gamma t_{(\alpha} \hat\xi_{\beta)}) +2\hat j^{\alpha\beta(\gamma\delta)}\hat\nabla_\delta t_\alpha\hat\nabla_\gamma\hat\xi_\beta \non\\
    &= \frac{1}{6} \mathsf{A}^{\beta\alpha}\hat v_{\beta}\hat \xi_\alpha-\frac{1}{2}\mathsf{A}^{\beta\alpha} \hv_\alpha\hat\xi_\beta + {\cal O}(\varepsilon^4)\non\\
    &=-\frac{1}{6}\hR_{\epsilon\mu\gamma}{}^{\beta}\hat J^{\alpha\gamma\epsilon\mu}\left(3 \hv_{\alpha}\xi_\beta - \hv_{\beta}\xi_{\alpha}\right) + {\cal O}(\varepsilon^4). 
\end{align}

Combining Eqs.~\eqref{term_pxi}, \eqref{term_sxi}, and \eqref{term_jvxi1}, we now obtain our desired result:
\beq
Q_{\xi}=P_{\xi} + {\cal O}(\varepsilon^4).
\eeq

\section{Teukolsky source and amplitudes}\label{App:source_term}

In this appendix, we derive formulas for the source term in the radial Teukolsky equation~\eqref{rad_Teuk} and for the resulting amplitudes~\eqref{eq:Teukolsky amplitudes} that determine the energy fluxes at infinity and the horizon. To avoid an overload of notation, we use $T^{\alpha\beta}$ to denote the test-body stress-energy tensor, denoted $T^{\alpha\beta}_{\rm test}$ in the body of the paper.

\subsection{Teukolsky source}

For a generic stress-energy tensor, the source term for the radial Teukolsky equation~\eqref{rad_Teuk} is given by~\cite{Piovano:2020zin}
\beq\label{source_rad_teuk}
\mathcal{J}_{l m\omega}=\int dt d\theta d\phi \Delta^2 e^{i(\omega t-m\phi)}\left(\mathcal{J}_{nn}+\mathcal{J}_{\overline{m}n}+\mathcal{J}_{\overline{m}\, \overline{m}}\right),
\eeq
where the integrands are constructed from projections of $T_{\alpha\beta}$ onto a complex null tetrad $\{l^{\mu},~n^{\mu},~m^{\mu},~\overline{m}^{\mu}\}$:
\beq\label{source_Teulkol}
\mathcal{J}_{nn} &=\left[f^{(0)}_{nn} T_{nn}\right]\,,\quad \mathcal{J}_{\overline mn} =\partial_r\left[f^{(1)}_{\overline mn} T_{\overline mn}\right]+\left[f^{(0)}_{\overline mn}T_{\overline mn}\right],\\
\mathcal{J}_{\overline{m}\, \overline{m}} &=\partial^2_r\left[f^{(2)}_{\overline{m}\, \overline{m}}T_{\overline{m}\, \overline{m}}\right]+\partial_r\left[f^{(1)}_{\overline{m}\, \overline{m}}T_{\overline{m}\, \overline{m}}\right]+\left[f^{(0)}_{\overline{m}\, \overline{m}}T_{\overline{m}\, \overline{m}}\right].
\eeq
In the context of a multiscale expansion, one does not work with true Fourier transforms and their inverses but instead with discrete Fourier series $\sum_{\vec k}e^{-i\vec k\cdot \vec \varphi}$ in the orbital phases $\vec{\varphi}=(\varphi^r,\varphi^\theta,\varphi^\phi)$~\cite{Pound:2021qin}, and the label $\omega$ is replaced by linear combinations of the orbital frequencies, $\omega_{\vec k} := \vec k\cdot\vec\Omega$. One should then replace $\int dt\, e^{i\omega t}$ in Eq.~\eqref{source_rad_teuk} with $\frac{1}{(2\pi)^3}\oint d^3\varphi\, e^{i\vec k\cdot \vec \varphi}$. For our case of quasicircular orbits, we have only one phase and frequency, such that $\omega$ becomes $\omega_m:= m\Omega$, and the integral becomes $\frac{1}{2\pi}\oint d\phi_p\, e^{im\phi_p}$. We return to this point later in the Appendix.

The coefficients in Eq.~\eqref{source_Teulkol} are given by 
\begingroup%
\allowdisplaybreaks%
\begin{subequations}    
\begin{align}
f^{(0)}_{nn} &=-\frac{2\sin\theta}{\Delta^2\rho^3\bar \rho}\left[\left(\mathcal L_{1}^{\dagger}-2i a \rho \sin\theta\right)\mathcal L_{2}^{\dagger}\Ang\right],\\
f^{(1)}_{\overline{m}n} &=\frac{4\sin \theta}{\sqrt{2}} \bigg[ 
\Big(\mathcal{L}^\dagger_2\Ang + i a \sin\theta(\bar{\rho}- \rho) \Ang \Big)\frac{1}{\rho^3 
\Delta}\bigg],\\
%
f^{(0)}_{\overline{m}n} &=\frac{4\sin \theta}{\sqrt{2}}\bigg[\bigg( \frac{i K}{\Delta} + \rho +\bar{\rho}\bigg)\mathcal{L}^\dagger_2\Ang \non\\*
&\quad - a 
\sin(\theta)\frac{K}{\Delta}(\bar{\rho}-\rho) \Ang \bigg] \frac{1}{\rho^3 \Delta},\\
f^{(2)}_{\overline{m}\, \overline{m}} &=- \left( \frac{\bar{\rho}}{\rho^3}  \right)\sin(\theta) \Ang,\\ 
%
f^{(1)}_{\overline{m}\, \overline{m}} &= - 2 \left[ \left( \frac{\bar{\rho}}{\rho^2} 
+\frac{\bar{\rho}}{\rho^3}\frac{i K}{\Delta} \right)\sin(\theta) \Ang   \right],\\
f^{(0)}_{\overline{m}\, \overline{m}} &=\frac{\bar{\rho}}{\rho^3}\!\left( \partial_{r}\!\left( \frac{i K}{\Delta} \right) \ - 2 \rho \frac{ i K}{\Delta} + 
\frac{K^2}{\Delta^2} \right)\sin(\theta) \Ang, 
\end{align}
\end{subequations}
\endgroup%
with
\begin{subequations}
\begin{align}
\rho &= \frac {1} {r - i a\cos \theta},  \quad \bar{\rho} = \frac {1}{r + i a\cos \theta},\\
\mathcal{L} _s &= \frac{\partial}{\partial\theta}+\frac{m}{\sin \theta} - a{\omega} \sin \theta + s \cot \theta ,\\
\mathcal{L} _s^\dagger &= \frac{\partial}{\partial\theta}-\frac{m}{\sin \theta} + a{\omega} \sin\theta + s 
\cot \theta.
\end{align}
\end{subequations}
The projected components of the stress-energy tensor, 
\begin{subequations}
    \begin{align}
T_{nn}&=T_{\alpha\beta}n^{\alpha}n^{\beta}\,,\\
      T_ {\overline{m}n}&=T_{\alpha\beta}\overline{m}^{\alpha}n^{\beta}\,,\\
      T_{\overline{m}\, \overline{m}}&=T_{\alpha\beta}\overline{m}^{\alpha}\overline{m}^{\beta}\,,
    \end{align}
\end{subequations}
are specifically given in terms of the Kinnersley tetrad,
\begin{subequations}
\begin{align}
l^\mu &= \frac{1}{\Delta}\left(r^2+a^2,\Delta,0,a\right),\\
n^{\mu} &= \frac{1}{2\Sigma}\left(r^2+a^2,-\Delta,0,a\right),\\
m^{\mu}&= \frac{\bar\rho}{\sqrt{2}}\left(i a \sin\theta,0,1,\frac{i}{\sin\theta}\right),
\end{align}
\end{subequations}
with $\overline{m}^{\mu}$ the complex conjugate of $m^{\mu}$.

The stress-energy tensor of a spinning, deformed object is given in Eqs.~\eqref{SET} and~\eqref{SET1}. Leaving the trajectory arbitrary but neglecting $h^{\rm R}_{\alpha\beta}$, we reduce those equations to 
\begin{align}
  T^{\alpha\beta} = \bigg[T_{(0)}^{\alpha\beta} \delta^3+ \nabla_\mu
  \Big(T_{(1)}^{\alpha\beta\mu} \delta^3 \Big) + \nabla_\mu \nabla_\nu
  \Big(T_{(2)}^{\alpha\beta\mu\nu} \delta^3 \Big)
\bigg] \,,
\end{align}
where we have defined
\begin{subequations}\label{eq:TmunuU}
\begin{align}
  &T_{(0)}^{\alpha\beta} = \frac{1}{v^t \sqrt{-g}} \Big( \, p^{(\alpha}v^{\beta)} +
  \frac{1}{3} R^{(\alpha}{}_{\,\lambda\mu\nu} J^{\beta)\lambda\mu\nu}\Big)\,,\\
  &T_{(1)}^{\alpha\beta\mu} = 
  \frac{1}{v^t \sqrt{-g}}  v^{(\alpha}S^{\beta)\mu}\,,\\
  &T_{(2)}^{\alpha\beta\mu\nu} =  \frac{1}{v^t \sqrt{-g}} \Big(-\frac{2}{3}
  J^{\mu(\alpha\beta)\nu}\Big) \,,
\end{align}
\end{subequations}
and $\delta^3:=\delta_r(r-r_p)\delta_{\theta}(\theta-\theta_p)\delta_{\phi}(\phi-\phi_p)$. 
We next rewrite the above equation in terms of partial derivatives, giving it the following form~\cite{Bohe:2015ana}:
\begin{align}
  \sqrt{-g}T^{\alpha\beta} =\bigg[\mathcal{P}^{\alpha\beta} \delta^3 + 
 \partial_\mu 
  \Big(\mathcal{S}^{\alpha\beta\mu} \delta^3 \Big) + 
  \partial_{\mu\nu} \Big(\mathcal{Q}^{\alpha\beta\mu\nu} \delta^3 \Big)
  \bigg] \,,
\end{align}
with
\begin{subequations}
\begin{align}
  &\mathcal{P}^{\alpha\beta} =\sqrt{-g}\left[T_{(0)}^{\alpha\beta} + 2
  \Gamma^{(\alpha}_{\,\mu\nu} T_{(1)}^{\beta)\mu\nu} + (\partial_\lambda
  \Gamma^{(\alpha}_{\mu\nu})\right]\, T_{(2)}^{\beta)\lambda\mu\nu}
  \nonumber \\ & \qquad + \Gamma^{(\alpha}_{\rho\lambda} (T_{(1)}^{\beta)\lambda\mu\nu}
  \Gamma^{\rho}_{\mu\nu} - \Gamma^{\beta)}_{\mu\nu}
  T_{(2)}^{\rho\lambda\mu\nu})\,,\\
  &\mathcal{S}^{\alpha\beta\mu} =  \sqrt{-g} \left[T_{(1)}^{\alpha\beta\mu} +
  \Gamma^{\mu}_{\nu\lambda} T_{(2)}^{\alpha\beta\nu\lambda} - 
  2 \Gamma^{(\alpha}_{\nu\lambda} T_{(2)}^{\beta)\mu\nu\lambda}\right]\,,\\
  &\mathcal{Q}^{\alpha\beta\mu\nu} =  \sqrt{-g} \, T_{(2)}^{\alpha\beta\mu\nu} \,.
\end{align}
\end{subequations}

From now on, we focus only on equatorial circular orbits, meaning $\theta_p=\pi/2$ within $\delta^3$ and $d\phi_p/dt=\Omega$. 
With this, the above equation simplifies to   
\bea\non
\sqrt{-g} \, T^{\alpha\beta}&=& \left(K_{000}^{\alpha\beta} \, \delta_{\phi} + K_{001}^{\alpha\beta} \, \delta^{\prime}_{\phi} + K_{002}^{\alpha\beta} \, \delta_{\phi}^{\prime \prime} \right) \delta_r\delta_\theta\\\non
&+& \left(K_{100}^{\alpha\beta} \, \delta_{\phi} + K_{101}^{\alpha\beta} \, \delta^{\prime}_{\phi} \right) \delta_r^{\prime} \delta_\theta + \left( K_{200}^{\alpha \beta} \delta_{\phi} \right) \delta_r^{\prime\prime} \delta_\theta\\\non
&+& \left(K_{020}^{\alpha \beta}\, \delta_{\phi} \right) \delta_r \delta_\theta^{\prime \prime} +  \left(K_{110}^{\alpha \beta}\, \delta_{\phi} \right) \delta_r^{\prime} \delta_\theta^{\prime}\\
&+& \left(K_{010}^{\alpha\beta} \, \delta_{\phi} + K_{011}^{\alpha\beta} \, \delta^{\prime}_{\phi} \right) \delta_r \delta_\theta^{\prime}\,,\label{SET_1}
\eea
where the symbol $'$ represents the derivative of the function with respect to its argument and 
\begin{subequations}
\begin{align}
K_{000}^{\alpha\beta} &=\left[\mathcal P^{\alpha\beta}\right]_{x_p},\,\quad K_{100}^{\alpha \beta} = \left[\mathcal{S}^{\alpha\beta r} \right]_{x_p},\\
K_{001}^{\alpha\beta}&=\left[\mathcal{S}^{\alpha\beta\phi}-\Omega \mathcal S^{\alpha\beta t}\right]_{x_p}, \\
K_{010}^{\alpha \beta} &= \left[\mathcal{S}^{\alpha\beta \theta}\right]_{x_p}, \quad K_{110}^{\alpha\beta} = \left[2 \mathcal{Q}^{\alpha \beta r \theta} \right]_{x_p},\\ 
K_{002}^{\alpha\beta}&=\left[\mathcal Q^{\alpha\beta\phi\phi}-2\Omega \mathcal Q^{\alpha\beta t\phi}+\Omega^2 \mathcal Q^{\alpha\beta tt}\right]_{x_p},\\
K_{101}^{\alpha\beta}&=\left[2\mathcal Q^{\alpha\beta r\phi}-2\Omega \mathcal Q^{\alpha\beta tr}\right]_{x_p},\\
K_{011}^{\alpha \beta} &= \left[2 \mathcal{Q}^{\alpha \beta \theta \phi} - 2 \Omega \mathcal{Q}^{\alpha \beta t \theta} \right]_{x_p}, \\
\quad K_{200}^{\alpha\beta} &= \left[\mathcal Q^{\alpha \beta r r} \right]_{x_p}, \quad K_{020}^{\alpha\beta} = \left[\mathcal Q^{\alpha \beta \theta \theta}\right]_{x_p},
\end{align}
\end{subequations}
where $[\cdot]_{x_p}$ indicates the function is evaluated at $x^{i}_{p}=(r_p,\pi/2,\phi_p)$. In the above equation, we have used the distributional properties of the Dirac delta function,
\beq\label{dirac_delta}
\mathfrak{f}(x)\delta^{(n)}(x-x_0)=\sum_{k=0}^{n}{}^n\mathbf{C}_k(-1)^k\delta^{(n-k)}(x-x_0)\mathfrak{f}^{(k)}(x_0),
\eeq
where $\mathfrak{f}(x)$ is an arbitrary smooth function,  ${}^n\mathbf{C}_k=n!/(k!(n-k)!)$, and $\mathfrak{f}^{(n)}(x)$ represents the $n$th derivative of the function with respect to its argument.

For our purpose, it is better to rewrite Eq.~\eqref{SET_1} as 
\begin{multline}
\sqrt{-g} \, T^{\alpha\beta} = K_{00}^{\alpha\beta}\delta_r \delta_\theta + K_{10}^{\alpha\beta}\delta_r^{\prime} \delta_{\theta} + K_{20}^{\alpha\beta}\delta_r^{\prime \prime} \delta_{\theta} \\ 
+ K_{01}^{\alpha \beta} \delta_{r} \delta_\theta^{'} + K_{11}^{\alpha \beta} \delta_{r}^{\prime} \delta_{\theta}^{\prime} + K_{02}^{\alpha \beta} \delta_{r} \delta_{\theta}^{\prime \prime}\label{SET_2},
\end{multline}
where we have defined the functions
\begin{subequations}\label{K_terms}
\begin{align}
K_{00}^{\alpha \beta} &= K^{\alpha \beta}_{000}\, \delta_{\phi} + K^{\alpha \beta}_{001} \delta_{\phi}^{\prime} + K^{\alpha \beta}_{002} \, \delta_{\phi}^{\prime \prime}, \\
K^{\alpha \beta}_{10} &= K^{\alpha \beta}_{100}\, \delta_{\phi} + K^{\alpha \beta}_{101} \delta_{\phi}^{\prime}, \quad K^{\alpha \beta}_{02} = K^{\alpha \beta}_{020}\, \delta_{\phi},\\
K^{\alpha \beta}_{20} &= K^{\alpha \beta}_{200}\, \delta_{\phi}, \quad K^{\alpha \beta}_{11} =  K^{\alpha \beta}_{110}\, \delta_{\phi},\\
K^{\alpha \beta}_{01} &= K^{\alpha \beta}_{010}\, \delta_{\phi} + K^{\alpha \beta}_{011} \delta_{\phi}^{\prime}.
\end{align}
\end{subequations}
Furthermore, using the distributional property~ \eqref{dirac_delta} of the Dirac delta function, we evaluate the quantities like  $f^{(i)}_{nn}T_{nn}$ appearing in Eq.~\eqref{source_Teulkol}. The explicit expressions for these quantities can be written as 
\beq\label{eq:fiATA}
f^{(i)}_{ab}T_{ab}=\mathcal{A}^{(i)}_{ab}\delta_r + \mathcal{B}^{(i)}_{ab} \delta_r^{\prime} + \mathcal{C}^{(i)}_{ab} \delta_r^{\prime \prime}, 
\eeq
where $a,b\in\{l,n,m,\overline{m}\}$,  $i\in \{ 0,1,2\}$, and there is no sum over $a,b$. Since all coefficients $\mathcal{A}^{(i)}_{ab}$, $\mathcal{B}^{(i)}_{ab}$, and $\mathcal{C}^{(i)}_{ab}$ share the same underlying structure, we present only the explicit expressions for $\mathcal{A}^{(0)}_{nn}$, $\mathcal{B}^{(0)}_{nn}$, and $\mathcal{C}^{(0)}_{nn}$ below. The corresponding coefficients for other choices of $a,b$ and $i$ can be obtained in a analogous manner. The coefficients $\mathcal{A}^{(0)}_{nn}$, $\mathcal{B}^{(0)}_{nn}$, and $\mathcal{C}^{(0)}_{nn}$ can be written as follows:
\begin{subequations}\label{eq:ABC}
\begin{align}
\mathcal{A}^{(0)}_{nn}&= \tilde{K}_{00}^{(nn0)}\,\delta_\theta + \tilde{K}_{01}^{(nn0)}\,\delta_\theta' +\tilde{K}_{02}^{(nn0)}\,\delta_\theta'', \,\\
\mathcal{B}^{(0)}_{nn}&= \tilde{K}_{10}^{(nn0)}\,\delta_\theta + \tilde{K}_{11}^{(nn0)}\,\delta_\theta' \, ,\\
\mathcal{C}^{(0)}_{nn}&= \tilde{K}_{20}^{(nn0)}\,\delta_\theta \,, 
\end{align}
\end{subequations}
where we have defined the quantities
\begin{subequations}
\begin{align}\non
\tilde{K}_{00}^{(nn0)} &= \Big[
K_{00}^{\alpha\beta}\,F_{\alpha\beta}^{(nn0)} - K_{10}^{\alpha\beta}\,\partial_r F_{\alpha\beta}^{(nn0)} \\\non &+ K_{20}^{\alpha\beta}\,\partial_r^2 F_{\alpha\beta}^{(nn0)} - K_{01}^{\alpha\beta}\,\partial_\theta F_{\alpha\beta}^{(nn0)} \\ &+ K_{02}^{\alpha\beta}\,\partial_\theta^2 F_{\alpha\beta}^{(nn0)} + K_{11}^{\alpha\beta}\,\partial_r \partial_\theta F_{\alpha\beta}^{(nn0)}
\Big]_{x_p},\\\non
\tilde{K}_{10}^{(nn0)} &= \Big[ K_{10}^{\alpha\beta}\,F_{\alpha\beta}^{(nn0)} - 2\,K_{20}^{\alpha\beta}\,\partial_r F_{\alpha\beta}^{(nn0)} \\ &- K_{11}^{\alpha\beta}\,\partial_\theta F_{\alpha\beta}^{(nn0)}
\Big]_{x_p},\\\non
\tilde{K}_{01}^{(nn0)} &= \Big[
K_{01}^{\alpha\beta}\,F_{\alpha\beta}^{(nn0)} - 2\,K_{02}^{\alpha\beta}\,\partial_\theta F_{\alpha\beta}^{(nn0)} \\ &- K_{11}^{\alpha\beta}\,\partial_r F_{\alpha\beta}^{(nn0)} \Big]_{x_p},\\
\tilde{K}_{20}^{(nn0)} &= \Big[ K_{20}^{\alpha\beta}\,F_{\alpha\beta}^{(nn0)}
\Big]_{x_p}, \\
\tilde{K}_{02}^{(nn0)} &= \Big[ K_{02}^{\alpha\beta}\,F_{\alpha\beta}^{(nn0)} \Big]_{x_p},\\
\tilde{K}_{11}^{(nn0)} &=\Big[K_{11}^{\alpha\beta}\,F_{\alpha\beta}^{(nn0)} \Big]_{x_p},
\end{align}
\end{subequations}
and
\beq
F_{\alpha\beta}^{(nn0)} = \frac{f_{nn}^{(0)}\, n_\alpha n_\beta}{\sqrt{-g}}.
\eeq

Substituting Eq.~\eqref{eq:fiATA} in Eq.~\eqref{source_Teulkol}, we obtain 
\begin{subequations}\label{eq:calJab}
\begin{align}
\mathcal{J}_{nn}&=\mathcal{J}^{(0)}_{nn}\delta_r+\mathcal{J}^{(1)}_{nn}\delta_r^{\prime}+\mathcal{J}^{(2)}_{nn}\delta_r^{\prime \prime},\\
\mathcal{J}_{\overline{m}n}&=\mathcal{J}^{(0)}_{\overline{m}n}\delta_r+\mathcal{J}^{(1)}_{\overline{m}n}\delta_r^{\prime}+\mathcal{J}^{(2)}_{\overline{m}n}\delta_r^{\prime \prime}+\mathcal{J}^{(3)}_{\overline{m}n}\delta_r^{(3)},\\\non
\mathcal{J}_{\overline{m}\,\overline{m}}&=\mathcal{J}^{(0)}_{\overline{m}\, \overline{m}}\delta_r + \mathcal{J}^{(1)}_{\overline{m}\, \overline{m}}\delta_r^{\prime}+\mathcal{J}^{(2)}_{\overline{m} \, \overline{m}}\delta_r^{\prime \prime} \\ & +\mathcal{J}^{(3)}_{\overline{m} \, \overline{m}}\delta_r^{(3)} + \mathcal{J}^{(4)}_{\overline{m}\, \overline{m}}\delta_r^{(4)},
\end{align}
\end{subequations}
where
\begin{subequations}
\begin{align}
\mathcal{J}^{(0)}_{nn}&=\mathcal{A}^{0}_{nn}\,,\quad \mathcal{J}^{(1)}_{nn}=\mathcal{B}^{0}_{nn}\,,\quad \mathcal{J}^{(2)}_{nn}=\mathcal{C}^{0}_{nn}\,,\\
\mathcal{J}^{(0)}_{\overline{m}n}&=\mathcal{A}^{0}_{\overline{m}n}\,,\quad \mathcal{J}^{(1)}_{\overline{m}n}=\mathcal{B}^{0}_{\overline{m}n}+\mathcal{A}^{1}_{\overline{m}n}\,,\\ \mathcal{J}^{(2)}_{\overline{m}n}&=\mathcal{C}^{0}_{\overline{m}n}+\mathcal{B}^{1}_{\overline{m}n}\,,\quad \mathcal{J}^{(3)}_{\overline{m}n}=\mathcal{C}^{1}_{\overline{m}n}\,,\\
\mathcal{J}^{(0)}_{\overline{m} \overline{m}}&=\mathcal{A}^{0}_{\overline{m}\, \overline{m}}\,,\quad \mathcal{J}^{(1)}_{\overline{m}\, \overline{m}}=\mathcal{B}^{0}_{\overline{m}\, \overline{m}}+\mathcal{A}^{1}_{\overline{m}\, \overline{m}}\,,\\ \mathcal{J}^{(2)}_{\overline{m}\, \overline{m}}&=\mathcal{C}^{0}_{\overline{m}\, \overline{m}}+\mathcal{B}^{1}_{\overline{m}\, \overline{m}}+\mathcal{A}^{2}_{\overline{m}\, \overline{m}}\,,\\\mathcal{J}^{(3)}_{\overline{m}\, \overline{m}}&=\mathcal{C}^{1}_{\overline{m}\, \overline{m}}+\mathcal{B}^{2}_{\overline{m}\, \overline{m}}\,,\quad \mathcal{J}^{(4)}_{\overline{m}\, \overline{m}}=\mathcal{C}^{2}_{\overline{m}\, \overline{m}}.
\end{align}
\end{subequations}
We now substitute these expressions in Eq.~\eqref{source_rad_teuk} and use the angular delta functions in Eq.~\eqref{eq:ABC} to perform the angular integration. We then find that the source term for the radial Teukolsky equation can be written as 
\beq\label{source_rad_teuk1}
\mathcal{J}_{\ell m\omega}=\int dt \Delta^2 ~e^{i(\omega t -m\phi_p)}\left(\tilde{\mathcal{J}}_{nn}+\tilde{\mathcal{J}}_{\overline{m}n}+\tilde{\mathcal{J}}_{\overline{m}\, \overline{m}}\right),
\eeq
where the quantities $\tilde{\mathcal{J}}_{ab}$ are given by Eq.~\eqref{eq:calJab} with the replacements $\delta_\phi\to 1$, $\delta_{\phi}^{\prime}\to im$, and $\delta_{\phi}^{\prime \prime}\to (im)^2$ in Eq.~\eqref{K_terms} and $\delta_\theta\to1$, $\delta'_\theta\to 0$ and $\delta''_\theta\to 0$ in Eq.~\eqref{eq:ABC}. 

If we fixed the particle to be on a test-body orbit with constant frequency, then the integrands $\tilde{\mathcal{J}}_{nn}$ would now all be independent of time, $\phi_p$ would become $\Omega t$, and the remaining integral over time would reduce to a Dirac delta, $\delta(\omega-\omega_m)$. In our multiscale expansion, the integrands are functions of the slowly evolving frequency, and $\phi_p\neq \Omega t$; but $\omega$ is replaced with $\omega_m$ and the integration $\int dt e^{i\omega t}$ is replaced with $\frac{1}{2\pi}\oint d\phi_p e^{im\phi_p}$, such that we are left with $\frac{1}{2\pi}\oint d\phi_p e^{im(\phi_p-\phi_p)}=1$. Hence, we obtain
\beq\label{source_rad_teuk1_multiscale}
\mathcal{J}_{\ell m\omega_m}= \Delta^2 \left(\tilde{\mathcal{J}}_{nn}+\tilde{\mathcal{J}}_{\overline{m}n}+\tilde{\mathcal{J}}_{\overline{m}\, \overline{m}}\right).
\eeq

\subsection{Teukolsky amplitudes}

Finally, the amplitudes $\mathcal{Z}_{\ell m\omega_m}^{\rm H,\infty}$ are given by the following relation: 
\begin{equation}\label{amp_def1}
		\begin{aligned}
\mathcal{Z}_{\ell m\omega_m}^{\rm H,\infty}&=\mathcal{C}^{\rm H,\infty}_{\ell m\omega_m}\int_{\hr_+}^{\infty
			}d\hr\frac{R_{\ell m\omega}^{\up,\In}\mathcal{J}_{\ell m\omega_m}}{\Delta^2}\,,\\
            &=\mathcal{C}^{\rm H,\infty}_{\ell m\omega_m}\mathcal{I}_{\ell m\omega_m}^{\rm H,\infty}\,,
		\end{aligned}
	\end{equation}
where
\begin{align}
\mathcal{I}_{\ell m \omega_m}^{\rm H,\infty}&=\int_{\hr_+}^{\infty}dr\Bigl[A_0\delta_r+A_1\delta_r' +A_2\delta_r''\notag\\*
&\qquad\quad +A_3\delta_r^{(3)}+A_4\delta_r^{(4)}\Bigr]R_{\ell m\omega_m}^{\up,\In}\\*
&=\sum_{j=0}^{4}\Big[{A}_j\delta_r^{(j)}\Big]R^{\up,\In}_{\ell m \omega_m}( r),
\end{align}
with
\begin{subequations}
\begin{align}
A_0 &= \tilde{\mathcal{J}}^{(0)}_{nn} + \tilde{\mathcal{J}}^{(0)}_{\overline{m} n} + \tilde{\mathcal{J}}^{(0)}_{\overline{m}\,\overline{m}}\,,\\
A_1 &= \tilde{\mathcal{J}}^{(1)}_{nn} + \tilde{\mathcal{J}}^{(1)}_{\overline{m} n} + \tilde{\mathcal{J}}^{(1)}_{\overline{m}\,\overline{m}}\,,\\
A_2 &= \tilde{\mathcal{J}}^{(2)}_{nn} + \tilde{\mathcal{J}}^{(2)}_{\overline{m} n} + \tilde{\mathcal{J}}^{(2)}_{\overline{m}\,\overline{m}}\,,\\
A_3 &= \tilde{\mathcal{J}}_{\overline{m} n} + \tilde{\mathcal{J}}^{(3)}_{\overline{m}\,\overline{m}}\,,\\
A_4 &= \tilde{\mathcal{J}}^{(4)}_{\overline{m}\,\overline{m}},
\end{align}
\end{subequations}
where the tildes denote replacement of delta functions as described below Eq.~\eqref{source_rad_teuk1}.

Up to this point, the expression for the amplitudes is quite general. In order to evaluate them in the fixed-frequency expansion, we note that $r_p=r_{0}+\e r_{1}+\e^2 r_{2}$. Now expanding the source coefficients as ${A}_j=\e {A}_{j0}+\e^2 {A}_{j1}+\e^3 {A}_{j2}$  and the Dirac delta function as 
\beq
\delta_r^{(j)}(r-r_p)&=\delta_r^{(j)}(r-r_{0})-\e r_1 \delta^{(j+1)}_{r}(r-r_{0})\\&-\e^2 r_2 \delta^{(j+1)}_{r}(r-r_{0})+\e^2 \frac{r_1}{2} \delta^{(j+2)}_{r}(r-r_{0}),
\eeq
we obtain 
\beq
\mathcal{I}_{\ell m\omega_m}^{\rm H,\infty}&= \sum_{j=0}^{4} (-1)^{j}\bigg[\e\left\{{A}_{j0}\frac{d^j R^{\textrm{in},\textrm{up}}_{l m \omega_m}}{dr^j}\right\}\\&\quad +\e^2\left\{r_1{A}_{j0}\frac{d^{j+1} R^{\textrm{in},\textrm{up}}_{\ell m \omega_m}}{dr^{j+1}}+{A}_{j1}\frac{d^j R^{\textrm{in},\textrm{up}}_{\ell m \omega_m}}{dr^j}\right\}\\
            &\quad
            +\e^3\bigg\{{A}_{j0}\left(r_2\frac{d^{j+1} R^{\textrm{in},\textrm{up}}_{\ell m \omega_m}}{dr^{j+1}}+\frac{r_1^2}{2}\frac{d^{j+2}R^{\textrm{in},\textrm{up}}_{\ell m \omega_m}}{d r^{j+2}}\right)\\&\quad +r_1{A}_{j1}\frac{d^{j+1} R^{\textrm{in},\textrm{up}}_{\ell m \omega_m}}{dr^{j+1}}+{A}_{j2}\frac{d^j R^{\textrm{in},\textrm{up}}_{\ell m \omega_m}}{dr^j}\bigg\}
            \bigg]_{r_0}\\
            &= \e \mathcal{I}^{\rm H,\infty}_{(0)}+\e^2 \mathcal{I}^{\rm H,\infty}_{(1)}+\e^3 \mathcal{I}^{\rm H,\infty}_{(2)},
\eeq
where in the last step we used the distributional property~\eqref{dirac_delta} of the Dirac delta function.\par
Finally, to define the mode amplitudes $\mathcal{Z}_{0}^{A}$, $\mathcal{Z}_{\chi}^{A}$, $\mathcal{Z}_{\chi\chi}^{A}$, $\mathcal{Z}_{C_Q}^{A}$, $\mathcal{Z}_{\mu_2}^{A}$ and $\mathcal{Z}_{\sigma_2}^{A}$ appearing in Eqs.~\eqref{0pa_flux}--\eqref{2pa_flux}, we introduce the parameter vector $\vec{\Theta}=\{\chi, C_Q,\mu_2,\sigma_2\}$. In terms of $\vec{\Theta}$, the mode amplitudes can be expanded as
\begin{subequations}\label{mode_amplitudes}
   \begin{align}
       \mathcal{Z}_{0}^{A}&=\mathcal{Z}_{\ell m\omega_m}^{A}\bigg|_{\vec{\Theta}=0}\label{amp_0pa},\\
       \mathcal{Z}_{\Theta_i}^{A}&=\frac{\partial\mathcal{Z}_{\ell m\omega_m}^{A}}{d\Theta_i}\bigg|_{\vec{\Theta}=0}\label{amp_1pa},\\
       \mathcal{Z}_{\Theta_i\Theta_i}^{A}&=\frac{1}{2}\frac{\partial^2\mathcal{Z}_{\ell m\omega_m}^{A}}{d\Theta_i^2}\bigg|_{\vec{\Theta}=0}\label{amp_2pa},
   \end{align} 
\end{subequations}
where $\textrm{A}\in\{\rm H,\infty\}$. The quantity $\mathcal{Z}_{0}^{A}$ corresponds to the point-particle mode amplitude. The amplitudes $\mathcal{Z}_{\Theta_i}^{A}$ represent the linear-order corrections arising from the spin and quadrupolar effects, namely $\mathcal{Z}_{\chi}^{A}$, $\mathcal{Z}_{C_Q}^{A}$, $\mathcal{Z}_{\mu_2}^{A}$ and $\mathcal{Z}_{\sigma_2}^{A}$, which enter the fluxes in Eqs.~\eqref{0pa_1pa_flux} and \eqref{2pa_flux}. Similarly, $\mathcal{Z}_{\Theta_i\Theta_i}^{A}$ denotes the corresponding quadratic-order contributions. In the present work, we retain only the quadratic-in-spin contribution, $\mathcal{Z}_{\chi\chi}^{A}$, as given in Eq.~\eqref{2pa_flux}.
\bibliography{Reference_1}

\begin{thebibliography}{205}%
\makeatletter
\providecommand \@ifxundefined [1]{%
 \@ifx{#1\undefined}
}%
\providecommand \@ifnum [1]{%
 \ifnum #1\expandafter \@firstoftwo
 \else \expandafter \@secondoftwo
 \fi
}%
\providecommand \@ifx [1]{%
 \ifx #1\expandafter \@firstoftwo
 \else \expandafter \@secondoftwo
 \fi
}%
\providecommand \natexlab [1]{#1}%
\providecommand \enquote  [1]{``#1''}%
\providecommand \bibnamefont  [1]{#1}%
\providecommand \bibfnamefont [1]{#1}%
\providecommand \citenamefont [1]{#1}%
\providecommand \href@noop [0]{\@secondoftwo}%
\providecommand \href [0]{\begingroup \@sanitize@url \@href}%
\providecommand \@href[1]{\@@startlink{#1}\@@href}%
\providecommand \@@href[1]{\endgroup#1\@@endlink}%
\providecommand \@sanitize@url [0]{\catcode `\\12\catcode `\$12\catcode `\&12\catcode `\#12\catcode `\^12\catcode `\_12\catcode `\%12\relax}%
\providecommand \@@startlink[1]{}%
\providecommand \@@endlink[0]{}%
\providecommand \url  [0]{\begingroup\@sanitize@url \@url }%
\providecommand \@url [1]{\endgroup\@href {#1}{\urlprefix }}%
\providecommand \urlprefix  [0]{URL }%
\providecommand \Eprint [0]{\href }%
\providecommand \doibase [0]{https://doi.org/}%
\providecommand \selectlanguage [0]{\@gobble}%
\providecommand \bibinfo  [0]{\@secondoftwo}%
\providecommand \bibfield  [0]{\@secondoftwo}%
\providecommand \translation [1]{[#1]}%
\providecommand \BibitemOpen [0]{}%
\providecommand \bibitemStop [0]{}%
\providecommand \bibitemNoStop [0]{.\EOS\space}%
\providecommand \EOS [0]{\spacefactor3000\relax}%
\providecommand \BibitemShut  [1]{\csname bibitem#1\endcsname}%
\let\auto@bib@innerbib\@empty
\bibitem [{\citenamefont {Abbott}\ \emph {et~al.}(2016)\citenamefont {Abbott} \emph {et~al.}}]{LIGOScientific:2016aoc}%
  \BibitemOpen
  \bibfield  {author} {\bibinfo {author} {\bibfnamefont {B.~P.}\ \bibnamefont {Abbott}} \emph {et~al.} (\bibinfo {collaboration} {LIGO Scientific, Virgo}),\ }\bibfield  {title} {\bibinfo {title} {{Observation of Gravitational Waves from a Binary Black Hole Merger}},\ }\href {https://doi.org/10.1103/PhysRevLett.116.061102} {\bibfield  {journal} {\bibinfo  {journal} {Phys. Rev. Lett.}\ }\textbf {\bibinfo {volume} {116}},\ \bibinfo {pages} {061102} (\bibinfo {year} {2016})},\ \Eprint {https://arxiv.org/abs/1602.03837} {arXiv:1602.03837 [gr-qc]} \BibitemShut {NoStop}%
\bibitem [{\citenamefont {Abac}\ \emph {et~al.}(2025{\natexlab{a}})\citenamefont {Abac} \emph {et~al.}}]{LIGOScientific:2025slb}%
  \BibitemOpen
  \bibfield  {author} {\bibinfo {author} {\bibfnamefont {A.~G.}\ \bibnamefont {Abac}} \emph {et~al.} (\bibinfo {collaboration} {LIGO Scientific, VIRGO, KAGRA}),\ }\bibfield  {title} {\bibinfo {title} {{GWTC-4.0: Updating the Gravitational-Wave Transient Catalog with Observations from the First Part of the Fourth LIGO-Virgo-KAGRA Observing Run}},\ }\href@noop {} {\  (\bibinfo {year} {2025}{\natexlab{a}})},\ \Eprint {https://arxiv.org/abs/2508.18082} {arXiv:2508.18082 [gr-qc]} \BibitemShut {NoStop}%
\bibitem [{\citenamefont {Abac}\ \emph {et~al.}(2026{\natexlab{a}})\citenamefont {Abac} \emph {et~al.}}]{LIGOScientific:2026sit}%
  \BibitemOpen
  \bibfield  {author} {\bibinfo {author} {\bibfnamefont {N.}~\bibnamefont {Abac}} \emph {et~al.} (\bibinfo {collaboration} {LIGO Scientific, VIRGO, KAGRA}),\ }\bibfield  {title} {\bibinfo {title} {{GWTC-5.0: An Introduction to Version 5.0 of the Gravitational-Wave Transient Catalog}},\ }\href@noop {} {\  (\bibinfo {year} {2026}{\natexlab{a}})},\ \Eprint {https://arxiv.org/abs/2605.27223} {arXiv:2605.27223 [gr-qc]} \BibitemShut {NoStop}%
\bibitem [{\citenamefont {Abbott}\ \emph {et~al.}(2017)\citenamefont {Abbott} \emph {et~al.}}]{LIGOScientific:2017vwq}%
  \BibitemOpen
  \bibfield  {author} {\bibinfo {author} {\bibfnamefont {B.~P.}\ \bibnamefont {Abbott}} \emph {et~al.} (\bibinfo {collaboration} {LIGO Scientific, Virgo}),\ }\bibfield  {title} {\bibinfo {title} {{GW170817: Observation of Gravitational Waves from a Binary Neutron Star Inspiral}},\ }\href {https://doi.org/10.1103/PhysRevLett.119.161101} {\bibfield  {journal} {\bibinfo  {journal} {Phys. Rev. Lett.}\ }\textbf {\bibinfo {volume} {119}},\ \bibinfo {pages} {161101} (\bibinfo {year} {2017})},\ \Eprint {https://arxiv.org/abs/1710.05832} {arXiv:1710.05832 [gr-qc]} \BibitemShut {NoStop}%
\bibitem [{\citenamefont {Annala}\ \emph {et~al.}(2018)\citenamefont {Annala}, \citenamefont {Gorda}, \citenamefont {Kurkela},\ and\ \citenamefont {Vuorinen}}]{Annala:2017llu}%
  \BibitemOpen
  \bibfield  {author} {\bibinfo {author} {\bibfnamefont {E.}~\bibnamefont {Annala}}, \bibinfo {author} {\bibfnamefont {T.}~\bibnamefont {Gorda}}, \bibinfo {author} {\bibfnamefont {A.}~\bibnamefont {Kurkela}},\ and\ \bibinfo {author} {\bibfnamefont {A.}~\bibnamefont {Vuorinen}},\ }\bibfield  {title} {\bibinfo {title} {{Gravitational-wave constraints on the neutron-star-matter Equation of State}},\ }\href {https://doi.org/10.1103/PhysRevLett.120.172703} {\bibfield  {journal} {\bibinfo  {journal} {Phys. Rev. Lett.}\ }\textbf {\bibinfo {volume} {120}},\ \bibinfo {pages} {172703} (\bibinfo {year} {2018})},\ \Eprint {https://arxiv.org/abs/1711.02644} {arXiv:1711.02644 [astro-ph.HE]} \BibitemShut {NoStop}%
\bibitem [{\citenamefont {Abac}\ \emph {et~al.}(2025{\natexlab{b}})\citenamefont {Abac} \emph {et~al.}}]{LIGOScientific:2025brd}%
  \BibitemOpen
  \bibfield  {author} {\bibinfo {author} {\bibfnamefont {A.~G.}\ \bibnamefont {Abac}} \emph {et~al.} (\bibinfo {collaboration} {LIGO Scientific, Virgo, KAGRA}),\ }\bibfield  {title} {\bibinfo {title} {{GW241011 and GW241110: Exploring Binary Formation and Fundamental Physics with Asymmetric, High-spin Black Hole Coalescences}},\ }\href {https://doi.org/10.3847/2041-8213/ae0d54} {\bibfield  {journal} {\bibinfo  {journal} {Astrophys. J. Lett.}\ }\textbf {\bibinfo {volume} {993}},\ \bibinfo {pages} {L21} (\bibinfo {year} {2025}{\natexlab{b}})},\ \Eprint {https://arxiv.org/abs/2510.26931} {arXiv:2510.26931 [astro-ph.HE]} \BibitemShut {NoStop}%
\bibitem [{\citenamefont {Das}\ \emph {et~al.}(2026)\citenamefont {Das}, \citenamefont {Krishnendu}, \citenamefont {Saleem}, \citenamefont {Mishra},\ and\ \citenamefont {Arun}}]{Das:2026tel}%
  \BibitemOpen
  \bibfield  {author} {\bibinfo {author} {\bibfnamefont {R.}~\bibnamefont {Das}}, \bibinfo {author} {\bibfnamefont {N.~V.}\ \bibnamefont {Krishnendu}}, \bibinfo {author} {\bibfnamefont {M.}~\bibnamefont {Saleem}}, \bibinfo {author} {\bibfnamefont {C.~K.}\ \bibnamefont {Mishra}},\ and\ \bibinfo {author} {\bibfnamefont {K.~G.}\ \bibnamefont {Arun}},\ }\bibfield  {title} {\bibinfo {title} {{Testing the Kerr hypothesis beyond the quadrupole with GW241011}},\ }\href@noop {} {\  (\bibinfo {year} {2026})},\ \Eprint {https://arxiv.org/abs/2604.09828} {arXiv:2604.09828 [gr-qc]} \BibitemShut {NoStop}%
\bibitem [{\citenamefont {Abac}\ \emph {et~al.}(2025{\natexlab{c}})\citenamefont {Abac} \emph {et~al.}}]{LIGOScientific:2025rid}%
  \BibitemOpen
  \bibfield  {author} {\bibinfo {author} {\bibfnamefont {A.~G.}\ \bibnamefont {Abac}} \emph {et~al.} (\bibinfo {collaboration} {LIGO Scientific, Virgo, KAGRA}),\ }\bibfield  {title} {\bibinfo {title} {{GW250114: Testing {H}awking{\textquoteright}s Area Law and the {K}err Nature of Black Holes}},\ }\href {https://doi.org/10.1103/kw5g-d732} {\bibfield  {journal} {\bibinfo  {journal} {Phys. Rev. Lett.}\ }\textbf {\bibinfo {volume} {135}},\ \bibinfo {pages} {111403} (\bibinfo {year} {2025}{\natexlab{c}})},\ \Eprint {https://arxiv.org/abs/2509.08054} {arXiv:2509.08054 [gr-qc]} \BibitemShut {NoStop}%
\bibitem [{\citenamefont {Arun}\ \emph {et~al.}(2022)\citenamefont {Arun} \emph {et~al.}}]{LISA:2022kgy}%
  \BibitemOpen
  \bibfield  {author} {\bibinfo {author} {\bibfnamefont {K.~G.}\ \bibnamefont {Arun}} \emph {et~al.} (\bibinfo {collaboration} {LISA}),\ }\bibfield  {title} {\bibinfo {title} {{New horizons for fundamental physics with LISA}},\ }\href {https://doi.org/10.1007/s41114-022-00036-9} {\bibfield  {journal} {\bibinfo  {journal} {Living Rev. Rel.}\ }\textbf {\bibinfo {volume} {25}},\ \bibinfo {pages} {4} (\bibinfo {year} {2022})},\ \Eprint {https://arxiv.org/abs/2205.01597} {arXiv:2205.01597 [gr-qc]} \BibitemShut {NoStop}%
\bibitem [{\citenamefont {Seoane}\ \emph {et~al.}(2023)\citenamefont {Seoane} \emph {et~al.}}]{LISA:2022yao}%
  \BibitemOpen
  \bibfield  {author} {\bibinfo {author} {\bibfnamefont {P.~A.}\ \bibnamefont {Seoane}} \emph {et~al.} (\bibinfo {collaboration} {LISA}),\ }\bibfield  {title} {\bibinfo {title} {{Astrophysics with the Laser Interferometer Space Antenna}},\ }\href {https://doi.org/10.1007/s41114-022-00041-y} {\bibfield  {journal} {\bibinfo  {journal} {Living Rev. Rel.}\ }\textbf {\bibinfo {volume} {26}},\ \bibinfo {pages} {2} (\bibinfo {year} {2023})},\ \Eprint {https://arxiv.org/abs/2203.06016} {arXiv:2203.06016 [gr-qc]} \BibitemShut {NoStop}%
\bibitem [{\citenamefont {Colpi}\ \emph {et~al.}(2024)\citenamefont {Colpi} \emph {et~al.}}]{LISA:2024hlh}%
  \BibitemOpen
  \bibfield  {author} {\bibinfo {author} {\bibfnamefont {M.}~\bibnamefont {Colpi}} \emph {et~al.} (\bibinfo {collaboration} {LISA}),\ }\bibfield  {title} {\bibinfo {title} {{LISA Definition Study Report}},\ }\href@noop {} {\  (\bibinfo {year} {2024})},\ \Eprint {https://arxiv.org/abs/2402.07571} {arXiv:2402.07571 [astro-ph.CO]} \BibitemShut {NoStop}%
\bibitem [{\citenamefont {Abac}\ \emph {et~al.}(2026{\natexlab{b}})\citenamefont {Abac} \emph {et~al.}}]{ET:2025xjr}%
  \BibitemOpen
  \bibfield  {author} {\bibinfo {author} {\bibfnamefont {A.}~\bibnamefont {Abac}} \emph {et~al.} (\bibinfo {collaboration} {ET}),\ }\bibfield  {title} {\bibinfo {title} {{The Science of the Einstein Telescope}},\ }\href {https://doi.org/10.1088/1475-7516/2026/03/081} {\bibfield  {journal} {\bibinfo  {journal} {JCAP}\ }\textbf {\bibinfo {volume} {03}},\ \bibinfo {pages} {081}},\ \Eprint {https://arxiv.org/abs/2503.12263} {arXiv:2503.12263 [gr-qc]} \BibitemShut {NoStop}%
\bibitem [{\citenamefont {P{\"u}rrer}\ and\ \citenamefont {Haster}(2020)}]{Purrer:2019jcp}%
  \BibitemOpen
  \bibfield  {author} {\bibinfo {author} {\bibfnamefont {M.}~\bibnamefont {P{\"u}rrer}}\ and\ \bibinfo {author} {\bibfnamefont {C.-J.}\ \bibnamefont {Haster}},\ }\bibfield  {title} {\bibinfo {title} {{Gravitational waveform accuracy requirements for future ground-based detectors}},\ }\href {https://doi.org/10.1103/PhysRevResearch.2.023151} {\bibfield  {journal} {\bibinfo  {journal} {Phys. Rev. Res.}\ }\textbf {\bibinfo {volume} {2}},\ \bibinfo {pages} {023151} (\bibinfo {year} {2020})},\ \Eprint {https://arxiv.org/abs/1912.10055} {arXiv:1912.10055 [gr-qc]} \BibitemShut {NoStop}%
\bibitem [{\citenamefont {Hu}\ and\ \citenamefont {Veitch}(2022)}]{Hu:2022rjq}%
  \BibitemOpen
  \bibfield  {author} {\bibinfo {author} {\bibfnamefont {Q.}~\bibnamefont {Hu}}\ and\ \bibinfo {author} {\bibfnamefont {J.}~\bibnamefont {Veitch}},\ }\bibfield  {title} {\bibinfo {title} {{Assessing the model waveform accuracy of gravitational waves}},\ }\href {https://doi.org/10.1103/PhysRevD.106.044042} {\bibfield  {journal} {\bibinfo  {journal} {Phys. Rev. D}\ }\textbf {\bibinfo {volume} {106}},\ \bibinfo {pages} {044042} (\bibinfo {year} {2022})},\ \Eprint {https://arxiv.org/abs/2205.08448} {arXiv:2205.08448 [gr-qc]} \BibitemShut {NoStop}%
\bibitem [{\citenamefont {Afshordi}\ \emph {et~al.}(2025)\citenamefont {Afshordi} \emph {et~al.}}]{LISAConsortiumWaveformWorkingGroup:2023arg}%
  \BibitemOpen
  \bibfield  {author} {\bibinfo {author} {\bibfnamefont {N.}~\bibnamefont {Afshordi}} \emph {et~al.} (\bibinfo {collaboration} {LISA Consortium Waveform Working Group}),\ }\bibfield  {title} {\bibinfo {title} {{Waveform modelling for the Laser Interferometer Space Antenna}},\ }\href {https://doi.org/10.1007/s41114-025-00056-1} {\bibfield  {journal} {\bibinfo  {journal} {Living Rev. Rel.}\ }\textbf {\bibinfo {volume} {28}},\ \bibinfo {pages} {9} (\bibinfo {year} {2025})},\ \Eprint {https://arxiv.org/abs/2311.01300} {arXiv:2311.01300 [gr-qc]} \BibitemShut {NoStop}%
\bibitem [{\citenamefont {Dhani}\ \emph {et~al.}(2025)\citenamefont {Dhani}, \citenamefont {V{\"o}lkel}, \citenamefont {Buonanno}, \citenamefont {Estelles}, \citenamefont {Gair}, \citenamefont {Pfeiffer}, \citenamefont {Pompili},\ and\ \citenamefont {Toubiana}}]{Dhani:2024jja}%
  \BibitemOpen
  \bibfield  {author} {\bibinfo {author} {\bibfnamefont {A.}~\bibnamefont {Dhani}}, \bibinfo {author} {\bibfnamefont {S.~H.}\ \bibnamefont {V{\"o}lkel}}, \bibinfo {author} {\bibfnamefont {A.}~\bibnamefont {Buonanno}}, \bibinfo {author} {\bibfnamefont {H.}~\bibnamefont {Estelles}}, \bibinfo {author} {\bibfnamefont {J.}~\bibnamefont {Gair}}, \bibinfo {author} {\bibfnamefont {H.~P.}\ \bibnamefont {Pfeiffer}}, \bibinfo {author} {\bibfnamefont {L.}~\bibnamefont {Pompili}},\ and\ \bibinfo {author} {\bibfnamefont {A.}~\bibnamefont {Toubiana}},\ }\bibfield  {title} {\bibinfo {title} {{Systematic Biases in Estimating the Properties of Black Holes Due to Inaccurate Gravitational-Wave Models}},\ }\href {https://doi.org/10.1103/5pks-qz6b} {\bibfield  {journal} {\bibinfo  {journal} {Phys. Rev. X}\ }\textbf {\bibinfo {volume} {15}},\ \bibinfo {pages} {031036} (\bibinfo {year} {2025})},\ \Eprint {https://arxiv.org/abs/2404.05811} {arXiv:2404.05811 [gr-qc]} \BibitemShut {NoStop}%
\bibitem [{\citenamefont {Chandramouli}\ \emph {et~al.}(2025)\citenamefont {Chandramouli}, \citenamefont {Prokup}, \citenamefont {Berti},\ and\ \citenamefont {Yunes}}]{Chandramouli:2024vhw}%
  \BibitemOpen
  \bibfield  {author} {\bibinfo {author} {\bibfnamefont {R.~S.}\ \bibnamefont {Chandramouli}}, \bibinfo {author} {\bibfnamefont {K.}~\bibnamefont {Prokup}}, \bibinfo {author} {\bibfnamefont {E.}~\bibnamefont {Berti}},\ and\ \bibinfo {author} {\bibfnamefont {N.}~\bibnamefont {Yunes}},\ }\bibfield  {title} {\bibinfo {title} {{Systematic biases due to waveform mismodeling in parametrized post-Einsteinian tests of general relativity: The impact of neglecting spin precession and higher modes}},\ }\href {https://doi.org/10.1103/PhysRevD.111.044026} {\bibfield  {journal} {\bibinfo  {journal} {Phys. Rev. D}\ }\textbf {\bibinfo {volume} {111}},\ \bibinfo {pages} {044026} (\bibinfo {year} {2025})},\ \Eprint {https://arxiv.org/abs/2410.06254} {arXiv:2410.06254 [gr-qc]} \BibitemShut {NoStop}%
\bibitem [{\citenamefont {Mahapatra}\ \emph {et~al.}(2026)\citenamefont {Mahapatra}, \citenamefont {Thompson}, \citenamefont {Fauchon-Jones},\ and\ \citenamefont {Hannam}}]{Mahapatra:2026wsp}%
  \BibitemOpen
  \bibfield  {author} {\bibinfo {author} {\bibfnamefont {P.}~\bibnamefont {Mahapatra}}, \bibinfo {author} {\bibfnamefont {J.~E.}\ \bibnamefont {Thompson}}, \bibinfo {author} {\bibfnamefont {E.}~\bibnamefont {Fauchon-Jones}},\ and\ \bibinfo {author} {\bibfnamefont {M.}~\bibnamefont {Hannam}},\ }\bibfield  {title} {\bibinfo {title} {{The High-Mass-Ratio Challenge in Gravitational Waveform Modelling}},\ }\href@noop {} {\  (\bibinfo {year} {2026})},\ \Eprint {https://arxiv.org/abs/2603.26521} {arXiv:2603.26521 [gr-qc]} \BibitemShut {NoStop}%
\bibitem [{\citenamefont {Marsat}(2015)}]{Marsat:2014xea}%
  \BibitemOpen
  \bibfield  {author} {\bibinfo {author} {\bibfnamefont {S.}~\bibnamefont {Marsat}},\ }\bibfield  {title} {\bibinfo {title} {{Cubic order spin effects in the dynamics and gravitational wave energy flux of compact object binaries}},\ }\href {https://doi.org/10.1088/0264-9381/32/8/085008} {\bibfield  {journal} {\bibinfo  {journal} {Class. Quant. Grav.}\ }\textbf {\bibinfo {volume} {32}},\ \bibinfo {pages} {085008} (\bibinfo {year} {2015})},\ \Eprint {https://arxiv.org/abs/1411.4118} {arXiv:1411.4118 [gr-qc]} \BibitemShut {NoStop}%
\bibitem [{\citenamefont {Boh\'e}\ \emph {et~al.}(2015)\citenamefont {Boh\'e}, \citenamefont {Faye}, \citenamefont {Marsat},\ and\ \citenamefont {Porter}}]{Bohe:2015ana}%
  \BibitemOpen
  \bibfield  {author} {\bibinfo {author} {\bibfnamefont {A.}~\bibnamefont {Boh\'e}}, \bibinfo {author} {\bibfnamefont {G.}~\bibnamefont {Faye}}, \bibinfo {author} {\bibfnamefont {S.}~\bibnamefont {Marsat}},\ and\ \bibinfo {author} {\bibfnamefont {E.~K.}\ \bibnamefont {Porter}},\ }\bibfield  {title} {\bibinfo {title} {{Quadratic-in-spin effects in the orbital dynamics and gravitational-wave energy flux of compact binaries at the 3PN order}},\ }\href {https://doi.org/10.1088/0264-9381/32/19/195010} {\bibfield  {journal} {\bibinfo  {journal} {Class. Quant. Grav.}\ }\textbf {\bibinfo {volume} {32}},\ \bibinfo {pages} {195010} (\bibinfo {year} {2015})},\ \Eprint {https://arxiv.org/abs/1501.01529} {arXiv:1501.01529 [gr-qc]} \BibitemShut {NoStop}%
\bibitem [{\citenamefont {Levi}\ and\ \citenamefont {Steinhoff}(2021)}]{Levi:2016ofk}%
  \BibitemOpen
  \bibfield  {author} {\bibinfo {author} {\bibfnamefont {M.}~\bibnamefont {Levi}}\ and\ \bibinfo {author} {\bibfnamefont {J.}~\bibnamefont {Steinhoff}},\ }\bibfield  {title} {\bibinfo {title} {{Complete conservative dynamics for inspiralling compact binaries with spins at the fourth post-Newtonian order}},\ }\href {https://doi.org/10.1088/1475-7516/2021/09/029} {\bibfield  {journal} {\bibinfo  {journal} {JCAP}\ }\textbf {\bibinfo {volume} {09}},\ \bibinfo {pages} {029}},\ \Eprint {https://arxiv.org/abs/1607.04252} {arXiv:1607.04252 [gr-qc]} \BibitemShut {NoStop}%
\bibitem [{\citenamefont {Siemonsen}\ \emph {et~al.}(2018)\citenamefont {Siemonsen}, \citenamefont {Steinhoff},\ and\ \citenamefont {Vines}}]{Siemonsen:2017yux}%
  \BibitemOpen
  \bibfield  {author} {\bibinfo {author} {\bibfnamefont {N.}~\bibnamefont {Siemonsen}}, \bibinfo {author} {\bibfnamefont {J.}~\bibnamefont {Steinhoff}},\ and\ \bibinfo {author} {\bibfnamefont {J.}~\bibnamefont {Vines}},\ }\bibfield  {title} {\bibinfo {title} {{Gravitational waves from spinning binary black holes at the leading post-Newtonian orders at all orders in spin}},\ }\href {https://doi.org/10.1103/PhysRevD.97.124046} {\bibfield  {journal} {\bibinfo  {journal} {Phys. Rev. D}\ }\textbf {\bibinfo {volume} {97}},\ \bibinfo {pages} {124046} (\bibinfo {year} {2018})},\ \Eprint {https://arxiv.org/abs/1712.08603} {arXiv:1712.08603 [gr-qc]} \BibitemShut {NoStop}%
\bibitem [{\citenamefont {Kastha}\ \emph {et~al.}(2018)\citenamefont {Kastha}, \citenamefont {Gupta}, \citenamefont {Arun}, \citenamefont {Sathyaprakash},\ and\ \citenamefont {Van Den~Broeck}}]{Kastha:2018bcr}%
  \BibitemOpen
  \bibfield  {author} {\bibinfo {author} {\bibfnamefont {S.}~\bibnamefont {Kastha}}, \bibinfo {author} {\bibfnamefont {A.}~\bibnamefont {Gupta}}, \bibinfo {author} {\bibfnamefont {K.~G.}\ \bibnamefont {Arun}}, \bibinfo {author} {\bibfnamefont {B.~S.}\ \bibnamefont {Sathyaprakash}},\ and\ \bibinfo {author} {\bibfnamefont {C.}~\bibnamefont {Van Den~Broeck}},\ }\bibfield  {title} {\bibinfo {title} {{Testing the multipole structure of compact binaries using gravitational wave observations}},\ }\href {https://doi.org/10.1103/PhysRevD.98.124033} {\bibfield  {journal} {\bibinfo  {journal} {Phys. Rev. D}\ }\textbf {\bibinfo {volume} {98}},\ \bibinfo {pages} {124033} (\bibinfo {year} {2018})},\ \Eprint {https://arxiv.org/abs/1809.10465} {arXiv:1809.10465 [gr-qc]} \BibitemShut {NoStop}%
\bibitem [{\citenamefont {Kastha}\ \emph {et~al.}(2019)\citenamefont {Kastha}, \citenamefont {Gupta}, \citenamefont {Arun}, \citenamefont {Sathyaprakash},\ and\ \citenamefont {Van Den~Broeck}}]{Kastha:2019brk}%
  \BibitemOpen
  \bibfield  {author} {\bibinfo {author} {\bibfnamefont {S.}~\bibnamefont {Kastha}}, \bibinfo {author} {\bibfnamefont {A.}~\bibnamefont {Gupta}}, \bibinfo {author} {\bibfnamefont {K.~G.}\ \bibnamefont {Arun}}, \bibinfo {author} {\bibfnamefont {B.~S.}\ \bibnamefont {Sathyaprakash}},\ and\ \bibinfo {author} {\bibfnamefont {C.}~\bibnamefont {Van Den~Broeck}},\ }\bibfield  {title} {\bibinfo {title} {{Testing the multipole structure and conservative dynamics of compact binaries using gravitational wave observations: The spinning case}},\ }\href {https://doi.org/10.1103/PhysRevD.100.044007} {\bibfield  {journal} {\bibinfo  {journal} {Phys. Rev. D}\ }\textbf {\bibinfo {volume} {100}},\ \bibinfo {pages} {044007} (\bibinfo {year} {2019})},\ \Eprint {https://arxiv.org/abs/1905.07277} {arXiv:1905.07277 [gr-qc]} \BibitemShut {NoStop}%
\bibitem [{\citenamefont {Henry}\ \emph {et~al.}(2020{\natexlab{a}})\citenamefont {Henry}, \citenamefont {Faye},\ and\ \citenamefont {Blanchet}}]{Henry:2019xhg}%
  \BibitemOpen
  \bibfield  {author} {\bibinfo {author} {\bibfnamefont {Q.}~\bibnamefont {Henry}}, \bibinfo {author} {\bibfnamefont {G.}~\bibnamefont {Faye}},\ and\ \bibinfo {author} {\bibfnamefont {L.}~\bibnamefont {Blanchet}},\ }\bibfield  {title} {\bibinfo {title} {{Tidal effects in the equations of motion of compact binary systems to next-to-next-to-leading post-Newtonian order}},\ }\href {https://doi.org/10.1103/PhysRevD.101.064047} {\bibfield  {journal} {\bibinfo  {journal} {Phys. Rev. D}\ }\textbf {\bibinfo {volume} {101}},\ \bibinfo {pages} {064047} (\bibinfo {year} {2020}{\natexlab{a}})},\ \Eprint {https://arxiv.org/abs/1912.01920} {arXiv:1912.01920 [gr-qc]} \BibitemShut {NoStop}%
\bibitem [{\citenamefont {Henry}\ \emph {et~al.}(2020{\natexlab{b}})\citenamefont {Henry}, \citenamefont {Faye},\ and\ \citenamefont {Blanchet}}]{Henry:2020ski}%
  \BibitemOpen
  \bibfield  {author} {\bibinfo {author} {\bibfnamefont {Q.}~\bibnamefont {Henry}}, \bibinfo {author} {\bibfnamefont {G.}~\bibnamefont {Faye}},\ and\ \bibinfo {author} {\bibfnamefont {L.}~\bibnamefont {Blanchet}},\ }\bibfield  {title} {\bibinfo {title} {{Tidal effects in the gravitational-wave phase evolution of compact binary systems to next-to-next-to-leading post-Newtonian order}},\ }\href {https://doi.org/10.1103/PhysRevD.102.044033} {\bibfield  {journal} {\bibinfo  {journal} {Phys. Rev. D}\ }\textbf {\bibinfo {volume} {102}},\ \bibinfo {pages} {044033} (\bibinfo {year} {2020}{\natexlab{b}})},\ \bibinfo {note} {[Erratum: Phys.Rev.D 108, 089901 (2023), Erratum: Phys.Rev.D 111, 029901 (2025)]},\ \Eprint {https://arxiv.org/abs/2005.13367} {arXiv:2005.13367 [gr-qc]} \BibitemShut {NoStop}%
\bibitem [{\citenamefont {Cho}\ \emph {et~al.}(2022)\citenamefont {Cho}, \citenamefont {Porto},\ and\ \citenamefont {Yang}}]{Cho:2022syn}%
  \BibitemOpen
  \bibfield  {author} {\bibinfo {author} {\bibfnamefont {G.}~\bibnamefont {Cho}}, \bibinfo {author} {\bibfnamefont {R.~A.}\ \bibnamefont {Porto}},\ and\ \bibinfo {author} {\bibfnamefont {Z.}~\bibnamefont {Yang}},\ }\bibfield  {title} {\bibinfo {title} {{Gravitational radiation from inspiralling compact objects: Spin effects to the fourth post-Newtonian order}},\ }\href {https://doi.org/10.1103/PhysRevD.106.L101501} {\bibfield  {journal} {\bibinfo  {journal} {Phys. Rev. D}\ }\textbf {\bibinfo {volume} {106}},\ \bibinfo {pages} {L101501} (\bibinfo {year} {2022})},\ \Eprint {https://arxiv.org/abs/2201.05138} {arXiv:2201.05138 [gr-qc]} \BibitemShut {NoStop}%
\bibitem [{\citenamefont {Blanchet}\ \emph {et~al.}(2023{\natexlab{a}})\citenamefont {Blanchet}, \citenamefont {Faye}, \citenamefont {Henry}, \citenamefont {Larrouturou},\ and\ \citenamefont {Trestini}}]{Blanchet:2023sbv}%
  \BibitemOpen
  \bibfield  {author} {\bibinfo {author} {\bibfnamefont {L.}~\bibnamefont {Blanchet}}, \bibinfo {author} {\bibfnamefont {G.}~\bibnamefont {Faye}}, \bibinfo {author} {\bibfnamefont {Q.}~\bibnamefont {Henry}}, \bibinfo {author} {\bibfnamefont {F.}~\bibnamefont {Larrouturou}},\ and\ \bibinfo {author} {\bibfnamefont {D.}~\bibnamefont {Trestini}},\ }\bibfield  {title} {\bibinfo {title} {{Gravitational-wave flux and quadrupole modes from quasicircular nonspinning compact binaries to the fourth post-Newtonian order}},\ }\href {https://doi.org/10.1103/PhysRevD.108.064041} {\bibfield  {journal} {\bibinfo  {journal} {Phys. Rev. D}\ }\textbf {\bibinfo {volume} {108}},\ \bibinfo {pages} {064041} (\bibinfo {year} {2023}{\natexlab{a}})},\ \Eprint {https://arxiv.org/abs/2304.11186} {arXiv:2304.11186 [gr-qc]} \BibitemShut {NoStop}%
\bibitem [{\citenamefont {Blanchet}\ \emph {et~al.}(2023{\natexlab{b}})\citenamefont {Blanchet}, \citenamefont {Faye}, \citenamefont {Henry}, \citenamefont {Larrouturou},\ and\ \citenamefont {Trestini}}]{Blanchet:2023bwj}%
  \BibitemOpen
  \bibfield  {author} {\bibinfo {author} {\bibfnamefont {L.}~\bibnamefont {Blanchet}}, \bibinfo {author} {\bibfnamefont {G.}~\bibnamefont {Faye}}, \bibinfo {author} {\bibfnamefont {Q.}~\bibnamefont {Henry}}, \bibinfo {author} {\bibfnamefont {F.}~\bibnamefont {Larrouturou}},\ and\ \bibinfo {author} {\bibfnamefont {D.}~\bibnamefont {Trestini}},\ }\bibfield  {title} {\bibinfo {title} {{Gravitational-Wave Phasing of Quasicircular Compact Binary Systems to the Fourth-and-a-Half Post-Newtonian Order}},\ }\href {https://doi.org/10.1103/PhysRevLett.131.121402} {\bibfield  {journal} {\bibinfo  {journal} {Phys. Rev. Lett.}\ }\textbf {\bibinfo {volume} {131}},\ \bibinfo {pages} {121402} (\bibinfo {year} {2023}{\natexlab{b}})},\ \Eprint {https://arxiv.org/abs/2304.11185} {arXiv:2304.11185 [gr-qc]} \BibitemShut {NoStop}%
\bibitem [{\citenamefont {Mandal}\ \emph {et~al.}(2025)\citenamefont {Mandal}, \citenamefont {Mastrolia}, \citenamefont {Patil},\ and\ \citenamefont {Steinhoff}}]{Mandal:2024iug}%
  \BibitemOpen
  \bibfield  {author} {\bibinfo {author} {\bibfnamefont {M.~K.}\ \bibnamefont {Mandal}}, \bibinfo {author} {\bibfnamefont {P.}~\bibnamefont {Mastrolia}}, \bibinfo {author} {\bibfnamefont {R.}~\bibnamefont {Patil}},\ and\ \bibinfo {author} {\bibfnamefont {J.}~\bibnamefont {Steinhoff}},\ }\bibfield  {title} {\bibinfo {title} {{Radiating Love: adiabatic tidal fluxes and modes up to next-to-next-to-leading post-Newtonian order}},\ }\href {https://doi.org/10.1007/JHEP05(2025)008} {\bibfield  {journal} {\bibinfo  {journal} {JHEP}\ }\textbf {\bibinfo {volume} {05}},\ \bibinfo {pages} {008}},\ \Eprint {https://arxiv.org/abs/2412.01706} {arXiv:2412.01706 [gr-qc]} \BibitemShut {NoStop}%
\bibitem [{\citenamefont {Bini}\ \emph {et~al.}(2020)\citenamefont {Bini}, \citenamefont {Damour},\ and\ \citenamefont {Geralico}}]{Bini:2020flp}%
  \BibitemOpen
  \bibfield  {author} {\bibinfo {author} {\bibfnamefont {D.}~\bibnamefont {Bini}}, \bibinfo {author} {\bibfnamefont {T.}~\bibnamefont {Damour}},\ and\ \bibinfo {author} {\bibfnamefont {A.}~\bibnamefont {Geralico}},\ }\bibfield  {title} {\bibinfo {title} {{Scattering of tidally interacting bodies in post-Minkowskian gravity}},\ }\href {https://doi.org/10.1103/PhysRevD.101.044039} {\bibfield  {journal} {\bibinfo  {journal} {Phys. Rev. D}\ }\textbf {\bibinfo {volume} {101}},\ \bibinfo {pages} {044039} (\bibinfo {year} {2020})},\ \Eprint {https://arxiv.org/abs/2001.00352} {arXiv:2001.00352 [gr-qc]} \BibitemShut {NoStop}%
\bibitem [{\citenamefont {Cheung}\ and\ \citenamefont {Solon}(2020)}]{Cheung:2020sdj}%
  \BibitemOpen
  \bibfield  {author} {\bibinfo {author} {\bibfnamefont {C.}~\bibnamefont {Cheung}}\ and\ \bibinfo {author} {\bibfnamefont {M.~P.}\ \bibnamefont {Solon}},\ }\bibfield  {title} {\bibinfo {title} {{Tidal Effects in the Post-Minkowskian Expansion}},\ }\href {https://doi.org/10.1103/PhysRevLett.125.191601} {\bibfield  {journal} {\bibinfo  {journal} {Phys. Rev. Lett.}\ }\textbf {\bibinfo {volume} {125}},\ \bibinfo {pages} {191601} (\bibinfo {year} {2020})},\ \Eprint {https://arxiv.org/abs/2006.06665} {arXiv:2006.06665 [hep-th]} \BibitemShut {NoStop}%
\bibitem [{\citenamefont {K{\"a}lin}\ \emph {et~al.}(2020)\citenamefont {K{\"a}lin}, \citenamefont {Liu},\ and\ \citenamefont {Porto}}]{Kalin:2020lmz}%
  \BibitemOpen
  \bibfield  {author} {\bibinfo {author} {\bibfnamefont {G.}~\bibnamefont {K{\"a}lin}}, \bibinfo {author} {\bibfnamefont {Z.}~\bibnamefont {Liu}},\ and\ \bibinfo {author} {\bibfnamefont {R.~A.}\ \bibnamefont {Porto}},\ }\bibfield  {title} {\bibinfo {title} {{Conservative Tidal Effects in Compact Binary Systems to Next-to-Leading Post-Minkowskian Order}},\ }\href {https://doi.org/10.1103/PhysRevD.102.124025} {\bibfield  {journal} {\bibinfo  {journal} {Phys. Rev. D}\ }\textbf {\bibinfo {volume} {102}},\ \bibinfo {pages} {124025} (\bibinfo {year} {2020})},\ \Eprint {https://arxiv.org/abs/2008.06047} {arXiv:2008.06047 [hep-th]} \BibitemShut {NoStop}%
\bibitem [{\citenamefont {Jakobsen}\ \emph {et~al.}(2024)\citenamefont {Jakobsen}, \citenamefont {Mogull}, \citenamefont {Plefka},\ and\ \citenamefont {Sauer}}]{Jakobsen:2023pvx}%
  \BibitemOpen
  \bibfield  {author} {\bibinfo {author} {\bibfnamefont {G.~U.}\ \bibnamefont {Jakobsen}}, \bibinfo {author} {\bibfnamefont {G.}~\bibnamefont {Mogull}}, \bibinfo {author} {\bibfnamefont {J.}~\bibnamefont {Plefka}},\ and\ \bibinfo {author} {\bibfnamefont {B.}~\bibnamefont {Sauer}},\ }\bibfield  {title} {\bibinfo {title} {{Tidal effects and renormalization at fourth post-Minkowskian order}},\ }\href {https://doi.org/10.1103/PhysRevD.109.L041504} {\bibfield  {journal} {\bibinfo  {journal} {Phys. Rev. D}\ }\textbf {\bibinfo {volume} {109}},\ \bibinfo {pages} {L041504} (\bibinfo {year} {2024})},\ \Eprint {https://arxiv.org/abs/2312.00719} {arXiv:2312.00719 [hep-th]} \BibitemShut {NoStop}%
\bibitem [{\citenamefont {Bernuzzi}\ \emph {et~al.}(2015)\citenamefont {Bernuzzi}, \citenamefont {Nagar}, \citenamefont {Dietrich},\ and\ \citenamefont {Damour}}]{Bernuzzi:2014owa}%
  \BibitemOpen
  \bibfield  {author} {\bibinfo {author} {\bibfnamefont {S.}~\bibnamefont {Bernuzzi}}, \bibinfo {author} {\bibfnamefont {A.}~\bibnamefont {Nagar}}, \bibinfo {author} {\bibfnamefont {T.}~\bibnamefont {Dietrich}},\ and\ \bibinfo {author} {\bibfnamefont {T.}~\bibnamefont {Damour}},\ }\bibfield  {title} {\bibinfo {title} {{Modeling the Dynamics of Tidally Interacting Binary Neutron Stars up to the Merger}},\ }\href {https://doi.org/10.1103/PhysRevLett.114.161103} {\bibfield  {journal} {\bibinfo  {journal} {Phys. Rev. Lett.}\ }\textbf {\bibinfo {volume} {114}},\ \bibinfo {pages} {161103} (\bibinfo {year} {2015})},\ \Eprint {https://arxiv.org/abs/1412.4553} {arXiv:1412.4553 [gr-qc]} \BibitemShut {NoStop}%
\bibitem [{\citenamefont {Steinhoff}\ \emph {et~al.}(2016)\citenamefont {Steinhoff}, \citenamefont {Hinderer}, \citenamefont {Buonanno},\ and\ \citenamefont {Taracchini}}]{Steinhoff:2016rfi}%
  \BibitemOpen
  \bibfield  {author} {\bibinfo {author} {\bibfnamefont {J.}~\bibnamefont {Steinhoff}}, \bibinfo {author} {\bibfnamefont {T.}~\bibnamefont {Hinderer}}, \bibinfo {author} {\bibfnamefont {A.}~\bibnamefont {Buonanno}},\ and\ \bibinfo {author} {\bibfnamefont {A.}~\bibnamefont {Taracchini}},\ }\bibfield  {title} {\bibinfo {title} {{Dynamical Tides in General Relativity: Effective Action and Effective-One-Body Hamiltonian}},\ }\href {https://doi.org/10.1103/PhysRevD.94.104028} {\bibfield  {journal} {\bibinfo  {journal} {Phys. Rev. D}\ }\textbf {\bibinfo {volume} {94}},\ \bibinfo {pages} {104028} (\bibinfo {year} {2016})},\ \Eprint {https://arxiv.org/abs/1608.01907} {arXiv:1608.01907 [gr-qc]} \BibitemShut {NoStop}%
\bibitem [{\citenamefont {Lackey}\ \emph {et~al.}(2019)\citenamefont {Lackey}, \citenamefont {P{\"u}rrer}, \citenamefont {Taracchini},\ and\ \citenamefont {Marsat}}]{Lackey:2018zvw}%
  \BibitemOpen
  \bibfield  {author} {\bibinfo {author} {\bibfnamefont {B.~D.}\ \bibnamefont {Lackey}}, \bibinfo {author} {\bibfnamefont {M.}~\bibnamefont {P{\"u}rrer}}, \bibinfo {author} {\bibfnamefont {A.}~\bibnamefont {Taracchini}},\ and\ \bibinfo {author} {\bibfnamefont {S.}~\bibnamefont {Marsat}},\ }\bibfield  {title} {\bibinfo {title} {{Surrogate model for an aligned-spin effective one body waveform model of binary neutron star inspirals using Gaussian process regression}},\ }\href {https://doi.org/10.1103/PhysRevD.100.024002} {\bibfield  {journal} {\bibinfo  {journal} {Phys. Rev. D}\ }\textbf {\bibinfo {volume} {100}},\ \bibinfo {pages} {024002} (\bibinfo {year} {2019})},\ \Eprint {https://arxiv.org/abs/1812.08643} {arXiv:1812.08643 [gr-qc]} \BibitemShut {NoStop}%
\bibitem [{\citenamefont {Nagar}\ \emph {et~al.}(2019{\natexlab{a}})\citenamefont {Nagar}, \citenamefont {Messina}, \citenamefont {Rettegno}, \citenamefont {Bini}, \citenamefont {Damour}, \citenamefont {Geralico}, \citenamefont {Akcay},\ and\ \citenamefont {Bernuzzi}}]{Nagar:2018plt}%
  \BibitemOpen
  \bibfield  {author} {\bibinfo {author} {\bibfnamefont {A.}~\bibnamefont {Nagar}}, \bibinfo {author} {\bibfnamefont {F.}~\bibnamefont {Messina}}, \bibinfo {author} {\bibfnamefont {P.}~\bibnamefont {Rettegno}}, \bibinfo {author} {\bibfnamefont {D.}~\bibnamefont {Bini}}, \bibinfo {author} {\bibfnamefont {T.}~\bibnamefont {Damour}}, \bibinfo {author} {\bibfnamefont {A.}~\bibnamefont {Geralico}}, \bibinfo {author} {\bibfnamefont {S.}~\bibnamefont {Akcay}},\ and\ \bibinfo {author} {\bibfnamefont {S.}~\bibnamefont {Bernuzzi}},\ }\bibfield  {title} {\bibinfo {title} {{Nonlinear-in-spin effects in effective-one-body waveform models of spin-aligned, inspiralling, neutron star binaries}},\ }\href {https://doi.org/10.1103/PhysRevD.99.044007} {\bibfield  {journal} {\bibinfo  {journal} {Phys. Rev. D}\ }\textbf {\bibinfo {volume} {99}},\ \bibinfo {pages} {044007} (\bibinfo {year} {2019}{\natexlab{a}})},\ \Eprint {https://arxiv.org/abs/1812.07923} {arXiv:1812.07923 [gr-qc]} \BibitemShut {NoStop}%
\bibitem [{\citenamefont {Thompson}\ \emph {et~al.}(2020)\citenamefont {Thompson}, \citenamefont {Fauchon-Jones}, \citenamefont {Khan}, \citenamefont {Nitoglia}, \citenamefont {Pannarale}, \citenamefont {Dietrich},\ and\ \citenamefont {Hannam}}]{Thompson:2020nei}%
  \BibitemOpen
  \bibfield  {author} {\bibinfo {author} {\bibfnamefont {J.~E.}\ \bibnamefont {Thompson}}, \bibinfo {author} {\bibfnamefont {E.}~\bibnamefont {Fauchon-Jones}}, \bibinfo {author} {\bibfnamefont {S.}~\bibnamefont {Khan}}, \bibinfo {author} {\bibfnamefont {E.}~\bibnamefont {Nitoglia}}, \bibinfo {author} {\bibfnamefont {F.}~\bibnamefont {Pannarale}}, \bibinfo {author} {\bibfnamefont {T.}~\bibnamefont {Dietrich}},\ and\ \bibinfo {author} {\bibfnamefont {M.}~\bibnamefont {Hannam}},\ }\bibfield  {title} {\bibinfo {title} {{Modeling the gravitational wave signature of neutron star black hole coalescences}},\ }\href {https://doi.org/10.1103/PhysRevD.101.124059} {\bibfield  {journal} {\bibinfo  {journal} {Phys. Rev. D}\ }\textbf {\bibinfo {volume} {101}},\ \bibinfo {pages} {124059} (\bibinfo {year} {2020})},\ \Eprint {https://arxiv.org/abs/2002.08383} {arXiv:2002.08383 [gr-qc]} \BibitemShut {NoStop}%
\bibitem [{\citenamefont {Abac}\ \emph {et~al.}(2024)\citenamefont {Abac}, \citenamefont {Dietrich}, \citenamefont {Buonanno}, \citenamefont {Steinhoff},\ and\ \citenamefont {Ujevic}}]{Abac:2023ujg}%
  \BibitemOpen
  \bibfield  {author} {\bibinfo {author} {\bibfnamefont {A.}~\bibnamefont {Abac}}, \bibinfo {author} {\bibfnamefont {T.}~\bibnamefont {Dietrich}}, \bibinfo {author} {\bibfnamefont {A.}~\bibnamefont {Buonanno}}, \bibinfo {author} {\bibfnamefont {J.}~\bibnamefont {Steinhoff}},\ and\ \bibinfo {author} {\bibfnamefont {M.}~\bibnamefont {Ujevic}},\ }\bibfield  {title} {\bibinfo {title} {{New and robust gravitational-waveform model for high-mass-ratio binary neutron star systems with dynamical tidal effects}},\ }\href {https://doi.org/10.1103/PhysRevD.109.024062} {\bibfield  {journal} {\bibinfo  {journal} {Phys. Rev. D}\ }\textbf {\bibinfo {volume} {109}},\ \bibinfo {pages} {024062} (\bibinfo {year} {2024})},\ \Eprint {https://arxiv.org/abs/2311.07456} {arXiv:2311.07456 [gr-qc]} \BibitemShut {NoStop}%
\bibitem [{\citenamefont {Williams}\ \emph {et~al.}(2024)\citenamefont {Williams}, \citenamefont {Schmidt},\ and\ \citenamefont {Pratten}}]{Williams:2024twp}%
  \BibitemOpen
  \bibfield  {author} {\bibinfo {author} {\bibfnamefont {N.}~\bibnamefont {Williams}}, \bibinfo {author} {\bibfnamefont {P.}~\bibnamefont {Schmidt}},\ and\ \bibinfo {author} {\bibfnamefont {G.}~\bibnamefont {Pratten}},\ }\bibfield  {title} {\bibinfo {title} {{Phenomenological model of gravitational self-force enhanced tides in inspiraling binary neutron stars}},\ }\href {https://doi.org/10.1103/PhysRevD.110.104013} {\bibfield  {journal} {\bibinfo  {journal} {Phys. Rev. D}\ }\textbf {\bibinfo {volume} {110}},\ \bibinfo {pages} {104013} (\bibinfo {year} {2024})},\ \Eprint {https://arxiv.org/abs/2407.08538} {arXiv:2407.08538 [gr-qc]} \BibitemShut {NoStop}%
\bibitem [{\citenamefont {Haberland}\ \emph {et~al.}(2025)\citenamefont {Haberland}, \citenamefont {Buonanno},\ and\ \citenamefont {Steinhoff}}]{Haberland:2025luz}%
  \BibitemOpen
  \bibfield  {author} {\bibinfo {author} {\bibfnamefont {M.}~\bibnamefont {Haberland}}, \bibinfo {author} {\bibfnamefont {A.}~\bibnamefont {Buonanno}},\ and\ \bibinfo {author} {\bibfnamefont {J.}~\bibnamefont {Steinhoff}},\ }\bibfield  {title} {\bibinfo {title} {{Modeling matter in seobnrv5thm: Generating fast and accurate effective-one-body waveforms for spin-aligned binary neutron stars}},\ }\href {https://doi.org/10.1103/d3ns-h77x} {\bibfield  {journal} {\bibinfo  {journal} {Phys. Rev. D}\ }\textbf {\bibinfo {volume} {112}},\ \bibinfo {pages} {084024} (\bibinfo {year} {2025})},\ \Eprint {https://arxiv.org/abs/2503.18934} {arXiv:2503.18934 [gr-qc]} \BibitemShut {NoStop}%
\bibitem [{\citenamefont {Albanesi}\ \emph {et~al.}(2025)\citenamefont {Albanesi}, \citenamefont {Gamba}, \citenamefont {Bernuzzi}, \citenamefont {Fontbut{\'e}}, \citenamefont {Gonzalez},\ and\ \citenamefont {Nagar}}]{Albanesi:2025txj}%
  \BibitemOpen
  \bibfield  {author} {\bibinfo {author} {\bibfnamefont {S.}~\bibnamefont {Albanesi}}, \bibinfo {author} {\bibfnamefont {R.}~\bibnamefont {Gamba}}, \bibinfo {author} {\bibfnamefont {S.}~\bibnamefont {Bernuzzi}}, \bibinfo {author} {\bibfnamefont {J.}~\bibnamefont {Fontbut{\'e}}}, \bibinfo {author} {\bibfnamefont {A.}~\bibnamefont {Gonzalez}},\ and\ \bibinfo {author} {\bibfnamefont {A.}~\bibnamefont {Nagar}},\ }\bibfield  {title} {\bibinfo {title} {{Effective-one-body modeling for generic compact binaries with arbitrary orbits}},\ }\href {https://doi.org/10.1103/3snf-w1x7} {\bibfield  {journal} {\bibinfo  {journal} {Phys. Rev. D}\ }\textbf {\bibinfo {volume} {112}},\ \bibinfo {pages} {L121503} (\bibinfo {year} {2025})},\ \Eprint {https://arxiv.org/abs/2503.14580} {arXiv:2503.14580 [gr-qc]} \BibitemShut {NoStop}%
\bibitem [{\citenamefont {Schulze}\ \emph {et~al.}(2026)\citenamefont {Schulze}, \citenamefont {Bernuzzi}, \citenamefont {Rettegno}, \citenamefont {Fontbut{\'e}}, \citenamefont {Placidi},\ and\ \citenamefont {Damour}}]{Schulze:2026ewu}%
  \BibitemOpen
  \bibfield  {author} {\bibinfo {author} {\bibfnamefont {M.}~\bibnamefont {Schulze}}, \bibinfo {author} {\bibfnamefont {S.}~\bibnamefont {Bernuzzi}}, \bibinfo {author} {\bibfnamefont {P.}~\bibnamefont {Rettegno}}, \bibinfo {author} {\bibfnamefont {J.}~\bibnamefont {Fontbut{\'e}}}, \bibinfo {author} {\bibfnamefont {A.}~\bibnamefont {Placidi}},\ and\ \bibinfo {author} {\bibfnamefont {T.}~\bibnamefont {Damour}},\ }\bibfield  {title} {\bibinfo {title} {{High-order effective-one-body tidal interactions and gravitational scattering}},\ }\href {https://doi.org/10.1103/j9pp-dlrz} {\bibfield  {journal} {\bibinfo  {journal} {Phys. Rev. D}\ }\textbf {\bibinfo {volume} {113}},\ \bibinfo {pages} {104027} (\bibinfo {year} {2026})},\ \Eprint {https://arxiv.org/abs/2603.22467} {arXiv:2603.22467 [gr-qc]} \BibitemShut {NoStop}%
\bibitem [{\citenamefont {Gamboa}\ \emph {et~al.}(2026)\citenamefont {Gamboa} \emph {et~al.}}]{Gamboa:2026jht}%
  \BibitemOpen
  \bibfield  {author} {\bibinfo {author} {\bibfnamefont {A.}~\bibnamefont {Gamboa}} \emph {et~al.},\ }\bibfield  {title} {\bibinfo {title} {{Accurate waveforms for generic planar-orbit binary black holes: The multipolar effective-one-body model SEOBNRv6EHM}},\ }\href@noop {} {\  (\bibinfo {year} {2026})},\ \Eprint {https://arxiv.org/abs/2605.28715} {arXiv:2605.28715 [gr-qc]} \BibitemShut {NoStop}%
\bibitem [{\citenamefont {Ramis~Vidal}\ \emph {et~al.}(2026)\citenamefont {Ramis~Vidal}, \citenamefont {Abac}, \citenamefont {Colleoni}, \citenamefont {Dietrich}, \citenamefont {Mourier}, \citenamefont {Gonzalez}, \citenamefont {Markin},\ and\ \citenamefont {Puecher}}]{RamisVidal:2026ycb}%
  \BibitemOpen
  \bibfield  {author} {\bibinfo {author} {\bibfnamefont {F.~A.}\ \bibnamefont {Ramis~Vidal}}, \bibinfo {author} {\bibfnamefont {A.}~\bibnamefont {Abac}}, \bibinfo {author} {\bibfnamefont {M.}~\bibnamefont {Colleoni}}, \bibinfo {author} {\bibfnamefont {T.}~\bibnamefont {Dietrich}}, \bibinfo {author} {\bibfnamefont {P.}~\bibnamefont {Mourier}}, \bibinfo {author} {\bibfnamefont {A.}~\bibnamefont {Gonzalez}}, \bibinfo {author} {\bibfnamefont {I.}~\bibnamefont {Markin}},\ and\ \bibinfo {author} {\bibfnamefont {A.}~\bibnamefont {Puecher}},\ }\bibfield  {title} {\bibinfo {title} {{Fast gravitational waveform models for quasi-circular coalescences of neutron star--black hole binaries}},\ }\href@noop {} {\  (\bibinfo {year} {2026})},\ \Eprint {https://arxiv.org/abs/2606.04810} {arXiv:2606.04810 [gr-qc]} \BibitemShut {NoStop}%
\bibitem [{\citenamefont {Owen}\ \emph {et~al.}(2023)\citenamefont {Owen}, \citenamefont {Haster}, \citenamefont {Perkins}, \citenamefont {Cornish},\ and\ \citenamefont {Yunes}}]{Owen:2023mid}%
  \BibitemOpen
  \bibfield  {author} {\bibinfo {author} {\bibfnamefont {C.~B.}\ \bibnamefont {Owen}}, \bibinfo {author} {\bibfnamefont {C.-J.}\ \bibnamefont {Haster}}, \bibinfo {author} {\bibfnamefont {S.}~\bibnamefont {Perkins}}, \bibinfo {author} {\bibfnamefont {N.~J.}\ \bibnamefont {Cornish}},\ and\ \bibinfo {author} {\bibfnamefont {N.}~\bibnamefont {Yunes}},\ }\bibfield  {title} {\bibinfo {title} {{Waveform accuracy and systematic uncertainties in current gravitational wave observations}},\ }\href {https://doi.org/10.1103/PhysRevD.108.044018} {\bibfield  {journal} {\bibinfo  {journal} {Phys. Rev. D}\ }\textbf {\bibinfo {volume} {108}},\ \bibinfo {pages} {044018} (\bibinfo {year} {2023})},\ \Eprint {https://arxiv.org/abs/2301.11941} {arXiv:2301.11941 [gr-qc]} \BibitemShut {NoStop}%
\bibitem [{\citenamefont {Estell{\'e}s}\ \emph {et~al.}(2022)\citenamefont {Estell{\'e}s} \emph {et~al.}}]{Estelles:2021jnz}%
  \BibitemOpen
  \bibfield  {author} {\bibinfo {author} {\bibfnamefont {H.}~\bibnamefont {Estell{\'e}s}} \emph {et~al.},\ }\bibfield  {title} {\bibinfo {title} {{A Detailed Analysis of GW190521 with Phenomenological Waveform Models}},\ }\href {https://doi.org/10.3847/1538-4357/ac33a0} {\bibfield  {journal} {\bibinfo  {journal} {Astrophys. J.}\ }\textbf {\bibinfo {volume} {924}},\ \bibinfo {pages} {79} (\bibinfo {year} {2022})},\ \Eprint {https://arxiv.org/abs/2105.06360} {arXiv:2105.06360 [gr-qc]} \BibitemShut {NoStop}%
\bibitem [{\citenamefont {Maggio}\ \emph {et~al.}(2023)\citenamefont {Maggio}, \citenamefont {Silva}, \citenamefont {Buonanno},\ and\ \citenamefont {Ghosh}}]{Maggio:2022hre}%
  \BibitemOpen
  \bibfield  {author} {\bibinfo {author} {\bibfnamefont {E.}~\bibnamefont {Maggio}}, \bibinfo {author} {\bibfnamefont {H.~O.}\ \bibnamefont {Silva}}, \bibinfo {author} {\bibfnamefont {A.}~\bibnamefont {Buonanno}},\ and\ \bibinfo {author} {\bibfnamefont {A.}~\bibnamefont {Ghosh}},\ }\bibfield  {title} {\bibinfo {title} {{Tests of general relativity in the nonlinear regime: A parametrized plunge-merger-ringdown gravitational waveform model}},\ }\href {https://doi.org/10.1103/PhysRevD.108.024043} {\bibfield  {journal} {\bibinfo  {journal} {Phys. Rev. D}\ }\textbf {\bibinfo {volume} {108}},\ \bibinfo {pages} {024043} (\bibinfo {year} {2023})},\ \Eprint {https://arxiv.org/abs/2212.09655} {arXiv:2212.09655 [gr-qc]} \BibitemShut {NoStop}%
\bibitem [{\citenamefont {Abac}\ \emph {et~al.}(2025{\natexlab{d}})\citenamefont {Abac} \emph {et~al.}}]{LIGOScientific:2025rsn}%
  \BibitemOpen
  \bibfield  {author} {\bibinfo {author} {\bibfnamefont {A.~G.}\ \bibnamefont {Abac}} \emph {et~al.} (\bibinfo {collaboration} {LIGO Scientific, VIRGO, KAGRA}),\ }\bibfield  {title} {\bibinfo {title} {{GW231123: A Binary Black Hole Merger with Total Mass 190{\textendash}265 M$_{\odot}$}},\ }\href {https://doi.org/10.3847/2041-8213/ae0c9c} {\bibfield  {journal} {\bibinfo  {journal} {Astrophys. J. Lett.}\ }\textbf {\bibinfo {volume} {993}},\ \bibinfo {pages} {L25} (\bibinfo {year} {2025}{\natexlab{d}})},\ \Eprint {https://arxiv.org/abs/2507.08219} {arXiv:2507.08219 [astro-ph.HE]} \BibitemShut {NoStop}%
\bibitem [{\citenamefont {van~de Meent}\ and\ \citenamefont {Pfeiffer}(2020)}]{vandeMeent:2020xgc}%
  \BibitemOpen
  \bibfield  {author} {\bibinfo {author} {\bibfnamefont {M.}~\bibnamefont {van~de Meent}}\ and\ \bibinfo {author} {\bibfnamefont {H.~P.}\ \bibnamefont {Pfeiffer}},\ }\bibfield  {title} {\bibinfo {title} {{Intermediate mass-ratio black hole binaries: Applicability of small mass-ratio perturbation theory}},\ }\href {https://doi.org/10.1103/PhysRevLett.125.181101} {\bibfield  {journal} {\bibinfo  {journal} {Phys. Rev. Lett.}\ }\textbf {\bibinfo {volume} {125}},\ \bibinfo {pages} {181101} (\bibinfo {year} {2020})},\ \Eprint {https://arxiv.org/abs/2006.12036} {arXiv:2006.12036 [gr-qc]} \BibitemShut {NoStop}%
\bibitem [{\citenamefont {Rink}\ \emph {et~al.}(2024)\citenamefont {Rink}, \citenamefont {Bachhar}, \citenamefont {Islam}, \citenamefont {Rifat}, \citenamefont {Gonzalez-Quesada}, \citenamefont {Field}, \citenamefont {Khanna}, \citenamefont {Hughes},\ and\ \citenamefont {Varma}}]{Rink:2024swg}%
  \BibitemOpen
  \bibfield  {author} {\bibinfo {author} {\bibfnamefont {K.}~\bibnamefont {Rink}}, \bibinfo {author} {\bibfnamefont {R.}~\bibnamefont {Bachhar}}, \bibinfo {author} {\bibfnamefont {T.}~\bibnamefont {Islam}}, \bibinfo {author} {\bibfnamefont {N.~E.~M.}\ \bibnamefont {Rifat}}, \bibinfo {author} {\bibfnamefont {K.}~\bibnamefont {Gonzalez-Quesada}}, \bibinfo {author} {\bibfnamefont {S.~E.}\ \bibnamefont {Field}}, \bibinfo {author} {\bibfnamefont {G.}~\bibnamefont {Khanna}}, \bibinfo {author} {\bibfnamefont {S.~A.}\ \bibnamefont {Hughes}},\ and\ \bibinfo {author} {\bibfnamefont {V.}~\bibnamefont {Varma}},\ }\bibfield  {title} {\bibinfo {title} {{Gravitational wave surrogate model for spinning, intermediate mass ratio binaries based on perturbation theory and numerical relativity}},\ }\href {https://doi.org/10.1103/PhysRevD.110.124069} {\bibfield  {journal} {\bibinfo  {journal} {Phys. Rev. D}\ }\textbf {\bibinfo {volume} {110}},\ \bibinfo {pages} {124069} (\bibinfo {year} {2024})},\ \Eprint
  {https://arxiv.org/abs/2407.18319} {arXiv:2407.18319 [gr-qc]} \BibitemShut {NoStop}%
\bibitem [{\citenamefont {Kumar}\ \emph {et~al.}(2026)\citenamefont {Kumar}, \citenamefont {Melching}, \citenamefont {Ohme}, \citenamefont {Narola}, \citenamefont {Dooney},\ and\ \citenamefont {Van Den~Broeck}}]{Kumar:2026ckp}%
  \BibitemOpen
  \bibfield  {author} {\bibinfo {author} {\bibfnamefont {S.}~\bibnamefont {Kumar}}, \bibinfo {author} {\bibfnamefont {M.}~\bibnamefont {Melching}}, \bibinfo {author} {\bibfnamefont {F.}~\bibnamefont {Ohme}}, \bibinfo {author} {\bibfnamefont {H.}~\bibnamefont {Narola}}, \bibinfo {author} {\bibfnamefont {T.}~\bibnamefont {Dooney}},\ and\ \bibinfo {author} {\bibfnamefont {C.}~\bibnamefont {Van Den~Broeck}},\ }\bibfield  {title} {\bibinfo {title} {{Mitigating Systematic Errors in Parameter Estimation of Binary Black Hole Mergers in O1-O3 LIGO-Virgo Data}},\ }\href@noop {} {\  (\bibinfo {year} {2026})},\ \Eprint {https://arxiv.org/abs/2604.21859} {arXiv:2604.21859 [astro-ph.HE]} \BibitemShut {NoStop}%
\bibitem [{\citenamefont {Mino}\ \emph {et~al.}(1997)\citenamefont {Mino}, \citenamefont {Sasaki},\ and\ \citenamefont {Tanaka}}]{Mino:1996nk}%
  \BibitemOpen
  \bibfield  {author} {\bibinfo {author} {\bibfnamefont {Y.}~\bibnamefont {Mino}}, \bibinfo {author} {\bibfnamefont {M.}~\bibnamefont {Sasaki}},\ and\ \bibinfo {author} {\bibfnamefont {T.}~\bibnamefont {Tanaka}},\ }\bibfield  {title} {\bibinfo {title} {{Gravitational radiation reaction to a particle motion}},\ }\href {https://doi.org/10.1103/PhysRevD.55.3457} {\bibfield  {journal} {\bibinfo  {journal} {Phys. Rev. D}\ }\textbf {\bibinfo {volume} {55}},\ \bibinfo {pages} {3457} (\bibinfo {year} {1997})},\ \Eprint {https://arxiv.org/abs/gr-qc/9606018} {arXiv:gr-qc/9606018} \BibitemShut {NoStop}%
\bibitem [{\citenamefont {Poisson}\ \emph {et~al.}(2011)\citenamefont {Poisson}, \citenamefont {Pound},\ and\ \citenamefont {Vega}}]{Poisson:2011nh}%
  \BibitemOpen
  \bibfield  {author} {\bibinfo {author} {\bibfnamefont {E.}~\bibnamefont {Poisson}}, \bibinfo {author} {\bibfnamefont {A.}~\bibnamefont {Pound}},\ and\ \bibinfo {author} {\bibfnamefont {I.}~\bibnamefont {Vega}},\ }\bibfield  {title} {\bibinfo {title} {{The Motion of point particles in curved spacetime}},\ }\href {https://doi.org/10.12942/lrr-2011-7} {\bibfield  {journal} {\bibinfo  {journal} {Living Rev. Rel.}\ }\textbf {\bibinfo {volume} {14}},\ \bibinfo {pages} {7} (\bibinfo {year} {2011})},\ \Eprint {https://arxiv.org/abs/1102.0529} {arXiv:1102.0529 [gr-qc]} \BibitemShut {NoStop}%
\bibitem [{\citenamefont {Barack}\ and\ \citenamefont {Pound}(2019)}]{Barack:2018yvs}%
  \BibitemOpen
  \bibfield  {author} {\bibinfo {author} {\bibfnamefont {L.}~\bibnamefont {Barack}}\ and\ \bibinfo {author} {\bibfnamefont {A.}~\bibnamefont {Pound}},\ }\bibfield  {title} {\bibinfo {title} {{Self-force and radiation reaction in general relativity}},\ }\href {https://doi.org/10.1088/1361-6633/aae552} {\bibfield  {journal} {\bibinfo  {journal} {Rept. Prog. Phys.}\ }\textbf {\bibinfo {volume} {82}},\ \bibinfo {pages} {016904} (\bibinfo {year} {2019})},\ \Eprint {https://arxiv.org/abs/1805.10385} {arXiv:1805.10385 [gr-qc]} \BibitemShut {NoStop}%
\bibitem [{\citenamefont {{Pound}}\ and\ \citenamefont {{Wardell}}(2022)}]{Pound:2021qin}%
  \BibitemOpen
  \bibfield  {author} {\bibinfo {author} {\bibfnamefont {A.}~\bibnamefont {{Pound}}}\ and\ \bibinfo {author} {\bibfnamefont {B.}~\bibnamefont {{Wardell}}},\ }\bibfield  {title} {\bibinfo {title} {{Black Hole Perturbation Theory and Gravitational Self-Force}},\ }in\ \href {https://doi.org/10.1007/978-981-15-4702-7_38-1} {\emph {\bibinfo {booktitle} {Handbook of Gravitational Wave Astronomy}}}\ (\bibinfo {year} {2022})\ p.~\bibinfo {pages} {38},\ \Eprint {https://arxiv.org/abs/2101.04592} {2101.04592} \BibitemShut {NoStop}%
\bibitem [{\citenamefont {Wardell}\ \emph {et~al.}(2023)\citenamefont {Wardell}, \citenamefont {Pound}, \citenamefont {Warburton}, \citenamefont {Miller}, \citenamefont {Durkan},\ and\ \citenamefont {Le~Tiec}}]{Wardell:2021fyy}%
  \BibitemOpen
  \bibfield  {author} {\bibinfo {author} {\bibfnamefont {B.}~\bibnamefont {Wardell}}, \bibinfo {author} {\bibfnamefont {A.}~\bibnamefont {Pound}}, \bibinfo {author} {\bibfnamefont {N.}~\bibnamefont {Warburton}}, \bibinfo {author} {\bibfnamefont {J.}~\bibnamefont {Miller}}, \bibinfo {author} {\bibfnamefont {L.}~\bibnamefont {Durkan}},\ and\ \bibinfo {author} {\bibfnamefont {A.}~\bibnamefont {Le~Tiec}},\ }\bibfield  {title} {\bibinfo {title} {{Gravitational Waveforms for Compact Binaries from Second-Order Self-Force Theory}},\ }\href {https://doi.org/10.1103/PhysRevLett.130.241402} {\bibfield  {journal} {\bibinfo  {journal} {Phys. Rev. Lett.}\ }\textbf {\bibinfo {volume} {130}},\ \bibinfo {pages} {241402} (\bibinfo {year} {2023})},\ \Eprint {https://arxiv.org/abs/2112.12265} {arXiv:2112.12265 [gr-qc]} \BibitemShut {NoStop}%
\bibitem [{\citenamefont {Mathews}\ \emph {et~al.}(2026)\citenamefont {Mathews}, \citenamefont {Wardell}, \citenamefont {Pound},\ and\ \citenamefont {Warburton}}]{Mathews:2025txc}%
  \BibitemOpen
  \bibfield  {author} {\bibinfo {author} {\bibfnamefont {J.}~\bibnamefont {Mathews}}, \bibinfo {author} {\bibfnamefont {B.}~\bibnamefont {Wardell}}, \bibinfo {author} {\bibfnamefont {A.}~\bibnamefont {Pound}},\ and\ \bibinfo {author} {\bibfnamefont {N.}~\bibnamefont {Warburton}},\ }\bibfield  {title} {\bibinfo {title} {{Postadiabatic self-force waveforms: Slowly spinning primary and precessing secondary}},\ }\href {https://doi.org/10.1103/ph3p-mscl} {\bibfield  {journal} {\bibinfo  {journal} {Phys. Rev. D}\ }\textbf {\bibinfo {volume} {113}},\ \bibinfo {pages} {064034} (\bibinfo {year} {2026})},\ \Eprint {https://arxiv.org/abs/2510.16113} {arXiv:2510.16113 [gr-qc]} \BibitemShut {NoStop}%
\bibitem [{\citenamefont {Katz}\ \emph {et~al.}(2021)\citenamefont {Katz}, \citenamefont {Chua}, \citenamefont {Speri}, \citenamefont {Warburton},\ and\ \citenamefont {Hughes}}]{Katz:2021yft}%
  \BibitemOpen
  \bibfield  {author} {\bibinfo {author} {\bibfnamefont {M.~L.}\ \bibnamefont {Katz}}, \bibinfo {author} {\bibfnamefont {A.~J.~K.}\ \bibnamefont {Chua}}, \bibinfo {author} {\bibfnamefont {L.}~\bibnamefont {Speri}}, \bibinfo {author} {\bibfnamefont {N.}~\bibnamefont {Warburton}},\ and\ \bibinfo {author} {\bibfnamefont {S.~A.}\ \bibnamefont {Hughes}},\ }\bibfield  {title} {\bibinfo {title} {{Fast extreme-mass-ratio-inspiral waveforms: New tools for millihertz gravitational-wave data analysis}},\ }\href {https://doi.org/10.1103/PhysRevD.104.064047} {\bibfield  {journal} {\bibinfo  {journal} {Phys. Rev. D}\ }\textbf {\bibinfo {volume} {104}},\ \bibinfo {pages} {064047} (\bibinfo {year} {2021})},\ \Eprint {https://arxiv.org/abs/2104.04582} {arXiv:2104.04582 [gr-qc]} \BibitemShut {NoStop}%
\bibitem [{\citenamefont {Speri}\ \emph {et~al.}(2023)\citenamefont {Speri}, \citenamefont {Katz}, \citenamefont {Chua}, \citenamefont {Hughes}, \citenamefont {Warburton}, \citenamefont {Thompson}, \citenamefont {Chapman-Bird},\ and\ \citenamefont {Gair}}]{Speri:2023jte}%
  \BibitemOpen
  \bibfield  {author} {\bibinfo {author} {\bibfnamefont {L.}~\bibnamefont {Speri}}, \bibinfo {author} {\bibfnamefont {M.~L.}\ \bibnamefont {Katz}}, \bibinfo {author} {\bibfnamefont {A.~J.~K.}\ \bibnamefont {Chua}}, \bibinfo {author} {\bibfnamefont {S.~A.}\ \bibnamefont {Hughes}}, \bibinfo {author} {\bibfnamefont {N.}~\bibnamefont {Warburton}}, \bibinfo {author} {\bibfnamefont {J.~E.}\ \bibnamefont {Thompson}}, \bibinfo {author} {\bibfnamefont {C.~E.~A.}\ \bibnamefont {Chapman-Bird}},\ and\ \bibinfo {author} {\bibfnamefont {J.~R.}\ \bibnamefont {Gair}},\ }\bibfield  {title} {\bibinfo {title} {{Fast and Fourier: Extreme Mass Ratio Inspiral Waveforms in the Frequency Domain}}\ }\href {https://doi.org/10.3389/fams.2023.1266739} {10.3389/fams.2023.1266739} (\bibinfo {year} {2023}),\ \Eprint {https://arxiv.org/abs/2307.12585} {arXiv:2307.12585 [gr-qc]} \BibitemShut {NoStop}%
\bibitem [{\citenamefont {Nasipak}(2024)}]{Nasipak:2023kuf}%
  \BibitemOpen
  \bibfield  {author} {\bibinfo {author} {\bibfnamefont {Z.}~\bibnamefont {Nasipak}},\ }\bibfield  {title} {\bibinfo {title} {{Adiabatic gravitational waveform model for compact objects undergoing quasicircular inspirals into rotating massive black holes}},\ }\href {https://doi.org/10.1103/PhysRevD.109.044020} {\bibfield  {journal} {\bibinfo  {journal} {Phys. Rev. D}\ }\textbf {\bibinfo {volume} {109}},\ \bibinfo {pages} {044020} (\bibinfo {year} {2024})},\ \Eprint {https://arxiv.org/abs/2310.19706} {arXiv:2310.19706 [gr-qc]} \BibitemShut {NoStop}%
\bibitem [{\citenamefont {Chapman-Bird}\ \emph {et~al.}(2025)\citenamefont {Chapman-Bird} \emph {et~al.}}]{Chapman-Bird:2025xtd}%
  \BibitemOpen
  \bibfield  {author} {\bibinfo {author} {\bibfnamefont {C.~E.~A.}\ \bibnamefont {Chapman-Bird}} \emph {et~al.},\ }\bibfield  {title} {\bibinfo {title} {{Efficient waveforms for asymmetric-mass eccentric equatorial inspirals into rapidly spinning black holes}},\ }\href {https://doi.org/10.1103/scbp-75pf} {\bibfield  {journal} {\bibinfo  {journal} {Phys. Rev. D}\ }\textbf {\bibinfo {volume} {112}},\ \bibinfo {pages} {104023} (\bibinfo {year} {2025})},\ \Eprint {https://arxiv.org/abs/2506.09470} {arXiv:2506.09470 [gr-qc]} \BibitemShut {NoStop}%
\bibitem [{\citenamefont {Ruangsri}\ \emph {et~al.}(2016)\citenamefont {Ruangsri}, \citenamefont {Vigeland},\ and\ \citenamefont {Hughes}}]{Ruangsri:2015cvg}%
  \BibitemOpen
  \bibfield  {author} {\bibinfo {author} {\bibfnamefont {U.}~\bibnamefont {Ruangsri}}, \bibinfo {author} {\bibfnamefont {S.~J.}\ \bibnamefont {Vigeland}},\ and\ \bibinfo {author} {\bibfnamefont {S.~A.}\ \bibnamefont {Hughes}},\ }\bibfield  {title} {\bibinfo {title} {{Gyroscopes orbiting black holes: A frequency-domain approach to precession and spin-curvature coupling for spinning bodies on generic Kerr orbits}},\ }\href {https://doi.org/10.1103/PhysRevD.94.044008} {\bibfield  {journal} {\bibinfo  {journal} {Phys. Rev. D}\ }\textbf {\bibinfo {volume} {94}},\ \bibinfo {pages} {044008} (\bibinfo {year} {2016})},\ \Eprint {https://arxiv.org/abs/1512.00376} {arXiv:1512.00376 [gr-qc]} \BibitemShut {NoStop}%
\bibitem [{\citenamefont {Harms}\ \emph {et~al.}(2016)\citenamefont {Harms}, \citenamefont {Lukes-Gerakopoulos}, \citenamefont {Bernuzzi},\ and\ \citenamefont {Nagar}}]{Harms:2015ixa}%
  \BibitemOpen
  \bibfield  {author} {\bibinfo {author} {\bibfnamefont {E.}~\bibnamefont {Harms}}, \bibinfo {author} {\bibfnamefont {G.}~\bibnamefont {Lukes-Gerakopoulos}}, \bibinfo {author} {\bibfnamefont {S.}~\bibnamefont {Bernuzzi}},\ and\ \bibinfo {author} {\bibfnamefont {A.}~\bibnamefont {Nagar}},\ }\bibfield  {title} {\bibinfo {title} {{Asymptotic gravitational wave fluxes from a spinning particle in circular equatorial orbits around a rotating black hole}},\ }\href {https://doi.org/10.1103/PhysRevD.93.044015} {\bibfield  {journal} {\bibinfo  {journal} {Phys. Rev. D}\ }\textbf {\bibinfo {volume} {93}},\ \bibinfo {pages} {044015} (\bibinfo {year} {2016})},\ \bibinfo {note} {[Addendum: Phys.Rev.D 100, 129901 (2019)]},\ \Eprint {https://arxiv.org/abs/1510.05548} {arXiv:1510.05548 [gr-qc]} \BibitemShut {NoStop}%
\bibitem [{\citenamefont {Warburton}\ \emph {et~al.}(2017)\citenamefont {Warburton}, \citenamefont {Osburn},\ and\ \citenamefont {Evans}}]{Warburton:2017sxk}%
  \BibitemOpen
  \bibfield  {author} {\bibinfo {author} {\bibfnamefont {N.}~\bibnamefont {Warburton}}, \bibinfo {author} {\bibfnamefont {T.}~\bibnamefont {Osburn}},\ and\ \bibinfo {author} {\bibfnamefont {C.~R.}\ \bibnamefont {Evans}},\ }\bibfield  {title} {\bibinfo {title} {{Evolution of small-mass-ratio binaries with a spinning secondary}},\ }\href {https://doi.org/10.1103/PhysRevD.96.084057} {\bibfield  {journal} {\bibinfo  {journal} {Phys. Rev. D}\ }\textbf {\bibinfo {volume} {96}},\ \bibinfo {pages} {084057} (\bibinfo {year} {2017})},\ \Eprint {https://arxiv.org/abs/1708.03720} {arXiv:1708.03720 [gr-qc]} \BibitemShut {NoStop}%
\bibitem [{\citenamefont {van~de Meent}(2020)}]{vandeMeent:2019cam}%
  \BibitemOpen
  \bibfield  {author} {\bibinfo {author} {\bibfnamefont {M.}~\bibnamefont {van~de Meent}},\ }\bibfield  {title} {\bibinfo {title} {{Analytic solutions for parallel transport along generic bound geodesics in Kerr spacetime}},\ }\href {https://doi.org/10.1088/1361-6382/ab79d5} {\bibfield  {journal} {\bibinfo  {journal} {Class. Quant. Grav.}\ }\textbf {\bibinfo {volume} {37}},\ \bibinfo {pages} {145007} (\bibinfo {year} {2020})},\ \Eprint {https://arxiv.org/abs/1906.05090} {arXiv:1906.05090 [gr-qc]} \BibitemShut {NoStop}%
\bibitem [{\citenamefont {Witzany}(2019)}]{Witzany:2019nml}%
  \BibitemOpen
  \bibfield  {author} {\bibinfo {author} {\bibfnamefont {V.}~\bibnamefont {Witzany}},\ }\bibfield  {title} {\bibinfo {title} {{Hamilton-Jacobi equation for spinning particles near black holes}},\ }\href {https://doi.org/10.1103/PhysRevD.100.104030} {\bibfield  {journal} {\bibinfo  {journal} {Phys. Rev. D}\ }\textbf {\bibinfo {volume} {100}},\ \bibinfo {pages} {104030} (\bibinfo {year} {2019})},\ \Eprint {https://arxiv.org/abs/1903.03651} {arXiv:1903.03651 [gr-qc]} \BibitemShut {NoStop}%
\bibitem [{\citenamefont {Piovano}\ \emph {et~al.}(2020)\citenamefont {Piovano}, \citenamefont {Maselli},\ and\ \citenamefont {Pani}}]{Piovano:2020zin}%
  \BibitemOpen
  \bibfield  {author} {\bibinfo {author} {\bibfnamefont {G.~A.}\ \bibnamefont {Piovano}}, \bibinfo {author} {\bibfnamefont {A.}~\bibnamefont {Maselli}},\ and\ \bibinfo {author} {\bibfnamefont {P.}~\bibnamefont {Pani}},\ }\bibfield  {title} {\bibinfo {title} {{Extreme mass ratio inspirals with spinning secondary: a detailed study of equatorial circular motion}},\ }\href {https://doi.org/10.1103/PhysRevD.102.024041} {\bibfield  {journal} {\bibinfo  {journal} {Phys. Rev. D}\ }\textbf {\bibinfo {volume} {102}},\ \bibinfo {pages} {024041} (\bibinfo {year} {2020})},\ \Eprint {https://arxiv.org/abs/2004.02654} {arXiv:2004.02654 [gr-qc]} \BibitemShut {NoStop}%
\bibitem [{\citenamefont {Mathews}\ \emph {et~al.}(2022)\citenamefont {Mathews}, \citenamefont {Pound},\ and\ \citenamefont {Wardell}}]{Mathews:2021rod}%
  \BibitemOpen
  \bibfield  {author} {\bibinfo {author} {\bibfnamefont {J.}~\bibnamefont {Mathews}}, \bibinfo {author} {\bibfnamefont {A.}~\bibnamefont {Pound}},\ and\ \bibinfo {author} {\bibfnamefont {B.}~\bibnamefont {Wardell}},\ }\bibfield  {title} {\bibinfo {title} {{Self-force calculations with a spinning secondary}},\ }\href {https://doi.org/10.1103/PhysRevD.105.084031} {\bibfield  {journal} {\bibinfo  {journal} {Phys. Rev. D}\ }\textbf {\bibinfo {volume} {105}},\ \bibinfo {pages} {084031} (\bibinfo {year} {2022})},\ \Eprint {https://arxiv.org/abs/2112.13069} {arXiv:2112.13069 [gr-qc]} \BibitemShut {NoStop}%
\bibitem [{\citenamefont {Skoup\'y}\ and\ \citenamefont {Lukes-Gerakopoulos}(2022)}]{Skoupy:2022adh}%
  \BibitemOpen
  \bibfield  {author} {\bibinfo {author} {\bibfnamefont {V.}~\bibnamefont {Skoup\'y}}\ and\ \bibinfo {author} {\bibfnamefont {G.}~\bibnamefont {Lukes-Gerakopoulos}},\ }\bibfield  {title} {\bibinfo {title} {{Adiabatic equatorial inspirals of a spinning body into a Kerr black hole}},\ }\href {https://doi.org/10.1103/PhysRevD.105.084033} {\bibfield  {journal} {\bibinfo  {journal} {Phys. Rev. D}\ }\textbf {\bibinfo {volume} {105}},\ \bibinfo {pages} {084033} (\bibinfo {year} {2022})},\ \Eprint {https://arxiv.org/abs/2201.07044} {arXiv:2201.07044 [gr-qc]} \BibitemShut {NoStop}%
\bibitem [{\citenamefont {Drummond}\ and\ \citenamefont {Hughes}(2022)}]{Drummond:2022efc}%
  \BibitemOpen
  \bibfield  {author} {\bibinfo {author} {\bibfnamefont {L.~V.}\ \bibnamefont {Drummond}}\ and\ \bibinfo {author} {\bibfnamefont {S.~A.}\ \bibnamefont {Hughes}},\ }\bibfield  {title} {\bibinfo {title} {{Precisely computing bound orbits of spinning bodies around black holes. II. Generic orbits}},\ }\href {https://doi.org/10.1103/PhysRevD.105.124041} {\bibfield  {journal} {\bibinfo  {journal} {Phys. Rev. D}\ }\textbf {\bibinfo {volume} {105}},\ \bibinfo {pages} {124041} (\bibinfo {year} {2022})},\ \Eprint {https://arxiv.org/abs/2201.13335} {arXiv:2201.13335 [gr-qc]} \BibitemShut {NoStop}%
\bibitem [{\citenamefont {Witzany}\ and\ \citenamefont {Piovano}(2024)}]{Witzany:2023bmq}%
  \BibitemOpen
  \bibfield  {author} {\bibinfo {author} {\bibfnamefont {V.}~\bibnamefont {Witzany}}\ and\ \bibinfo {author} {\bibfnamefont {G.~A.}\ \bibnamefont {Piovano}},\ }\bibfield  {title} {\bibinfo {title} {{Analytic Solutions for the Motion of Spinning Particles near Spherically Symmetric Black Holes and Exotic Compact Objects}},\ }\href {https://doi.org/10.1103/PhysRevLett.132.171401} {\bibfield  {journal} {\bibinfo  {journal} {Phys. Rev. Lett.}\ }\textbf {\bibinfo {volume} {132}},\ \bibinfo {pages} {171401} (\bibinfo {year} {2024})},\ \Eprint {https://arxiv.org/abs/2308.00021} {arXiv:2308.00021 [gr-qc]} \BibitemShut {NoStop}%
\bibitem [{\citenamefont {Drummond}\ \emph {et~al.}(2024)\citenamefont {Drummond}, \citenamefont {Lynch}, \citenamefont {Hanselman}, \citenamefont {Becker},\ and\ \citenamefont {Hughes}}]{Drummond:2023wqc}%
  \BibitemOpen
  \bibfield  {author} {\bibinfo {author} {\bibfnamefont {L.~V.}\ \bibnamefont {Drummond}}, \bibinfo {author} {\bibfnamefont {P.}~\bibnamefont {Lynch}}, \bibinfo {author} {\bibfnamefont {A.~G.}\ \bibnamefont {Hanselman}}, \bibinfo {author} {\bibfnamefont {D.~R.}\ \bibnamefont {Becker}},\ and\ \bibinfo {author} {\bibfnamefont {S.~A.}\ \bibnamefont {Hughes}},\ }\bibfield  {title} {\bibinfo {title} {{Extreme mass-ratio inspiral and waveforms for a spinning body into a Kerr black hole via osculating geodesics and near-identity transformations}},\ }\href {https://doi.org/10.1103/PhysRevD.109.064030} {\bibfield  {journal} {\bibinfo  {journal} {Phys. Rev. D}\ }\textbf {\bibinfo {volume} {109}},\ \bibinfo {pages} {064030} (\bibinfo {year} {2024})},\ \Eprint {https://arxiv.org/abs/2310.08438} {arXiv:2310.08438 [gr-qc]} \BibitemShut {NoStop}%
\bibitem [{\citenamefont {Skoupy}\ \emph {et~al.}(2023)\citenamefont {Skoupy}, \citenamefont {Lukes-Gerakopoulos}, \citenamefont {Drummond},\ and\ \citenamefont {Hughes}}]{Skoupy:2023lih}%
  \BibitemOpen
  \bibfield  {author} {\bibinfo {author} {\bibfnamefont {V.}~\bibnamefont {Skoupy}}, \bibinfo {author} {\bibfnamefont {G.}~\bibnamefont {Lukes-Gerakopoulos}}, \bibinfo {author} {\bibfnamefont {L.~V.}\ \bibnamefont {Drummond}},\ and\ \bibinfo {author} {\bibfnamefont {S.~A.}\ \bibnamefont {Hughes}},\ }\bibfield  {title} {\bibinfo {title} {{Asymptotic gravitational-wave fluxes from a spinning test body on generic orbits around a Kerr black hole}},\ }\href {https://doi.org/10.1103/PhysRevD.108.044041} {\bibfield  {journal} {\bibinfo  {journal} {Phys. Rev. D}\ }\textbf {\bibinfo {volume} {108}},\ \bibinfo {pages} {044041} (\bibinfo {year} {2023})},\ \Eprint {https://arxiv.org/abs/2303.16798} {arXiv:2303.16798 [gr-qc]} \BibitemShut {NoStop}%
\bibitem [{\citenamefont {Kerachian}\ \emph {et~al.}(2023)\citenamefont {Kerachian}, \citenamefont {Polcar}, \citenamefont {Skoup{\'y}}, \citenamefont {Efthymiopoulos},\ and\ \citenamefont {Lukes-Gerakopoulos}}]{Kerachian:2023oiw}%
  \BibitemOpen
  \bibfield  {author} {\bibinfo {author} {\bibfnamefont {M.}~\bibnamefont {Kerachian}}, \bibinfo {author} {\bibfnamefont {L.}~\bibnamefont {Polcar}}, \bibinfo {author} {\bibfnamefont {V.}~\bibnamefont {Skoup{\'y}}}, \bibinfo {author} {\bibfnamefont {C.}~\bibnamefont {Efthymiopoulos}},\ and\ \bibinfo {author} {\bibfnamefont {G.}~\bibnamefont {Lukes-Gerakopoulos}},\ }\bibfield  {title} {\bibinfo {title} {{Action-angle formalism for extreme mass ratio inspirals in Kerr spacetime}},\ }\href {https://doi.org/10.1103/PhysRevD.108.044004} {\bibfield  {journal} {\bibinfo  {journal} {Phys. Rev. D}\ }\textbf {\bibinfo {volume} {108}},\ \bibinfo {pages} {044004} (\bibinfo {year} {2023})},\ \Eprint {https://arxiv.org/abs/2301.08150} {arXiv:2301.08150 [gr-qc]} \BibitemShut {NoStop}%
\bibitem [{\citenamefont {Ramond}\ and\ \citenamefont {Isoyama}(2025)}]{Ramond:2024sfp}%
  \BibitemOpen
  \bibfield  {author} {\bibinfo {author} {\bibfnamefont {P.}~\bibnamefont {Ramond}}\ and\ \bibinfo {author} {\bibfnamefont {S.}~\bibnamefont {Isoyama}},\ }\bibfield  {title} {\bibinfo {title} {{Symplectic mechanics of relativistic spinning compact bodies: Canonical formalism in the Schwarzschild spacetime}},\ }\href {https://doi.org/10.1103/PhysRevD.111.064027} {\bibfield  {journal} {\bibinfo  {journal} {Phys. Rev. D}\ }\textbf {\bibinfo {volume} {111}},\ \bibinfo {pages} {064027} (\bibinfo {year} {2025})},\ \Eprint {https://arxiv.org/abs/2402.05049} {arXiv:2402.05049 [gr-qc]} \BibitemShut {NoStop}%
\bibitem [{\citenamefont {Grant}(2025)}]{Grant:2024ivt}%
  \BibitemOpen
  \bibfield  {author} {\bibinfo {author} {\bibfnamefont {A.~M.}\ \bibnamefont {Grant}},\ }\bibfield  {title} {\bibinfo {title} {{Flux-balance laws for spinning bodies under the gravitational self-force}},\ }\href {https://doi.org/10.1103/PhysRevD.111.084015} {\bibfield  {journal} {\bibinfo  {journal} {Phys. Rev. D}\ }\textbf {\bibinfo {volume} {111}},\ \bibinfo {pages} {084015} (\bibinfo {year} {2025})},\ \Eprint {https://arxiv.org/abs/2406.10343} {arXiv:2406.10343 [gr-qc]} \BibitemShut {NoStop}%
\bibitem [{\citenamefont {Piovano}\ \emph {et~al.}(2025)\citenamefont {Piovano}, \citenamefont {Pantelidou}, \citenamefont {Mac~Uilliam},\ and\ \citenamefont {Witzany}}]{Piovano:2024yks}%
  \BibitemOpen
  \bibfield  {author} {\bibinfo {author} {\bibfnamefont {G.~A.}\ \bibnamefont {Piovano}}, \bibinfo {author} {\bibfnamefont {C.}~\bibnamefont {Pantelidou}}, \bibinfo {author} {\bibfnamefont {J.}~\bibnamefont {Mac~Uilliam}},\ and\ \bibinfo {author} {\bibfnamefont {V.}~\bibnamefont {Witzany}},\ }\bibfield  {title} {\bibinfo {title} {{Spinning particles near Kerr black holes: Orbits and gravitational-wave fluxes through the Hamilton-Jacobi formalism}},\ }\href {https://doi.org/10.1103/PhysRevD.111.044009} {\bibfield  {journal} {\bibinfo  {journal} {Phys. Rev. D}\ }\textbf {\bibinfo {volume} {111}},\ \bibinfo {pages} {044009} (\bibinfo {year} {2025})},\ \Eprint {https://arxiv.org/abs/2410.05769} {arXiv:2410.05769 [gr-qc]} \BibitemShut {NoStop}%
\bibitem [{\citenamefont {Witzany}\ \emph {et~al.}(2025)\citenamefont {Witzany}, \citenamefont {Skoup{\'y}}, \citenamefont {Stein},\ and\ \citenamefont {Tanay}}]{Witzany:2024ttz}%
  \BibitemOpen
  \bibfield  {author} {\bibinfo {author} {\bibfnamefont {V.}~\bibnamefont {Witzany}}, \bibinfo {author} {\bibfnamefont {V.}~\bibnamefont {Skoup{\'y}}}, \bibinfo {author} {\bibfnamefont {L.~C.}\ \bibnamefont {Stein}},\ and\ \bibinfo {author} {\bibfnamefont {S.}~\bibnamefont {Tanay}},\ }\bibfield  {title} {\bibinfo {title} {{Actions of spinning compact binaries: Spinning particle in Kerr matched to dynamics at 1.5 post-Newtonian order}},\ }\href {https://doi.org/10.1103/PhysRevD.111.044032} {\bibfield  {journal} {\bibinfo  {journal} {Phys. Rev. D}\ }\textbf {\bibinfo {volume} {111}},\ \bibinfo {pages} {044032} (\bibinfo {year} {2025})},\ \Eprint {https://arxiv.org/abs/2411.09742} {arXiv:2411.09742 [gr-qc]} \BibitemShut {NoStop}%
\bibitem [{\citenamefont {Skoup{\'y}}\ and\ \citenamefont {Witzany}(2024)}]{Skoupy:2024jsi}%
  \BibitemOpen
  \bibfield  {author} {\bibinfo {author} {\bibfnamefont {V.}~\bibnamefont {Skoup{\'y}}}\ and\ \bibinfo {author} {\bibfnamefont {V.}~\bibnamefont {Witzany}},\ }\bibfield  {title} {\bibinfo {title} {{Post-Newtonian expansions of extreme mass ratio inspirals of spinning bodies into Schwarzschild black holes}},\ }\href {https://doi.org/10.1103/PhysRevD.110.084061} {\bibfield  {journal} {\bibinfo  {journal} {Phys. Rev. D}\ }\textbf {\bibinfo {volume} {110}},\ \bibinfo {pages} {084061} (\bibinfo {year} {2024})},\ \Eprint {https://arxiv.org/abs/2406.14291} {arXiv:2406.14291 [gr-qc]} \BibitemShut {NoStop}%
\bibitem [{\citenamefont {Mathews}\ and\ \citenamefont {Pound}(2025)}]{Mathews:2025nyb}%
  \BibitemOpen
  \bibfield  {author} {\bibinfo {author} {\bibfnamefont {J.}~\bibnamefont {Mathews}}\ and\ \bibinfo {author} {\bibfnamefont {A.}~\bibnamefont {Pound}},\ }\bibfield  {title} {\bibinfo {title} {{Postadiabatic waveform-generation framework for asymmetric precessing binaries}},\ }\href {https://doi.org/10.1103/rbkb-qnxv} {\bibfield  {journal} {\bibinfo  {journal} {Phys. Rev. D}\ }\textbf {\bibinfo {volume} {112}},\ \bibinfo {pages} {104078} (\bibinfo {year} {2025})},\ \Eprint {https://arxiv.org/abs/2501.01413} {arXiv:2501.01413 [gr-qc]} \BibitemShut {NoStop}%
\bibitem [{\citenamefont {Skoup{\'y}}\ \emph {et~al.}(2025)\citenamefont {Skoup{\'y}}, \citenamefont {Piovano},\ and\ \citenamefont {Witzany}}]{Skoupy:2025nie}%
  \BibitemOpen
  \bibfield  {author} {\bibinfo {author} {\bibfnamefont {V.}~\bibnamefont {Skoup{\'y}}}, \bibinfo {author} {\bibfnamefont {G.~A.}\ \bibnamefont {Piovano}},\ and\ \bibinfo {author} {\bibfnamefont {V.}~\bibnamefont {Witzany}},\ }\bibfield  {title} {\bibinfo {title} {{Spherical inspirals of spinning bodies into Kerr black holes}},\ }\href {https://doi.org/10.1103/x9yy-c2jq} {\bibfield  {journal} {\bibinfo  {journal} {Phys. Rev. D}\ }\textbf {\bibinfo {volume} {112}},\ \bibinfo {pages} {124054} (\bibinfo {year} {2025})},\ \Eprint {https://arxiv.org/abs/2506.20726} {arXiv:2506.20726 [gr-qc]} \BibitemShut {NoStop}%
\bibitem [{\citenamefont {Honet}\ \emph {et~al.}(2025)\citenamefont {Honet}, \citenamefont {Mathews}, \citenamefont {Comp{\`e}re}, \citenamefont {Pound}, \citenamefont {Wardell}, \citenamefont {Piovano}, \citenamefont {van~de Meent},\ and\ \citenamefont {Warburton}}]{Honet:2025lmk}%
  \BibitemOpen
  \bibfield  {author} {\bibinfo {author} {\bibfnamefont {L.}~\bibnamefont {Honet}}, \bibinfo {author} {\bibfnamefont {J.}~\bibnamefont {Mathews}}, \bibinfo {author} {\bibfnamefont {G.}~\bibnamefont {Comp{\`e}re}}, \bibinfo {author} {\bibfnamefont {A.}~\bibnamefont {Pound}}, \bibinfo {author} {\bibfnamefont {B.}~\bibnamefont {Wardell}}, \bibinfo {author} {\bibfnamefont {G.~A.}\ \bibnamefont {Piovano}}, \bibinfo {author} {\bibfnamefont {M.}~\bibnamefont {van~de Meent}},\ and\ \bibinfo {author} {\bibfnamefont {N.}~\bibnamefont {Warburton}},\ }\bibfield  {title} {\bibinfo {title} {{Spin-aligned inspiral waveforms from self-force and post-Newtonian theory}},\ }\href@noop {} {\  (\bibinfo {year} {2025})},\ \Eprint {https://arxiv.org/abs/2510.16112} {arXiv:2510.16112 [gr-qc]} \BibitemShut {NoStop}%
\bibitem [{\citenamefont {Drummond}\ \emph {et~al.}(2026)\citenamefont {Drummond}, \citenamefont {Hughes}, \citenamefont {Skoup{\'y}}, \citenamefont {Lynch},\ and\ \citenamefont {Piovano}}]{Drummond:2026haw}%
  \BibitemOpen
  \bibfield  {author} {\bibinfo {author} {\bibfnamefont {L.~V.}\ \bibnamefont {Drummond}}, \bibinfo {author} {\bibfnamefont {S.~A.}\ \bibnamefont {Hughes}}, \bibinfo {author} {\bibfnamefont {V.}~\bibnamefont {Skoup{\'y}}}, \bibinfo {author} {\bibfnamefont {P.}~\bibnamefont {Lynch}},\ and\ \bibinfo {author} {\bibfnamefont {G.~A.}\ \bibnamefont {Piovano}},\ }\bibfield  {title} {\bibinfo {title} {{Shifted-geodesic approximation for spinning-body gravitational wave fluxes}},\ }\href@noop {} {\  (\bibinfo {year} {2026})},\ \Eprint {https://arxiv.org/abs/2603.12189} {arXiv:2603.12189 [gr-qc]} \BibitemShut {NoStop}%
\bibitem [{\citenamefont {Skoup{\'y}}(2026)}]{Skoupy:2026ewu}%
  \BibitemOpen
  \bibfield  {author} {\bibinfo {author} {\bibfnamefont {V.}~\bibnamefont {Skoup{\'y}}},\ }\bibfield  {title} {\bibinfo {title} {{A new approach to the calculation of extreme-mass-ratio inspirals with a spinning secondary}},\ }\href@noop {} {\  (\bibinfo {year} {2026})},\ \Eprint {https://arxiv.org/abs/2603.13482} {arXiv:2603.13482 [gr-qc]} \BibitemShut {NoStop}%
\bibitem [{\citenamefont {{Cui}}\ and\ \citenamefont {{Han}}(2026)}]{Cui:2026qsk}%
  \BibitemOpen
  \bibfield  {author} {\bibinfo {author} {\bibfnamefont {Q.}~\bibnamefont {{Cui}}}\ and\ \bibinfo {author} {\bibfnamefont {W.-B.}\ \bibnamefont {{Han}}},\ }\bibfield  {title} {\bibinfo {title} {{Waveforms and Fluxes of Generic Extreme-Mass-Ratio Inspirals with a Spinning Secondary}},\ }\href {https://doi.org/10.48550/arXiv.2603.18075} {\bibfield  {journal} {\bibinfo  {journal} {arXiv e-prints}\ ,\ \bibinfo {eid} {arXiv:2603.18075}} (\bibinfo {year} {2026})},\ \Eprint {https://arxiv.org/abs/2603.18075} {arXiv:2603.18075 [astro-ph.HE]} \BibitemShut {NoStop}%
\bibitem [{\citenamefont {Steinhoff}\ and\ \citenamefont {Puetzfeld}(2012)}]{Steinhoff:2012rw}%
  \BibitemOpen
  \bibfield  {author} {\bibinfo {author} {\bibfnamefont {J.}~\bibnamefont {Steinhoff}}\ and\ \bibinfo {author} {\bibfnamefont {D.}~\bibnamefont {Puetzfeld}},\ }\bibfield  {title} {\bibinfo {title} {{Influence of internal structure on the motion of test bodies in extreme mass ratio situations}},\ }\href {https://doi.org/10.1103/PhysRevD.86.044033} {\bibfield  {journal} {\bibinfo  {journal} {Phys. Rev. D}\ }\textbf {\bibinfo {volume} {86}},\ \bibinfo {pages} {044033} (\bibinfo {year} {2012})},\ \Eprint {https://arxiv.org/abs/1205.3926} {arXiv:1205.3926 [gr-qc]} \BibitemShut {NoStop}%
\bibitem [{\citenamefont {Rahman}\ and\ \citenamefont {Bhattacharyya}(2023)}]{Rahman:2021eay}%
  \BibitemOpen
  \bibfield  {author} {\bibinfo {author} {\bibfnamefont {M.}~\bibnamefont {Rahman}}\ and\ \bibinfo {author} {\bibfnamefont {A.}~\bibnamefont {Bhattacharyya}},\ }\bibfield  {title} {\bibinfo {title} {{Prospects for determining the nature of the secondaries of extreme mass-ratio inspirals using the spin-induced quadrupole deformation}},\ }\href {https://doi.org/10.1103/PhysRevD.107.024006} {\bibfield  {journal} {\bibinfo  {journal} {Phys. Rev. D}\ }\textbf {\bibinfo {volume} {107}},\ \bibinfo {pages} {024006} (\bibinfo {year} {2023})},\ \Eprint {https://arxiv.org/abs/2112.13869} {arXiv:2112.13869 [gr-qc]} \BibitemShut {NoStop}%
\bibitem [{\citenamefont {Timogiannis}\ \emph {et~al.}(2023)\citenamefont {Timogiannis}, \citenamefont {Lukes-Gerakopoulos},\ and\ \citenamefont {Apostolatos}}]{Timogiannis:2023pop}%
  \BibitemOpen
  \bibfield  {author} {\bibinfo {author} {\bibfnamefont {I.}~\bibnamefont {Timogiannis}}, \bibinfo {author} {\bibfnamefont {G.}~\bibnamefont {Lukes-Gerakopoulos}},\ and\ \bibinfo {author} {\bibfnamefont {T.~A.}\ \bibnamefont {Apostolatos}},\ }\bibfield  {title} {\bibinfo {title} {{Extended bodies with spin induced quadrupoles on circular equatorial orbits in Kerr spacetime}},\ }\href {https://doi.org/10.1103/PhysRevD.108.084023} {\bibfield  {journal} {\bibinfo  {journal} {Phys. Rev. D}\ }\textbf {\bibinfo {volume} {108}},\ \bibinfo {pages} {084023} (\bibinfo {year} {2023})},\ \Eprint {https://arxiv.org/abs/2306.11558} {arXiv:2306.11558 [gr-qc]} \BibitemShut {NoStop}%
\bibitem [{\citenamefont {Ramond}(2025)}]{Ramond:2024ozy}%
  \BibitemOpen
  \bibfield  {author} {\bibinfo {author} {\bibfnamefont {P.}~\bibnamefont {Ramond}},\ }\bibfield  {title} {\bibinfo {title} {{On the integrability of extended test body dynamics around black holes}},\ }\href {https://doi.org/10.1088/1361-6382/adb197} {\bibfield  {journal} {\bibinfo  {journal} {Class. Quant. Grav.}\ }\textbf {\bibinfo {volume} {42}},\ \bibinfo {pages} {065019} (\bibinfo {year} {2025})},\ \Eprint {https://arxiv.org/abs/2402.02670} {arXiv:2402.02670 [gr-qc]} \BibitemShut {NoStop}%
\bibitem [{\citenamefont {Shahzadi}\ \emph {et~al.}(2025)\citenamefont {Shahzadi}, \citenamefont {Lukes-Gerakopoulos},\ and\ \citenamefont {Kolo{\v{s}}}}]{Shahzadi:2025ebj}%
  \BibitemOpen
  \bibfield  {author} {\bibinfo {author} {\bibfnamefont {M.}~\bibnamefont {Shahzadi}}, \bibinfo {author} {\bibfnamefont {G.}~\bibnamefont {Lukes-Gerakopoulos}},\ and\ \bibinfo {author} {\bibfnamefont {M.}~\bibnamefont {Kolo{\v{s}}}},\ }\bibfield  {title} {\bibinfo {title} {{Circular equatorial orbits of extended bodies with spin-induced quadrupole around a Kerr black hole: Comparing spin-supplementary conditions}},\ }\href {https://doi.org/10.1103/wkcy-38fh} {\bibfield  {journal} {\bibinfo  {journal} {Phys. Rev. D}\ }\textbf {\bibinfo {volume} {112}},\ \bibinfo {pages} {064013} (\bibinfo {year} {2025})},\ \Eprint {https://arxiv.org/abs/2505.16783} {arXiv:2505.16783 [gr-qc]} \BibitemShut {NoStop}%
\bibitem [{\citenamefont {Ramond}\ \emph {et~al.}(2026)\citenamefont {Ramond}, \citenamefont {Isoyama},\ and\ \citenamefont {Druart}}]{Ramond:2026fpi}%
  \BibitemOpen
  \bibfield  {author} {\bibinfo {author} {\bibfnamefont {P.}~\bibnamefont {Ramond}}, \bibinfo {author} {\bibfnamefont {S.}~\bibnamefont {Isoyama}},\ and\ \bibinfo {author} {\bibfnamefont {A.}~\bibnamefont {Druart}},\ }\bibfield  {title} {\bibinfo {title} {{Symplectic mechanics of relativistic spinning compact bodies. III. quadratic-in-spin integrability in Type-D Einstein spacetimes: persistence and breakdown}},\ }\href@noop {} {\  (\bibinfo {year} {2026})},\ \Eprint {https://arxiv.org/abs/2601.06416} {arXiv:2601.06416 [gr-qc]} \BibitemShut {NoStop}%
\bibitem [{\citenamefont {Gourgoulhon}\ \emph {et~al.}(2019)\citenamefont {Gourgoulhon}, \citenamefont {Le~Tiec}, \citenamefont {Vincent},\ and\ \citenamefont {Warburton}}]{Gourgoulhon:2019iyu}%
  \BibitemOpen
  \bibfield  {author} {\bibinfo {author} {\bibfnamefont {E.}~\bibnamefont {Gourgoulhon}}, \bibinfo {author} {\bibfnamefont {A.}~\bibnamefont {Le~Tiec}}, \bibinfo {author} {\bibfnamefont {F.~H.}\ \bibnamefont {Vincent}},\ and\ \bibinfo {author} {\bibfnamefont {N.}~\bibnamefont {Warburton}},\ }\bibfield  {title} {\bibinfo {title} {{Gravitational waves from bodies orbiting the Galactic Center black hole and their detectability by LISA}},\ }\href {https://doi.org/10.1051/0004-6361/201935406} {\bibfield  {journal} {\bibinfo  {journal} {Astron. Astrophys.}\ }\textbf {\bibinfo {volume} {627}},\ \bibinfo {pages} {A92} (\bibinfo {year} {2019})},\ \Eprint {https://arxiv.org/abs/1903.02049} {arXiv:1903.02049 [gr-qc]} \BibitemShut {NoStop}%
\bibitem [{\citenamefont {Amaro~Seoane}(2022)}]{Amaro-Seoane:2020zbo}%
  \BibitemOpen
  \bibfield  {author} {\bibinfo {author} {\bibfnamefont {P.}~\bibnamefont {Amaro~Seoane}},\ }\bibinfo {title} {The gravitational capture of compact objects by massive black holes},\ in\ \href {https://doi.org/10.1007/978-981-16-4306-4_17} {\emph {\bibinfo {booktitle} {Handbook of Gravitational Wave Astronomy}}},\ \bibinfo {editor} {edited by\ \bibinfo {editor} {\bibfnamefont {C.}~\bibnamefont {Bambi}}, \bibinfo {editor} {\bibfnamefont {S.}~\bibnamefont {Katsanevas}},\ and\ \bibinfo {editor} {\bibfnamefont {K.~D.}\ \bibnamefont {Kokkotas}}}\ (\bibinfo  {publisher} {Springer Nature Singapore},\ \bibinfo {address} {Singapore},\ \bibinfo {year} {2022})\ pp.\ \bibinfo {pages} {771--849}\BibitemShut {NoStop}%
\bibitem [{\citenamefont {Albertini}\ \emph {et~al.}(2022)\citenamefont {Albertini}, \citenamefont {Nagar}, \citenamefont {Pound}, \citenamefont {Warburton}, \citenamefont {Wardell}, \citenamefont {Durkan},\ and\ \citenamefont {Miller}}]{Albertini:2022rfe}%
  \BibitemOpen
  \bibfield  {author} {\bibinfo {author} {\bibfnamefont {A.}~\bibnamefont {Albertini}}, \bibinfo {author} {\bibfnamefont {A.}~\bibnamefont {Nagar}}, \bibinfo {author} {\bibfnamefont {A.}~\bibnamefont {Pound}}, \bibinfo {author} {\bibfnamefont {N.}~\bibnamefont {Warburton}}, \bibinfo {author} {\bibfnamefont {B.}~\bibnamefont {Wardell}}, \bibinfo {author} {\bibfnamefont {L.}~\bibnamefont {Durkan}},\ and\ \bibinfo {author} {\bibfnamefont {J.}~\bibnamefont {Miller}},\ }\bibfield  {title} {\bibinfo {title} {{Comparing second-order gravitational self-force, numerical relativity, and effective one body waveforms from inspiralling, quasicircular, and nonspinning black hole binaries}},\ }\href {https://doi.org/10.1103/PhysRevD.106.084061} {\bibfield  {journal} {\bibinfo  {journal} {Phys. Rev. D}\ }\textbf {\bibinfo {volume} {106}},\ \bibinfo {pages} {084061} (\bibinfo {year} {2022})},\ \Eprint {https://arxiv.org/abs/2208.01049} {arXiv:2208.01049 [gr-qc]} \BibitemShut {NoStop}%
\bibitem [{\citenamefont {Ramos-Buades}\ \emph {et~al.}(2022)\citenamefont {Ramos-Buades}, \citenamefont {van~de Meent}, \citenamefont {Pfeiffer}, \citenamefont {R{\"u}ter}, \citenamefont {Scheel}, \citenamefont {Boyle},\ and\ \citenamefont {Kidder}}]{Ramos-Buades:2022lgf}%
  \BibitemOpen
  \bibfield  {author} {\bibinfo {author} {\bibfnamefont {A.}~\bibnamefont {Ramos-Buades}}, \bibinfo {author} {\bibfnamefont {M.}~\bibnamefont {van~de Meent}}, \bibinfo {author} {\bibfnamefont {H.~P.}\ \bibnamefont {Pfeiffer}}, \bibinfo {author} {\bibfnamefont {H.~R.}\ \bibnamefont {R{\"u}ter}}, \bibinfo {author} {\bibfnamefont {M.~A.}\ \bibnamefont {Scheel}}, \bibinfo {author} {\bibfnamefont {M.}~\bibnamefont {Boyle}},\ and\ \bibinfo {author} {\bibfnamefont {L.~E.}\ \bibnamefont {Kidder}},\ }\bibfield  {title} {\bibinfo {title} {{Eccentric binary black holes: Comparing numerical relativity and small mass-ratio perturbation theory}},\ }\href {https://doi.org/10.1103/PhysRevD.106.124040} {\bibfield  {journal} {\bibinfo  {journal} {Phys. Rev. D}\ }\textbf {\bibinfo {volume} {106}},\ \bibinfo {pages} {124040} (\bibinfo {year} {2022})},\ \Eprint {https://arxiv.org/abs/2209.03390} {arXiv:2209.03390 [gr-qc]} \BibitemShut {NoStop}%
\bibitem [{\citenamefont {Damour}(2010)}]{Damour:2009sm}%
  \BibitemOpen
  \bibfield  {author} {\bibinfo {author} {\bibfnamefont {T.}~\bibnamefont {Damour}},\ }\bibfield  {title} {\bibinfo {title} {{Gravitational Self Force in a Schwarzschild Background and the Effective One Body Formalism}},\ }\href {https://doi.org/10.1103/PhysRevD.81.024017} {\bibfield  {journal} {\bibinfo  {journal} {Phys. Rev. D}\ }\textbf {\bibinfo {volume} {81}},\ \bibinfo {pages} {024017} (\bibinfo {year} {2010})},\ \Eprint {https://arxiv.org/abs/0910.5533} {arXiv:0910.5533 [gr-qc]} \BibitemShut {NoStop}%
\bibitem [{\citenamefont {Nagar}\ and\ \citenamefont {Albanesi}(2022)}]{Nagar:2022fep}%
  \BibitemOpen
  \bibfield  {author} {\bibinfo {author} {\bibfnamefont {A.}~\bibnamefont {Nagar}}\ and\ \bibinfo {author} {\bibfnamefont {S.}~\bibnamefont {Albanesi}},\ }\bibfield  {title} {\bibinfo {title} {{Toward a gravitational self-force-informed effective-one-body waveform model for nonprecessing, eccentric, large-mass-ratio inspirals}},\ }\href {https://doi.org/10.1103/PhysRevD.106.064049} {\bibfield  {journal} {\bibinfo  {journal} {Phys. Rev. D}\ }\textbf {\bibinfo {volume} {106}},\ \bibinfo {pages} {064049} (\bibinfo {year} {2022})},\ \Eprint {https://arxiv.org/abs/2207.14002} {arXiv:2207.14002 [gr-qc]} \BibitemShut {NoStop}%
\bibitem [{\citenamefont {van~de Meent}\ \emph {et~al.}(2023)\citenamefont {van~de Meent}, \citenamefont {Buonanno}, \citenamefont {Mihaylov}, \citenamefont {Ossokine}, \citenamefont {Pompili}, \citenamefont {Warburton}, \citenamefont {Pound}, \citenamefont {Wardell}, \citenamefont {Durkan},\ and\ \citenamefont {Miller}}]{vandeMeent:2023ols}%
  \BibitemOpen
  \bibfield  {author} {\bibinfo {author} {\bibfnamefont {M.}~\bibnamefont {van~de Meent}}, \bibinfo {author} {\bibfnamefont {A.}~\bibnamefont {Buonanno}}, \bibinfo {author} {\bibfnamefont {D.~P.}\ \bibnamefont {Mihaylov}}, \bibinfo {author} {\bibfnamefont {S.}~\bibnamefont {Ossokine}}, \bibinfo {author} {\bibfnamefont {L.}~\bibnamefont {Pompili}}, \bibinfo {author} {\bibfnamefont {N.}~\bibnamefont {Warburton}}, \bibinfo {author} {\bibfnamefont {A.}~\bibnamefont {Pound}}, \bibinfo {author} {\bibfnamefont {B.}~\bibnamefont {Wardell}}, \bibinfo {author} {\bibfnamefont {L.}~\bibnamefont {Durkan}},\ and\ \bibinfo {author} {\bibfnamefont {J.}~\bibnamefont {Miller}},\ }\bibfield  {title} {\bibinfo {title} {{Enhancing the SEOBNRv5 effective-one-body waveform model with second-order gravitational self-force fluxes}},\ }\href {https://doi.org/10.1103/PhysRevD.108.124038} {\bibfield  {journal} {\bibinfo  {journal} {Phys. Rev. D}\ }\textbf {\bibinfo {volume} {108}},\ \bibinfo {pages} {124038} (\bibinfo {year} {2023})},\
  \Eprint {https://arxiv.org/abs/2303.18026} {arXiv:2303.18026 [gr-qc]} \BibitemShut {NoStop}%
\bibitem [{\citenamefont {Albertini}\ \emph {et~al.}(2024)\citenamefont {Albertini}, \citenamefont {Nagar}, \citenamefont {Mathews},\ and\ \citenamefont {Lukes-Gerakopoulos}}]{Albertini:2024rrs}%
  \BibitemOpen
  \bibfield  {author} {\bibinfo {author} {\bibfnamefont {A.}~\bibnamefont {Albertini}}, \bibinfo {author} {\bibfnamefont {A.}~\bibnamefont {Nagar}}, \bibinfo {author} {\bibfnamefont {J.}~\bibnamefont {Mathews}},\ and\ \bibinfo {author} {\bibfnamefont {G.}~\bibnamefont {Lukes-Gerakopoulos}},\ }\bibfield  {title} {\bibinfo {title} {{Comparing second-order gravitational self-force and effective-one-body waveforms from inspiralling, quasicircular black hole binaries with a nonspinning primary and a spinning secondary}},\ }\href {https://doi.org/10.1103/PhysRevD.110.044034} {\bibfield  {journal} {\bibinfo  {journal} {Phys. Rev. D}\ }\textbf {\bibinfo {volume} {110}},\ \bibinfo {pages} {044034} (\bibinfo {year} {2024})},\ \Eprint {https://arxiv.org/abs/2406.04108} {arXiv:2406.04108 [gr-qc]} \BibitemShut {NoStop}%
\bibitem [{\citenamefont {Leather}\ \emph {et~al.}(2025)\citenamefont {Leather}, \citenamefont {Buonanno},\ and\ \citenamefont {van~de Meent}}]{Leather:2025nhu}%
  \BibitemOpen
  \bibfield  {author} {\bibinfo {author} {\bibfnamefont {B.}~\bibnamefont {Leather}}, \bibinfo {author} {\bibfnamefont {A.}~\bibnamefont {Buonanno}},\ and\ \bibinfo {author} {\bibfnamefont {M.}~\bibnamefont {van~de Meent}},\ }\bibfield  {title} {\bibinfo {title} {{Inspiral-merger-ringdown waveforms with gravitational self-force results within the effective-one-body formalism}},\ }\href {https://doi.org/10.1103/6qc3-xn17} {\bibfield  {journal} {\bibinfo  {journal} {Phys. Rev. D}\ }\textbf {\bibinfo {volume} {112}},\ \bibinfo {pages} {044012} (\bibinfo {year} {2025})},\ \Eprint {https://arxiv.org/abs/2505.11242} {arXiv:2505.11242 [gr-qc]} \BibitemShut {NoStop}%
\bibitem [{\citenamefont {Honet}\ \emph {et~al.}(2026)\citenamefont {Honet}, \citenamefont {Pound},\ and\ \citenamefont {Comp{\`e}re}}]{Honet:2025gge}%
  \BibitemOpen
  \bibfield  {author} {\bibinfo {author} {\bibfnamefont {L.}~\bibnamefont {Honet}}, \bibinfo {author} {\bibfnamefont {A.}~\bibnamefont {Pound}},\ and\ \bibinfo {author} {\bibfnamefont {G.}~\bibnamefont {Comp{\`e}re}},\ }\bibfield  {title} {\bibinfo {title} {{Hybrid waveform model for asymmetric spinning binaries: Self-force meets post-Newtonian theory}},\ }\href {https://doi.org/10.1103/rhwy-59y2} {\bibfield  {journal} {\bibinfo  {journal} {Phys. Rev. D}\ }\textbf {\bibinfo {volume} {113}},\ \bibinfo {pages} {064035} (\bibinfo {year} {2026})},\ \Eprint {https://arxiv.org/abs/2510.16114} {arXiv:2510.16114 [gr-qc]} \BibitemShut {NoStop}%
\bibitem [{\citenamefont {Hinderer}\ and\ \citenamefont {Flanagan}(2008{\natexlab{a}})}]{PhysRevD.78.064028}%
  \BibitemOpen
  \bibfield  {author} {\bibinfo {author} {\bibfnamefont {T.}~\bibnamefont {Hinderer}}\ and\ \bibinfo {author} {\bibfnamefont {E.~E.}\ \bibnamefont {Flanagan}},\ }\bibfield  {title} {\bibinfo {title} {Two-timescale analysis of extreme mass ratio inspirals in kerr spacetime: Orbital motion},\ }\href {https://doi.org/10.1103/PhysRevD.78.064028} {\bibfield  {journal} {\bibinfo  {journal} {Phys. Rev. D}\ }\textbf {\bibinfo {volume} {78}},\ \bibinfo {pages} {064028} (\bibinfo {year} {2008}{\natexlab{a}})}\BibitemShut {NoStop}%
\bibitem [{\citenamefont {Miller}\ and\ \citenamefont {Pound}(2021)}]{Miller:2020bft}%
  \BibitemOpen
  \bibfield  {author} {\bibinfo {author} {\bibfnamefont {J.}~\bibnamefont {Miller}}\ and\ \bibinfo {author} {\bibfnamefont {A.}~\bibnamefont {Pound}},\ }\bibfield  {title} {\bibinfo {title} {{Two-timescale evolution of extreme-mass-ratio inspirals: waveform generation scheme for quasicircular orbits in Schwarzschild spacetime}},\ }\href {https://doi.org/10.1103/PhysRevD.103.064048} {\bibfield  {journal} {\bibinfo  {journal} {Phys. Rev. D}\ }\textbf {\bibinfo {volume} {103}},\ \bibinfo {pages} {064048} (\bibinfo {year} {2021})},\ \Eprint {https://arxiv.org/abs/2006.11263} {arXiv:2006.11263 [gr-qc]} \BibitemShut {NoStop}%
\bibitem [{\citenamefont {Wei}\ \emph {et~al.}(2025)\citenamefont {Wei}, \citenamefont {Zhu}, \citenamefont {Zhang},\ and\ \citenamefont {Mei}}]{Wei:2025lva}%
  \BibitemOpen
  \bibfield  {author} {\bibinfo {author} {\bibfnamefont {Y.-X.}\ \bibnamefont {Wei}}, \bibinfo {author} {\bibfnamefont {X.-L.}\ \bibnamefont {Zhu}}, \bibinfo {author} {\bibfnamefont {J.-d.}\ \bibnamefont {Zhang}},\ and\ \bibinfo {author} {\bibfnamefont {J.}~\bibnamefont {Mei}},\ }\bibfield  {title} {\bibinfo {title} {{Toward second-order self-force for eccentric extreme-mass ratio inspirals in Schwarzschild spacetimes}},\ }\href {https://doi.org/10.1103/42fh-sw5h} {\bibfield  {journal} {\bibinfo  {journal} {Phys. Rev. D}\ }\textbf {\bibinfo {volume} {112}},\ \bibinfo {pages} {064048} (\bibinfo {year} {2025})},\ \Eprint {https://arxiv.org/abs/2504.09640} {arXiv:2504.09640 [gr-qc]} \BibitemShut {NoStop}%
\bibitem [{\citenamefont {Lewis}\ \emph {et~al.}(2026)\citenamefont {Lewis}, \citenamefont {Kakehi}, \citenamefont {Pound},\ and\ \citenamefont {Tanaka}}]{Lewis:2025ydo}%
  \BibitemOpen
  \bibfield  {author} {\bibinfo {author} {\bibfnamefont {J.}~\bibnamefont {Lewis}}, \bibinfo {author} {\bibfnamefont {T.}~\bibnamefont {Kakehi}}, \bibinfo {author} {\bibfnamefont {A.}~\bibnamefont {Pound}},\ and\ \bibinfo {author} {\bibfnamefont {T.}~\bibnamefont {Tanaka}},\ }\bibfield  {title} {\bibinfo {title} {{Postadiabatic dynamics and waveform generation in self-force theory: An invariant pseudo-Hamiltonian framework}},\ }\href {https://doi.org/10.1103/jllr-kl86} {\bibfield  {journal} {\bibinfo  {journal} {Phys. Rev. D}\ }\textbf {\bibinfo {volume} {113}},\ \bibinfo {pages} {064046} (\bibinfo {year} {2026})},\ \Eprint {https://arxiv.org/abs/2507.08081} {arXiv:2507.08081 [gr-qc]} \BibitemShut {NoStop}%
\bibitem [{\citenamefont {Bini}\ \emph {et~al.}(2015)\citenamefont {Bini}, \citenamefont {Faye},\ and\ \citenamefont {Geralico}}]{Bini:2015zya}%
  \BibitemOpen
  \bibfield  {author} {\bibinfo {author} {\bibfnamefont {D.}~\bibnamefont {Bini}}, \bibinfo {author} {\bibfnamefont {G.}~\bibnamefont {Faye}},\ and\ \bibinfo {author} {\bibfnamefont {A.}~\bibnamefont {Geralico}},\ }\bibfield  {title} {\bibinfo {title} {{Dynamics of extended bodies in a Kerr spacetime with spin-induced quadrupole tensor}},\ }\href {https://doi.org/10.1103/PhysRevD.92.104003} {\bibfield  {journal} {\bibinfo  {journal} {Phys. Rev. D}\ }\textbf {\bibinfo {volume} {92}},\ \bibinfo {pages} {104003} (\bibinfo {year} {2015})},\ \Eprint {https://arxiv.org/abs/1507.07441} {arXiv:1507.07441 [gr-qc]} \BibitemShut {NoStop}%
\bibitem [{\citenamefont {Mathisson}(2010)}]{Mathisson:2010opl}%
  \BibitemOpen
  \bibfield  {author} {\bibinfo {author} {\bibfnamefont {M.}~\bibnamefont {Mathisson}},\ }\bibfield  {title} {\bibinfo {title} {{Republication of: New mechanics of material systems}},\ }\href {https://doi.org/10.1007/s10714-010-0939-y} {\bibfield  {journal} {\bibinfo  {journal} {Gen. Rel. Grav.}\ }\textbf {\bibinfo {volume} {42}},\ \bibinfo {pages} {1011} (\bibinfo {year} {2010})}\BibitemShut {NoStop}%
\bibitem [{\citenamefont {Dixon}(1970)}]{Dixon1970DynamicsOE}%
  \BibitemOpen
  \bibfield  {author} {\bibinfo {author} {\bibfnamefont {W.~G.}\ \bibnamefont {Dixon}},\ }\bibfield  {title} {\bibinfo {title} {Dynamics of extended bodies in general relativity. i. momentum and angular momentum},\ }\href@noop {} {\bibfield  {journal} {\bibinfo  {journal} {Proceedings of the Royal Society of London. A. Mathematical and Physical Sciences}\ }\textbf {\bibinfo {volume} {314}},\ \bibinfo {pages} {499 } (\bibinfo {year} {1970})}\BibitemShut {NoStop}%
\bibitem [{\citenamefont {Dixon}(1973)}]{Dixon1973TheDO}%
  \BibitemOpen
  \bibfield  {author} {\bibinfo {author} {\bibfnamefont {W.~G.}\ \bibnamefont {Dixon}},\ }\bibfield  {title} {\bibinfo {title} {The definition of multipole moments for extended bodies},\ }\href@noop {} {\bibfield  {journal} {\bibinfo  {journal} {General Relativity and Gravitation}\ }\textbf {\bibinfo {volume} {4}},\ \bibinfo {pages} {199} (\bibinfo {year} {1973})}\BibitemShut {NoStop}%
\bibitem [{\citenamefont {Dixon}(1974)}]{Dixon1974DynamicsOE}%
  \BibitemOpen
  \bibfield  {author} {\bibinfo {author} {\bibfnamefont {W.~G.}\ \bibnamefont {Dixon}},\ }\bibfield  {title} {\bibinfo {title} {Dynamics of extended bodies in general relativity iii. equations of motion},\ }\href@noop {} {\bibfield  {journal} {\bibinfo  {journal} {Philosophical Transactions of the Royal Society of London. Series A, Mathematical and Physical Sciences}\ }\textbf {\bibinfo {volume} {277}},\ \bibinfo {pages} {119 } (\bibinfo {year} {1974})}\BibitemShut {NoStop}%
\bibitem [{\citenamefont {Harte}(2012)}]{Harte:2011ku}%
  \BibitemOpen
  \bibfield  {author} {\bibinfo {author} {\bibfnamefont {A.~I.}\ \bibnamefont {Harte}},\ }\bibfield  {title} {\bibinfo {title} {{Mechanics of extended masses in general relativity}},\ }\href {https://doi.org/10.1088/0264-9381/29/5/055012} {\bibfield  {journal} {\bibinfo  {journal} {Class. Quant. Grav.}\ }\textbf {\bibinfo {volume} {29}},\ \bibinfo {pages} {055012} (\bibinfo {year} {2012})},\ \Eprint {https://arxiv.org/abs/1103.0543} {arXiv:1103.0543 [gr-qc]} \BibitemShut {NoStop}%
\bibitem [{\citenamefont {Harte}\ \emph {et~al.}(2025)\citenamefont {Harte}, \citenamefont {Blanco},\ and\ \citenamefont {Flanagan}}]{Harte:2025tmd}%
  \BibitemOpen
  \bibfield  {author} {\bibinfo {author} {\bibfnamefont {A.~I.}\ \bibnamefont {Harte}}, \bibinfo {author} {\bibfnamefont {F.~M.}\ \bibnamefont {Blanco}},\ and\ \bibinfo {author} {\bibfnamefont {E.~E.}\ \bibnamefont {Flanagan}},\ }\bibfield  {title} {\bibinfo {title} {{Nonlinearly Self-Interacting Extended Bodies Move as Test Bodies in Effective External Fields}},\ }\href {https://doi.org/10.1103/k74d-dj7y} {\bibfield  {journal} {\bibinfo  {journal} {Phys. Rev. Lett.}\ }\textbf {\bibinfo {volume} {135}},\ \bibinfo {pages} {151401} (\bibinfo {year} {2025})},\ \Eprint {https://arxiv.org/abs/2504.11912} {arXiv:2504.11912 [gr-qc]} \BibitemShut {NoStop}%
\bibitem [{\citenamefont {Tulczyjew}(1959)}]{Tulczyjew:1959}%
  \BibitemOpen
  \bibfield  {author} {\bibinfo {author} {\bibfnamefont {W.}~\bibnamefont {Tulczyjew}},\ }\bibfield  {title} {\bibinfo {title} {Motion of multipole particles in general relativity theory},\ }\href@noop {} {\bibfield  {journal} {\bibinfo  {journal} {Acta Phys. Pol.}\ }\textbf {\bibinfo {volume} {18}},\ \bibinfo {pages} {393} (\bibinfo {year} {1959})}\BibitemShut {NoStop}%
\bibitem [{\citenamefont {Steinhoff}\ and\ \citenamefont {Puetzfeld}(2010)}]{Steinhoff:2009tk}%
  \BibitemOpen
  \bibfield  {author} {\bibinfo {author} {\bibfnamefont {J.}~\bibnamefont {Steinhoff}}\ and\ \bibinfo {author} {\bibfnamefont {D.}~\bibnamefont {Puetzfeld}},\ }\bibfield  {title} {\bibinfo {title} {{Multipolar equations of motion for extended test bodies in General Relativity}},\ }\href {https://doi.org/10.1103/PhysRevD.81.044019} {\bibfield  {journal} {\bibinfo  {journal} {Phys. Rev. D}\ }\textbf {\bibinfo {volume} {81}},\ \bibinfo {pages} {044019} (\bibinfo {year} {2010})},\ \Eprint {https://arxiv.org/abs/0909.3756} {arXiv:0909.3756 [gr-qc]} \BibitemShut {NoStop}%
\bibitem [{\citenamefont {D'Eath}(1975)}]{DEath:1975jps}%
  \BibitemOpen
  \bibfield  {author} {\bibinfo {author} {\bibfnamefont {P.~D.}\ \bibnamefont {D'Eath}},\ }\bibfield  {title} {\bibinfo {title} {{Dynamics of a small black hole in a background universe}},\ }\href {https://doi.org/10.1103/PhysRevD.11.1387} {\bibfield  {journal} {\bibinfo  {journal} {Phys. Rev. D}\ }\textbf {\bibinfo {volume} {11}},\ \bibinfo {pages} {1387} (\bibinfo {year} {1975})}\BibitemShut {NoStop}%
\bibitem [{\citenamefont {Thorne}\ and\ \citenamefont {Hartle}(1984)}]{Thorne:1984mz}%
  \BibitemOpen
  \bibfield  {author} {\bibinfo {author} {\bibfnamefont {K.~S.}\ \bibnamefont {Thorne}}\ and\ \bibinfo {author} {\bibfnamefont {J.~B.}\ \bibnamefont {Hartle}},\ }\bibfield  {title} {\bibinfo {title} {{Laws of motion and precession for black holes and other bodies}},\ }\href {https://doi.org/10.1103/PhysRevD.31.1815} {\bibfield  {journal} {\bibinfo  {journal} {Phys. Rev. D}\ }\textbf {\bibinfo {volume} {31}},\ \bibinfo {pages} {1815} (\bibinfo {year} {1984})}\BibitemShut {NoStop}%
\bibitem [{\citenamefont {Zhang}(1985)}]{Zhang:1985qz}%
  \BibitemOpen
  \bibfield  {author} {\bibinfo {author} {\bibfnamefont {X.~H.}\ \bibnamefont {Zhang}},\ }\bibfield  {title} {\bibinfo {title} {{HIGHER ORDER CORRECTIONS TO THE LAWS OF MOTION AND PRECESSION FOR BLACK HOLES AND OTHER BODIES}},\ }\href {https://doi.org/10.1103/PhysRevD.31.3130} {\bibfield  {journal} {\bibinfo  {journal} {Phys. Rev. D}\ }\textbf {\bibinfo {volume} {31}},\ \bibinfo {pages} {3130} (\bibinfo {year} {1985})}\BibitemShut {NoStop}%
\bibitem [{\citenamefont {Detweiler}(2001)}]{Detweiler:2000gt}%
  \BibitemOpen
  \bibfield  {author} {\bibinfo {author} {\bibfnamefont {S.~L.}\ \bibnamefont {Detweiler}},\ }\bibfield  {title} {\bibinfo {title} {{Radiation reaction and the selfforce for a point mass in general relativity}},\ }\href {https://doi.org/10.1103/PhysRevLett.86.1931} {\bibfield  {journal} {\bibinfo  {journal} {Phys. Rev. Lett.}\ }\textbf {\bibinfo {volume} {86}},\ \bibinfo {pages} {1931} (\bibinfo {year} {2001})},\ \Eprint {https://arxiv.org/abs/gr-qc/0011039} {arXiv:gr-qc/0011039} \BibitemShut {NoStop}%
\bibitem [{\citenamefont {Gralla}\ and\ \citenamefont {Wald}(2008)}]{Gralla:2008fg}%
  \BibitemOpen
  \bibfield  {author} {\bibinfo {author} {\bibfnamefont {S.~E.}\ \bibnamefont {Gralla}}\ and\ \bibinfo {author} {\bibfnamefont {R.~M.}\ \bibnamefont {Wald}},\ }\bibfield  {title} {\bibinfo {title} {{A Rigorous Derivation of Gravitational Self-force}},\ }\href {https://doi.org/10.1088/0264-9381/25/20/205009} {\bibfield  {journal} {\bibinfo  {journal} {Class. Quant. Grav.}\ }\textbf {\bibinfo {volume} {25}},\ \bibinfo {pages} {205009} (\bibinfo {year} {2008})},\ \bibinfo {note} {[Erratum: Class.Quant.Grav. 28, 159501 (2011)]},\ \Eprint {https://arxiv.org/abs/0806.3293} {arXiv:0806.3293 [gr-qc]} \BibitemShut {NoStop}%
\bibitem [{\citenamefont {Pound}(2012{\natexlab{a}})}]{Pound:2012dk}%
  \BibitemOpen
  \bibfield  {author} {\bibinfo {author} {\bibfnamefont {A.}~\bibnamefont {Pound}},\ }\bibfield  {title} {\bibinfo {title} {{Nonlinear gravitational self-force. I. Field outside a small body}},\ }\href {https://doi.org/10.1103/PhysRevD.86.084019} {\bibfield  {journal} {\bibinfo  {journal} {Phys. Rev. D}\ }\textbf {\bibinfo {volume} {86}},\ \bibinfo {pages} {084019} (\bibinfo {year} {2012}{\natexlab{a}})},\ \Eprint {https://arxiv.org/abs/1206.6538} {arXiv:1206.6538 [gr-qc]} \BibitemShut {NoStop}%
\bibitem [{\citenamefont {Pound}(2015)}]{Pound:2015tma}%
  \BibitemOpen
  \bibfield  {author} {\bibinfo {author} {\bibfnamefont {A.}~\bibnamefont {Pound}},\ }\bibfield  {title} {\bibinfo {title} {{Motion of small objects in curved spacetimes: An introduction to gravitational self-force}},\ }\href {https://doi.org/10.1007/978-3-319-18335-0_13} {\bibfield  {journal} {\bibinfo  {journal} {Fund. Theor. Phys.}\ }\textbf {\bibinfo {volume} {179}},\ \bibinfo {pages} {399} (\bibinfo {year} {2015})},\ \Eprint {https://arxiv.org/abs/1506.06245} {arXiv:1506.06245 [gr-qc]} \BibitemShut {NoStop}%
\bibitem [{\citenamefont {Upton}\ and\ \citenamefont {Pound}(2021)}]{Upton:2021oxf}%
  \BibitemOpen
  \bibfield  {author} {\bibinfo {author} {\bibfnamefont {S.~D.}\ \bibnamefont {Upton}}\ and\ \bibinfo {author} {\bibfnamefont {A.}~\bibnamefont {Pound}},\ }\bibfield  {title} {\bibinfo {title} {{Second-order gravitational self-force in a highly regular gauge}},\ }\href {https://doi.org/10.1103/PhysRevD.103.124016} {\bibfield  {journal} {\bibinfo  {journal} {Phys. Rev. D}\ }\textbf {\bibinfo {volume} {103}},\ \bibinfo {pages} {124016} (\bibinfo {year} {2021})},\ \Eprint {https://arxiv.org/abs/2101.11409} {arXiv:2101.11409 [gr-qc]} \BibitemShut {NoStop}%
\bibitem [{\citenamefont {Musaeus}\ \emph {et~al.}()\citenamefont {Musaeus}, \citenamefont {Pound},\ and\ \citenamefont {Upton}}]{Musaeus:InPrep}%
  \BibitemOpen
  \bibfield  {author} {\bibinfo {author} {\bibfnamefont {J.}~\bibnamefont {Musaeus}}, \bibinfo {author} {\bibfnamefont {A.}~\bibnamefont {Pound}},\ and\ \bibinfo {author} {\bibfnamefont {S.~D.}\ \bibnamefont {Upton}},\ }\bibfield  {title} {\bibinfo {title} {Point particles in general relativity: beyond linear perturbation theory},\ }\bibinfo {note} {in preparation}\BibitemShut {NoStop}%
\bibitem [{\citenamefont {Pound}(2010)}]{Pound:2009sm}%
  \BibitemOpen
  \bibfield  {author} {\bibinfo {author} {\bibfnamefont {A.}~\bibnamefont {Pound}},\ }\bibfield  {title} {\bibinfo {title} {{Self-consistent gravitational self-force}},\ }\href {https://doi.org/10.1103/PhysRevD.81.024023} {\bibfield  {journal} {\bibinfo  {journal} {Phys. Rev. D}\ }\textbf {\bibinfo {volume} {81}},\ \bibinfo {pages} {024023} (\bibinfo {year} {2010})},\ \Eprint {https://arxiv.org/abs/0907.5197} {arXiv:0907.5197 [gr-qc]} \BibitemShut {NoStop}%
\bibitem [{\citenamefont {{Nauenberg}}(1972)}]{1972ApJ...175..417N}%
  \BibitemOpen
  \bibfield  {author} {\bibinfo {author} {\bibfnamefont {M.}~\bibnamefont {{Nauenberg}}},\ }\bibfield  {title} {\bibinfo {title} {{Analytic Approximations to the Mass-Radius Relation and Energy of Zero-Temperature Stars}},\ }\href {https://doi.org/10.1086/151568} {\bibfield  {journal} {\bibinfo  {journal} {\apj}\ }\textbf {\bibinfo {volume} {175}},\ \bibinfo {pages} {417} (\bibinfo {year} {1972})}\BibitemShut {NoStop}%
\bibitem [{\citenamefont {{\"O}zel}\ and\ \citenamefont {Freire}(2016)}]{Ozel:2016oaf}%
  \BibitemOpen
  \bibfield  {author} {\bibinfo {author} {\bibfnamefont {F.}~\bibnamefont {{\"O}zel}}\ and\ \bibinfo {author} {\bibfnamefont {P.}~\bibnamefont {Freire}},\ }\bibfield  {title} {\bibinfo {title} {{Masses, Radii, and the Equation of State of Neutron Stars}},\ }\href {https://doi.org/10.1146/annurev-astro-081915-023322} {\bibfield  {journal} {\bibinfo  {journal} {Ann. Rev. Astron. Astrophys.}\ }\textbf {\bibinfo {volume} {54}},\ \bibinfo {pages} {401} (\bibinfo {year} {2016})},\ \Eprint {https://arxiv.org/abs/1603.02698} {arXiv:1603.02698 [astro-ph.HE]} \BibitemShut {NoStop}%
\bibitem [{\citenamefont {Papapetrou}(1951)}]{Papapetrou:1951pa}%
  \BibitemOpen
  \bibfield  {author} {\bibinfo {author} {\bibfnamefont {A.}~\bibnamefont {Papapetrou}},\ }\bibfield  {title} {\bibinfo {title} {{Spinning test particles in general relativity. 1.}},\ }\href {https://doi.org/10.1098/rspa.1951.0200} {\bibfield  {journal} {\bibinfo  {journal} {Proc. Roy. Soc. Lond. A}\ }\textbf {\bibinfo {volume} {209}},\ \bibinfo {pages} {248} (\bibinfo {year} {1951})}\BibitemShut {NoStop}%
\bibitem [{\citenamefont {Pound}(2012{\natexlab{b}})}]{Pound:2012nt}%
  \BibitemOpen
  \bibfield  {author} {\bibinfo {author} {\bibfnamefont {A.}~\bibnamefont {Pound}},\ }\bibfield  {title} {\bibinfo {title} {{Second-order gravitational self-force}},\ }\href {https://doi.org/10.1103/PhysRevLett.109.051101} {\bibfield  {journal} {\bibinfo  {journal} {Phys. Rev. Lett.}\ }\textbf {\bibinfo {volume} {109}},\ \bibinfo {pages} {051101} (\bibinfo {year} {2012}{\natexlab{b}})},\ \Eprint {https://arxiv.org/abs/1201.5089} {arXiv:1201.5089 [gr-qc]} \BibitemShut {NoStop}%
\bibitem [{\citenamefont {Pound}(2017)}]{Pound:2017psq}%
  \BibitemOpen
  \bibfield  {author} {\bibinfo {author} {\bibfnamefont {A.}~\bibnamefont {Pound}},\ }\bibfield  {title} {\bibinfo {title} {{Nonlinear gravitational self-force: second-order equation of motion}},\ }\href {https://doi.org/10.1103/PhysRevD.95.104056} {\bibfield  {journal} {\bibinfo  {journal} {Phys. Rev. D}\ }\textbf {\bibinfo {volume} {95}},\ \bibinfo {pages} {104056} (\bibinfo {year} {2017})},\ \Eprint {https://arxiv.org/abs/1703.02836} {arXiv:1703.02836 [gr-qc]} \BibitemShut {NoStop}%
\bibitem [{\citenamefont {Gralla}(2012)}]{Gralla:2012db}%
  \BibitemOpen
  \bibfield  {author} {\bibinfo {author} {\bibfnamefont {S.~E.}\ \bibnamefont {Gralla}},\ }\bibfield  {title} {\bibinfo {title} {{Second Order Gravitational Self Force}},\ }\href {https://doi.org/10.1103/PhysRevD.85.124011} {\bibfield  {journal} {\bibinfo  {journal} {Phys. Rev. D}\ }\textbf {\bibinfo {volume} {85}},\ \bibinfo {pages} {124011} (\bibinfo {year} {2012})},\ \Eprint {https://arxiv.org/abs/1203.3189} {arXiv:1203.3189 [gr-qc]} \BibitemShut {NoStop}%
\bibitem [{\citenamefont {{Mino}}\ \emph {et~al.}(1997)\citenamefont {{Mino}}, \citenamefont {{Sasaki}},\ and\ \citenamefont {{Tanaka}}}]{Mino:1997wh}%
  \BibitemOpen
  \bibfield  {author} {\bibinfo {author} {\bibfnamefont {Y.}~\bibnamefont {{Mino}}}, \bibinfo {author} {\bibfnamefont {M.}~\bibnamefont {{Sasaki}}},\ and\ \bibinfo {author} {\bibfnamefont {T.}~\bibnamefont {{Tanaka}}},\ }\bibfield  {title} {\bibinfo {title} {{Gravitational Radiation Reaction to a Particle Motion II: Spinning Particle}},\ }\href {https://doi.org/10.48550/arXiv.gr-qc/9705073} {\bibfield  {journal} {\bibinfo  {journal} {arXiv e-prints}\ ,\ \bibinfo {eid} {gr-qc/9705073}} (\bibinfo {year} {1997})},\ \Eprint {https://arxiv.org/abs/gr-qc/9705073} {arXiv:gr-qc/9705073 [gr-qc]} \BibitemShut {NoStop}%
\bibitem [{\citenamefont {Hinderer}\ \emph {et~al.}(2013)\citenamefont {Hinderer} \emph {et~al.}}]{Hinderer:2013uwa}%
  \BibitemOpen
  \bibfield  {author} {\bibinfo {author} {\bibfnamefont {T.}~\bibnamefont {Hinderer}} \emph {et~al.},\ }\bibfield  {title} {\bibinfo {title} {{Periastron advance in spinning black hole binaries: comparing effective-one-body and Numerical Relativity}},\ }\href {https://doi.org/10.1103/PhysRevD.88.084005} {\bibfield  {journal} {\bibinfo  {journal} {Phys. Rev. D}\ }\textbf {\bibinfo {volume} {88}},\ \bibinfo {pages} {084005} (\bibinfo {year} {2013})},\ \Eprint {https://arxiv.org/abs/1309.0544} {arXiv:1309.0544 [gr-qc]} \BibitemShut {NoStop}%
\bibitem [{\citenamefont {Dones}\ \emph {et~al.}(2025)\citenamefont {Dones}, \citenamefont {Henry},\ and\ \citenamefont {Bernard}}]{Dones:2024odv}%
  \BibitemOpen
  \bibfield  {author} {\bibinfo {author} {\bibfnamefont {E.}~\bibnamefont {Dones}}, \bibinfo {author} {\bibfnamefont {Q.}~\bibnamefont {Henry}},\ and\ \bibinfo {author} {\bibfnamefont {L.}~\bibnamefont {Bernard}},\ }\bibfield  {title} {\bibinfo {title} {{Tidal contributions to the full gravitational waveform to the second-and-a-half post-Newtonian order}},\ }\href {https://doi.org/10.1103/PhysRevD.111.084043} {\bibfield  {journal} {\bibinfo  {journal} {Phys. Rev. D}\ }\textbf {\bibinfo {volume} {111}},\ \bibinfo {pages} {084043} (\bibinfo {year} {2025})},\ \Eprint {https://arxiv.org/abs/2412.14249} {arXiv:2412.14249 [gr-qc]} \BibitemShut {NoStop}%
\bibitem [{\citenamefont {Hansen}(1974)}]{Hansen:1974zz}%
  \BibitemOpen
  \bibfield  {author} {\bibinfo {author} {\bibfnamefont {R.~O.}\ \bibnamefont {Hansen}},\ }\bibfield  {title} {\bibinfo {title} {{Multipole moments of stationary space-times}},\ }\href {https://doi.org/10.1063/1.1666501} {\bibfield  {journal} {\bibinfo  {journal} {J. Math. Phys.}\ }\textbf {\bibinfo {volume} {15}},\ \bibinfo {pages} {46} (\bibinfo {year} {1974})}\BibitemShut {NoStop}%
\bibitem [{\citenamefont {Poisson}(1998)}]{PhysRevD.57.5287}%
  \BibitemOpen
  \bibfield  {author} {\bibinfo {author} {\bibfnamefont {E.}~\bibnamefont {Poisson}},\ }\bibfield  {title} {\bibinfo {title} {Gravitational waves from inspiraling compact binaries: The quadrupole-moment term},\ }\href {https://doi.org/10.1103/PhysRevD.57.5287} {\bibfield  {journal} {\bibinfo  {journal} {Phys. Rev. D}\ }\textbf {\bibinfo {volume} {57}},\ \bibinfo {pages} {5287} (\bibinfo {year} {1998})}\BibitemShut {NoStop}%
\bibitem [{\citenamefont {Binnington}\ and\ \citenamefont {Poisson}(2009)}]{Binnington:2009bb}%
  \BibitemOpen
  \bibfield  {author} {\bibinfo {author} {\bibfnamefont {T.}~\bibnamefont {Binnington}}\ and\ \bibinfo {author} {\bibfnamefont {E.}~\bibnamefont {Poisson}},\ }\bibfield  {title} {\bibinfo {title} {{Relativistic theory of tidal Love numbers}},\ }\href {https://doi.org/10.1103/PhysRevD.80.084018} {\bibfield  {journal} {\bibinfo  {journal} {Phys. Rev. D}\ }\textbf {\bibinfo {volume} {80}},\ \bibinfo {pages} {084018} (\bibinfo {year} {2009})},\ \Eprint {https://arxiv.org/abs/0906.1366} {arXiv:0906.1366 [gr-qc]} \BibitemShut {NoStop}%
\bibitem [{\citenamefont {Damour}\ and\ \citenamefont {Nagar}(2009)}]{Damour:2009vw}%
  \BibitemOpen
  \bibfield  {author} {\bibinfo {author} {\bibfnamefont {T.}~\bibnamefont {Damour}}\ and\ \bibinfo {author} {\bibfnamefont {A.}~\bibnamefont {Nagar}},\ }\bibfield  {title} {\bibinfo {title} {{Relativistic tidal properties of neutron stars}},\ }\href {https://doi.org/10.1103/PhysRevD.80.084035} {\bibfield  {journal} {\bibinfo  {journal} {Phys. Rev. D}\ }\textbf {\bibinfo {volume} {80}},\ \bibinfo {pages} {084035} (\bibinfo {year} {2009})},\ \Eprint {https://arxiv.org/abs/0906.0096} {arXiv:0906.0096 [gr-qc]} \BibitemShut {NoStop}%
\bibitem [{\citenamefont {Le~Tiec}\ and\ \citenamefont {Casals}(2021)}]{LeTiec:2020spy}%
  \BibitemOpen
  \bibfield  {author} {\bibinfo {author} {\bibfnamefont {A.}~\bibnamefont {Le~Tiec}}\ and\ \bibinfo {author} {\bibfnamefont {M.}~\bibnamefont {Casals}},\ }\bibfield  {title} {\bibinfo {title} {{Spinning Black Holes Fall in Love}},\ }\href {https://doi.org/10.1103/PhysRevLett.126.131102} {\bibfield  {journal} {\bibinfo  {journal} {Phys. Rev. Lett.}\ }\textbf {\bibinfo {volume} {126}},\ \bibinfo {pages} {131102} (\bibinfo {year} {2021})},\ \Eprint {https://arxiv.org/abs/2007.00214} {arXiv:2007.00214 [gr-qc]} \BibitemShut {NoStop}%
\bibitem [{\citenamefont {Chia}(2021)}]{Chia:2020yla}%
  \BibitemOpen
  \bibfield  {author} {\bibinfo {author} {\bibfnamefont {H.~S.}\ \bibnamefont {Chia}},\ }\bibfield  {title} {\bibinfo {title} {{Tidal deformation and dissipation of rotating black holes}},\ }\href {https://doi.org/10.1103/PhysRevD.104.024013} {\bibfield  {journal} {\bibinfo  {journal} {Phys. Rev. D}\ }\textbf {\bibinfo {volume} {104}},\ \bibinfo {pages} {024013} (\bibinfo {year} {2021})},\ \Eprint {https://arxiv.org/abs/2010.07300} {arXiv:2010.07300 [gr-qc]} \BibitemShut {NoStop}%
\bibitem [{\citenamefont {Hinderer}(2008)}]{Hinderer:2007mb}%
  \BibitemOpen
  \bibfield  {author} {\bibinfo {author} {\bibfnamefont {T.}~\bibnamefont {Hinderer}},\ }\bibfield  {title} {\bibinfo {title} {{Tidal Love numbers of neutron stars}},\ }\href {https://doi.org/10.1086/533487} {\bibfield  {journal} {\bibinfo  {journal} {Astrophys. J.}\ }\textbf {\bibinfo {volume} {677}},\ \bibinfo {pages} {1216} (\bibinfo {year} {2008})},\ \bibinfo {note} {[Erratum: Astrophys.J. 697, 964 (2009)]},\ \Eprint {https://arxiv.org/abs/0711.2420} {arXiv:0711.2420 [astro-ph]} \BibitemShut {NoStop}%
\bibitem [{\citenamefont {{Cook}}\ \emph {et~al.}(1994)\citenamefont {{Cook}}, \citenamefont {{Shapiro}},\ and\ \citenamefont {{Teukolsky}}}]{1994ApJ...424..823C}%
  \BibitemOpen
  \bibfield  {author} {\bibinfo {author} {\bibfnamefont {G.~B.}\ \bibnamefont {{Cook}}}, \bibinfo {author} {\bibfnamefont {S.~L.}\ \bibnamefont {{Shapiro}}},\ and\ \bibinfo {author} {\bibfnamefont {S.~A.}\ \bibnamefont {{Teukolsky}}},\ }\bibfield  {title} {\bibinfo {title} {{Rapidly Rotating Neutron Stars in General Relativity: Realistic Equations of State}},\ }\href {https://doi.org/10.1086/173934} {\bibfield  {journal} {\bibinfo  {journal} {\apj}\ }\textbf {\bibinfo {volume} {424}},\ \bibinfo {pages} {823} (\bibinfo {year} {1994})}\BibitemShut {NoStop}%
\bibitem [{\citenamefont {Lo}\ and\ \citenamefont {Lin}(2011)}]{Lo:2010bj}%
  \BibitemOpen
  \bibfield  {author} {\bibinfo {author} {\bibfnamefont {K.-W.}\ \bibnamefont {Lo}}\ and\ \bibinfo {author} {\bibfnamefont {L.-M.}\ \bibnamefont {Lin}},\ }\bibfield  {title} {\bibinfo {title} {{The spin parameter of uniformly rotating compact stars}},\ }\href {https://doi.org/10.1088/0004-637X/728/1/12} {\bibfield  {journal} {\bibinfo  {journal} {Astrophys. J.}\ }\textbf {\bibinfo {volume} {728}},\ \bibinfo {pages} {12} (\bibinfo {year} {2011})},\ \Eprint {https://arxiv.org/abs/1011.3563} {arXiv:1011.3563 [astro-ph.HE]} \BibitemShut {NoStop}%
\bibitem [{\citenamefont {Laarakkers}\ and\ \citenamefont {Poisson}(1999)}]{Laarakkers:1997hb}%
  \BibitemOpen
  \bibfield  {author} {\bibinfo {author} {\bibfnamefont {W.~G.}\ \bibnamefont {Laarakkers}}\ and\ \bibinfo {author} {\bibfnamefont {E.}~\bibnamefont {Poisson}},\ }\bibfield  {title} {\bibinfo {title} {{Quadrupole moments of rotating neutron stars}},\ }\href {https://doi.org/10.1086/306732} {\bibfield  {journal} {\bibinfo  {journal} {Astrophys. J.}\ }\textbf {\bibinfo {volume} {512}},\ \bibinfo {pages} {282} (\bibinfo {year} {1999})},\ \Eprint {https://arxiv.org/abs/gr-qc/9709033} {arXiv:gr-qc/9709033} \BibitemShut {NoStop}%
\bibitem [{\citenamefont {Harry}\ and\ \citenamefont {Hinderer}(2018)}]{Harry:2018hke}%
  \BibitemOpen
  \bibfield  {author} {\bibinfo {author} {\bibfnamefont {I.}~\bibnamefont {Harry}}\ and\ \bibinfo {author} {\bibfnamefont {T.}~\bibnamefont {Hinderer}},\ }\bibfield  {title} {\bibinfo {title} {{Observing and measuring the neutron-star equation-of-state in spinning binary neutron star systems}},\ }\href {https://doi.org/10.1088/1361-6382/aac7e3} {\bibfield  {journal} {\bibinfo  {journal} {Class. Quant. Grav.}\ }\textbf {\bibinfo {volume} {35}},\ \bibinfo {pages} {145010} (\bibinfo {year} {2018})},\ \Eprint {https://arxiv.org/abs/1801.09972} {arXiv:1801.09972 [gr-qc]} \BibitemShut {NoStop}%
\bibitem [{\citenamefont {Pappas}\ and\ \citenamefont {Apostolatos}(2012)}]{Pappas:2012qg}%
  \BibitemOpen
  \bibfield  {author} {\bibinfo {author} {\bibfnamefont {G.}~\bibnamefont {Pappas}}\ and\ \bibinfo {author} {\bibfnamefont {T.~A.}\ \bibnamefont {Apostolatos}},\ }\href@noop {} {\bibinfo {title} {{Multipole Moments of numerical spacetimes}}} (\bibinfo {year} {2012}),\ \Eprint {https://arxiv.org/abs/1211.6299} {arXiv:1211.6299 [gr-qc]} \BibitemShut {NoStop}%
\bibitem [{\citenamefont {Hinderer}\ \emph {et~al.}(2010)\citenamefont {Hinderer}, \citenamefont {Lackey}, \citenamefont {Lang},\ and\ \citenamefont {Read}}]{Hinderer:2009ca}%
  \BibitemOpen
  \bibfield  {author} {\bibinfo {author} {\bibfnamefont {T.}~\bibnamefont {Hinderer}}, \bibinfo {author} {\bibfnamefont {B.~D.}\ \bibnamefont {Lackey}}, \bibinfo {author} {\bibfnamefont {R.~N.}\ \bibnamefont {Lang}},\ and\ \bibinfo {author} {\bibfnamefont {J.~S.}\ \bibnamefont {Read}},\ }\bibfield  {title} {\bibinfo {title} {{Tidal deformability of neutron stars with realistic equations of state and their gravitational wave signatures in binary inspiral}},\ }\href {https://doi.org/10.1103/PhysRevD.81.123016} {\bibfield  {journal} {\bibinfo  {journal} {Phys. Rev. D}\ }\textbf {\bibinfo {volume} {81}},\ \bibinfo {pages} {123016} (\bibinfo {year} {2010})},\ \Eprint {https://arxiv.org/abs/0911.3535} {arXiv:0911.3535 [astro-ph.HE]} \BibitemShut {NoStop}%
\bibitem [{\citenamefont {Piekarewicz}\ and\ \citenamefont {Fattoyev}(2019)}]{Piekarewicz:2018sgy}%
  \BibitemOpen
  \bibfield  {author} {\bibinfo {author} {\bibfnamefont {J.}~\bibnamefont {Piekarewicz}}\ and\ \bibinfo {author} {\bibfnamefont {F.~J.}\ \bibnamefont {Fattoyev}},\ }\bibfield  {title} {\bibinfo {title} {{Impact of the neutron star crust on the tidal polarizability}},\ }\href {https://doi.org/10.1103/PhysRevC.99.045802} {\bibfield  {journal} {\bibinfo  {journal} {Phys. Rev. C}\ }\textbf {\bibinfo {volume} {99}},\ \bibinfo {pages} {045802} (\bibinfo {year} {2019})},\ \Eprint {https://arxiv.org/abs/1812.09974} {arXiv:1812.09974 [nucl-th]} \BibitemShut {NoStop}%
\bibitem [{\citenamefont {Gagnon-Bischoff}\ \emph {et~al.}(2018)\citenamefont {Gagnon-Bischoff}, \citenamefont {Green}, \citenamefont {Landry},\ and\ \citenamefont {Ortiz}}]{Gagnon-Bischoff:2017tnz}%
  \BibitemOpen
  \bibfield  {author} {\bibinfo {author} {\bibfnamefont {J.}~\bibnamefont {Gagnon-Bischoff}}, \bibinfo {author} {\bibfnamefont {S.~R.}\ \bibnamefont {Green}}, \bibinfo {author} {\bibfnamefont {P.}~\bibnamefont {Landry}},\ and\ \bibinfo {author} {\bibfnamefont {N.}~\bibnamefont {Ortiz}},\ }\bibfield  {title} {\bibinfo {title} {{Extended I-Love relations for slowly rotating neutron stars}},\ }\href {https://doi.org/10.1103/PhysRevD.97.064042} {\bibfield  {journal} {\bibinfo  {journal} {Phys. Rev. D}\ }\textbf {\bibinfo {volume} {97}},\ \bibinfo {pages} {064042} (\bibinfo {year} {2018})},\ \Eprint {https://arxiv.org/abs/1711.05694} {arXiv:1711.05694 [gr-qc]} \BibitemShut {NoStop}%
\bibitem [{\citenamefont {Landry}\ and\ \citenamefont {Poisson}(2015)}]{Landry:2015cva}%
  \BibitemOpen
  \bibfield  {author} {\bibinfo {author} {\bibfnamefont {P.}~\bibnamefont {Landry}}\ and\ \bibinfo {author} {\bibfnamefont {E.}~\bibnamefont {Poisson}},\ }\bibfield  {title} {\bibinfo {title} {{Gravitomagnetic response of an irrotational body to an applied tidal field}},\ }\href {https://doi.org/10.1103/PhysRevD.91.104026} {\bibfield  {journal} {\bibinfo  {journal} {Phys. Rev. D}\ }\textbf {\bibinfo {volume} {91}},\ \bibinfo {pages} {104026} (\bibinfo {year} {2015})},\ \Eprint {https://arxiv.org/abs/1504.06606} {arXiv:1504.06606 [gr-qc]} \BibitemShut {NoStop}%
\bibitem [{\citenamefont {Sennett}\ \emph {et~al.}(2017)\citenamefont {Sennett}, \citenamefont {Hinderer}, \citenamefont {Steinhoff}, \citenamefont {Buonanno},\ and\ \citenamefont {Ossokine}}]{Sennett:2017etc}%
  \BibitemOpen
  \bibfield  {author} {\bibinfo {author} {\bibfnamefont {N.}~\bibnamefont {Sennett}}, \bibinfo {author} {\bibfnamefont {T.}~\bibnamefont {Hinderer}}, \bibinfo {author} {\bibfnamefont {J.}~\bibnamefont {Steinhoff}}, \bibinfo {author} {\bibfnamefont {A.}~\bibnamefont {Buonanno}},\ and\ \bibinfo {author} {\bibfnamefont {S.}~\bibnamefont {Ossokine}},\ }\bibfield  {title} {\bibinfo {title} {{Distinguishing Boson Stars from Black Holes and Neutron Stars from Tidal Interactions in Inspiraling Binary Systems}},\ }\href {https://doi.org/10.1103/PhysRevD.96.024002} {\bibfield  {journal} {\bibinfo  {journal} {Phys. Rev. D}\ }\textbf {\bibinfo {volume} {96}},\ \bibinfo {pages} {024002} (\bibinfo {year} {2017})},\ \Eprint {https://arxiv.org/abs/1704.08651} {arXiv:1704.08651 [gr-qc]} \BibitemShut {NoStop}%
\bibitem [{\citenamefont {Ryan}(1997)}]{PhysRevD.55.6081}%
  \BibitemOpen
  \bibfield  {author} {\bibinfo {author} {\bibfnamefont {F.~D.}\ \bibnamefont {Ryan}},\ }\bibfield  {title} {\bibinfo {title} {Spinning boson stars with large self-interaction},\ }\href {https://doi.org/10.1103/PhysRevD.55.6081} {\bibfield  {journal} {\bibinfo  {journal} {Phys. Rev. D}\ }\textbf {\bibinfo {volume} {55}},\ \bibinfo {pages} {6081} (\bibinfo {year} {1997})}\BibitemShut {NoStop}%
\bibitem [{\citenamefont {Siemonsen}\ and\ \citenamefont {East}(2021)}]{Siemonsen:2020hcg}%
  \BibitemOpen
  \bibfield  {author} {\bibinfo {author} {\bibfnamefont {N.}~\bibnamefont {Siemonsen}}\ and\ \bibinfo {author} {\bibfnamefont {W.~E.}\ \bibnamefont {East}},\ }\bibfield  {title} {\bibinfo {title} {{Stability of rotating scalar boson stars with nonlinear interactions}},\ }\href {https://doi.org/10.1103/PhysRevD.103.044022} {\bibfield  {journal} {\bibinfo  {journal} {Phys. Rev. D}\ }\textbf {\bibinfo {volume} {103}},\ \bibinfo {pages} {044022} (\bibinfo {year} {2021})},\ \Eprint {https://arxiv.org/abs/2011.08247} {arXiv:2011.08247 [gr-qc]} \BibitemShut {NoStop}%
\bibitem [{\citenamefont {Vaglio}\ \emph {et~al.}(2022)\citenamefont {Vaglio}, \citenamefont {Pacilio}, \citenamefont {Maselli},\ and\ \citenamefont {Pani}}]{Vaglio:2022flq}%
  \BibitemOpen
  \bibfield  {author} {\bibinfo {author} {\bibfnamefont {M.}~\bibnamefont {Vaglio}}, \bibinfo {author} {\bibfnamefont {C.}~\bibnamefont {Pacilio}}, \bibinfo {author} {\bibfnamefont {A.}~\bibnamefont {Maselli}},\ and\ \bibinfo {author} {\bibfnamefont {P.}~\bibnamefont {Pani}},\ }\bibfield  {title} {\bibinfo {title} {{Multipolar structure of rotating boson stars}},\ }\href {https://doi.org/10.1103/PhysRevD.105.124020} {\bibfield  {journal} {\bibinfo  {journal} {Phys. Rev. D}\ }\textbf {\bibinfo {volume} {105}},\ \bibinfo {pages} {124020} (\bibinfo {year} {2022})},\ \Eprint {https://arxiv.org/abs/2203.07442} {arXiv:2203.07442 [gr-qc]} \BibitemShut {NoStop}%
\bibitem [{\citenamefont {Hartl}(2003)}]{Hartl:2002ig}%
  \BibitemOpen
  \bibfield  {author} {\bibinfo {author} {\bibfnamefont {M.~D.}\ \bibnamefont {Hartl}},\ }\bibfield  {title} {\bibinfo {title} {{Dynamics of spinning test particles in Kerr space-time}},\ }\href {https://doi.org/10.1103/PhysRevD.67.024005} {\bibfield  {journal} {\bibinfo  {journal} {Phys. Rev. D}\ }\textbf {\bibinfo {volume} {67}},\ \bibinfo {pages} {024005} (\bibinfo {year} {2003})},\ \Eprint {https://arxiv.org/abs/gr-qc/0210042} {arXiv:gr-qc/0210042} \BibitemShut {NoStop}%
\bibitem [{\citenamefont {{Geroyannis}}\ and\ \citenamefont {{Papasotiriou}}(2000)}]{2000ApJ...534..359G}%
  \BibitemOpen
  \bibfield  {author} {\bibinfo {author} {\bibfnamefont {V.~S.}\ \bibnamefont {{Geroyannis}}}\ and\ \bibinfo {author} {\bibfnamefont {P.~J.}\ \bibnamefont {{Papasotiriou}}},\ }\bibfield  {title} {\bibinfo {title} {{Spin-up and Spin-down of Rotating Magnetic White Dwarfs: A Straightforward Numerical Approach}},\ }\href {https://doi.org/10.1086/308728} {\bibfield  {journal} {\bibinfo  {journal} {\apj}\ }\textbf {\bibinfo {volume} {534}},\ \bibinfo {pages} {359} (\bibinfo {year} {2000})}\BibitemShut {NoStop}%
\bibitem [{\citenamefont {Taylor}\ \emph {et~al.}(2020)\citenamefont {Taylor}, \citenamefont {Yagi},\ and\ \citenamefont {Arras}}]{Taylor:2019hle}%
  \BibitemOpen
  \bibfield  {author} {\bibinfo {author} {\bibfnamefont {A.}~\bibnamefont {Taylor}}, \bibinfo {author} {\bibfnamefont {K.}~\bibnamefont {Yagi}},\ and\ \bibinfo {author} {\bibfnamefont {P.~L.}\ \bibnamefont {Arras}},\ }\bibfield  {title} {\bibinfo {title} {{I\textendash{}Love\textendash{}Q relations for realistic white dwarfs}},\ }\href {https://doi.org/10.1093/mnras/stz3519} {\bibfield  {journal} {\bibinfo  {journal} {Mon. Not. Roy. Astron. Soc.}\ }\textbf {\bibinfo {volume} {492}},\ \bibinfo {pages} {978} (\bibinfo {year} {2020})},\ \Eprint {https://arxiv.org/abs/1912.09557} {arXiv:1912.09557 [gr-qc]} \BibitemShut {NoStop}%
\bibitem [{\citenamefont {{Rodr{\'\i}guez}}\ \emph {et~al.}(2026)\citenamefont {{Rodr{\'\i}guez}}, \citenamefont {{Santoni}},\ and\ \citenamefont {{Solomon}}}]{Rodriguez:2026iot}%
  \BibitemOpen
  \bibfield  {author} {\bibinfo {author} {\bibfnamefont {M.~J.}\ \bibnamefont {{Rodr{\'\i}guez}}}, \bibinfo {author} {\bibfnamefont {L.}~\bibnamefont {{Santoni}}},\ and\ \bibinfo {author} {\bibfnamefont {A.~R.}\ \bibnamefont {{Solomon}}},\ }\bibfield  {title} {\bibinfo {title} {{Love numbers of black holes and compact objects}},\ }\href {https://doi.org/10.48550/arXiv.2604.08653} {\bibfield  {journal} {\bibinfo  {journal} {arXiv e-prints}\ ,\ \bibinfo {eid} {arXiv:2604.08653}} (\bibinfo {year} {2026})},\ \Eprint {https://arxiv.org/abs/2604.08653} {arXiv:2604.08653 [gr-qc]} \BibitemShut {NoStop}%
\bibitem [{\citenamefont {{Chakraborty}}\ and\ \citenamefont {{Pani}}(2026)}]{Chakraborty:2026qru}%
  \BibitemOpen
  \bibfield  {author} {\bibinfo {author} {\bibfnamefont {S.}~\bibnamefont {{Chakraborty}}}\ and\ \bibinfo {author} {\bibfnamefont {P.}~\bibnamefont {{Pani}}},\ }\bibfield  {title} {\bibinfo {title} {{Tidal Response of Compact Objects}},\ }\href {https://doi.org/10.48550/arXiv.2604.08679} {\bibfield  {journal} {\bibinfo  {journal} {arXiv e-prints}\ ,\ \bibinfo {eid} {arXiv:2604.08679}} (\bibinfo {year} {2026})},\ \Eprint {https://arxiv.org/abs/2604.08679} {arXiv:2604.08679 [gr-qc]} \BibitemShut {NoStop}%
\bibitem [{\citenamefont {Akcay}\ \emph {et~al.}(2020)\citenamefont {Akcay}, \citenamefont {Dolan}, \citenamefont {Kavanagh}, \citenamefont {Moxon}, \citenamefont {Warburton},\ and\ \citenamefont {Wardell}}]{Akcay:2019bvk}%
  \BibitemOpen
  \bibfield  {author} {\bibinfo {author} {\bibfnamefont {S.}~\bibnamefont {Akcay}}, \bibinfo {author} {\bibfnamefont {S.~R.}\ \bibnamefont {Dolan}}, \bibinfo {author} {\bibfnamefont {C.}~\bibnamefont {Kavanagh}}, \bibinfo {author} {\bibfnamefont {J.}~\bibnamefont {Moxon}}, \bibinfo {author} {\bibfnamefont {N.}~\bibnamefont {Warburton}},\ and\ \bibinfo {author} {\bibfnamefont {B.}~\bibnamefont {Wardell}},\ }\bibfield  {title} {\bibinfo {title} {{Dissipation in extreme-mass ratio binaries with a spinning secondary}},\ }\href {https://doi.org/10.1103/PhysRevD.102.064013} {\bibfield  {journal} {\bibinfo  {journal} {Phys. Rev. D}\ }\textbf {\bibinfo {volume} {102}},\ \bibinfo {pages} {064013} (\bibinfo {year} {2020})},\ \Eprint {https://arxiv.org/abs/1912.09461} {arXiv:1912.09461 [gr-qc]} \BibitemShut {NoStop}%
\bibitem [{\citenamefont {Warburton}\ \emph {et~al.}(2026)\citenamefont {Warburton}, \citenamefont {Wardell}, \citenamefont {Trestini}, \citenamefont {Henry}, \citenamefont {Pound}, \citenamefont {Blanchet}, \citenamefont {Durkan}, \citenamefont {Faye},\ and\ \citenamefont {Miller}}]{Warburton:2024xnr}%
  \BibitemOpen
  \bibfield  {author} {\bibinfo {author} {\bibfnamefont {N.}~\bibnamefont {Warburton}}, \bibinfo {author} {\bibfnamefont {B.}~\bibnamefont {Wardell}}, \bibinfo {author} {\bibfnamefont {D.}~\bibnamefont {Trestini}}, \bibinfo {author} {\bibfnamefont {Q.}~\bibnamefont {Henry}}, \bibinfo {author} {\bibfnamefont {A.}~\bibnamefont {Pound}}, \bibinfo {author} {\bibfnamefont {L.}~\bibnamefont {Blanchet}}, \bibinfo {author} {\bibfnamefont {L.}~\bibnamefont {Durkan}}, \bibinfo {author} {\bibfnamefont {G.}~\bibnamefont {Faye}},\ and\ \bibinfo {author} {\bibfnamefont {J.}~\bibnamefont {Miller}},\ }\bibfield  {title} {\bibinfo {title} {{Comparison of 4.5PN and 2SF gravitational energy fluxes from quasicircular compact binaries}},\ }\href {https://doi.org/10.1103/tzsw-kcyt} {\bibfield  {journal} {\bibinfo  {journal} {Phys. Rev. D}\ }\textbf {\bibinfo {volume} {113}},\ \bibinfo {pages} {084050} (\bibinfo {year} {2026})},\ \Eprint {https://arxiv.org/abs/2407.00366} {arXiv:2407.00366 [gr-qc]} \BibitemShut {NoStop}%
\bibitem [{\citenamefont {Miller}\ \emph {et~al.}(2024)\citenamefont {Miller}, \citenamefont {Leather}, \citenamefont {Pound},\ and\ \citenamefont {Warburton}}]{Miller:2023ers}%
  \BibitemOpen
  \bibfield  {author} {\bibinfo {author} {\bibfnamefont {J.}~\bibnamefont {Miller}}, \bibinfo {author} {\bibfnamefont {B.}~\bibnamefont {Leather}}, \bibinfo {author} {\bibfnamefont {A.}~\bibnamefont {Pound}},\ and\ \bibinfo {author} {\bibfnamefont {N.}~\bibnamefont {Warburton}},\ }\bibfield  {title} {\bibinfo {title} {{Worldtube puncture scheme for first- and second-order self-force calculations in the Fourier domain}},\ }\href {https://doi.org/10.1103/PhysRevD.109.104010} {\bibfield  {journal} {\bibinfo  {journal} {Phys. Rev. D}\ }\textbf {\bibinfo {volume} {109}},\ \bibinfo {pages} {104010} (\bibinfo {year} {2024})},\ \Eprint {https://arxiv.org/abs/2401.00455} {arXiv:2401.00455 [gr-qc]} \BibitemShut {NoStop}%
\bibitem [{\citenamefont {Hinderer}\ and\ \citenamefont {Flanagan}(2008{\natexlab{b}})}]{Hinderer:2008dm}%
  \BibitemOpen
  \bibfield  {author} {\bibinfo {author} {\bibfnamefont {T.}~\bibnamefont {Hinderer}}\ and\ \bibinfo {author} {\bibfnamefont {E.~E.}\ \bibnamefont {Flanagan}},\ }\bibfield  {title} {\bibinfo {title} {{Two timescale analysis of extreme mass ratio inspirals in Kerr. I. Orbital Motion}},\ }\href {https://doi.org/10.1103/PhysRevD.78.064028} {\bibfield  {journal} {\bibinfo  {journal} {Phys. Rev. D}\ }\textbf {\bibinfo {volume} {78}},\ \bibinfo {pages} {064028} (\bibinfo {year} {2008}{\natexlab{b}})},\ \Eprint {https://arxiv.org/abs/0805.3337} {arXiv:0805.3337 [gr-qc]} \BibitemShut {NoStop}%
\bibitem [{\citenamefont {Shah}\ and\ \citenamefont {Pound}(2015)}]{Shah:2015nva}%
  \BibitemOpen
  \bibfield  {author} {\bibinfo {author} {\bibfnamefont {A.~G.}\ \bibnamefont {Shah}}\ and\ \bibinfo {author} {\bibfnamefont {A.}~\bibnamefont {Pound}},\ }\bibfield  {title} {\bibinfo {title} {{Linear-in-mass-ratio contribution to spin precession and tidal invariants in Schwarzschild spacetime at very high post-Newtonian order}},\ }\href {https://doi.org/10.1103/PhysRevD.91.124022} {\bibfield  {journal} {\bibinfo  {journal} {Phys. Rev. D}\ }\textbf {\bibinfo {volume} {91}},\ \bibinfo {pages} {124022} (\bibinfo {year} {2015})},\ \Eprint {https://arxiv.org/abs/1503.02414} {arXiv:1503.02414 [gr-qc]} \BibitemShut {NoStop}%
\bibitem [{\citenamefont {Gal'tsov}(1982)}]{Galtsov:1982hwm}%
  \BibitemOpen
  \bibfield  {author} {\bibinfo {author} {\bibfnamefont {D.~V.}\ \bibnamefont {Gal'tsov}},\ }\bibfield  {title} {\bibinfo {title} {{Radiation reaction in the Kerr gravitational field}},\ }\href {https://doi.org/10.1088/0305-4470/15/12/025} {\bibfield  {journal} {\bibinfo  {journal} {J. Phys. A}\ }\textbf {\bibinfo {volume} {15}},\ \bibinfo {pages} {3737} (\bibinfo {year} {1982})}\BibitemShut {NoStop}%
\bibitem [{\citenamefont {Sago}\ \emph {et~al.}(2006)\citenamefont {Sago}, \citenamefont {Tanaka}, \citenamefont {Hikida}, \citenamefont {Ganz},\ and\ \citenamefont {Nakano}}]{Sago:2005fn}%
  \BibitemOpen
  \bibfield  {author} {\bibinfo {author} {\bibfnamefont {N.}~\bibnamefont {Sago}}, \bibinfo {author} {\bibfnamefont {T.}~\bibnamefont {Tanaka}}, \bibinfo {author} {\bibfnamefont {W.}~\bibnamefont {Hikida}}, \bibinfo {author} {\bibfnamefont {K.}~\bibnamefont {Ganz}},\ and\ \bibinfo {author} {\bibfnamefont {H.}~\bibnamefont {Nakano}},\ }\bibfield  {title} {\bibinfo {title} {{The Adiabatic evolution of orbital parameters in the Kerr spacetime}},\ }\href {https://doi.org/10.1143/PTP.115.873} {\bibfield  {journal} {\bibinfo  {journal} {Prog. Theor. Phys.}\ }\textbf {\bibinfo {volume} {115}},\ \bibinfo {pages} {873} (\bibinfo {year} {2006})},\ \Eprint {https://arxiv.org/abs/gr-qc/0511151} {arXiv:gr-qc/0511151} \BibitemShut {NoStop}%
\bibitem [{\citenamefont {Isoyama}\ \emph {et~al.}(2019)\citenamefont {Isoyama}, \citenamefont {Fujita}, \citenamefont {Nakano}, \citenamefont {Sago},\ and\ \citenamefont {Tanaka}}]{Isoyama:2018sib}%
  \BibitemOpen
  \bibfield  {author} {\bibinfo {author} {\bibfnamefont {S.}~\bibnamefont {Isoyama}}, \bibinfo {author} {\bibfnamefont {R.}~\bibnamefont {Fujita}}, \bibinfo {author} {\bibfnamefont {H.}~\bibnamefont {Nakano}}, \bibinfo {author} {\bibfnamefont {N.}~\bibnamefont {Sago}},\ and\ \bibinfo {author} {\bibfnamefont {T.}~\bibnamefont {Tanaka}},\ }\bibfield  {title} {\bibinfo {title} {{{\textquotedblleft}Flux-balance formulae{\textquotedblright} for extreme mass-ratio inspirals}},\ }\href {https://doi.org/10.1093/ptep/pty136} {\bibfield  {journal} {\bibinfo  {journal} {PTEP}\ }\textbf {\bibinfo {volume} {2019}},\ \bibinfo {pages} {013E01} (\bibinfo {year} {2019})},\ \Eprint {https://arxiv.org/abs/1809.11118} {arXiv:1809.11118 [gr-qc]} \BibitemShut {NoStop}%
\bibitem [{\citenamefont {Detweiler}\ and\ \citenamefont {Whiting}(2003)}]{Detweiler:2002mi}%
  \BibitemOpen
  \bibfield  {author} {\bibinfo {author} {\bibfnamefont {S.~L.}\ \bibnamefont {Detweiler}}\ and\ \bibinfo {author} {\bibfnamefont {B.~F.}\ \bibnamefont {Whiting}},\ }\bibfield  {title} {\bibinfo {title} {{Selfforce via a Green's function decomposition}},\ }\href {https://doi.org/10.1103/PhysRevD.67.024025} {\bibfield  {journal} {\bibinfo  {journal} {Phys. Rev. D}\ }\textbf {\bibinfo {volume} {67}},\ \bibinfo {pages} {024025} (\bibinfo {year} {2003})},\ \Eprint {https://arxiv.org/abs/gr-qc/0202086} {arXiv:gr-qc/0202086} \BibitemShut {NoStop}%
\bibitem [{\citenamefont {Drasco}\ \emph {et~al.}(2005)\citenamefont {Drasco}, \citenamefont {Flanagan},\ and\ \citenamefont {Hughes}}]{Drasco:2005is}%
  \BibitemOpen
  \bibfield  {author} {\bibinfo {author} {\bibfnamefont {S.}~\bibnamefont {Drasco}}, \bibinfo {author} {\bibfnamefont {E.~E.}\ \bibnamefont {Flanagan}},\ and\ \bibinfo {author} {\bibfnamefont {S.~A.}\ \bibnamefont {Hughes}},\ }\bibfield  {title} {\bibinfo {title} {{Computing inspirals in Kerr in the adiabatic regime. I. The Scalar case}},\ }\href {https://doi.org/10.1088/0264-9381/22/15/011} {\bibfield  {journal} {\bibinfo  {journal} {Class. Quant. Grav.}\ }\textbf {\bibinfo {volume} {22}},\ \bibinfo {pages} {S801} (\bibinfo {year} {2005})},\ \Eprint {https://arxiv.org/abs/gr-qc/0505075} {arXiv:gr-qc/0505075} \BibitemShut {NoStop}%
\bibitem [{\citenamefont {Chrzanowski}\ and\ \citenamefont {Misner}(1974)}]{Chrzanowski:1974nr}%
  \BibitemOpen
  \bibfield  {author} {\bibinfo {author} {\bibfnamefont {P.~L.}\ \bibnamefont {Chrzanowski}}\ and\ \bibinfo {author} {\bibfnamefont {C.~W.}\ \bibnamefont {Misner}},\ }\bibfield  {title} {\bibinfo {title} {{Geodesic synchrotron radiation in the Kerr geometry by the method of asymptotically factorized Green's functions}},\ }\href {https://doi.org/10.1103/PhysRevD.10.1701} {\bibfield  {journal} {\bibinfo  {journal} {Phys. Rev. D}\ }\textbf {\bibinfo {volume} {10}},\ \bibinfo {pages} {1701} (\bibinfo {year} {1974})}\BibitemShut {NoStop}%
\bibitem [{\citenamefont {Trestini}\ \emph {et~al.}(2026)\citenamefont {Trestini}, \citenamefont {Nasipak},\ and\ \citenamefont {Pound}}]{Trestini:2026tky}%
  \BibitemOpen
  \bibfield  {author} {\bibinfo {author} {\bibfnamefont {D.}~\bibnamefont {Trestini}}, \bibinfo {author} {\bibfnamefont {Z.}~\bibnamefont {Nasipak}},\ and\ \bibinfo {author} {\bibfnamefont {A.}~\bibnamefont {Pound}},\ }\bibfield  {title} {\bibinfo {title} {{Constants of motion in gravitational self-force theory}},\ }\href {https://doi.org/10.1103/5qb1-9f7c} {\bibfield  {journal} {\bibinfo  {journal} {Phys. Rev. D}\ }\textbf {\bibinfo {volume} {113}},\ \bibinfo {pages} {124049} (\bibinfo {year} {2026})},\ \Eprint {https://arxiv.org/abs/2601.05223} {arXiv:2601.05223 [gr-qc]} \BibitemShut {NoStop}%
\bibitem [{\citenamefont {Comp{\`e}re}\ \emph {et~al.}(2020)\citenamefont {Comp{\`e}re}, \citenamefont {Oliveri},\ and\ \citenamefont {Seraj}}]{Compere:2019gft}%
  \BibitemOpen
  \bibfield  {author} {\bibinfo {author} {\bibfnamefont {G.}~\bibnamefont {Comp{\`e}re}}, \bibinfo {author} {\bibfnamefont {R.}~\bibnamefont {Oliveri}},\ and\ \bibinfo {author} {\bibfnamefont {A.}~\bibnamefont {Seraj}},\ }\bibfield  {title} {\bibinfo {title} {{The Poincar{\'e} and BMS flux-balance laws with application to binary systems}},\ }\href {https://doi.org/10.1007/JHEP10(2020)116} {\bibfield  {journal} {\bibinfo  {journal} {JHEP}\ }\textbf {\bibinfo {volume} {10}},\ \bibinfo {pages} {116}},\ \bibinfo {note} {[Erratum: JHEP 06, 045 (2024)]},\ \Eprint {https://arxiv.org/abs/1912.03164} {arXiv:1912.03164 [gr-qc]} \BibitemShut {NoStop}%
\bibitem [{\citenamefont {Ashtekar}\ and\ \citenamefont {Krishnan}(2004)}]{Ashtekar:2004cn}%
  \BibitemOpen
  \bibfield  {author} {\bibinfo {author} {\bibfnamefont {A.}~\bibnamefont {Ashtekar}}\ and\ \bibinfo {author} {\bibfnamefont {B.}~\bibnamefont {Krishnan}},\ }\bibfield  {title} {\bibinfo {title} {{Isolated and dynamical horizons and their applications}},\ }\href {https://doi.org/10.12942/lrr-2004-10} {\bibfield  {journal} {\bibinfo  {journal} {Living Rev. Rel.}\ }\textbf {\bibinfo {volume} {7}},\ \bibinfo {pages} {10} (\bibinfo {year} {2004})},\ \Eprint {https://arxiv.org/abs/gr-qc/0407042} {arXiv:gr-qc/0407042} \BibitemShut {NoStop}%
\bibitem [{\citenamefont {{Ashtekar}}(2023)}]{Ashtekar:2023kei}%
  \BibitemOpen
  \bibfield  {author} {\bibinfo {author} {\bibfnamefont {A.}~\bibnamefont {{Ashtekar}}},\ }\bibfield  {title} {\bibinfo {title} {{Black Hole Horizons and their Mechanics}},\ }\href {https://doi.org/10.48550/arXiv.2308.08729} {\bibfield  {journal} {\bibinfo  {journal} {arXiv e-prints}\ ,\ \bibinfo {eid} {arXiv:2308.08729}} (\bibinfo {year} {2023})},\ \Eprint {https://arxiv.org/abs/2308.08729} {arXiv:2308.08729 [gr-qc]} \BibitemShut {NoStop}%
\bibitem [{\citenamefont {Chandrasekaran}\ \emph {et~al.}(2018)\citenamefont {Chandrasekaran}, \citenamefont {Flanagan},\ and\ \citenamefont {Prabhu}}]{Chandrasekaran:2018aop}%
  \BibitemOpen
  \bibfield  {author} {\bibinfo {author} {\bibfnamefont {V.}~\bibnamefont {Chandrasekaran}}, \bibinfo {author} {\bibfnamefont {{\'E}.~{\'E}.}\ \bibnamefont {Flanagan}},\ and\ \bibinfo {author} {\bibfnamefont {K.}~\bibnamefont {Prabhu}},\ }\bibfield  {title} {\bibinfo {title} {{Symmetries and charges of general relativity at null boundaries}},\ }\href {https://doi.org/10.1007/JHEP11(2018)125} {\bibfield  {journal} {\bibinfo  {journal} {JHEP}\ }\textbf {\bibinfo {volume} {11}},\ \bibinfo {pages} {125}},\ \bibinfo {note} {[Erratum: JHEP 07, 224 (2023)]},\ \Eprint {https://arxiv.org/abs/1807.11499} {arXiv:1807.11499 [hep-th]} \BibitemShut {NoStop}%
\bibitem [{\citenamefont {Pound}\ \emph {et~al.}(2020)\citenamefont {Pound}, \citenamefont {Wardell}, \citenamefont {Warburton},\ and\ \citenamefont {Miller}}]{Pound:2019lzj}%
  \BibitemOpen
  \bibfield  {author} {\bibinfo {author} {\bibfnamefont {A.}~\bibnamefont {Pound}}, \bibinfo {author} {\bibfnamefont {B.}~\bibnamefont {Wardell}}, \bibinfo {author} {\bibfnamefont {N.}~\bibnamefont {Warburton}},\ and\ \bibinfo {author} {\bibfnamefont {J.}~\bibnamefont {Miller}},\ }\bibfield  {title} {\bibinfo {title} {{Second-Order Self-Force Calculation of Gravitational Binding Energy in Compact Binaries}},\ }\href {https://doi.org/10.1103/PhysRevLett.124.021101} {\bibfield  {journal} {\bibinfo  {journal} {Phys. Rev. Lett.}\ }\textbf {\bibinfo {volume} {124}},\ \bibinfo {pages} {021101} (\bibinfo {year} {2020})},\ \Eprint {https://arxiv.org/abs/1908.07419} {arXiv:1908.07419 [gr-qc]} \BibitemShut {NoStop}%
\bibitem [{\citenamefont {Trestini}(2025{\natexlab{a}})}]{Trestini:2025nzr}%
  \BibitemOpen
  \bibfield  {author} {\bibinfo {author} {\bibfnamefont {D.}~\bibnamefont {Trestini}},\ }\bibfield  {title} {\bibinfo {title} {{Schott term in the binding energy for compact binaries on circular orbits at fourth post-Newtonian order}},\ }\href {https://doi.org/10.1103/lsbb-sv45} {\bibfield  {journal} {\bibinfo  {journal} {Phys. Rev. D}\ }\textbf {\bibinfo {volume} {112}},\ \bibinfo {pages} {024076} (\bibinfo {year} {2025}{\natexlab{a}})},\ \Eprint {https://arxiv.org/abs/2504.13245} {arXiv:2504.13245 [gr-qc]} \BibitemShut {NoStop}%
\bibitem [{\citenamefont {Tanaka}\ \emph {et~al.}(1996)\citenamefont {Tanaka}, \citenamefont {Mino}, \citenamefont {Sasaki},\ and\ \citenamefont {Shibata}}]{Tanaka:1996ht}%
  \BibitemOpen
  \bibfield  {author} {\bibinfo {author} {\bibfnamefont {T.}~\bibnamefont {Tanaka}}, \bibinfo {author} {\bibfnamefont {Y.}~\bibnamefont {Mino}}, \bibinfo {author} {\bibfnamefont {M.}~\bibnamefont {Sasaki}},\ and\ \bibinfo {author} {\bibfnamefont {M.}~\bibnamefont {Shibata}},\ }\bibfield  {title} {\bibinfo {title} {{Gravitational waves from a spinning particle in circular orbits around a rotating black hole}},\ }\href {https://doi.org/10.1103/PhysRevD.54.3762} {\bibfield  {journal} {\bibinfo  {journal} {Phys. Rev. D}\ }\textbf {\bibinfo {volume} {54}},\ \bibinfo {pages} {3762} (\bibinfo {year} {1996})},\ \Eprint {https://arxiv.org/abs/gr-qc/9602038} {arXiv:gr-qc/9602038} \BibitemShut {NoStop}%
\bibitem [{\citenamefont {Piovano}\ \emph {et~al.}(2021)\citenamefont {Piovano}, \citenamefont {Brito}, \citenamefont {Maselli},\ and\ \citenamefont {Pani}}]{Piovano:2021iwv}%
  \BibitemOpen
  \bibfield  {author} {\bibinfo {author} {\bibfnamefont {G.~A.}\ \bibnamefont {Piovano}}, \bibinfo {author} {\bibfnamefont {R.}~\bibnamefont {Brito}}, \bibinfo {author} {\bibfnamefont {A.}~\bibnamefont {Maselli}},\ and\ \bibinfo {author} {\bibfnamefont {P.}~\bibnamefont {Pani}},\ }\bibfield  {title} {\bibinfo {title} {{Assessing the detectability of the secondary spin in extreme mass-ratio inspirals with fully relativistic numerical waveforms}},\ }\href {https://doi.org/10.1103/PhysRevD.104.124019} {\bibfield  {journal} {\bibinfo  {journal} {Phys. Rev. D}\ }\textbf {\bibinfo {volume} {104}},\ \bibinfo {pages} {124019} (\bibinfo {year} {2021})},\ \Eprint {https://arxiv.org/abs/2105.07083} {arXiv:2105.07083 [gr-qc]} \BibitemShut {NoStop}%
\bibitem [{\citenamefont {Skoup\'y}\ and\ \citenamefont {Lukes-Gerakopoulos}(2021)}]{Skoupy:2021asz}%
  \BibitemOpen
  \bibfield  {author} {\bibinfo {author} {\bibfnamefont {V.}~\bibnamefont {Skoup\'y}}\ and\ \bibinfo {author} {\bibfnamefont {G.}~\bibnamefont {Lukes-Gerakopoulos}},\ }\bibfield  {title} {\bibinfo {title} {{Spinning test body orbiting around a Kerr black hole: Eccentric equatorial orbits and their asymptotic gravitational-wave fluxes}},\ }\href {https://doi.org/10.1103/PhysRevD.103.104045} {\bibfield  {journal} {\bibinfo  {journal} {Phys. Rev. D}\ }\textbf {\bibinfo {volume} {103}},\ \bibinfo {pages} {104045} (\bibinfo {year} {2021})},\ \Eprint {https://arxiv.org/abs/2102.04819} {arXiv:2102.04819 [gr-qc]} \BibitemShut {NoStop}%
\bibitem [{\citenamefont {Shah}(2014)}]{Shah:2014tka}%
  \BibitemOpen
  \bibfield  {author} {\bibinfo {author} {\bibfnamefont {A.~G.}\ \bibnamefont {Shah}},\ }\bibfield  {title} {\bibinfo {title} {{Gravitational-wave flux for a particle orbiting a Kerr black hole to 20th post-Newtonian order: a numerical approach}},\ }\href {https://doi.org/10.1103/PhysRevD.90.044025} {\bibfield  {journal} {\bibinfo  {journal} {Phys. Rev. D}\ }\textbf {\bibinfo {volume} {90}},\ \bibinfo {pages} {044025} (\bibinfo {year} {2014})},\ \Eprint {https://arxiv.org/abs/1403.2697} {arXiv:1403.2697 [gr-qc]} \BibitemShut {NoStop}%
\bibitem [{\citenamefont {Boh{\'e}}\ \emph {et~al.}(2013)\citenamefont {Boh{\'e}}, \citenamefont {Marsat},\ and\ \citenamefont {Blanchet}}]{Bohe:2013cla}%
  \BibitemOpen
  \bibfield  {author} {\bibinfo {author} {\bibfnamefont {A.}~\bibnamefont {Boh{\'e}}}, \bibinfo {author} {\bibfnamefont {S.}~\bibnamefont {Marsat}},\ and\ \bibinfo {author} {\bibfnamefont {L.}~\bibnamefont {Blanchet}},\ }\bibfield  {title} {\bibinfo {title} {{Next-to-next-to-leading order spin{\textendash}orbit effects in the gravitational wave flux and orbital phasing of compact binaries}},\ }\href {https://doi.org/10.1088/0264-9381/30/13/135009} {\bibfield  {journal} {\bibinfo  {journal} {Class. Quant. Grav.}\ }\textbf {\bibinfo {volume} {30}},\ \bibinfo {pages} {135009} (\bibinfo {year} {2013})},\ \Eprint {https://arxiv.org/abs/1303.7412} {arXiv:1303.7412 [gr-qc]} \BibitemShut {NoStop}%
\bibitem [{\citenamefont {Abdelsalhin}\ \emph {et~al.}(2018)\citenamefont {Abdelsalhin}, \citenamefont {Gualtieri},\ and\ \citenamefont {Pani}}]{Abdelsalhin:2018reg}%
  \BibitemOpen
  \bibfield  {author} {\bibinfo {author} {\bibfnamefont {T.}~\bibnamefont {Abdelsalhin}}, \bibinfo {author} {\bibfnamefont {L.}~\bibnamefont {Gualtieri}},\ and\ \bibinfo {author} {\bibfnamefont {P.}~\bibnamefont {Pani}},\ }\bibfield  {title} {\bibinfo {title} {{Post-Newtonian spin-tidal couplings for compact binaries}},\ }\href {https://doi.org/10.1103/PhysRevD.98.104046} {\bibfield  {journal} {\bibinfo  {journal} {Phys. Rev. D}\ }\textbf {\bibinfo {volume} {98}},\ \bibinfo {pages} {104046} (\bibinfo {year} {2018})},\ \Eprint {https://arxiv.org/abs/1805.01487} {arXiv:1805.01487 [gr-qc]} \BibitemShut {NoStop}%
\bibitem [{\citenamefont {Saketh}\ \emph {et~al.}(2023)\citenamefont {Saketh}, \citenamefont {Steinhoff}, \citenamefont {Vines},\ and\ \citenamefont {Buonanno}}]{Saketh:2022xjb}%
  \BibitemOpen
  \bibfield  {author} {\bibinfo {author} {\bibfnamefont {M.~V.~S.}\ \bibnamefont {Saketh}}, \bibinfo {author} {\bibfnamefont {J.}~\bibnamefont {Steinhoff}}, \bibinfo {author} {\bibfnamefont {J.}~\bibnamefont {Vines}},\ and\ \bibinfo {author} {\bibfnamefont {A.}~\bibnamefont {Buonanno}},\ }\bibfield  {title} {\bibinfo {title} {{Modeling horizon absorption in spinning binary black holes using effective worldline theory}},\ }\href {https://doi.org/10.1103/PhysRevD.107.084006} {\bibfield  {journal} {\bibinfo  {journal} {Phys. Rev. D}\ }\textbf {\bibinfo {volume} {107}},\ \bibinfo {pages} {084006} (\bibinfo {year} {2023})},\ \Eprint {https://arxiv.org/abs/2212.13095} {arXiv:2212.13095 [gr-qc]} \BibitemShut {NoStop}%
\bibitem [{\citenamefont {Fujita}(2012)}]{Fujita:2012cm}%
  \BibitemOpen
  \bibfield  {author} {\bibinfo {author} {\bibfnamefont {R.}~\bibnamefont {Fujita}},\ }\bibfield  {title} {\bibinfo {title} {{Gravitational Waves from a Particle in Circular Orbits around a Schwarzschild Black Hole to the 22nd Post-Newtonian Order}},\ }\href {https://doi.org/10.1143/PTP.128.971} {\bibfield  {journal} {\bibinfo  {journal} {Prog. Theor. Phys.}\ }\textbf {\bibinfo {volume} {128}},\ \bibinfo {pages} {971} (\bibinfo {year} {2012})},\ \Eprint {https://arxiv.org/abs/1211.5535} {arXiv:1211.5535 [gr-qc]} \BibitemShut {NoStop}%
\bibitem [{\citenamefont {Cipriani}\ \emph {et~al.}(2025)\citenamefont {Cipriani}, \citenamefont {Di~Russo}, \citenamefont {Fucito}, \citenamefont {Morales}, \citenamefont {Poghosyan},\ and\ \citenamefont {Poghossian}}]{Cipriani:2025ikx}%
  \BibitemOpen
  \bibfield  {author} {\bibinfo {author} {\bibfnamefont {A.}~\bibnamefont {Cipriani}}, \bibinfo {author} {\bibfnamefont {G.}~\bibnamefont {Di~Russo}}, \bibinfo {author} {\bibfnamefont {F.}~\bibnamefont {Fucito}}, \bibinfo {author} {\bibfnamefont {J.~F.}\ \bibnamefont {Morales}}, \bibinfo {author} {\bibfnamefont {H.}~\bibnamefont {Poghosyan}},\ and\ \bibinfo {author} {\bibfnamefont {R.}~\bibnamefont {Poghossian}},\ }\bibfield  {title} {\bibinfo {title} {{Resumming post-Minkowskian and post-Newtonian gravitational waveform expansions}},\ }\href {https://doi.org/10.21468/SciPostPhys.19.2.057} {\bibfield  {journal} {\bibinfo  {journal} {SciPost Phys.}\ }\textbf {\bibinfo {volume} {19}},\ \bibinfo {pages} {057} (\bibinfo {year} {2025})},\ \Eprint {https://arxiv.org/abs/2501.19257} {arXiv:2501.19257 [gr-qc]} \BibitemShut {NoStop}%
\bibitem [{\citenamefont {Mano}\ \emph {et~al.}(1996)\citenamefont {Mano}, \citenamefont {Suzuki},\ and\ \citenamefont {Takasugi}}]{10.1143/PTP.95.1079}%
  \BibitemOpen
  \bibfield  {author} {\bibinfo {author} {\bibfnamefont {S.}~\bibnamefont {Mano}}, \bibinfo {author} {\bibfnamefont {H.}~\bibnamefont {Suzuki}},\ and\ \bibinfo {author} {\bibfnamefont {E.}~\bibnamefont {Takasugi}},\ }\bibfield  {title} {\bibinfo {title} {{Analytic Solutions of the Teukolsky Equation and Their Low Frequency Expansions}},\ }\href {https://doi.org/10.1143/PTP.95.1079} {\bibfield  {journal} {\bibinfo  {journal} {Progress of Theoretical Physics}\ }\textbf {\bibinfo {volume} {95}},\ \bibinfo {pages} {1079} (\bibinfo {year} {1996})}\BibitemShut {NoStop}%
\bibitem [{\citenamefont {Sasaki}\ and\ \citenamefont {Tagoshi}(2003)}]{Sasaki:2003xr}%
  \BibitemOpen
  \bibfield  {author} {\bibinfo {author} {\bibfnamefont {M.}~\bibnamefont {Sasaki}}\ and\ \bibinfo {author} {\bibfnamefont {H.}~\bibnamefont {Tagoshi}},\ }\bibfield  {title} {\bibinfo {title} {{Analytic black hole perturbation approach to gravitational radiation}},\ }\href {https://doi.org/10.12942/lrr-2003-6} {\bibfield  {journal} {\bibinfo  {journal} {Living Rev. Rel.}\ }\textbf {\bibinfo {volume} {6}},\ \bibinfo {pages} {6} (\bibinfo {year} {2003})},\ \Eprint {https://arxiv.org/abs/gr-qc/0306120} {arXiv:gr-qc/0306120} \BibitemShut {NoStop}%
\bibitem [{\citenamefont {Nagar}\ \emph {et~al.}(2019{\natexlab{b}})\citenamefont {Nagar}, \citenamefont {Messina}, \citenamefont {Kavanagh}, \citenamefont {Lukes-Gerakopoulos}, \citenamefont {Warburton}, \citenamefont {Bernuzzi},\ and\ \citenamefont {Harms}}]{Nagar:2019wrt}%
  \BibitemOpen
  \bibfield  {author} {\bibinfo {author} {\bibfnamefont {A.}~\bibnamefont {Nagar}}, \bibinfo {author} {\bibfnamefont {F.}~\bibnamefont {Messina}}, \bibinfo {author} {\bibfnamefont {C.}~\bibnamefont {Kavanagh}}, \bibinfo {author} {\bibfnamefont {G.}~\bibnamefont {Lukes-Gerakopoulos}}, \bibinfo {author} {\bibfnamefont {N.}~\bibnamefont {Warburton}}, \bibinfo {author} {\bibfnamefont {S.}~\bibnamefont {Bernuzzi}},\ and\ \bibinfo {author} {\bibfnamefont {E.}~\bibnamefont {Harms}},\ }\bibfield  {title} {\bibinfo {title} {{Factorization and resummation: A new paradigm to improve gravitational wave amplitudes. III: the spinning test-body terms}},\ }\href {https://doi.org/10.1103/PhysRevD.100.104056} {\bibfield  {journal} {\bibinfo  {journal} {Phys. Rev. D}\ }\textbf {\bibinfo {volume} {100}},\ \bibinfo {pages} {104056} (\bibinfo {year} {2019}{\natexlab{b}})},\ \Eprint {https://arxiv.org/abs/1907.12233} {arXiv:1907.12233 [gr-qc]} \BibitemShut {NoStop}%
\bibitem [{Sup()}]{SupplementalMaterial}%
  \BibitemOpen
  \href@noop {} {}\bibinfo {note} {{See Supplemental Material at [URL will be inserted by publisher] for the complete dataset.}}\BibitemShut {Stop}%
\bibitem [{\citenamefont {{\relax Black Hole Perturbation Toolkit,}}()}]{BHPT}%
  \BibitemOpen
  \bibfield  {author} {\bibinfo {author} {\bibnamefont {{\relax Black Hole Perturbation Toolkit,}}},\ }\href@noop {} {}\bibinfo {howpublished} {\href{http://bhptoolkit.org/}{(http://bhptoolkit.org/})}\BibitemShut {NoStop}%
\bibitem [{\citenamefont {Wardell}\ \emph {et~al.}(2025{\natexlab{a}})\citenamefont {Wardell}, \citenamefont {Warburton}, \citenamefont {Cunningham}, \citenamefont {Durkan}, \citenamefont {Leather}, \citenamefont {Nasipak}, \citenamefont {Kavanagh}, \citenamefont {Ottewill}, \citenamefont {Casals}, \citenamefont {Torres}, \citenamefont {Neef},\ and\ \citenamefont {Barsanti}}]{bhpt_teukolsky}%
  \BibitemOpen
  \bibfield  {author} {\bibinfo {author} {\bibfnamefont {B.}~\bibnamefont {Wardell}}, \bibinfo {author} {\bibfnamefont {N.}~\bibnamefont {Warburton}}, \bibinfo {author} {\bibfnamefont {K.}~\bibnamefont {Cunningham}}, \bibinfo {author} {\bibfnamefont {L.}~\bibnamefont {Durkan}}, \bibinfo {author} {\bibfnamefont {B.}~\bibnamefont {Leather}}, \bibinfo {author} {\bibfnamefont {Z.}~\bibnamefont {Nasipak}}, \bibinfo {author} {\bibfnamefont {C.}~\bibnamefont {Kavanagh}}, \bibinfo {author} {\bibfnamefont {A.}~\bibnamefont {Ottewill}}, \bibinfo {author} {\bibfnamefont {M.}~\bibnamefont {Casals}}, \bibinfo {author} {\bibfnamefont {T.}~\bibnamefont {Torres}}, \bibinfo {author} {\bibfnamefont {J.}~\bibnamefont {Neef}},\ and\ \bibinfo {author} {\bibfnamefont {S.}~\bibnamefont {Barsanti}},\ }\href {https://doi.org/10.5281/zenodo.15745889} {\bibinfo {title} {Teukolsky}} (\bibinfo {year} {2025}{\natexlab{a}})\BibitemShut {NoStop}%
\bibitem [{\citenamefont {Wardell}\ \emph {et~al.}(2024)\citenamefont {Wardell}, \citenamefont {Warburton}, \citenamefont {Fransen}, \citenamefont {Upton}, \citenamefont {Cunningham}, \citenamefont {Ottewill},\ and\ \citenamefont {Casals}}]{bhpt_swsh}%
  \BibitemOpen
  \bibfield  {author} {\bibinfo {author} {\bibfnamefont {B.}~\bibnamefont {Wardell}}, \bibinfo {author} {\bibfnamefont {N.}~\bibnamefont {Warburton}}, \bibinfo {author} {\bibfnamefont {K.}~\bibnamefont {Fransen}}, \bibinfo {author} {\bibfnamefont {S.~D.}\ \bibnamefont {Upton}}, \bibinfo {author} {\bibfnamefont {K.}~\bibnamefont {Cunningham}}, \bibinfo {author} {\bibfnamefont {A.}~\bibnamefont {Ottewill}},\ and\ \bibinfo {author} {\bibfnamefont {M.}~\bibnamefont {Casals}},\ }\href {https://doi.org/10.5281/zenodo.11199019} {\bibinfo {title} {Spinweightedspheroidalharmonics}} (\bibinfo {year} {2024})\BibitemShut {NoStop}%
\bibitem [{\citenamefont {van~de Meent}\ and\ \citenamefont {Shah}(2015)}]{vandeMeent:2015lxa}%
  \BibitemOpen
  \bibfield  {author} {\bibinfo {author} {\bibfnamefont {M.}~\bibnamefont {van~de Meent}}\ and\ \bibinfo {author} {\bibfnamefont {A.~G.}\ \bibnamefont {Shah}},\ }\bibfield  {title} {\bibinfo {title} {{Metric perturbations produced by eccentric equatorial orbits around a Kerr black hole}},\ }\href {https://doi.org/10.1103/PhysRevD.92.064025} {\bibfield  {journal} {\bibinfo  {journal} {Phys. Rev. D}\ }\textbf {\bibinfo {volume} {92}},\ \bibinfo {pages} {064025} (\bibinfo {year} {2015})},\ \Eprint {https://arxiv.org/abs/1506.04755} {arXiv:1506.04755 [gr-qc]} \BibitemShut {NoStop}%
\bibitem [{\citenamefont {Wardell}\ \emph {et~al.}(2025{\natexlab{b}})\citenamefont {Wardell}, \citenamefont {Warburton}, \citenamefont {Kavanagh}, \citenamefont {Munna}, \citenamefont {Evans},\ and\ \citenamefont {Trestini}}]{bhpt_postnewtonianselfforce}%
  \BibitemOpen
  \bibfield  {author} {\bibinfo {author} {\bibfnamefont {B.}~\bibnamefont {Wardell}}, \bibinfo {author} {\bibfnamefont {N.}~\bibnamefont {Warburton}}, \bibinfo {author} {\bibfnamefont {C.}~\bibnamefont {Kavanagh}}, \bibinfo {author} {\bibfnamefont {C.}~\bibnamefont {Munna}}, \bibinfo {author} {\bibfnamefont {C.~R.}\ \bibnamefont {Evans}},\ and\ \bibinfo {author} {\bibfnamefont {D.}~\bibnamefont {Trestini}},\ }\href {https://doi.org/10.5281/zenodo.15969633} {\bibinfo {title} {Postnewtonianselfforce}} (\bibinfo {year} {2025}{\natexlab{b}})\BibitemShut {NoStop}%
\bibitem [{BHP()}]{BHPToolkit}%
  \BibitemOpen
  \href@noop {} {\bibinfo {title} {{Black Hole Perturbation Toolkit}}},\ \bibinfo {howpublished} {(\href{http://bhptoolkit.org/}{bhptoolkit.org})}\BibitemShut {NoStop}%
\bibitem [{\citenamefont {Yunes}\ and\ \citenamefont {Berti}(2008)}]{Yunes:2008tw}%
  \BibitemOpen
  \bibfield  {author} {\bibinfo {author} {\bibfnamefont {N.}~\bibnamefont {Yunes}}\ and\ \bibinfo {author} {\bibfnamefont {E.}~\bibnamefont {Berti}},\ }\bibfield  {title} {\bibinfo {title} {{Accuracy of the post-Newtonian approximation: Optimal asymptotic expansion for quasicircular, extreme-mass ratio inspirals}},\ }\href {https://doi.org/10.1103/PhysRevD.77.124006} {\bibfield  {journal} {\bibinfo  {journal} {Phys. Rev. D}\ }\textbf {\bibinfo {volume} {77}},\ \bibinfo {pages} {124006} (\bibinfo {year} {2008})},\ \bibinfo {note} {[Erratum: Phys.Rev.D 83, 109901 (2011)]},\ \Eprint {https://arxiv.org/abs/0803.1853} {arXiv:0803.1853 [gr-qc]} \BibitemShut {NoStop}%
\bibitem [{\citenamefont {Sun}\ \emph {et~al.}(2026)\citenamefont {Sun}, \citenamefont {Bonga}, \citenamefont {Stein},\ and\ \citenamefont {Da~Re}}]{Sun:2026lfd}%
  \BibitemOpen
  \bibfield  {author} {\bibinfo {author} {\bibfnamefont {D.}~\bibnamefont {Sun}}, \bibinfo {author} {\bibfnamefont {B.}~\bibnamefont {Bonga}}, \bibinfo {author} {\bibfnamefont {L.~C.}\ \bibnamefont {Stein}},\ and\ \bibinfo {author} {\bibfnamefont {G.}~\bibnamefont {Da~Re}},\ }\bibfield  {title} {\bibinfo {title} {{Convergence of post-Newtonian for quasi-circular non-precessing comparable mass ratios BBHs}},\ }\href@noop {} {\  (\bibinfo {year} {2026})},\ \Eprint {https://arxiv.org/abs/2605.20562} {arXiv:2605.20562 [gr-qc]} \BibitemShut {NoStop}%
\bibitem [{\citenamefont {Oliynyk}(2007)}]{Oliynyk:2007uno}%
  \BibitemOpen
  \bibfield  {author} {\bibinfo {author} {\bibfnamefont {T.~A.}\ \bibnamefont {Oliynyk}},\ }\bibfield  {title} {\bibinfo {title} {{The Newtonian limit for perfect fluids}},\ }\href {https://doi.org/10.1007/s00220-007-0334-z} {\bibfield  {journal} {\bibinfo  {journal} {Commun. Math. Phys.}\ }\textbf {\bibinfo {volume} {276}},\ \bibinfo {pages} {131} (\bibinfo {year} {2007})},\ \Eprint {https://arxiv.org/abs/0810.3744} {arXiv:0810.3744 [gr-qc]} \BibitemShut {NoStop}%
\bibitem [{\citenamefont {Oliynyk}(2009)}]{Oliynyk:2008vx}%
  \BibitemOpen
  \bibfield  {author} {\bibinfo {author} {\bibfnamefont {T.~A.}\ \bibnamefont {Oliynyk}},\ }\bibfield  {title} {\bibinfo {title} {{Post-Newtonian expansions for perfect fluids}},\ }\href {https://doi.org/10.1007/s00220-009-0738-z} {\bibfield  {journal} {\bibinfo  {journal} {Commun. Math. Phys.}\ }\textbf {\bibinfo {volume} {288}},\ \bibinfo {pages} {847} (\bibinfo {year} {2009})},\ \Eprint {https://arxiv.org/abs/0810.3752} {arXiv:0810.3752 [gr-qc]} \BibitemShut {NoStop}%
\bibitem [{\citenamefont {Fujita}\ \emph {et~al.}(2018)\citenamefont {Fujita}, \citenamefont {Sago},\ and\ \citenamefont {Nakano}}]{Fujita:2017wjq}%
  \BibitemOpen
  \bibfield  {author} {\bibinfo {author} {\bibfnamefont {R.}~\bibnamefont {Fujita}}, \bibinfo {author} {\bibfnamefont {N.}~\bibnamefont {Sago}},\ and\ \bibinfo {author} {\bibfnamefont {H.}~\bibnamefont {Nakano}},\ }\bibfield  {title} {\bibinfo {title} {{Note on accuracy of the post-Newtonian approximation for extreme-mass ratio inspirals: retrograde orbits}},\ }\href {https://doi.org/10.1088/1361-6382/aa9ad5} {\bibfield  {journal} {\bibinfo  {journal} {Class. Quant. Grav.}\ }\textbf {\bibinfo {volume} {35}},\ \bibinfo {pages} {027001} (\bibinfo {year} {2018})},\ \Eprint {https://arxiv.org/abs/1707.09309} {arXiv:1707.09309 [gr-qc]} \BibitemShut {NoStop}%
\bibitem [{\citenamefont {Sago}\ \emph {et~al.}(2026)\citenamefont {Sago}, \citenamefont {Fujita}, \citenamefont {Isoyama},\ and\ \citenamefont {Nakano}}]{Sago:2026gxb}%
  \BibitemOpen
  \bibfield  {author} {\bibinfo {author} {\bibfnamefont {N.}~\bibnamefont {Sago}}, \bibinfo {author} {\bibfnamefont {R.}~\bibnamefont {Fujita}}, \bibinfo {author} {\bibfnamefont {S.}~\bibnamefont {Isoyama}},\ and\ \bibinfo {author} {\bibfnamefont {H.}~\bibnamefont {Nakano}},\ }\bibfield  {title} {\bibinfo {title} {{Secular evolution of orbital parameters for general bound orbits in Kerr spacetime}},\ }\href@noop {} {\  (\bibinfo {year} {2026})},\ \Eprint {https://arxiv.org/abs/2603.27941} {arXiv:2603.27941 [gr-qc]} \BibitemShut {NoStop}%
\bibitem [{\citenamefont {{ Jos\`{e} M. Mart\`{i}n-Garc\`{i}a et. al.,}}(2025)}]{xACT}%
  \BibitemOpen
  \bibfield  {author} {\bibinfo {author} {\bibnamefont {{ Jos\`{e} M. Mart\`{i}n-Garc\`{i}a et. al.,}}},\ }\href@noop {} {}\bibinfo {howpublished} {\href{http://www.xact.es/}{xACT, {V}ersion 1.3.0}} (\bibinfo {year} {2025})\BibitemShut {NoStop}%
\bibitem [{\citenamefont {Trestini}(2025{\natexlab{b}})}]{PNpedia}%
  \BibitemOpen
  \bibfield  {author} {\bibinfo {author} {\bibfnamefont {D.}~\bibnamefont {Trestini}},\ }\href {https://doi.org/10.5281/zenodo.15002834} {\bibinfo {title} {{PNpedia}}} (\bibinfo {year} {2025}{\natexlab{b}})\BibitemShut {NoStop}%
\end{thebibliography}%

\end{document}